\documentclass[aps,rmp,twocolumn,showpacs]{revtex4-1}
\pdfoutput=1
\usepackage{graphicx}
\usepackage{amsmath}
\usepackage{amsfonts}
\usepackage{amssymb,mathrsfs}

\usepackage{cleveref}
\crefformat{footnote}{#2\footnotemark[#1]#3}

\newcommand{\bs}[1]{\boldsymbol{#1}}

\begin{document}

\title{Coulomb drag}

\author{B.N. Narozhny}
\affiliation{Institut f\"ur Theorie der Kondensierten Materie,
  Karlsruhe Institute of Technology, 76128 Karlsruhe, Germany}
\affiliation{National Research Nuclear University MEPhI (Moscow
  Engineering Physics Institute), Kashirskoe shosse 31, 115409 Moscow,
  Russia }

\author{A. Levchenko} 
\affiliation{Department of Physics, University of Wisconsin-Madison,
  Madison, Wisconsin 53706, USA} 
\affiliation{Institut fur Nanotechnologie, Karlsruhe Institute of
  Technology, 76021 Karlsruhe, Germany}

\date{\today}

\begin{abstract}

Coulomb drag is a transport phenomenon whereby long-range Coulomb
interaction between charge carriers in two closely spaced but
electrically isolated conductors induces a voltage (or, in a closed
circuit, a current) in one of the conductors when an electrical
current is passed through the other. The magnitude of the effect
depends on the exact nature of the charge carriers and microscopic,
many-body structure of the electronic systems in the two conductors.
Drag measurements have become part of the standard toolbox in
condensed matter physics that can be used to study fundamental
properties of diverse physical systems including semiconductor
heterostructures, graphene, quantum wires, quantum dots, and optical
cavities.

\end{abstract}

\pacs{73.63.-b,73.43.Cd, 81.05.ue}

\maketitle

\tableofcontents

%
%
%
%


\section{Frictional drag}
\label{intro}


Inner workings of solids are often studied with the help of transport
measurements. Within linear response, the outcome of such measurements
is determined by the properties of the unperturbed system, which are
often the object of study. In a typical experiment a current is driven
through a conductor and the voltage drop along the conductor is
measured. In conventional conductors at low temperatures the resulting
Ohmic resistance is mostly determined by disorder (which is always
present in any sample) \cite{ziman1965,dau10}, while interactions
between charge carriers lead to corrections that affect the
temperature dependence of transport coefficients \cite{aar}.

\begin{figure}
\begin{center}
\includegraphics[width=0.7\linewidth]{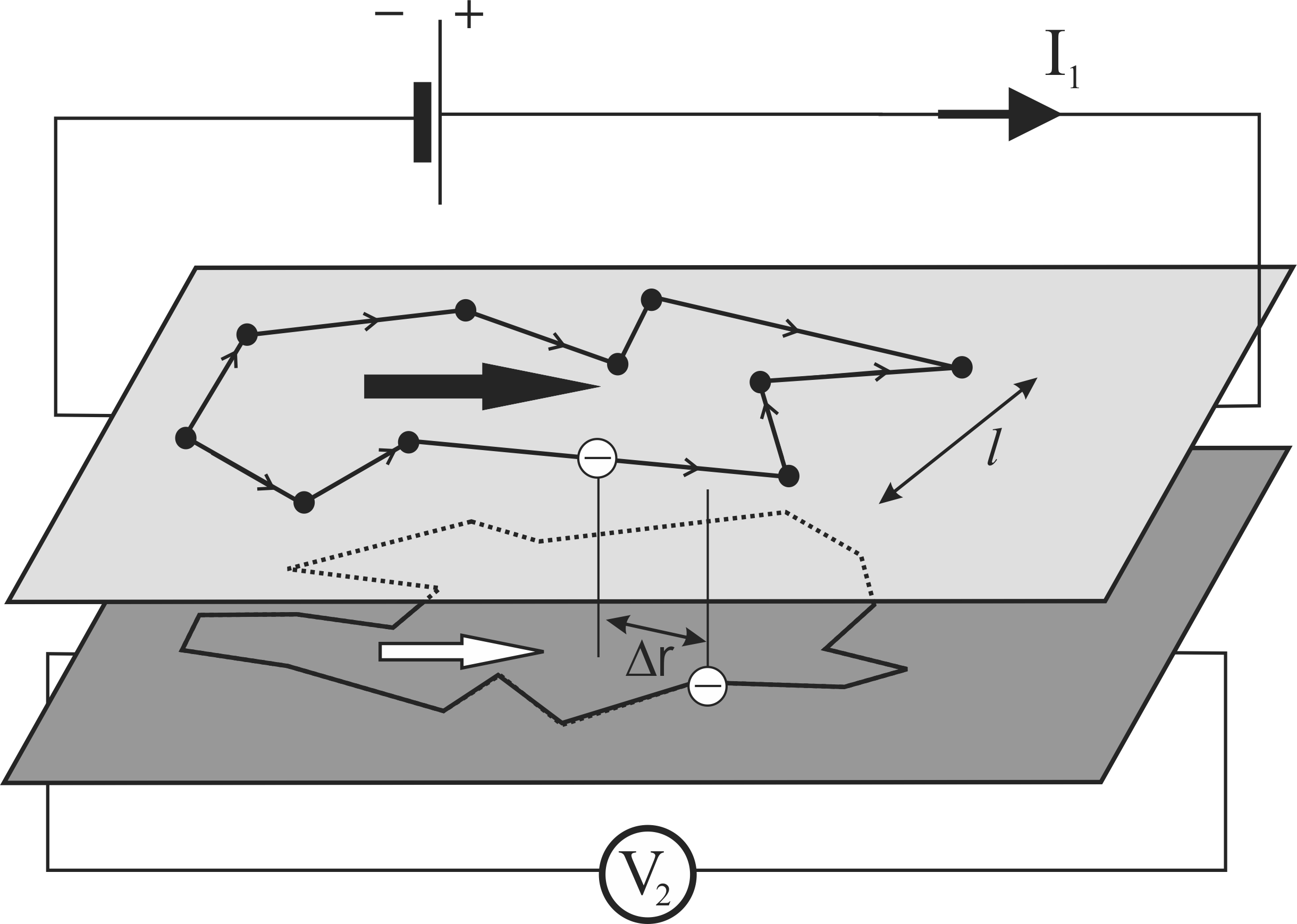}
\end{center}
\caption{\label{fig_d0} Schematic showing the drag signal $V_2$
  induced by the current $I_1$. [From \textcite{mes}. Reprinted with
    permission from AAAS.] }
\end{figure}

In his pioneering work, \textcite{pog} has suggested an
alternative measurement that involves two closely spaced, but
electrically isolated conductors (hereafter referred to as
``layers''). In such a system, an electric current $I_1$ flowing
through one of the layers, known as the ``active'' layer, induces a
current (or, in an open circuit, a voltage $V_2$, see
Fig.~\ref{fig_d0}) in the other, ``passive'' layer by means of
``mutual friction''. By this one typically understands scattering
between charge carriers belonging to different layers due to
long-range interactions. These scattering events are accompanied by
energy and momentum transfer from the carriers in the active layer to
the carriers in the passive layer, effectively ``dragging'' them
along. At the simplest level, such friction effects can be described
by introducing a phenomenological relaxation rate.  In the case of
frictional drag, the corresponding rate $\tau_D^{-1}$ generally
depends on the exact nature of the charge carriers, interlayer
interaction, and microscopic structure of the electronic system. Thus,
measurements of this relaxation rate provide additional insight into
microscopic properties of interacting many-body systems.

A related phenomenon, where a quasiparticle flow instigates a partial
transfer of energy and momentum between separate, but interacting
subsystems of quasiparticles, is known as ``phonon drag''
\cite{lgu1,lgu2,her} and manifests itself in a rising
thermoelectric power in semiconductors at low temperatures
\cite{fre1,fre2,geb1,geb2}. In the presence of a temperature gradient,
lattice vibrations become anisotropic since the phonons travel
preferentially from hot to cold (providing a mechanism for thermal
conduction). Interacting with electrons, the phonons effectively
drag them towards the cold end of the sample, creating an excess
charge density (this process will continue until the electrostatic
field created by the accumulated charge will counterbalance the drag
effect). In a nonequilibrium system of electrons and phonons, their
mutual drag is intertwined with heating effects and affects charge
transport \cite{gum}. The resulting correction to the standard
transport theory is important in thermoelectric measurements.

In contrast, frictional drag in double-layer systems is {\it not a
  correction}: in the absence of the interlayer interaction charge
carriers in two disjoined conductors are insensitive to each other
(therefore, any drag effect should necessarily vanish in the limit of
infinitely remote layers). In other words, {\it the drag phenomenon
  simply does not exist in noninteracting systems}! Consequently,
initial experimental work on mutual drag was devoted to quantitative
measurement of the strength of interactions between quasiparticle
subsystems in various semiconductor devices including
$p$-modulation-doped GaAs quantum wells \cite{ex0,ex01}, capacitively
coupled two- and three-dimensional (2D-3D) electron systems in
AlGaAs/GaAs heterostructures \cite{ex1,ex11}, 2D electron systems in
AlGaAs/GaAs double quantum wells \cite{eis,ex11,ex2,ex22,ex24}, and
electron-hole bilayers \cite{siv}. Drag between 3D systems was
numerically simulated in \textcite{jac}. At low temperatures and for
closely spaced layers, the interlayer scattering rate $\tau_D^{-1}$
appeared to be dominated by the Coulomb interaction \cite{pri,pr1}.

Coulomb drag between spatially separated electron systems is
ultimately caused by fluctuations (or inhomogeneities) of the charge
density in the two layers \cite{mac}. Indeed, an infinite layer with
the uniformly distributed electric charge creates a uniform electric
field in the normal direction that does not exert any lateral force
upon the carriers in another layer. If both layers are in the
Fermi-liquid state, then the usual phase-space argument \cite{ex2}
yields the quadratic temperature dependence
${\tau_D^{-1}\propto{T}^2}$ in qualitative agreement with the observed
behavior at low enough temperatures. More detailed analysis of the
experimental data revealed the presence of additional mechanisms
leading to frictional drag, such as the indirect interlayer
interaction mediated by phonons \cite{ex23,ru3,ru32,noh}, plasmon
effects \cite{no1,hil}, and thermoelectric phenomena \cite{ex11}.

Theoretically, it was realized early on that mutual Coulomb scattering
between electrons in the two layers results in the {\it exchange of
  both energy and momentum} \cite{pri,bo1,mas}. Initial calculations
aimed at energy and momentum relaxation in a nondegenerate 2D electron
gas (2DEG) due to proximity to a 3D conductor \cite{bo1,bo12} were
followed by investigation of transport properties in coupled 2D and 3D
systems \cite{boi,las}, 1D systems coupled to conductors of arbitrary
dimensionality \cite{si2}, coupled 1D wires \cite{ts4,ta22,ta2,pev},
and quantum Hall edge states \cite{org}. Following the groundbreaking
experiments in AlGaAs/GaAs double quantum wells \cite{eis,ex2}, a lot
of work was devoted to drag between two degenerate 2DEGs. While the
purely Coulomb mechanism \cite{jho,mac,kor,fl2} does capture the most
qualitative features of the effect, other mechanisms of momentum
transfer may also contribute to the observed behavior. In samples with
larger interlayer spacing (${d\sim{50}}$-$500${nm}) as much as $30$\%
of the measured signal was attributed to phonon-mediated interactions
\cite{ex23}. These measurements appeared to be consistent with the
virtual-phonon exchange mechanism \cite{tso,tso2}. Other suggested
scattering mechanisms involved acoustic \cite{bo2} and optical
\cite{hu3} phonons, plasmon effects \cite{fln} and coupled
plasmon-phonon modes \cite{tan}. Bilayers subject to strong magnetic
field were shown to form interlayer correlated states
\cite{var,gir}. For superconducting layers, interlayer
magnetic interaction due to spontaneously created vortices has also
been suggested \cite{sh1}.

Mutual Coulomb scattering has been studied also in a hybrid device
\cite{hua} comprising normal (Au/Ti) and superconducting (AlO$_x$) 2D
films separated by an insulating (Al$_2$O$_3$) layer. In that case, as
well as in ``cross-talk'' measurements in
superconductor--insulator--normal-metal trilayers \cite{gio}, the
phenomenological Drude-like description of drag in terms of
$\tau_D^{-1}$ does not apply. The Drude description also fails when
the system is subjected to a strong magnetic field: in contrast to the
naive description, numerous experiments
\cite{hi3,rub,pat,ru2,fen,hi2,jo1,lok,lo1} show significant dependence
of the measured drag resistivity $\rho_D$ on the applied field,
especially in the extreme quantum regime \cite{lil}. More
sophisticated theoretical calculations on Coulomb drag in quantum Hall
states \cite{shi}, superfluid condensates in paired electron-hole
layers \cite{vi2}, drag of composite fermions \cite{us1,ust,kim,zho},
vortex drag \cite{vit}, non-dissipative drag \cite{rom}, supercurrent
drag \cite{dua}, as well as drag between charged Bose gases \cite{ta3}
and mesoscopic rings \cite{sha,sha1,ba4} have confirmed the
expectation that {\it the drag resistivity reflects not only the exact
  character of interlayer interaction, but also the nature of
  elementary excitations} in each layer and their fundamental
properties.

After the turn of the century, drag measurements became part of the
standard toolbox in condensed matter physics. They have been used to
investigate properties of electron-electron scattering in low-density
2D electron systems \cite{ke1,san}; signatures of metal-insulator
transition in dilute 2D hole systems \cite{jor,jo2,pi2,pil,pi23};
quantum coherence of electrons \cite{mes,sav,tu1} and composite
fermions \cite{prs}; exciton effects in electron-hole bilayers
\cite{keo,cro,sea,mor}; exotic bilayer collective states \cite{eis2},
especially the quantum Hall effect (QHE) at the total filling factor
$\nu_T=1$ \cite{ke2,kel,spi,tut,fin,std}; compressible quantum Hall
(QH) states at half-integer filling factor \cite{zel,mur}; integer QH
regime \cite{lkd}; Luttinger liquid effects \cite{deb,la2,lar2};
Wigner crystallization in quantum wires \cite{ya1,yam,ya2}; and
one-dimensional (1D) sub-bands in quasi 1D wires \cite{de2,lar}. More
generally, interlayer interaction and corresponding transport
properties have been studied in hybrid devices comprising a quantum
wire and a quantum dot \cite{kri}; a SC film and a 2D electron gas
\cite{far}; Si metal-oxide-semiconductor systems \cite{la3}; quantum
point contacts \cite{khr}; insulating a-SiNb films \cite{els};
ferromagnetic-antiferromagnetic-SC trilayers \cite{cuo}; nanosize
CdSe-CdS semiconductor tetrapods \cite{mau}; electron-hole scattering
in quantum wells \cite{pru,tak,yan}; graphene monolayers
\cite{tu1,exg,tu2,meg}; and hybrid graphene-semiconductor systems
\cite{gam}.

On the theory side, the variety of suggested extensions and
generalizations of the original drag problem is even richer. The
theory of Coulomb drag between two 2DEGs was extended to dilute 2D
hole systems \cite{hbs} and to the cases where one allows for certain
tunneling processes between the layers \cite{or2,ore}, interlayer
disorder correlations \cite{go2,hu2}, in-plane potential modulation
\cite{alk}, and disorder inhomogeneities
\cite{apa,spk,zou,zou2}. Theory of Coulomb drag between composite
fermions was generalized to include phonon-mediated coupling
\cite{khv,bon}. Mutual friction was also suggested to occur between
non-Fermi-Liquid phases including Luttinger liquids
\cite{fle,naz,kle}, Wigner crystals \cite{bra,bak}, and strongly
localized electrons \cite{rai}. Drag or similar measurements of
interlayer interactions were also considered for composite (or hybrid)
systems comprising ballistic quantum wires \cite{ra2,wmd,gur,gur2},
coupled 2D-1D systems \cite{lyo}, nonequilibrium charged gases
\cite{wan}, multi-wall nanotubes \cite{luj,lfj}, quantum point
contacts \cite{le2}, few level quantum dots \cite{mol}, optical
cavities \cite{loz,ber}, coupled mesoscopic rings \cite{ma2},
superconductors \cite{lev}, and
normal-metal--ferromagnet--normal-metal structures \cite{zh2}. Other
developments include mesoscopic fluctuations of Coulomb drag
\cite{me0,me2}, frictional drag mediated by virtual photons \cite{don}
and plasmons \cite{ba3}, exciton effects in semiconductors \cite{lai}
and topological insulators \cite{min}, interlayer Seebeck effect
\cite{lun} and spin drag
\cite{dam,vig,sta,tse,du2,bav,du1,dui,gla}. Recently, the focus of the
theoretical work was shifted towards the drag effect in graphene-based
devices \cite{da1,me1,mef,sl3} and strongly interacting high-mobility
double-layers with low-density carrier concentration \cite{apos,chaa}.

Given the rather large amount of literature devoted to frictional
drag, it seems unreasonable to cover all possible angles in a single
paper. Early work on frictional drag was reviewed by
\textcite{roj}. Various experimental aspects were discussed in reviews
on exciton condensates \cite{sno,eis2}, electron-hole bilayers
\cite{gup}, strongly-correlated 2D electron fluids \cite{spiv}, and 1D
ballistic electron systems \cite{dzg}. A discussion of drag in strong
magnetic fields was included in a review of magnetotransport in 2D
electron systems \cite{dmi}. In the present review, we limit ourselves
to the discussion of standard (``electrical'') Coulomb
drag. Spin-related phenomena and thermoelectric effects are beyond the
scope of this review.


\section{Coulomb drag in semiconductor heterostructures}
\label{drag0}


In an idealized experiment, a constant ({\it dc}) current $I_1$ is
passed through the active layer, keeping the passive layer isolated at
the same time (such that no current is allowed to flow in it), see
Fig.~\ref{fig_d0}. The voltage $V_2$ induced in the passive layer is
proportional to $I_1$ and the coefficient\footnote{The minus sign in
  Eq.~(\ref{rd_def}) is motivated by Eq.~(\ref{rd0}). An alternative
  definition without the explicit minus sign is also widely used in
  literature.}
\begin{equation}
\label{rd_def}
R_D = - V_2/I_1,
\end{equation}
is a direct measure of interlayer interactions.

In his original paper, \textcite{pog} derived the Drude-like
description of transport in double-layer systems comprising two
coupled equations of motion
\begin{subequations}
\label{dt}
\begin{eqnarray}
\label{de1}
\frac{d\bs{v}_1}{dt} = \frac{e}{m_1}\bs{E}_1 
+ \frac{e}{m_1c} [\bs{v}_1\times\bs{B}]
- \frac{\bs{v}_1}{\tau_1} - 
\frac{\bs{v}_1-\bs{v}_2}{\tau_D} , \qquad
\end{eqnarray}
\begin{eqnarray}
\label{de2}
\frac{d\bs{v}_2}{dt} = \frac{e}{m_2}\bs{E}_2 
+ \frac{e}{m_2c} [\bs{v}_2\times\bs{B}]
- \frac{\bs{v}_2}{\tau_2} - 
\frac{\bs{v}_2-\bs{v}_1}{\tau_D} , \qquad
\end{eqnarray}
\end{subequations}
where $e$ is the electric charge, $\bs{v}_i$, $m_i$, and $\bs{E}_i$
are the drift velocities, effective masses, and electric fields in the
two layers, and the nonquantizing magnetic field $\bs{B}$ is assumed
to be uniform. Intralayer impurity scattering processes yielding the
Drude resistivity in the two layers are described by the mean free
times $\tau_i$. The last term in each of Eqs.~(\ref{dt}) describes the
mutual friction between the charge carriers in the two layers that
tends to equalize drift velocities. If treated phenomenologically, the
model (\ref{dt}) describes two distinct types of carriers coupled by
the friction term, but does not explicitly require them to be
spatially separated \cite{cui,mah,sod}.

Solving the equations (\ref{dt}), one finds the resistivity matrix
$\rho_{\alpha\beta}^{(ij)}$ [hereafter the indices $i,j=1,2$ denote
  the two layers and $\alpha,\beta=x,y$ -- spatial coordinates
  orthogonal to $\bs{B}=B\bs{e}_z$; the ``layers'' described by
  Eqs.~(\ref{dt}) can represent 2D or 3D conductors, see Sec.~\ref{nw}
  for the 1D case]. The ``drag resistivity'' (also called the
transresistivity or the drag coefficient) is given by the Drude-like
formula
\begin{subequations}
\label{dres}
\begin{equation}
\label{rd0}
\rho_D = - \rho_{xx}^{(12)} = m_2/(e^2 n_1 \tau_D).
\end{equation}
The expression (\ref{rd0}) is {\it independent of the magnetic
  field}. This statement has the same status as the absence of the
classical magnetoresistance\footnote{In this Section we are discussing
  the simplest situation, where both $\tau_D$ and $\tau$ are
  unaffected by weak enough magnetic fields.}. Indeed, the
single-layer longitudinal resistivity derived from Eqs.~(\ref{dt})
is given by
\begin{equation}
\label{r0}
\rho_{xx}^{(11)} = \frac{m_1}{e^2 n_1} 
\left(\frac{1}{\tau}+\frac{1}{\tau_D}\right).
\end{equation}
In most cases, drag is rather weak (${\tau_D\gg\tau}$) and the usual
Drude formula remains a good approximation for ${\rho_{xx}^{(11)}}$
\cite{eis,roj}. The single-layer Hall coefficient is unaffected by the
presence of the second layer and is determined solely by the carrier
density
\begin{equation}
\label{hall0}
\rho_{yx}^{(11)} = B/(n_1 e c).
\end{equation}
Within the applicability of the Drude model, frictional drag is purely
longitudinal: ``Hall drag'' does not occur\footnote{\label{fn_no_mag}
  Under the assumptions of the present Section, magnetic field has no
  effect on drag. Hence, up until Sec.~\ref{wmf} we focus on the
  zero-field, longitudinal transport. Drag in magnetic field is
  discussed in Secs.~\ref{wmf}, \ref{hdg}, and \ref{qhe1}.}
\begin{equation}
\label{rH0} 
\rho_D^H = \rho_{yx}^{(12)} = 0.
\end{equation}
\end{subequations}

At the phenomenological level, the drag resistivity (\ref{rd0}) is
independent of the disorder strength. Moreover, in the ``clean'' limit
${\tau\rightarrow\infty}$ the inter- and intralayer resistivities tend
to the same value and the resistivity matrix becomes degenerate (the
corresponding conductivities diverge):
\begin{equation}
\label{rc}
\rho_{xx}^{(11)}(\tau\rightarrow\infty) = \rho_D(\tau\rightarrow\infty).
\end{equation}
Thus a system comprising two capacitively coupled, ideal conductors is
characterized by non-zero resistivity and exhibits {\it perfect drag!}


\subsection{Interlayer Coulomb interaction}
\label{int}

\begin{figure}
\begin{center}
\includegraphics[width=0.9\linewidth]{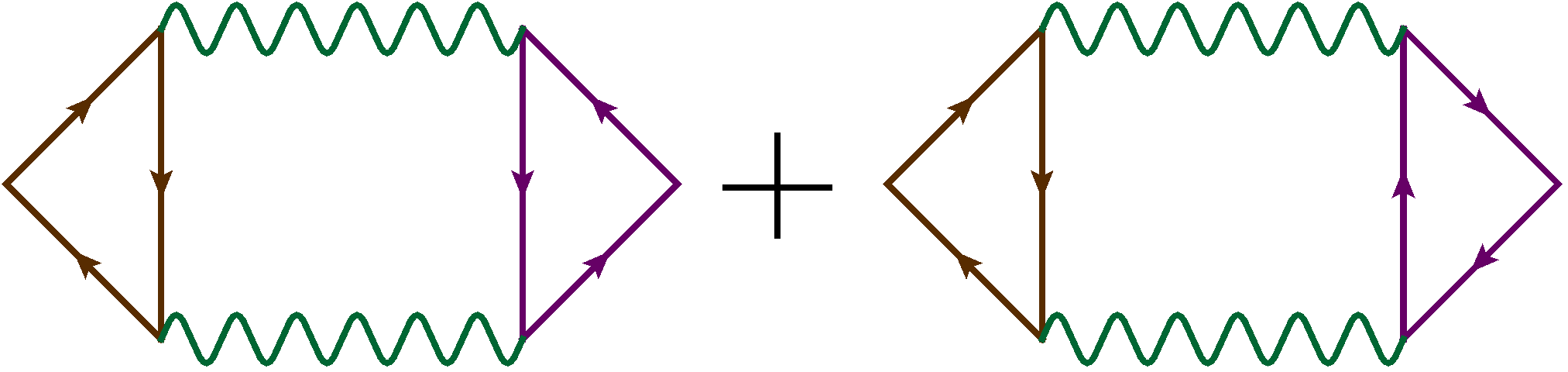}
\end{center}
\caption{\label{fig_fd} (Color online) Aslamazov-Larkin diagrams
  describing the lowest-order contribution to drag. The solid lines
  refer to quasiparticle Green's functions and the wavy lines describe
  the interlayer interaction. The left and right triangles correspond
  to non-linear susceptibilities of the two layers.}
\end{figure}

The ``Drude formula'' (\ref{rd0}) for the drag resistivity becomes
falsifiable provided that something is known about the properties of
the ``drag rate'' $\tau_D^{-1}$ (i.e. its dependence on temperature,
carrier density, interlayer separation, and other experimentally
relevant parameters). To the leading order, the contribution of the
interlayer Coulomb interaction to $\tau_D^{-1}$ can be calculated
within the Born approximation (or equivalently, using the Fermi's
Golden Rule) \cite{las,jho}. In the language of Feynman diagrams, the
corresponding process \cite{kor,mac,fl2} is described by the
Aslamazov-Larkin diagrams \cite{asl} shown in Fig.~\ref{fig_fd}.

The effective interlayer interaction can be found as a solution to the
Poisson equation for the potential of a point source belonging to one
of the layers. In principle, this can be done for any system of
coupled conductors. Coupling between a 2DEG and a 3DEG was considered
in \textcite{las}. A double quantum well system was discussed in
\textcite{jho} where the finite width of the wells was taken into
account by assuming a specific form of the electron wave function in
the direction perpendicular to the layers. However, the obtained
results are qualitatively the same as in the simplest case of purely
two-dimensional layers.

If electrons in each layer are confined to move in a 2D plane, the
``bare'' Coulomb potential has the form\footnote{Although electrons
  are confined to move in two dimensions, they interact by means of
  ``real, 3D'' Coulomb interaction.}
\begin{equation}
\label{coulomb}
V_{11}=V_{22} = 2\pi e^2/q, \quad
V_{12}(q) = (2\pi e^2/q) e^{-qd}.
\end{equation}
Here $e$ is the electron charge and $d$ is the interlayer separation,
that determines the maximum value (or rather, the order of magnitude
thereof) of the momentum $q$ that can be transferred between the
layers\footnote{While discussing the theory, we use the natural units
  with ${\hbar=k_B=1}$. We attempt to restore the Planck's constant in
  final answers and while discussing experimental findings.}:
\begin{equation}
\label{qd}
q \ll 1/d.
\end{equation}
Taking into account dynamical screening within the usual Random Phase
Approximation (RPA) modifies the interlayer interaction propagator
\cite{ady}, but does not change the exponential decay at large $q$:
\begin{eqnarray}
\label{d12}
{\cal D}^R_{12}\!=\!-
\frac{1}{\Pi^R_{1}\Pi^R_{2}\frac{4\pi{e}^2}{q}\!\sinh{qd}\!+
\!\left[\frac{q}{2\pi e^2}+\Pi^R_{1}+\Pi^R_{2}\right]\!e^{qd}}.
\quad\quad
\end{eqnarray}
Here $\Pi^R_i$ is the single-layer retarded polarization operator. It
is quite common [see, e.g., \textcite{las,jho}], to include the
dielectric constant $\epsilon$ of the insulating spacer into the
``bare'' potential. Since the same $\epsilon$ should enter
the expression for the inverse Thomas-Fermi screening length
\begin{equation}
\label{tf}
\varkappa = 2\pi e^2 \nu = 2\pi e^2 \Pi^R(q<2k_F, \omega=T=0),
\end{equation}
the dielectric environment can be taken into account by expressing the
results in terms of $\varkappa$ ($\nu$ denotes the thermodynamic
density of states of the 2DEG). For high carrier densities \cite{ex2},
Eq.~(\ref{d12}) can be simplified \cite{kor} by assuming the small
screening length, ${\varkappa{d}\gg{1}}$ [see Eqs.~(\ref{db}) and
  (\ref{ddkor}) below].

The condition (\ref{qd}) allows one to distinguish the following two
regimes \cite{kor}:
\begin{itemize}
\item[(i)] 
if the interlayer separation is large compared to the mean-free path
${d\gg\ell}$, then it follows from Eq.~(\ref{qd}) that ${q\ll1/\ell}$;
in this case the motion of charge carriers is {\it diffusive};
\item[(ii)] in the opposite case, ${d\ll\ell}$, transport is dominated
  by {\it ballistic} propagation of charge carriers with ${1/d\gg
    q\gg1/\ell}$, see Eq.~(\ref{baldif}) below. Most measurements
  \cite{ex2,exg} are performed on ballistic samples.
\end{itemize}
The majority of analytic \cite{roj} and numerical \cite{mos} work on
Coulomb drag in semiconductor heterostructures was performed treating
the interaction (\ref{d12}) in the lowest order of perturbation
theory. For generalizations see Secs.~\ref{third} and \ref{exp1}.


\subsection{Kinetic theory of ballistic drag}
\label{kt}


Ballistic motion of charge carriers in semiconductors can be described
by using the kinetic equation approach, where impurity scattering is
taken into account within the simplest $\tau$-approximation
\cite{pog,las,jho}. One starts with the generic Boltzmann equation
\begin{equation}
\label{beq0}
\frac{\partial f_i}{\partial t} + \bs{v}_i \bs{\nabla}f_i
+\left(e\bs{E}_i +\frac{e}{c} \left[\bs{v}_i\times\bs{B}\right]\right)
\frac{\partial f_i}{\partial \bs{p}} = -\frac{\delta f_i}{\tau} + {\cal I}_{ij},
\end{equation}
where ${f_i}$ is the distribution function (in layer ${i=1,2}$),
${{\cal{I}}_{ij}}$ is the collision integral due to interlayer Coulomb
interaction, $\tau$ is the transport impurity scattering time, and
${\delta{f_i}}$ is the nonequilibrium correction to the distribution
function. Here we will only consider {\it degenerate} electron systems
[as realized in semiconductor heterostructures \cite{ex2}]. Weak
deviations from the equilibrium Fermi-Dirac distribution function
${f^{(0)}_i}$ (as appropriate within linear response) are described by
\begin{equation}
\label{df}
\delta f_i \!=\! f_i \!-\! f^{(0)}_i \!=\! f^{(0)}_i\!\left[1-f^{(0)}_i\right]\!
h_i\!=\!-\! T 
[\partial f^{(0)}_i\!/\partial \epsilon]
h_i.
\end{equation}
Here we only consider the steady state and uniform
fields
\begin{equation}
\label{stst}
\partial f_i/\partial t = 0, \quad
\bs{\nabla} f_i = 0.
\end{equation}
The latter condition physically means that the sample size is large
compared to the length scale of typical relaxation processes in the
system, see also Sec.~\ref{gmd}.

In the absence of interlayer interaction, the task of finding
linear-response transport coefficients from Eq.~(\ref{beq0}) is a
textbook problem \cite{ziman1965,smith1989,seeger2002}. Under the
above assumptions, the theory is qualitatively equivalent to the Drude
theory (\ref{dt}) yielding the standard results (\ref{r0}) and
(\ref{hall0}). Not surprisingly, taking into account the collision
integral ${\cal I}_{ij}$ leads to the Drude-like description of the
drag resistivity (\ref{rd0}) and (\ref{rH0}). The advantage of the
present ``microscopic'' calculation is that now we can determine the
phenomenological relaxation time $\tau_D$ in terms of the model
parameters.

The standard perturbative calculation \cite{dau10,las,boi,jho} amounts
to finding the nonequilibrium distribution functions ${h_i}$ in the
two layers to the leading order in the interlayer interaction and the
electric field ${\bs{E}_1}$ applied to the active layer. Then one uses
the definition of the electric current (here the sum runs over all of
the single-particle states)
\begin{equation}
\label{jdef}
\bs{j}_i = e \sum \bs{v} \delta f_i,
\end{equation}
and finds the current $\bs{j}_2$ in the passive layers. The
coefficient of proportionality between $j_{2x}$ and $E_{1x}$ defines
the drag conductivity $\sigma_D$. The drag coefficient $\rho_D$ can
then be obtained by inverting the 2$\times$2 conductivity
matrix\cref{fn_no_mag}
\begin{equation}
\label{rd1}
\rho_D = \frac{\sigma_D}{\sigma_1\sigma_2-\sigma_D^2}
\approx \frac{\sigma_D}{\sigma_1\sigma_2},
\end{equation}
where $\sigma_i$ is the longitudinal conductivity in layer $i$; the
latter relation follows from the smallness of the effect
\begin{equation}
\label{small}
\sigma_D \ll \sigma_i,
\end{equation}
as observed in experiment \cite{eis,roj}. This way, one finds
for the phenomenological ``drag rate''
\begin{widetext}
\begin{equation}
\label{td}
\tau_D^{-1}=\frac{m_1}{16 \pi e^2 \tau^2 n_2 T}
\int\limits_{-\infty}^\infty\! \frac{d\omega}{\sinh^2[\omega/(2T)]}
\int\!\frac{d^2q}{(2\pi)^2} |{\cal D}_{12}(\omega, \bs{q})|^2 
\Gamma^x_1(\omega, \bs{q})
\Gamma^x_2(\omega, \bs{q}).
\end{equation}
\end{widetext}
The nonlinear susceptibility (also known as the rectification
function) ${\bs{\Gamma}_i(\omega,\bs{q})}$ (in layer $i$) is a
response function relating a voltage ${V(r_i)e^{i\omega{t}}}$ to a
$dc$ current it induces by the quadratic response:
\begin{equation}
\bs{J}=\int\!d{\bf r}_1\!\int\!d{\bf r}_2\bs{\Gamma}(\omega; {\bf r}_1, {\bf r}_2) 
V({\bf r}_1)V({\bf r}_2),
\label{gamma-el}
\end{equation}
\noindent
with $\bs{J}$ being the induced $dc$ current. From  
gauge invariance $\int\!d{\bf r}_1\bs{\Gamma}(\omega) = 
\int\!d{\bf r}_2\bs{\Gamma}(\omega) = 0$.

The same result follows from the standard Kubo formula approach within
the diagrammatic perturbation theory \cite{kor,fl2}, memory
function formalism \cite{mac}, and more general Boltzmann-Langevin
theory of stochastic kinetic equation \cite{chaa}.

\subsubsection{Electron-hole asymmetry and rectification}
\hfill

The rectification function ${\bs{\Gamma}(\omega,\bs{q})}$ is the
central object in the perturbative theory of Coulomb drag. The
expression (\ref{td}) of the interlayer relaxation rate in terms of
${\bs{\Gamma}(\omega,\bs{q})}$ explicitly demonstrates the key role of
electron-hole asymmetry in the leading-order drag
effect\footnote{\label{fnm} Another known effect of the electron-hole
  asymmetry in electronic systems is the thermopower described by the
  Mott formula \cite{Mott,gl1,gl2}.}.

\begin{figure}
\begin{center}
\includegraphics[width=0.7\linewidth]{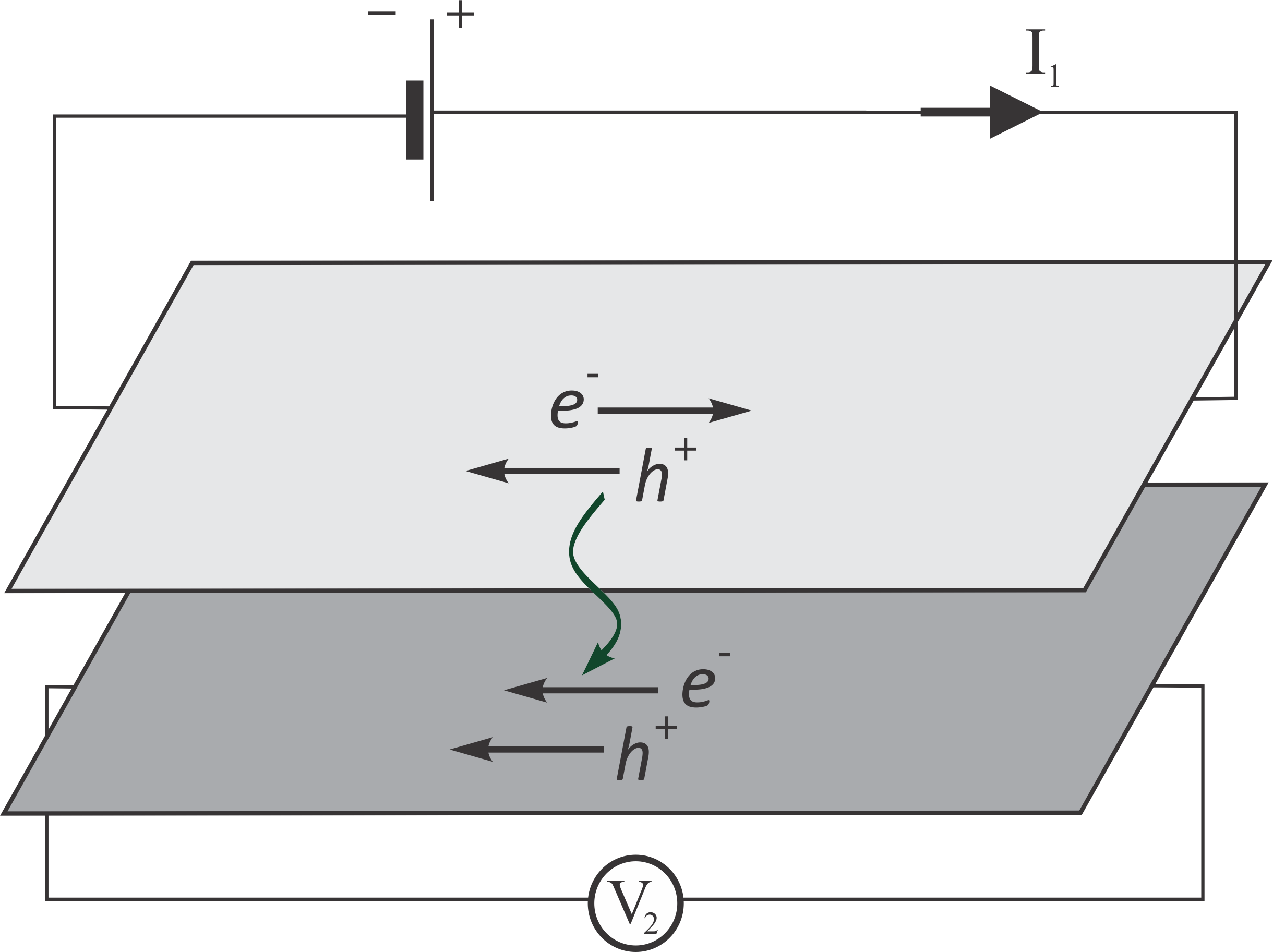}
\end{center}
\caption{\label{fig_as} Schematic illustration of the momentum
  transfer due to interlayer interaction. As the current $I_1$ is
  driven through the active layer, electrons and holes are moving in
  the opposite directions since they carry the opposite charge. Such a
  state has non-zero total momentum only due to electron-hole
  asymmetry. Once the momentum is transferred to the passive layer,
  the electrons and holes there are pushed in the same direction. This
  process can induce a voltage again only due to electron-hole
  asymmetry.}
\end{figure}

Indeed, in order to induce a voltage (or generate a current) in the
passive layer, one needs to somehow move the charge carriers. This is
achieved by transferring momentum from the active layer. The
macroscopic state of the electronic system in the active layer is
characterized by the finite electric current driven by an external
source. In a typical electron gas, there are two kinds of excitations
- ``electron-like'', with energies ${\epsilon>E_F}$ above the Fermi
energy (i.e. the occupied states outside the Fermi surface), and
``hole-like'', with ${\epsilon<E_F}$. These quasiparticles are
oppositely charged. As the current is driven through the active layer
they move in opposite directions, see Fig.~\ref{fig_as}. Then the
active layer can be characterized by a nonzero total momentum only if
there is some asymmetry between electron-like and hole-like
quasiparticles. Likewise, in the passive layer the momentum is
transferred equally to electrons and holes, such that the resulting
state can carry current only in the case of electron-hole
asymmetry. In conventional semiconductors \cite{kor}, the
electron-hole asymmetry appears due to curvature of the conduction
band spectrum [leading to the energy dependence of the density of
  states (DoS) and/or diffusion coefficient]. Consequently, in the
Fermi-liquid theory the electron-hole asymmetry can be expressed
\cite{me2} as a derivative of the single-layer conductivity
$\sigma_{1(2)}$ with respect to the chemical potential (assuming
either a constant impurity scattering time or diffusive
transport). The simple estimate
$\partial\sigma_{1(2)}/\partial\mu\sim\sigma_{1(2)}/\mu$ then explains
the typical smallness of the effect \cite{ex1,ex2,siv}, see
Eq.~(\ref{small}).

Same arguments can be applied to any system containing carriers with
opposite signs of the electric charge. For instance, one can consider
semimetals (or even band insulators at high enough temperature) where
the electric current can be carried by electrons from the conduction
band and holes from the valence band. A particularly interesting
example is graphene (see Sec.~\ref{dig}), which exhibits exact
particle-hole symmetry at the charge neutrality point \cite{kats2012}.
At that point, the non-linear susceptibility of graphene (\ref{gg})
vanishes \cite{da1,me1} implying the absence of the drag effect.  In
contrast, experiment \cite{exg} shows nonzero drag resistivity at
charge neutrality, which in addition is greatly enhanced by the
external magnetic field \cite{meg}.

Indeed, the outlined physical picture is not universal. In fact, it
only describes a particular (although often dominant) scattering
process, where momentum is transferred from an electron-hole pair in
the active layer to another electron-hole pair in the passive
layer. Technically, this process is described by the leading-order
perturbation theory, see Fig.~\ref{fig_fd}, yielding Eq.~(\ref{td}).
Higher-order processes may result in additional contributions which
are less sensitive to electron-hole symmetry. These include the
so-called ``third-order'' drag \cite{le1,mem} and the effect of the
correlated disorder \cite{go2,go22,mem,sl3}, see Secs.~\ref{third} and
\ref{tod}.

In conventional heterostructures, higher-order processes remain
subleading at least within the temperature range where most of the
experiments are performed, see Sec.~\ref{third}. Specifically in the
ballistic regime, the dominant contribution to drag is indeed given by
Eq.~(\ref{td}) [with the corresponding drag resistivity (\ref{rd0})]
and is determined by the nonlinear susceptibility, which in the
simplest case of energy-independent impurity-scattering time $\tau$ is
given by \cite{kor}
\begin{eqnarray}
\label{gb}
\bs{\Gamma}(\bs{q},\omega) = 
\frac{2}{\pi} \; e\tau \bs{q} \frac{\omega}{v_F q} \,
\theta(v_Fq-\omega).
\end{eqnarray}
As shown in \textcite{kor}, the resulting expression (\ref{gb}) for the
nonlinear susceptibility is proportional to the imaginary part of the
single-layer polarization operator
\begin{eqnarray}
\label{gip}
\bs{\Gamma}(\bs{q},\omega) = \frac{2e\tau\bs{q}}{m}
{\rm Im}\Pi^R(\bs{q},\omega),
\end{eqnarray}
where (for two-dimensional, noninteracting electron gas in the
ballistic regime)
\begin{eqnarray}
\label{pb}
{\rm Im}\Pi^R(\bs{q},\omega) = \nu \frac{\omega}{v_F q} \,
\theta(v_Fq-\omega).
\end{eqnarray}

Within the kinetic theory, one can observe Eq.~(\ref{gip}) already at
the level of the collision integral \cite{qui}; hence many authors
[see, e.g. \textcite{jho,mac,shi,us1}] proceed to express
Eq.~(\ref{td}) in terms of ${{\rm{Im}}\Pi^R(\bs{q},\omega)}$ instead
of the nonlinear susceptibility. Under the assumption of
energy-independent impurity-scattering time $\tau$ and neglecting
intralayer correlations \cite{kor,fl2}, such calculations lead to the
correct result [see Eq.~(\ref{rdb}) below]. At the same time, within
such an approach the physics of electron-hole asymmetry remains
hidden. Generalization to more general settings is also nontrivial:
the relation (\ref{gip}) is by no means a general theorem
\cite{kor,fl2,me0,met}; for explicit examples of the two quantities
being inequivalent see Secs.~\ref{mf} and \ref{dig}.

\subsubsection{Drag resistivity in ballistic samples.}
\hfill

In the limit of strong screening, ${\varkappa{d}\gg{1}}$, one can
approximate \cite{kor} the interlayer interaction propagator
(\ref{d12}) by the expression
\begin{equation}
\label{db}
{\cal D}^R_{12} = -\frac{\pi e^2}{\varkappa_1 \varkappa_2}
\frac{q}{\sinh qd}.
\end{equation}
Combining the nonlinear susceptibility (\ref{gip}) with the
interlayer relaxation rate (\ref{td}) and Eq.~(\ref{rd0}), one finds
the following expression for the drag resistivity
\cite{jho,mac,fln,kor}
\begin{subequations}
\label{rdb}
\begin{equation}
\rho_D = \frac{\hbar}{e^2} \frac{\pi^2\zeta(3)}{16} \frac{T^2}{E_{F1} E_{F2}}
\frac{1}{\varkappa_1\varkappa_2k_{F1}k_{F2}d^4}.
\end{equation}
The same result can be also expressed\cref{fn1} in terms of the
interlayer relaxation rate (\ref{td}) [e.g., using Eq.~(\ref{rd0})]
\begin{equation}
\label{tdb}
\tau_D^{-1} = \frac{\pi^2\zeta(3)}{16} \frac{n_1}{m_2}
\frac{T^2}{E_{F1} E_{F2}}
\frac{1}{\varkappa_1\varkappa_2k_{F1}k_{F2}d^4}.
\end{equation}
\end{subequations}
Physically, these expressions\footnote{\label{fn1} Most expressions
  for $\rho_D$ \cite{jho,mac,fln,fl2,roj} can be reduced to
  Eqs.~(\ref{rdb}) using the following simple relations, valid under
  the assumptions of this Section: ${E_F=\pi{n}/m}$, ${n=E_F\nu}$,
  ${\nu D=E_F\tau/\pi}$, where ${D=v_F^2\tau/2}$ is the diffusion
  constant and ${\nu=m/\pi}$ is the density of states.} can be
understood based on the Fermi Golden Rule [which was explicitly used
  in the solution of the kinetic equation \cite{jho,las}]. Indeed,
there are three basic elements that combine into the result
(\ref{rdb}): (i) the phase space available for electron-hole pairs in
the two layers, which is limited by temperature, hence
${\tau_D^{-1}\propto{T}^2}$; (ii) the electron-hole asymmetry, which
results in the overall smallness of the effect,
${\tau_D^{-1}\propto(E_{F1}E_{F2})^{-1}}$; and (iii) the matrix
element of the interlayer interaction, determining the dependence on
the interlayer separation; in the ballistic case the matrix element is
dominated by small-angle scattering \cite{ex2}.

The drag resistivity (\ref{rdb}) -- and especially the quadratic
temperature dependence -- is often quoted as the ``Fermi-liquid''
result. However, Eq.~(\ref{rdb}) was obtained under a number of
assumptions: (i) ${\varkappa{d}\gg{1}}$, (ii) ${d\gg\ell}$, and (iii)
${T\ll{T}_d\sim{v}_F/d\sim{E}_F/(k_Fd)}$. The latter assumption
appears only implicitly and is often overlooked.

Indeed, substituting the interaction propagator (\ref{db}) and the
nonlinear susceptibility (\ref{gb}) into Eq.~(\ref{td}), one finds
that except for the $\theta$-function in Eq.~(\ref{gb}) the frequency
and momentum integrals factorize. The exponential decay of the
corresponding integrands allows one to estimate the typical values of
transferred energy ${\omega\sim{T}}$ and momentum ${q\sim1/d}$.
Assuming ${T\ll{T}_d}$, this yields ${\omega<v_Fq}$, which
automatically satisfies the $\theta$-function. Based on this
observation, one may omit the $\theta$-function and subsequently
extend the integration limits in both integrals in Eq.~(\ref{td}) to
infinity. The remaining integration is straightforward and yields
Eq.~(\ref{rdb}).

At higher temperatures, ${T\gg{T}_d}$, the $\theta$-function in
Eq.~(\ref{gb}) is not satisfied automatically. Physically, it
represents kinematic restrictions on the phase space available to
electron-hole pairs associated with predominantly small-angle
scattering \cite{ex2}. The frequency integration is now cut off at
${v_Fq}$ (or $T_d$), rather than $T$, which leads to the {\it linear}
temperature dependence [first reported in \textcite{ex2,ex11}, see
  also \textcite{jho}, and recently rediscovered in \textcite{chaa}],
\begin{equation}
\label{rdph}
\rho_D(T\gg T_d)=\frac{\hbar}{e^2}
\frac{\pi^3}{360}\frac{1}{(k_Fd)^3(\varkappa d)^2}\frac{T}{E_F}.
\end{equation}
This behavior may be observable in samples with either nondegenerate
2DEGs or large interlayer separation. In the latter case,
${T_d\ll{T}\ll{E}_F}$, both layers are perfectly described by the
Fermi liquid theory which is not synonymous with quadratic temperature
dependence of transport coefficients.

\subsubsection{Plasmon contribution}
\hfill

The approximate form of the interlayer Coulomb interaction (\ref{db})
appears justified in the ``ballistic'' regime where the dominant
interlayer relaxation processes are characterized by relatively large
momentum transfers ${\omega<v_Fq}$. The imaginary part of the
single-particle polarization operator (\ref{pb}) vanishes at smaller
momenta (or larger frequencies) making the above calculations
consistent. At the same time, approximating the interlayer interaction
propagator (\ref{d12}) by Eq.~(\ref{db}) one completely neglects a
possible contribution of plasmon modes, that (within the simplest RPA
approach) can be found by setting the denominator of Eq.~(\ref{d12})
to zero. At zero temperature and for ${\omega\gg{v}_Fq}$ (where
${\rm{Im}\Pi^R=0}$), the polarization operator is known to be given by
\cite{ste}
\[
\Pi(\bs{q},\omega) \simeq - nq^2/(m\omega^2).
\]
Using this expression and expanding the bare Coulomb potential in
small momenta, yields the acoustic (``$-$'') and optical (``$+$'')
plasmon modes with dispersions
\[
\omega_{-} = e q \sqrt{2\pi nd/m}, \quad
\omega_{+} = e\sqrt{4\pi nq/m}.
\]
Both of these modes lie outside of the particle-hole continuum and in
the parameter region, where the nonlinear susceptibility (\ref{gb})
vanishes. Hence, one may conclude that the plasmons cannot
contribute to frictional drag.

However \cite{fln}, at finite temperatures thermally excited
quasiparticles and plasmons may coexist in the same parameter region,
which may result in an additional contribution to drag. In order to
accurately describe the plasmon contribution to $\rho_D$, one has to
consider intralayer equilibration due to electron-electron collisions
\cite{chaa} which gives rise to two important features: (i) the
polarization operator acquires nonvanishing spectral weight within the
high frequency part of the spectrum at $\omega>v_Fq$ \cite{fln}; and
(ii) the plasmons acquire finite life-time \cite{reiz,hru} that
regularizes the pole in the interaction propagator.

The theory discussed in the preceding Sections in based on the
implicit assumption that the intralayer equilibration is the fastest
process in the system. Characterizing inelastic electron-electron
scattering by the quasiparticle lifetime ${\tau_{ee}}$, one finds
that the standard theory -- and hence Eq.~(\ref{rdph}) -- is valid as
long as the length $\tau_{ee}$ is much smaller than the interlayer
scattering time, ${\tau_{ee}\ll\tau_D}$ and at temperatures below the
corresponding threshold
${T\ll{T}_c\sim{E}_F\sqrt{k_F/(\varkappa^2d)}}$.

\begin{figure}
\begin{center}
\includegraphics[width=0.9\linewidth]{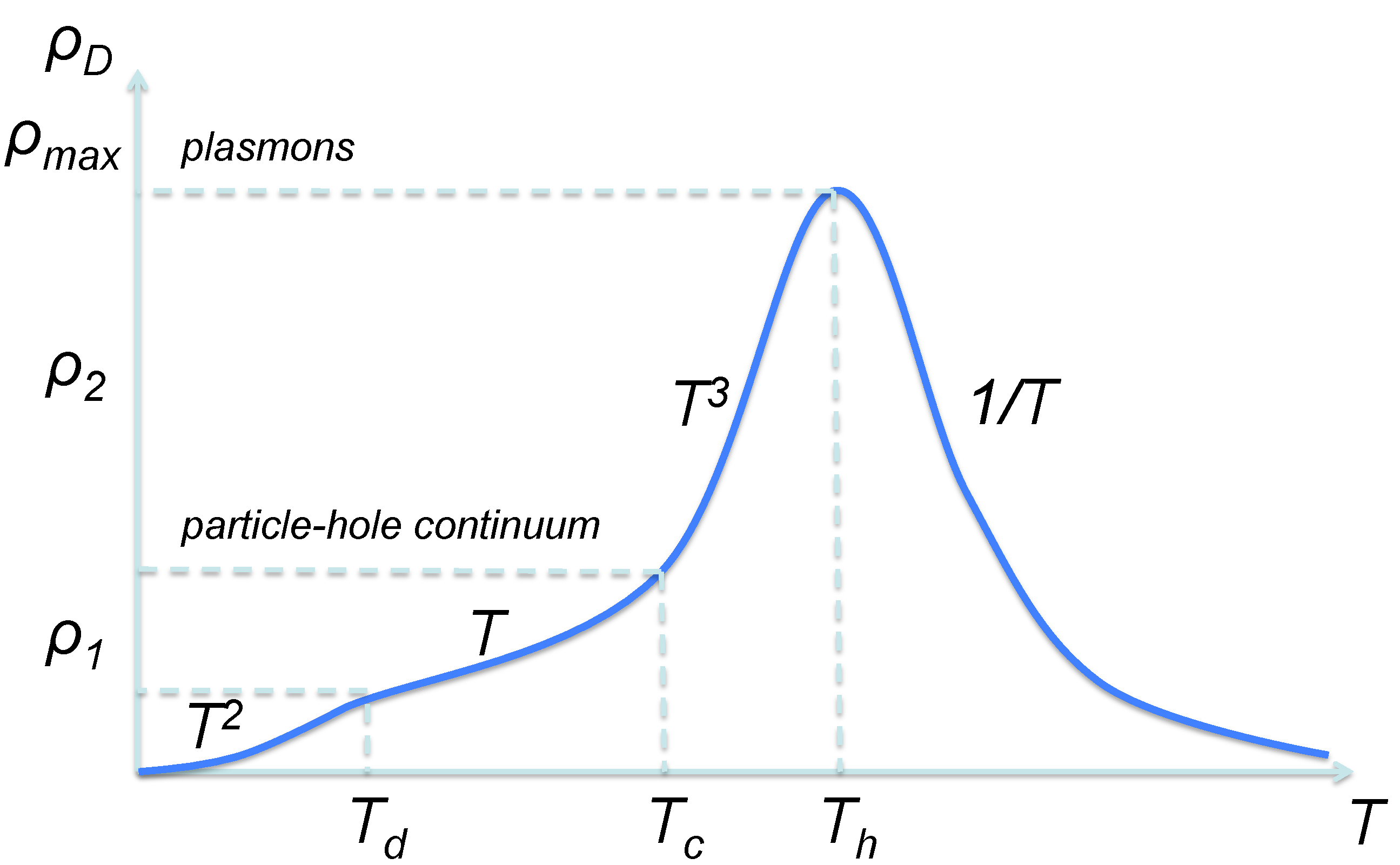}
\end{center}
\caption{(color online) Schematic illustration for the drag
  resistivity at high temperatures showing the plasmon peak at
  ${T\sim{T}_h}$. The asymptotic dependences are exaggerated for
  clarity. Definitions of the three crossover scales are given in the
  main text. [Reproduced from \textcite{chaa}.]}
\label{plas}
\end{figure}

At higher temperatures, ${T>T_c}$, the system enters the
collision-dominated regime, where Coulomb drag is dominated by
plasmons. In this regime, \textcite{chaa} find a stronger temperature
dependence
\begin{equation}
\label{rdp1}
\rho_D\simeq\frac{\hbar}{e^2}\frac{1}{(k_Fd)^4}\frac{T^3}{E^3_F}. 
\end{equation} 
The rise of the plasmon contribution to the drag resistivity persists
so long as the quasiparticle decay rate remains small compared to the
plasma frequency (at the wave vector ${1/d}$), i.e. up to the third
crossover temperature
${T_h\sim{E}_F\sqrt{k_F/\varkappa}\sqrt[4]{1/\varkappa{d}}}$. At
temperatures above the crossover, ${T>T_h}$, the electronic system
enters the hydrodynamic regime of density fluctuations, where 
\begin{equation}
\label{rdh}
\rho_D\simeq\frac{\hbar}{e^2}\frac{1}{(k_Fd)^2(\varkappa d)^3}\frac{E_F}{T}. 
\end{equation}
This limit can be understood on the basis of the classical
Navier-Stokes hydrodynamics \cite{apos}.

The resulting temperature dependence of the drag coefficient is
summarized in Fig.~\ref{plas}. The nonmonotonicity of $\rho_D$
originates from the delicate interplay of various scattering channels
in the electronic system. Perhaps the most striking feature of the
theory of \textcite{chaa} is that intralayer collisions promote
stronger drag. Indeed, should one naively continue Eq.~(\ref{rdph}) up
to the temperatures of the order $T_h$ one would underestimate the
actual maximum value of $\rho_D$ by $\sqrt{k_Fd}\gg1$.


\subsection{Effects of potential disorder}
\label{dis}


In ballistic samples, potential disorder played a very limited role.
In fact, the resulting drag resistivity (\ref{rdb}) is independent of
the impurity scattering time $\tau$. In diffusive samples with
${d\gg\ell}$ only small momenta ${q\ll1/\ell}$ can be transferred
between the layers. Typically this results in a small contribution to
the drag resistivity, which in ballistic samples can be neglected.
This is not always the case -- at low enough temperatures drag is
dominated by mesoscopic fluctuations which are mostly due to processes
with small momentum transfers, see Sec.~\ref{mf}.

Coulomb drag in diffusive systems was considered in \textcite{mac}
using the memory function formalism and in \textcite{kor} using the
diagrammatic technique. To the lowest order in interlayer interaction,
one can use the Kubo formula analysis \cite{kor,me0} to derive the
expression for the drag conductivity ($S$ is the area of the sample)
\begin{equation}
\sigma_D= \frac{1}{16\pi T S} 
\int\!\frac{d\omega}{\sinh^2\frac{\omega}{2T}}
{D}^R_{12}\varGamma^{x}_{23}
{D}^A_{34}\varGamma^{x\; *}_{41},
\label{dia}
\end{equation}
where numerical subscripts indicate spatial coordinates, and are
implied to be integrated over. Averaging over disorder restores
translational invariance. In the absence of interlayer disorder
correlations [this special case was considered in \textcite{go2}], the
nonlinear susceptibilities in each layer have to be averaged
independently of each other. Then one recovers the drag relaxation
rate (\ref{td}), where each quantity should be understood as
disorder-averaged,
i.e. ${\langle\varGamma^{x}_{23}\rangle\rightarrow\Gamma^x(\bs{q})}$.

\subsubsection{Drag resistivity in diffusive regime}
\hfill

In the diffusive regime, the nonlinear susceptibility $\bs{\Gamma}$
can be found from the Ohm's law \cite{dau8},
\begin{equation}
\bs{j} = {\hat\sigma }\bs{E} - e D \bs{\nabla} n,
\label{o}
\end{equation}
where $\hat\sigma$ is the conductivity matrix and $D$ is the diffusion
coefficient (in two dimensions ${D=v_F^2\tau/2}$). Combining
Eq.~(\ref{o}) with the continuity equation, one finds the linear
response of the carrier density $n$ to the electric field $\bs{E}$
\begin{equation}
\langle n(\bs{q},\omega)\rangle = \frac{1}{e} 
\frac{iq^\alpha\sigma^{\alpha\beta}E^\beta(\bs{q},\omega)}
{-i\omega + D q^2}.
\label{den}
\end{equation}
where $\langle\dots\rangle$ indicates averaging over disorder.
Nonlinear response follows from the density dependence of the
conductivity
${j_{dc}={\rm{Re}}\left(\partial\sigma/\partial{n}\right)n(\bs{q},\omega)E(-\bs{q},-\omega)}$,
and yields
\begin{equation}
\langle\varGamma^\gamma\rangle = 
\frac{2\nu}{e}\, 
\frac{\partial\langle\sigma^{\gamma\delta}\rangle}{\partial n}\, q^\delta 
\frac{\omega Dq^2}{\omega^2+D^2q^4}.
\label{gd}
\end{equation}

\begin{figure}
\begin{center}
\includegraphics[width=0.6\linewidth]{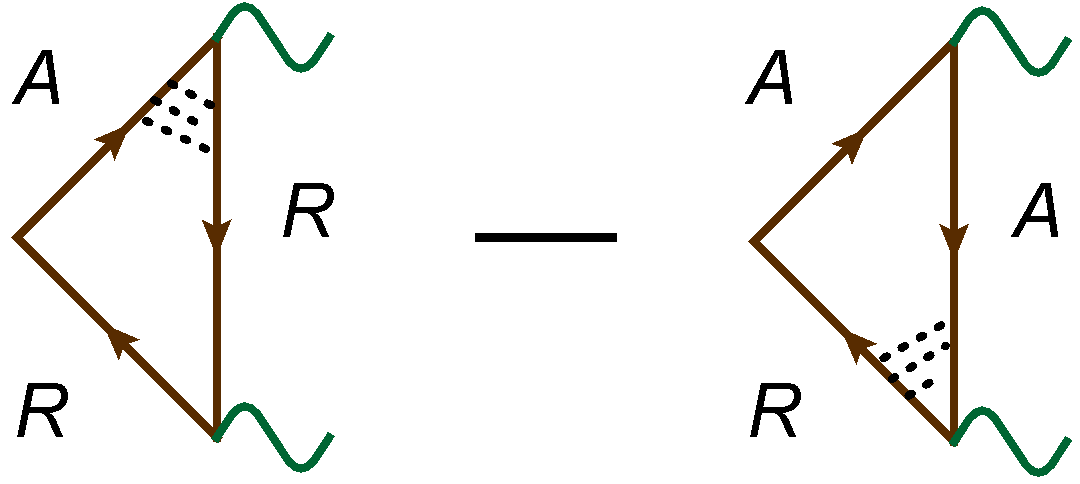}
\end{center}
\caption{(Color online) Disorder averaging of the nonlinear
  susceptibility \cite{kor}. The dotted lines represent the diffuson
  ladder \cite{aar}.}
\label{fig_gd} 
\end{figure}

In the absence of a magnetic field
${\langle\sigma^{\alpha\beta}\rangle=\sigma\delta^{\alpha\beta}}$, and
the nonlinear susceptibility (\ref{gd}) is parallel to $\bs{q}$.  The
disorder-averaged conductivity is linear in the carrier density,
${\partial\sigma^{\alpha\beta}/\partial{n}\approx\sigma^{\alpha\beta}/n}$.
As result,
\begin{equation}
\langle\bs{\varGamma}\rangle = 2\bs{q} \frac{e\nu D}{E_F}
\frac{\omega Dq^2}{\omega^2+D^2q^4}.
\label{gd1}
\end{equation}
This expression can be recast into two equivalent forms. Noting the
similarity between Eq.~(\ref{gd1}) and the standard diffusive form of
the polarization operator \cite{smith1989,aar}
\begin{equation}
\label{prd}
\Pi^R(\bs{q}, \omega)=\nu\frac{Dq^2}{-i\omega+Dq^2},
\end{equation}
one finds \cite{kor}
\begin{equation}
\label{gdr}
\langle\bs{\varGamma}\rangle = 2\bs{q} \frac{eD}{E_F} {\rm Im}\Pi^R(\bs{q}, \omega).
\end{equation}
Furthermore, one can emphasize the fact, that the density dependence
of the conductivity $\sigma$ is a manifestation of electron-hole
asymmetry by rewriting the fraction in Eq.~(\ref{gdr}) as \cite{me0}
\begin{equation}
\label{gdr2}
\langle \bs{\varGamma} \rangle = 2e\bs{q} D \frac{\partial\ln(\nu D)}{\partial\mu}
{\rm Im}\Pi^R(\bs{q}, \omega).
\end{equation}
In the simplest case\cref{fn1}, this expression can be obtained
directly from Eq.~(\ref{gd}) by noticing that
$\partial\sigma^{\alpha\beta}/\partial
n=(\partial\sigma^{\alpha\beta}/\partial\mu) (\partial\mu/\partial n)
= (\partial\sigma^{\alpha\beta}/\partial\mu)(1/\nu)$ and using the
Einstein relation. The same result can be found evaluating diagrams
shown in Fig.~\ref{fig_gd}.

The diffusive approximation for the interlayer interaction follows
from Eqs.~(\ref{d12}) and (\ref{prd}). Focusing on small momenta
${q\ll{1}/d}$, one can obtain alternative expressions for the
interaction propagator by either expanding the bare matrix element
(\ref{coulomb}) in small $qd$ (and subsequently limiting the momentum
integration from above) or keeping the exponential in
Eq.~(\ref{coulomb}) intact, leaving the momentum integral converging
in the ultraviolet. The former approach was taken in \textcite{me0}.
Generalizing to inequivalent layers one finds
\begin{subequations}
\label{dd}
\begin{eqnarray}
\label{ddme}
{\cal D}^R_{12} = -\frac{1}{q^2}
\frac{(-i\omega+D_1q^2)(-i\omega+D_2q^2)}
{(\nu_1D_1+\nu_2D_2)[-i\omega + (1+\varkappa^*d)D^*q^2]}, \qquad\,\,
\end{eqnarray}
where
\begin{eqnarray*}
\varkappa^* = 4\pi e^2 \frac{\nu_1\nu_2}{\nu_1+\nu_2}, \quad
D^*=\frac{(\nu_1+\nu_2)D_1D_2}{\nu_1D_1+\nu_2D_2}.
\end{eqnarray*}
The latter alternative was taken in \textcite{kor}, where in addition
(just as in the ballistic case) the limit ${\varkappa{d}\gg{1}}$ was
used. As a result, the interaction propagator takes the form
\begin{eqnarray}
\label{ddkor}
{\cal D}^R_{12} = - \frac{\pi e^2 q}{\varkappa_1\varkappa_2 \sinh qd}
\frac{-i\omega+D_1q^2}{D_1q^2} \frac{-i\omega+D_2q^2}{D_2q^2}. \qquad
\end{eqnarray}
\end{subequations}

With logarithmic accuracy, the resulting drag coefficient is
independent of the distinction between the two and can be written as
\cite{kor}
\begin{subequations}
\label{rdd}
\begin{equation}
\langle\rho_D\rangle
= \frac{\hbar}{e^2} \frac{\pi^2T^2}{12E_{F1} E_{F2}}
\frac{1}{\varkappa_1\varkappa_2k_{F1}k_{F2}\ell_1\ell_2d^2}
\ln\frac{T_0}{2T}.
\end{equation}
The only difference between using the two expressions for the
interaction propagator in Eq.~(\ref{dd}) is the exact value of
$T_0$. Using Eq.~(\ref{ddme}) in the limit $\varkappa^*d\gg 1$, one
finds
\begin{eqnarray*}
T_0 = \frac{4\pi e^2 \nu_1D_1 \nu_2D_2}{(\nu_1D_1 + \nu_2D_2)d},
\end{eqnarray*}
while Eq.~(\ref{ddkor}) leads to \cite{kor}
\begin{eqnarray*}
T_0 = {\rm min}\{\varkappa_1D_1, \varkappa_2D_2\}/d.
\end{eqnarray*}
Both expressions are of the same order of magnitude and coincide for
the case of identical layers.

The result can be expressed also in terms of the interlayer relaxation
rate \cite{mac}
\begin{eqnarray}
\label{tdd}
\frac{1}{\tau_D} =  \frac{\pi^2}{12} \frac{n_1}{m_2} \frac{T^2}{E_{F1} E_{F2}}
\ln \frac{T_0}{2T}
\frac{1}{\varkappa_1\varkappa_2k_{F1}k_{F2}\ell_1\ell_2d^2}. \qquad
\end{eqnarray}
\end{subequations}
Equivalently, one can use Eq.~(\ref{gdr2}) and express the drag
conductivity \cite{me0} as (here the layers are assumed to be
identical for simplicity)
\begin{eqnarray}
\label{sdd}
\sigma_D = \frac{e^2}{\hbar} \frac{\pi^2}{3}
\frac{(\hbar T)^2}{g^2(\varkappa d)^2}
\left(\frac{\partial}{\partial\mu}(\nu D)\right)^2
\ln\frac{T_0}{2T},
\end{eqnarray}
where the derivative highlights the crucial role of the electron-hole
asymmetry in the leading-order drag effect.

The diffusive result for the drag resistivity (\ref{rdd}) appears to
be rather similar to its ballistic counterpart
Eq.~(\ref{rdb}). Indeed, disregarding the numerical prefactors and the
logarithm in Eq.~(\ref{rdd}), one finds
\begin{equation}
\label{baldif}
\rho_D^{diff}/\rho_D^{bal} \sim d^2/(\ell_1\ell_2).
\end{equation}
This relation may serve as an {\it a posteriori} justification for the
statement that the drag effect in samples with ${d\ll\ell}$ is
dominated by ballistic propagation of carriers with momenta
${\ell^{-1}\ll{q}\ll{d}^{-1}}$. Carriers with small momenta
${q\ll\ell^{-1}}$ also participate in drag, but their contribution is
small [according to Eq.~(\ref{baldif})] and is typically neglected.

\subsubsection{Weak localization corrections}
\hfill

The nonlinear susceptibility (\ref{gd1}) and drag coefficient
(\ref{rdd}) were obtained as the leading approximation in the standard
perturbation theory of disordered metals \cite{aar}, controlled by the
large parameter ${g=25.8}$k${\Omega/R_\Box}$ representing the
dimensionless conductance of the layers (with $R_\Box$ being the layer
(sheet) resistance). Within the assumptions adopted in this
Section\cref{fn1} ${g\sim\nu{D}\sim{k}_F\ell\sim{E}_F\tau\gg{1}}$.

\begin{figure}
\begin{center}
\includegraphics[width=0.7\linewidth]{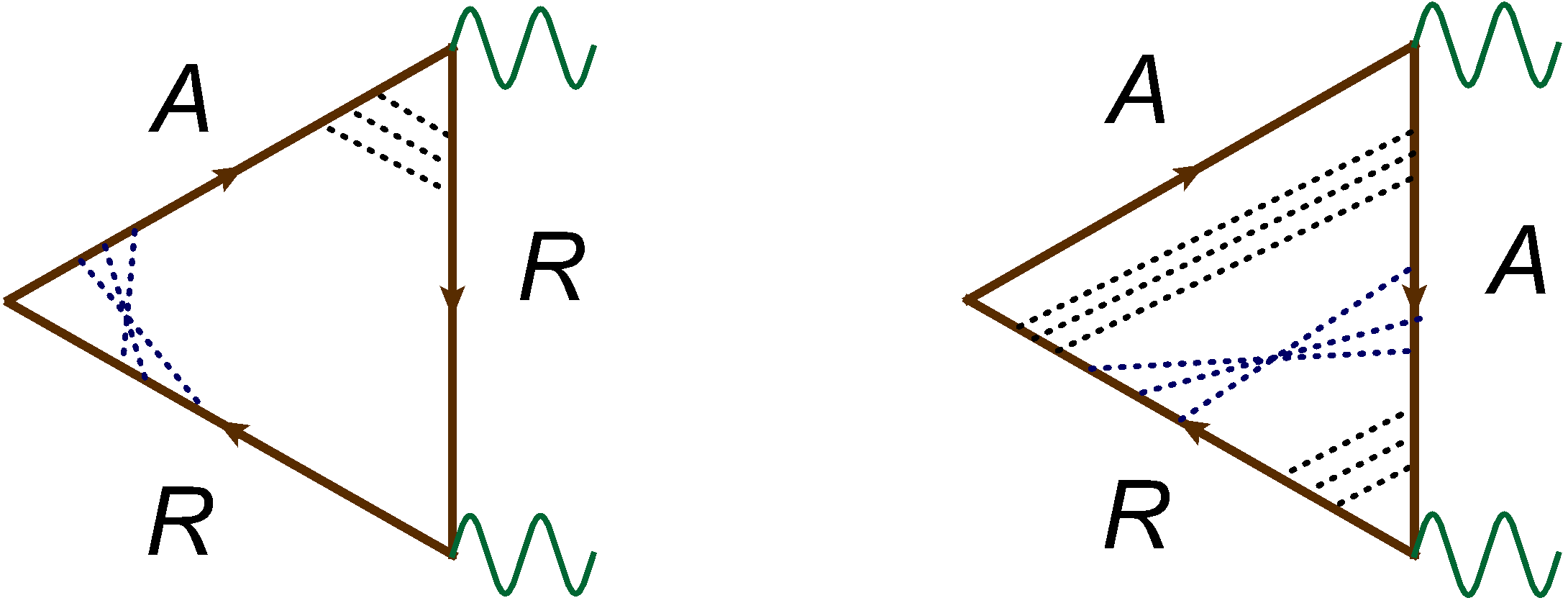}
\end{center}
\caption{\label{fig_wl} (Color online) Leading weak localization
  corrections to the nonlinear susceptibility \cite{kor}. The black,
  parallel dotted lines represent the diffuson ladder
  \cite{aar,aag}. The blue, crossing lines represent the Cooperon
  \cite{glk}.}
\end{figure}

The next-order terms in the perturbation theory are known as quantum
corrections to transport coefficients \cite{aar,aag}. Physically, they
describe leading interference processes that arise in the course of
subsequent scattering events. Although the resulting contribution to
transport is proportional to a small factor ${1/g}$, quantum
corrections dominate the temperature and magnetic field dependence of
transport coefficients at low temperatures.

To the leading order in ${1/g}$, one may distinguish three types of
corrections: (i) interference between self-intersecting, time-reversed
scattering paths leads to a positive correction to resistivity, known
as the weak localization correction \cite{glk,akl,gang4}; (ii)
coherent scattering off Friedel oscillations yields the
Altshuler-Aronov correction \cite{aa1,zna,fnk,fnk1}; and (iii) in
small, mesoscopic samples interference between scattering paths gives
rise to universal conductance fluctuations \cite{lees,alts}. The
latter effect has a direct counterpart in double-layer systems, namely
mesoscopic fluctuations of Coulomb drag discussed in Sec.~\ref{mf}. At
the time of writing, no qualitative interference effect due to
electron-electron interaction has been identified for drag
measurements. At the technical level, the third-order drag effect (see
the following Section) bears certain resemblance to the
Altshuler-Aronov diagrams \cite{gorn}. Here we discuss
the weak localization correction to Coulomb drag \cite{kor,fl2}.

In the absence of interlayer disorder correlations [such effects were
  discussed in \textcite{go2}], impurity scattering is confined to
each individual layer. It should come as no surprise, that the same
mechanism behind the weak localization correction to single-layer
conductivity (i.e., interference between time-reversed,
self-intersecting paths) yields a correction to the nonlinear
susceptibility. Technically, this interference mechanism is described
by a ``maximally crossed'' element of the diagram technique known as
the Cooperon \cite{glk}. Diagrams for the corresponding corrections to
the nonlinear susceptibility are shown in Fig.~\ref{fig_wl} [further
  corrections, e.g. two-Cooperon diagrams, considered in
  \textcite{kor,fl2} were found to be subleading]. The resulting 
nonlinear susceptibility is given by 
\begin{equation}
\label{gwl}
\left\langle\bs{\varGamma}\right\rangle = 
2\bs{q} \frac{e\nu D(\tau_\varphi^{-1},0)}{E_F}
\frac{\omega D(\omega,\bs{q})q^2}{\omega^2+D^2(\omega,\bs{q})q^4},
\end{equation}
where the renormalized diffusion coefficient in two dimensions in
\cite{glk}
\begin{equation}
\label{dwl}
D(\omega,\bs{q}) = 
D\left(1-\frac{1}{\pi k_F\ell}\ln\frac{1}{\omega\tau}\right),
\end{equation}
and $\tau_\varphi$ is the dephasing time \cite{akl}.
The result (\ref{gwl}) is valid in the first order in
${\delta{D}=D(\omega,\bs{q})-D}$.

The resulting leading-order weak localization correction to Coulomb
drag is \cite{kor}
\begin{equation}
\label{rdwl}
\frac{\delta \rho_D}{\rho_D} = 
- \frac{1}{\pi k_{F1}\ell_1}\ln\frac{1}{2T\tau_1}
- \frac{1}{\pi k_{F2}\ell_2}\ln\frac{1}{2T\tau_2},
\end{equation}
where $\rho_D$ is given by Eq.~(\ref{rdd}). The result (\ref{rdwl}) is
similar to the weak localization corrections in 2D \cite{akl,glk},
except that in Eq.~(\ref{rdwl}) the logarithmic singularity is cut by
temperature rather than by the dephasing time.

In conventional 2DEG, weak localization effects result in a dependence
on a weak magnetic field \cite{akl}. Here, the characteristic scale of
the magnetic field would be ${H_c\sim{T}/(eD)}$. Similar scale describes
intralayer interaction corrections to magnetoresistance \cite{aar},
making the weak localization corrections to the drag coefficient hard
to observe experimentally \cite{kor}.


\subsection{Third-order drag effect}
\label{third}


The leading contribution to Coulomb drag, Eqs.~(\ref{td}),
(\ref{dia}), describes the effect to the lowest order in the
inter-layer Coulomb interaction, see Fig.~\ref{fig_fd}. Since the
particles belonging to different layers interact through a layer of an
insulating material, certain weakness of the effective interaction is
intuitively expected. In many-body electron systems the Coulomb
interaction is usually screened and the perturbative analysis gives
reasonable account of most basic observable quantities
\cite{ziman1965,aar}. Consequently, the vast majority of theoretical
studies of Coulomb drag are devoted to the investigation of the
lowest-order effect. Notable exceptions are given by the studies of
the interlayer correlated states, either in the context of the quantum
Hall devices \cite{gir,ya3,ma2,ad2,ads,dem} or quantum wires
\cite{naz,kle}, as well as strongly correlated intralayer states, such
as Wigner crystals \cite{bra,bak} or Anderson insulators \cite{rai}.

The ``single-particle'' drag resistivity, Eqs.~(\ref{rdb}),
(\ref{rdd}), is determined (besides the interlayer interaction) by the
quasiparticle phase space, electron-hole asymmetry (see
Sec.~\ref{kt}), and disorder effects (see Secs.~\ref{dis} and
\ref{mf}). At ${T=0}$ or at a point of exact electron-hole symmetry
(e.g., in neutral graphene, see Sec.~\ref{dig}), these factors may
conspire to nullify the effect. Then $\rho_D$ may be determined by
higher orders of the perturbation theory, implying that saturation of
drag resistivity at low temperatures should not necessarily point
towards a strongly correlated state.

\begin{figure}
\begin{center}
\includegraphics[width=0.9\linewidth]{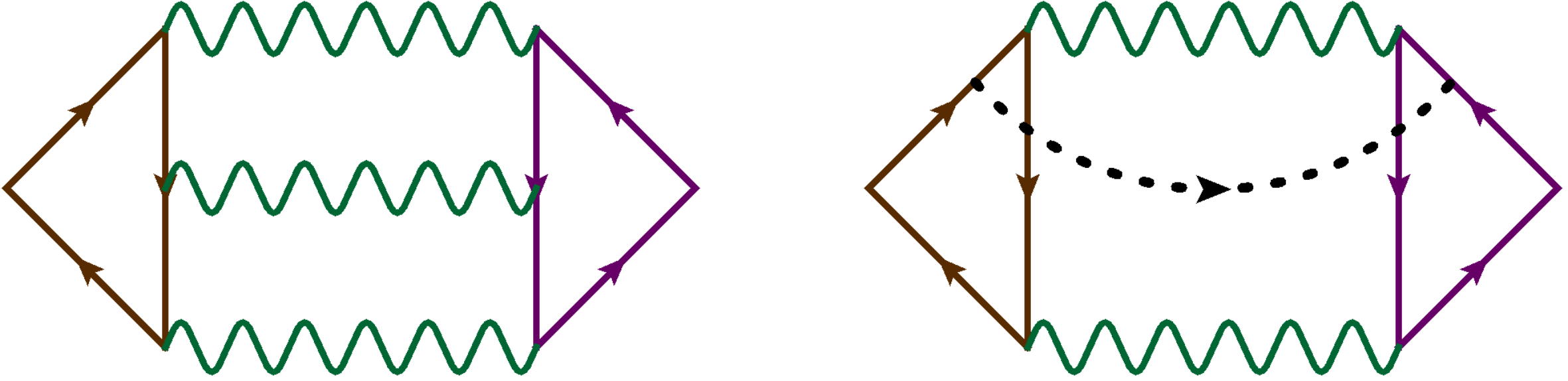}
\end{center}
\caption{(color online) Typical diagrams describing higher-order drag
  effects. Left: third-order drag \cite{le1}. Right: the effect of
  interlayer disorder correlations \cite{go2,hu2}.}
\label{todiag}
\end{figure}

To the third order in interlayer interaction (see Fig.~\ref{todiag}
for the ``skeleton'' diagram), Coulomb drag was first discussed in
\textcite{le1} in the diffusive regime, ${T\tau\ll{1}}$. It was shown
that the third-order drag contribution remains finite at zero
temperature\footnote{More precisely, the result is valid down to
  lowest temperatures ${T\sim\tau^{-1}e^{-\pi g}}$. Below this scale
  the diffusive approximation breaks down.}:
\begin{equation}
\label{tord1}
\rho_D^{(3)}(T<h/\tau)=0.27 (h/e^2)g^{-3}(\varkappa d)^{-2}.
\end{equation}

This surprising result was attributed \cite{le1} to the singular
behavior of matrix elements in the diffusive regime. In single-layer
systems, similar enhancement of the matrix elements leads to singular
interaction effects \cite{aar}. Here, the divergence of the matrix
elements is compensated by the smallness of the phase space yielding
the $T$-independent contribution to the drag resistivity.

The third-order effect (\ref{tord1}) does not rely on electron-hole
asymmetry (technically, the third-order diagram in Fig.~\ref{todiag}
contains four-point vertices instead of the triangular vertices in
Fig.~\ref{fig_fd}). Hence, $\rho_D^{(3)}$ is independent of
$E_F$. This provides an additional explanation of the $T$-independent
result (\ref{tord1}): in the diffusive regime there is no other scale
for a temperature dependence.

Another contribution to drag that is insensitive to electron-hole
symmetry is due to interlayer disorder correlations \cite{go2,hu2}.
For temperatures higher than the inverse interlayer coherence time,
but still in the diffusive regime, ${\tau_g^{-1}\ll{T}\ll\tau}$, one
finds
\begin{equation}
\label{corrd}
\rho_D (\tau_g^{-1}\!\!\ll T\!\ll\tau_{tr}^{-1})
\sim (h/e^2) (k_F^2d^2\varkappa\ell)^{-2}\ln(T\tau_g),
\end{equation}
which might dominate over Eq.~(\ref{rdd}).

While the above higher-order effect have not been observed in
semiconductor samples, they may provide an explanation of the observed
nonzero drag resistivity in neutral graphene \cite{exg}, see
Sec.~\ref{dig}.


\subsection{Transconductance due to tunneling bridges}
\label{tunnel}


A qualitatively different mechanism of transconductance takes place in
the double layer systems with point-like shortages (or bridges) --
places where electrons may tunnel between the two layers
\cite{or2,ore}. Such bridges can be present in metallic double-layer
systems due to device fabrication imperfections, or they can be
introduced on purpose \cite{gio}. In sharp contrast to pure Coulomb
drag, the interplay between interlayer tunneling and Coulomb
interaction leads to a transconductance that is nonvanishing at zero
temperature and \textit{negative} for carriers of the same charge: the
induced current flows in the direction opposite to the driving
one. The origin of the effect is in the interlayer \textit{exchange}
interaction, which is possible due to wave functions overlap at the
bridges.

\begin{figure}
\begin{center}
\includegraphics[width=0.9\linewidth]{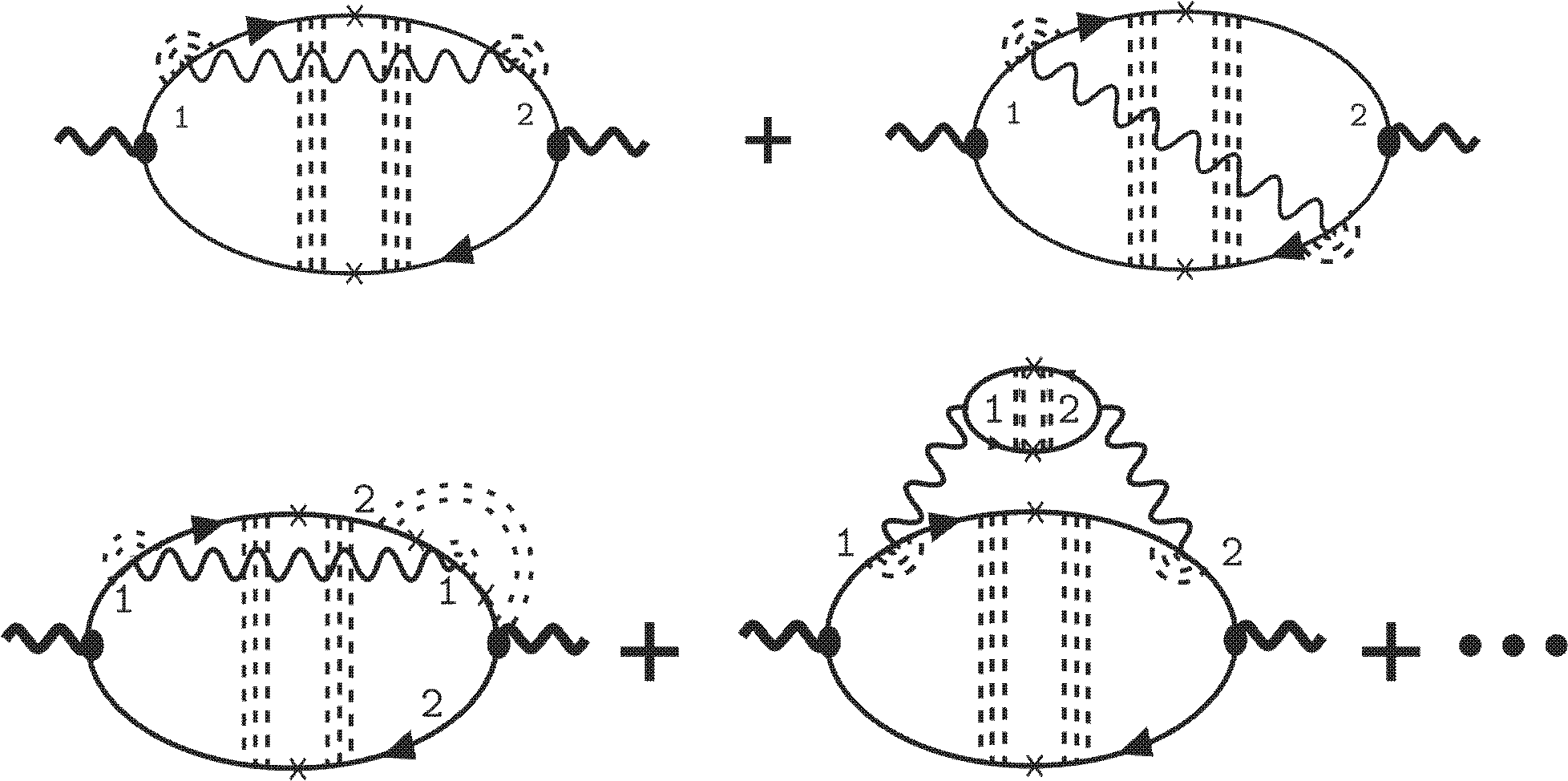}
\end{center}
\caption{(a) Two diagrams contributing to the transconductance that
  are second order in tunneling matrix element denoted by a
  cross. Full lines with arrows are electron Green functions, dashed
  lines represent diffusons, and wavy lines screened interactions. Two
  additional diagrams with arrows in the opposite direction should be
  included. The numbers indicate the layer index. (b) Examples of
  diagrams contributing to the transconductance that are fourth order
  in tunneling. Diagrams with interaction lines connecting ``upper"
  and ``lower" Green functions, as well as diagrams with an opposite
  direction of electron lines are also implicit. [Reproduced from
    \textcite{or2}.]}
\label{Fig-Drag-Tunnel}
\end{figure}

To illustrate the physics of the effect consider a bunch of electrons
injected into the active layer. Some of them tunnel through a bridge
to the passive layer, while the others continue to propagate in the
initial direction. Without interlayer electron-electron interaction,
the tunnelled electrons ``forget" about the direction of their initial
momentum: the wave front propagating in the passive layer is
spherically symmetric; thus no net current is induced. The repulsive
interaction between the tunneled electrons and the electrons remaining
in the active layer makes the former move in a direction opposite to
the injected current. This leads to a negative transconductance. This
lowest order in tunneling mechanism, which is illustrated
diagrammatically in Fig.~(\ref{Fig-Drag-Tunnel}a) is dominant only at
temperatures that are not too small. At lower temperatures the leading
mechanism involves coherent tunneling to the second layer and back to
the first one accompanied by intralayer Coulomb interactions, see
Fig.~(\ref{Fig-Drag-Tunnel}b). Since the exchange contribution is not
proportional to a small electron-hole asymmetry factor, its absolute
value may well overcome the standard Coulomb drag even for a small
tunneling rate.

The interlayer tunneling rate may be described by a mean intralayer
lifetime $\tau_{12}$, which is related to the interlayer tunneling
conductance per unit area ${\sigma_\perp=e^2\nu/\tau_{12}}$. Assuming
weak tunneling, $\tau_{12}\gg\tau$ ($\tau$ is the mean elastic
scattering time within each layer), one finds the following regimes
determined by the three energy scales
${\tau^{-1}_{12}<\varkappa{d}\tau^{-1}_{12}<\tau^{-1}}$.

At high temperatures, ${T>\tau^{-1}}$, the motion of the wave packets
is ballistic and the exchange contribution to the transconductance is
negative and independent of $T$
\begin{equation}
\sigma_D=-\frac{e^2}{\hbar}\frac{\pi}{32}
\frac{1}{\varkappa d}\frac{v_F\tau^2}{d\tau_{12}}. 
\end{equation}

At temperatures ${T<\tau^{-1}}$ the diffusive character of the
electron motion should be taken into account. To the leading order
in the tunneling amplitude, one finds the following temperature
dependent contribution:
\begin{equation}
\label{sdt2}
\sigma_D=-\frac{e^2}{\hbar}\frac{1}{24\pi}
\frac{\ln(\varkappa d)}{\varkappa d}\frac{1}{T\tau_{12}}.
\end{equation}
Singular temperature dependences are not uncommon in diffusive
systems. What is less common is the fact that the divergence is so
pronounced, ${\sim1/T}$, instead of being logarithmic (cf. the
Altshuler-Aronov corrections to the conductivity of 2D systems).

The temperature dependence becomes even more singular at smaller
temperatures, ${T<\varkappa d\tau^{-1}_{12}}$, where
\begin{equation}
\label{sdt3}
\sigma_D=-\frac{e^2}{\hbar}\frac{3\zeta(3)}{8\pi^4}
\frac{\ln(T\tau_{12})}{(T\tau_{12})^2}. 
\end{equation}

The divergences in Eqs.~(\ref{sdt2}) and (\ref{sdt3}) are cut off by
the Thouless energy ${E_T=D/L^2}$ ($L$ is the system size). For large
systems, ${L\gg\sqrt{D\tau_{12}}}$, there is an additional temperature
range ${E_T<T<\tau^{-1}_{12}}$. Here the exchange contribution to the
$\sigma_D$ is dominated by multiple tunneling processes, leading
to the form
\begin{equation}
\sigma_D=-\frac{e^2}{\hbar}\frac{1}{8\pi^2}\ln\frac{1}{T\tau_{12}}. 
\end{equation}
This expression differs from the Altshuler-Aronov correction to the 2D
conductivity only by a factor of $1/4$, which reflects the peculiarity
of the drag measurement setup: the current flows only in one half of
the system and the potential is measured in the other half.

To reveal above quantum corrections $\sigma_D$ in an actual
experimental situation one should
take into account a classical
contribution to the transresistance. It originates from currents which
may flow in the second layer in the presence of interlayer
tunneling. These currents lead to an additional voltage drop in
the
second layer, which modifies the measured transresistance. Such
currents exist either in the presence of many point-like bridges, or in
the case of a uniform tunneling through a sufficiently thin
barrier. In the latter case the tunneling rate is sensitive to a
Fermi-surface mismatch between the layers, and thus may be affected by
a gate voltage or an in-plane magnetic field \cite{boeb,berk}. The
case of many point-like bridges may be visualized by a network of
classical resistors. Calculations done in the continuous limit
\cite{raichev} show that for relatively long samples,
$L>\sqrt{D\tau_{12}}$, approximately half of the current applied to
the first layer eventually leaks out to the second one. In this case
the transresistance practically reduces to the resistance of a single
layer of a doubled width. Such classical effect however leads to a
weakly temperature dependent transresistance. Thus the exchange
contribution, although possibly smaller in its absolute value, may be
distinguished due to its singular temperature dependence.


\subsection{Comparison to experiment}
\label{exp1}


The theory outlined in the preceding sections describes an idealized
phenomenon of mutual friction between two two-dimensional electron
systems. The electrons were assumed to belong to a parabolic band,
with energy-independent impurity-scattering time and negligible
intralayer correlations. Clearly, such assumptions can be realized in
any experimental sample only approximately.

\begin{figure}
\begin{center}
\includegraphics[width=0.95\linewidth]{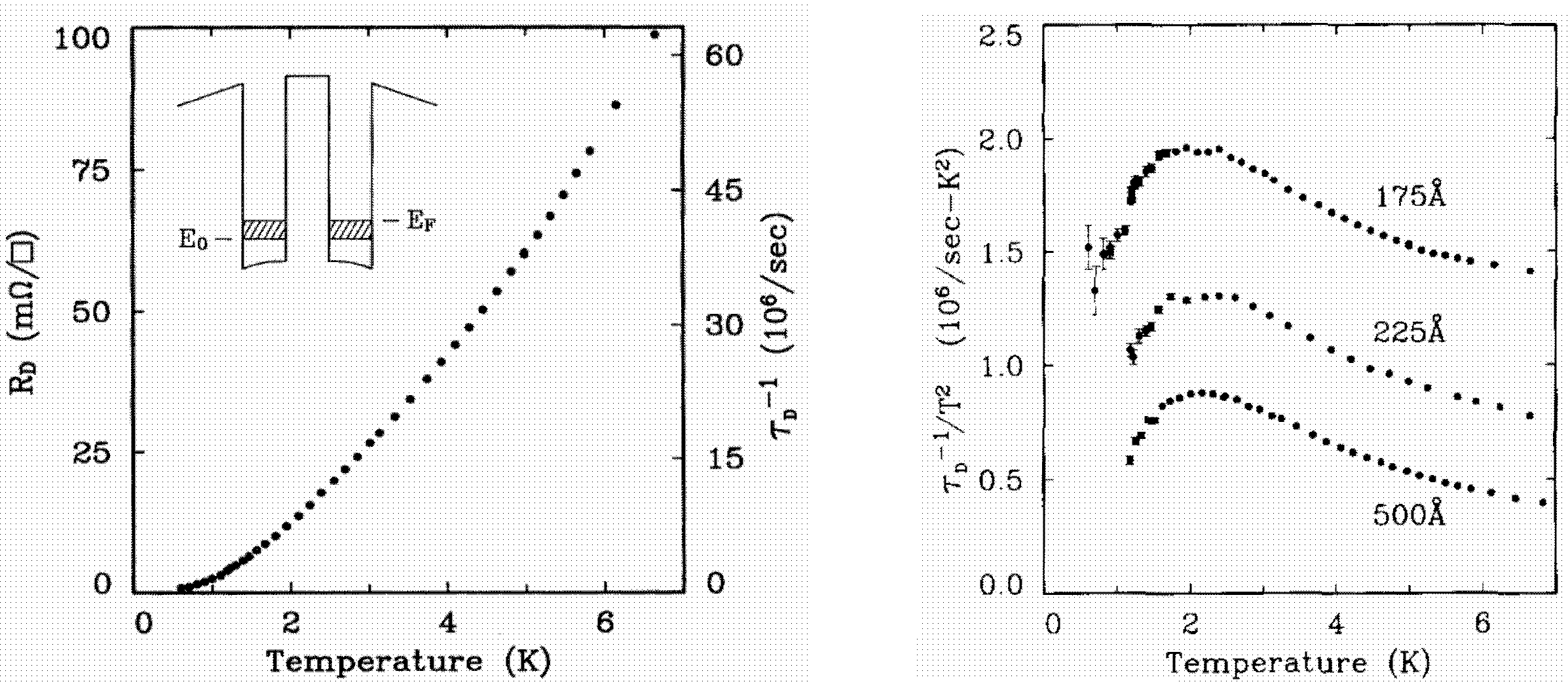}
\end{center}
\caption{Left panel: temperature dependence of the drag resistivity
  GaAs double-quantum-wells. The additional scale on the right
  provides the corresponding values of the momentum-transfer rate (see
  main text for more details). The inset shows an idealized energy
  diagram for a double-quantum-well structure indicating the ground
  subband energy $E_0$ and the Fermi energy $E_F$. [Reproduced from
    \textcite{ex2}.] Right panel: Temperature dependence of the
  interlayer momentum transfer rate divided by $T^2$. The three sets
  of data were measured in samples with interwell barrier widths of
  $175${\AA}, $225$\AA, and $500$\AA. [Reproduced from
    \textcite{ex22}.]}
\label{data_ex2}
\end{figure}

Coulomb drag between two two-dimensional electron gases was first
observed by the group of J. Eisenstein \cite{ex2,eis} in GaAs
double-quantum-wells, see Fig.~\ref{data_ex2}. Detailed comparison of
the experimental data to the quantitative predictions of the Coulomb
drag theory showed that the latter accounts for about 50\% of the
measured values\footnote{The momentum relaxation time reported in
  \textcite{ex2} is twice smaller than Eq.~(\ref{tdb}). In addition,
  the paper cited unpublished calculations of MacDonald, Gramila, and
  Eisenstein involving a more realistic modeling of finite-width
  quantum wells. In particular, these calculations were reported to
  include vertex corrections to the RPA interaction propagator
  (\ref{d12}).  Hence, it is difficult to judge whether that factor of
  $2$ has played any role in the actual analysis of \textcite{ex2}.}.
This was judged as sufficient evidence of the relevance of the Coulomb
mechanism of frictional drag. Also, the overall reduction of the drag
resistance with the increase of the interwell barrier width (see
Fig.~\ref{data_ex2}) was in rough agreement with Eq.~(\ref{tdb}). At
the same time, the data (see the right panel in Fig.~\ref{data_ex2})
show noticeable deviations from the $T^2$ behavior predicted by
Eqs.~(\ref{rdb}) and (\ref{rdd}), indicating that
other scattering mechanisms might also be important.

\begin{figure}
\begin{center}
\includegraphics[width=0.95\linewidth]{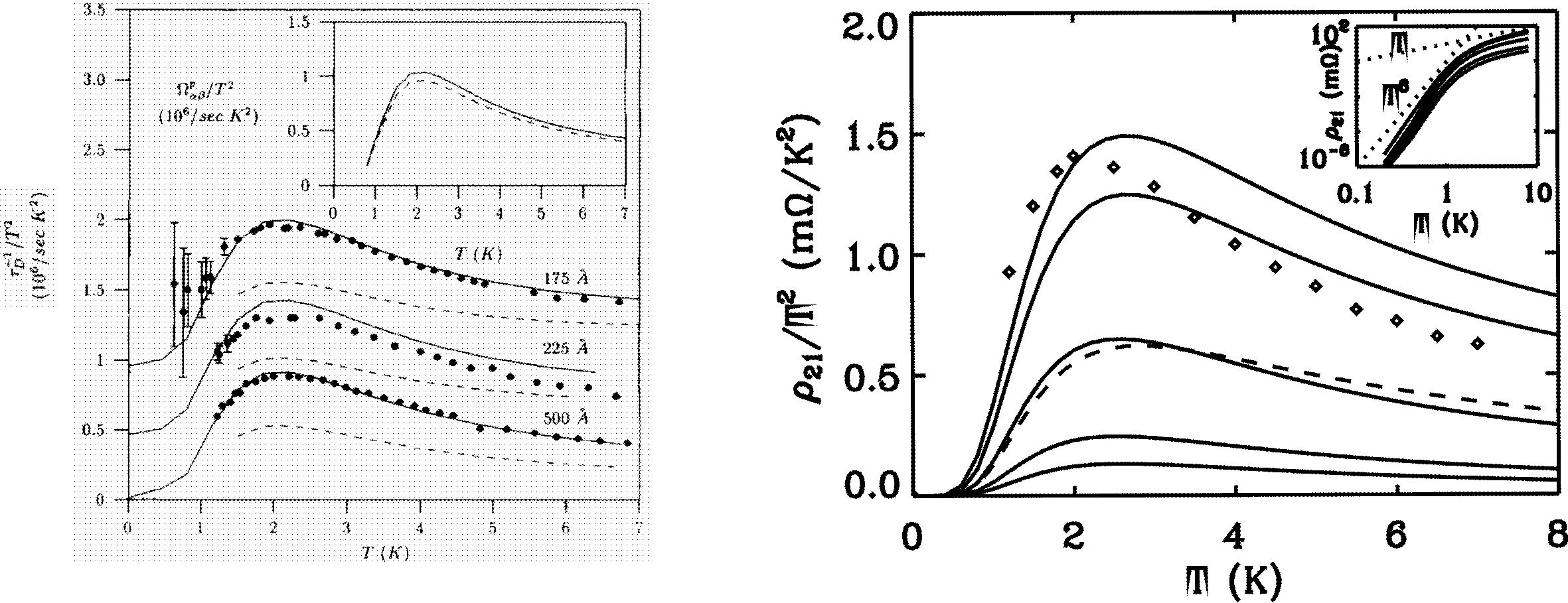}
\end{center}
\caption{Left panel: calculated ${\tau_D^{-1}/T^2}$ (solid curves)
  compared to the data of \textcite{ex2}. A less optimal choice of a
  fitting parameter yields results shown by the dashed curves. Inset:
  calculated contribution of virtual-phonon exchange processes to
  ${\tau_D^{-1}/T^2}$). [Reproduced from \textcite{tso}.] Right panel:
  calculated ${\rho_D/T^2}$ for various values of the phonon mean free
  path and ${d=500}$\AA (solid curves). The dots show the data of
  \textcite{ex22}. The dotted line represents the contribution of the
  modified plasmon pole. Inset: the crossover of the $T^6$ to $T$
  temperature dependence. [Reproduced from \textcite{bo2}.]}
\label{data3}
\end{figure}

One additional mechanism \cite{ex2} is due to electron-phonon
interaction. This suggestion was developed theoretically in
\textcite{tso,bo2,bo22,ba2} and experimentally in
\textcite{noh,ru3,jo2}, see Fig.~\ref{data3}.

\begin{figure}
\begin{center}
\includegraphics[width=0.95\linewidth]{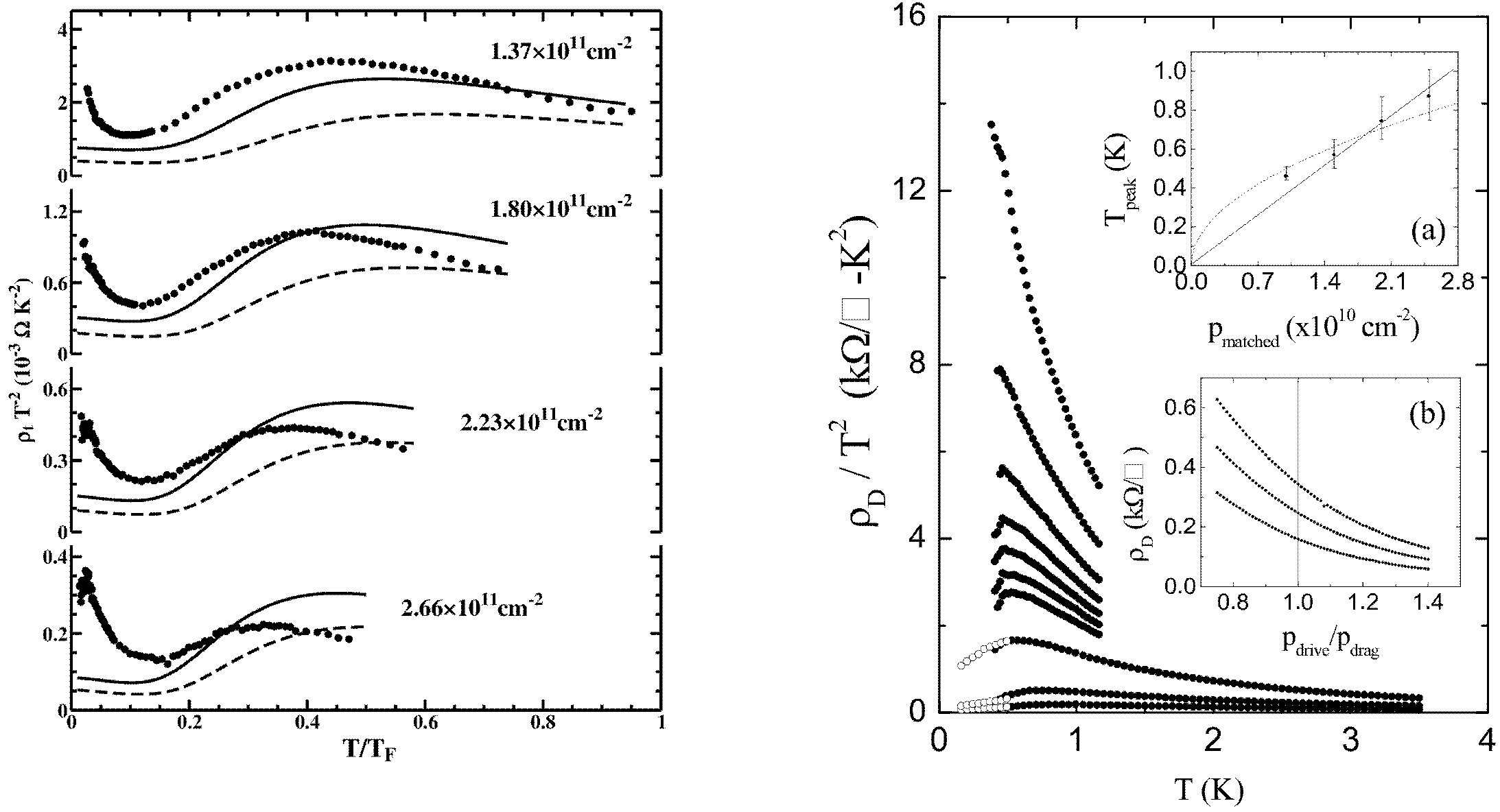}
\end{center}
\caption{Left panel: measured ${\rho_D/T^2}$ for various values of the
  carrier density, ${n_1=n_2}$. Dashed lines represent the results of
  \textcite{fln} adjusted for the sample parameters of the
  experiment. Solid lines show the results of additional calculations
  taking into account intralayer many-body correlations within the
  Hubbard approximation. [Reproduced from \textcite{hil}.] Right
  panel: measured ${\rho_D/T^2}$ for different carrier
  densities. Inset: (a) peak position temperature vs matched layer
  density; (b) $\rho_D$ vs density ratio for $T=860$, $730$, and
  $600$mK. [Reproduced from \textcite{pi2}.]}
\label{data_pl}
\end{figure}

Further corrections to the single-particle Coulomb mechanism are
associated with the plasmon contribution. As shown in \textcite{fln},
plasmons are expected to be most important at intermediate
temperatures, ${T\sim{0.5}T_F}$. This prediction was tested
experimentally in \textcite{hil}, see the left panel in
Fig.~\ref{data_pl}, and in \textcite{no1}. While the theoretical
results show qualitative agreement with the data, discrepancies
persist. Taking into account many-body correlations [see, e.g.,
  \textcite{ssg}] improves the agreement, but further advances in
many-body theory are necessary before a more precise quantitative
description of the correlation effects in double-layer structures is
achieved.

The discrepancies between the simple single-particle description and
laboratory experiments are by no means universal, especially since
many measurements were performed in very different systems. One of the
first drag experiments \cite{ex1} has been performed on a hybrid 2D-3D
system. This device showed considerable thermoelectric effects masking
the purely Coulomb contribution to drag. Experiments on electron-hole
systems \cite{siv} showed behavior that could not be accounted by
neither the phonon, nor plasmon corrections. Instead, generalized RPA
(taking into account exchange processes to all orders) \cite{tvp}
appears to yield satisfactory agreement with observations of
\textcite{siv} at low temperatures. Apparently, the traditional RPA
overestimates screening which results in the underestimated drag
resistivity.

Experiments on dilute 2D hole systems \cite{pi2,pi21} show marked
enhancement of the drag resistivity, along with the stronger
temperature dependence (empirically, ${\rho_D\propto{T}^{2.5}}$ at low
temperatures, followed by a crossover towards a sublinear temperature
dependence at ${T\simeq{E_F}}$). These systems are characterized by
rather high values of the interaction parameter
${r_s\simeq{20}}$--$40$ and also exhibit signs of a metal-insulator
transition in single-layer measurements \cite{pil}. The data obtained
in \textcite{pi2} are not explained by taking into account corrections
due to phonons \cite{bo2}, plasmons \cite{fln}, or many-body effects
\cite{ssg}, as follows from the density dependence of the measured
drag, illustrated in the right panel of Fig.~\ref{data_pl}. The lack
of adequate theoretical description of these experiments is not
surprising, given that the regime of relatively high $r_s$ remains an
unsolved problem in single-layer (bulk) systems as well.

\textcite{cro,gup2} report anomalous drag in electron-hole
bilayers. Below ${T=1}$K, the measured drag resistivity exhibits an
upturn that may be followed by a downturn, although $\rho_D$ does not
seem to vanish for ${T\rightarrow0}$. The observed upturn may indicate
exciton formation \cite{vi2,hu5}, however neither the observed
violation of Onsager reciprocity, nor the apparent downturn at lower
temperatures are anticipated by the theory. The effect of density
imbalance on the drag upturn was studied in \textcite{mor}. The data
were interpreted in terms of a pairing-fluctuations mechanism based on
the theory of \textcite{hwa2}. The theory accounts for most
qualitative features of the effect, however the predicted peak in
$\rho_D$ at equal layer densities was not observed in experiment
\cite{mor}.

Further experiments demonstrate interesting correlation effects such
as Wigner crystallization in quantum wires \cite{ya1,yam}, exciton
formation in electron-hole bilayers \cite{sea}, or quantum Hall effect
\cite{lil,gir}, see Sec.~\ref{qhe1}. Clearly these phenomena cannot be
described by the simple theory presented in this Section. At the same
time, single-particle effects are still important at relatively low
temperatures (${T\lesssim{0.2}T_F}$) in traditional semiconductor
heterostructures hosting two-dimensional electron systems and even
more so in graphene (see Sec.~\ref{dig}), where interlayer separation
can be as small as several interatomic distances \cite{exg}.

\subsubsection{Phonon effects}
\hfill

Electrical resistivity due to electron-phonon scattering is a standard
topic in condensed matter physics \cite{ziman1965}. At temperatures
higher than the Debye frequency ${T\gg\omega_D}$, it exhibits linear
behavior ${\rho\propto{T}}$, that is observed in a wide class of
materials including high-mobility 2DEG \cite{storm} and graphene
\cite{efk}. At low temperatures ${T\ll\omega_D}$ [in low density
  electron systems the crossover occurs at a lower scale, the
  so-called Bloch-Gr\"uneisen temperature ${T\ll{T}_{BG}<\omega_D}$
  \cite{storm}] the phonon contribution is rapidly decreasing as
${\rho\propto{T}^5}$ in metals \cite{bloch,gru} and heterostructures
\cite{price,storm} and as ${\rho\propto{T}^4}$ in graphene
\cite{hwang,efk}.

Qualitative physics of electron-phonon interaction in semiconductor
double-quantum-well heterostructures is captured by the following
interaction Hamiltonian
\begin{eqnarray}
&&
{\cal H}_{ep} = \frac{1}{\sqrt{V}} 
\sum_{\lambda, \lambda'; \bs{k}}\sum_{\bs{Q}; \eta}
M_{\lambda, \lambda'}^\eta(\bs{Q}) F_{\lambda, \lambda'}(q_z) 
\nonumber\\
&&
\nonumber\\
&&
\qquad
\times
\left[\hat{b}_{\eta}^\dagger(-\bs{Q}) + \hat{b}_{\eta}(\bs{Q})\right]
\hat{c}^\dagger_\lambda(\bs{k}) \hat{c}^{\,}_{\lambda'}(\bs{k}+\bs{q}).
\label{heph}
\end{eqnarray}
Here $\bs{Q}=(\bs{q},q_z)$ is the 3D wave vector of a phonon with
polarization $\eta$, $\bs{k}$ is the 2D electron wave vector,
$M_{\lambda, \lambda'}^\nu$ is the bulk electron-phonon matrix element
corrected by the subband form-factor
\begin{equation}
\label{ff}
F_{\lambda, \lambda'}(q_z) = \int\limits_{-\infty}^\infty\!dz 
\xi_\lambda(z)\xi^*_{\lambda'} e^{iq_zz},
\end{equation}
where $\xi_\lambda(z)$ is the bound state wave function associated
with the quantized motion in the subband $\lambda$. This Hamiltonian
was used to study effects of interaction between electrons and
longitudinal optical phonons in \textcite{sdas} and to calculate
quasiparticle properties in weakly polar 2DEG in \textcite{sdas2}. In
double-layer systems, the Hamiltonian (\ref{heph}) was used to
describe interlayer interaction mediated by acoustic phonons in
\textcite{zha,bo2} and by optical phonons in \textcite{hu3}.

Electrons experience the phonon-mediated interaction (\ref{heph})
alongside the Coulomb interaction. The propagator of the effective
interlayer interaction can be obtained within the RPA
\cite{zha,bo2,sdas2} similarly to Eq.~(\ref{d12}). The result can be
represented in the form
${{\cal{D}}_{12}=(V_{12}+D_{12})/\epsilon(\bs{q},\omega)}$, where
$D_{12}$ is the propagator of the phonon-mediated interaction and
${\epsilon(\bs{q},\omega)}$ is the effective dielectric function for
interlayer interactions that is also determined by the sum of the
Coulomb interaction (\ref{coulomb}) and the phonon propagator. Thus
the phonon and Coulomb mechanisms are generally not independent of
each other. However, the Coulomb interaction contributes only to small
momentum transfers (\ref{qd}), while the phonon contribution peaks at
${q\sim{2}k_F}$ \cite{bo2}. Neglecting interference between the two,
one can estimate the effect of phonon-mediated interaction by
considering only the phonon part
${{\cal{D}}_{12}\rightarrow{D}_{12}/\epsilon(\bs{q},\omega)}$.

A simple analytical estimate for the strength of the phonon-mediated
interaction in GaAs/AlGaAs systems was suggested in \textcite{bo2}.
In this material, electron-phonon interaction is due to the
deformation potential and piezoelectric effect. It turns out, that the
deformation mechanism dominates (except for very low electron
densities). Assuming infinite phonon mean free path, the corresponding
(unscreened) effective interaction has the form
\begin{equation}
\label{d12ph}
{\cal D}_{12} = -
\frac{C_{DP}\omega^2e^{-d\sqrt{q^2-\omega^2c_l^{-2}}}}
{\nu k_F c_l\sqrt{c_l^2q^2-\omega^2}},
\end{equation}
where $c_l$ is the velocity of longitudinal acoustic phonons and
${C_{DP}\approx2.7\times10^{-3}k_F/(10^6{\rm{cm}}^{-1})}$. The
smallness of electron-phonon coupling constants implies weakness of
the phonon-mediated interlayer interaction as compared to the Coulomb
interaction. However, the effective interaction (\ref{d12ph}) diverges
near ${\omega\approx{c}_lq}$ leading to a logarithmic divergence in
the drag resistivity. Although this divergence is removed by either
dynamic screening or phonon relaxation, the above argument illustrates
the reason behind the relative strength of the phonon-mediated
interlayer interaction.

Detailed calculations of the phonon-mediated drag resistivity have
been performed numerically by several authors. \textcite{tso} showed
that combining the phonon and Coulomb mechanisms of mutual friction
accounts for the nonparabolic temperature dependence observed in
GaAs/AlGaAs devices \cite{ex2}, see the left panel of
Fig.~\ref{data3}. A refined discussion of the phonon mechanism was
given in \textcite{bo2}, see the right panel of Fig.~\ref{data3}.  It
was shown, that the temperature dependence of the phonon contribution
to drag exhibits a crossover from linear to $T^6$ behavior around the
Bloch-Gr\"uneisen temperature (see the inset in Fig.~\ref{data3}),
explaining the peak in the drag resistivity, Fig.~\ref{data_ex2}. In
addition, it was shown that there exists a collective mode that can be
found setting ${\epsilon(\bs{q},\omega)=0}$. This mode is similar to
the usual plasmon and results from coupling of the electrons from both
layers to the phonons with ${\omega\sim{c}_lq}$. A similar mode
resulting from interaction between electrons and optical phonons was
discussed in \textcite{tan,tan2}. A detailed analysis of the mutual
friction due to optical phonons is given in \textcite{hu3}.

\subsubsection{Interlayer interaction beyond RPA}
\hfill

The expression (\ref{d12}) for the dynamically screened interlayer
Coulomb interaction has been obtained within the RPA. While capturing
the qualitative physics of the effect, this representation is by no
means exact. In particular, RPA-based calculations seem to
underestimate the value of $\rho_D$ as compared to experimental data
\cite{siv}. A pedagogical discussion of the RPA and possible
approaches to interacting many-body systems that go ``beyond'' the RPA
can be found in \textcite{Giuliani}. Most of these approaches are not
parametrically justified. The results of the calculations are
typically compared to either experimental data or computer
simulations.

Coulomb drag between electron and hole layers within the generalized
RPA approach was considered in \textcite{tvp}. The resulting $\rho_D$
is about twice larger than that calculated within RPA, but still about
twice smaller than the experimental data. Furthermore, it was
understood in \textcite{ssg} that the true temperature dependence of
$\rho_D$ should exhibit a crossover from the $T^2$ dependence at low
temperatures to a power-law at higher temperatures.  However, the
local field approach [or the Singwi-Tosi-Land-Sj\"olander method
  \cite{stls}] used in this work still fails to reproduce $\rho_D(T)$
measured in \textcite{siv}, although yields roughly the same magnitude
of the effect (in contrast the above RPA and generalized RPA
calculations). This approach was further extended to drag between two
2DEG in \textcite{ssg2,ssg3}. The results of that work suggest that
many-body correlations enhance interlayer interaction and improve
agreement with experiments. Nevertheless, experiments [see, e.g.,
  \textcite{hil} and Fig.~\ref{data_pl}] show, that existing
theoretical methods are still incapable of providing precise
quantitative description of real systems.

A detailed consideration of Coulomb drag resistivity based on an
extrapolation of Fermi-liquid-based formulas to the region where
intralayer correlations are strong has been carried out by
\textcite{hbs} in an attempt to address the striking data of
\textcite{pi2} in low density and high mobility hole bilayers. The
observed drag was two to three orders of magnitude larger than
previously reported values. The calculations of \textcite{hbs} were
different from that leading to Eq. (21a) in several points, all of
them leading to an increase of the drag resistivity: (i) Hubbard
approximation was employed to obtain polarization operator, which
accounts for the exchange-driven local field corrections; (ii)
experimentally measured dependence of conductivity on density was used
to extract electron-hole asymmetry factor; (iii) large-momentum
transfer component was included to calculate drag; (iv) finite
thickness of quantum wells was included to calculate form-factors of
Coulomb matrix elements; (v) lastly, phonon contribution was
added. Combining all these factors, \textcite{hbs} were able to
account for most of the results of the measurements within a Fermi
liquid approach.


\subsection{Single-particle drag in magnetic field}
\label{wmf}


The semiclassical Drude model described by Eqs.(\ref{dt}) predicts
that the drag resistivity is independent of the magnetic
field. Moreover, there is no ``Hall drag'': the direction of the
induced motion of charge carriers in the passive layer is expected to
coincide with that of the driving current. These predictions
contradict numerous experiments [see, e.g.,
  \textcite{lil,rub,mur,fin,nan,eim}] showing that Coulomb drag is not
only sensitive to magnetic field, but in fact the drag resistivity can
be greatly enhanced, once the field is applied.

In single-layer measurements, magnetoresistance is usually associated
with either (i) multi-band systems, or (ii) quantum effects. A close
analog of the former can be found in graphene-based systems, see
\textcite{exg}, \textcite{meg} and Sec.~\ref{dig}. The latter effects
are manifest in strong, quantizing magnetic fields leading to
emergence of a qualitatively different behavior \cite{gir,eim}
discussed in Sec.~\ref{qhe}.

The situation somewhat simplifies if the field is tuned close enough
to the point where the Landau level filling factor in the two layers
is about ${\nu=1/2}$. In this case, the many-body state in each layer
can be viewed as a Fermi liquid of composite fermions \cite{hlr}.  In
some sense composite fermions are similar to usual electronic
quasiparticle in metals.  Long-range, interlayer interaction between
such excitations could lead either to a ``single-particle'' drag
effect \cite{us1,ust,kim,sak}, or to novel correlated states, see
Sec.~\ref{qhe}. Alternative approaches include magnetodrag due to
electron-phonon interaction \cite{bkm}, semiclassical theory
\cite{mez}, diagrammatic theory in high Landau levels
\cite{gor,opp,bo3,bo32}, self-consistent Hartree approximation
\cite{ts2}, and the effect of magnetoplasmons \cite{mat,kha}.

\subsubsection{Hall drag in weak (classical) magnetic field}
\hfill

Recall that the standard single-band Drude theory (\ref{dt}) does not
allow for any dependence of the drag resistivity on the magnetic field
and in particular predicts zero Hall drag, see Eq.~(\ref{rH0}). Same
conclusion can be reached using the diagrammatic perturbation theory
\cite{kor}. This result is justified by the assumption of
energy-independent impurity scattering time $\tau$. Lifting this
assumption \cite{hu0}, one can show that a weak Hall drag signal may
appear
\begin{equation}
\label{rh_hu}
\rho_D^H\propto s T^4, \quad s= \frac{\partial\tau(\epsilon)}{\partial\epsilon}
\frac{E_F}{\tau(E_F)}.
\end{equation}
As argued in \textcite{hu0}, this effect is hard to observe in
conventional semiconductor heterostructures where intralayer
relaxation processes are dominated by electron-electron interaction:
in this case the nonequilibrium distribution function quickly relaxes
to a drifted Fermi-Dirac distribution and hence the impurity
scattering time is effectively almost independent of energy,
i.e. ${s\ll{1}}$. 

Hall drag in weak magnetic fields was studied in \textcite{pat}. The
experimental device comprised two $180${\AA}-wide quantum wells
separated by $100${\AA} and exhibited measurable tunneling between the
layers, contrary to the assumptions of \textcite{hu0}. Hall drag in
graphene \cite{meg} was attributed to a different mechanism, see
Sec.~\ref{dig}. Other observations of Hall drag were performed in the
quantum Hall regime (see Sec.~\ref{qhe}), where the effect is much
stronger \cite{opp} than Eq.~(\ref{rh_hu}).

\subsubsection{Coulomb drag of composite fermions}
\hfill
\label{cfdrag}

All of the previous discussion was based on the underlying physical
picture of weakly interacting fermions. Typically, this picture
becomes invalid in a strong, quantizing magnetic field. The only
exception to this statement is the peculiar state at the half-filled
Landau level. This state can be described as a Fermi liquid of
composite fermions \cite{hlr}. Each composite fermion is an electron
with two attached flux quanta \cite{jai}, that interacts with the
others both electrostatically and by means of a Chern-Simons
interaction.

Composite fermions can be characterized by linear response functions
similar to those of electrons. In particular, their respective
single-layer resistivities related to each other by \cite{hlr}
\begin{equation}
\label{rcf}
\hat\rho_{el} = \hat\rho_{cf} + \frac{2h}{e^2}
\begin{pmatrix}
0 & 1 \cr
-1 & 0
\end{pmatrix}.
\end{equation}
If one is interested in the relation between conductivities of the
electrons and composite fermions, then one has to invert the
resistivity matrices in Eq.~(\ref{rcf}). Clearly, the electronic
conductivity is not identical to that of the composite fermions.

Extending Eq.~(\ref{rcf}) to the case of a double-layer system, one
obtains a similar relation for the ${4\times{4}}$ resistivity matrices
\cite{us1}. If interlayer interaction is weak enough, so that
composite fermions in a given layer are not sensitive to the
Chern-Simons field of the other layer, then similarly to
Eq.~(\ref{rcf}), longitudinal resistivities of the electrons and
composite fermions are the same and hence
\begin{equation}
\label{rdcf}
\rho_D^{el} = \rho_D^{cf}.
\end{equation}
Again, conductivities (in particular, drag conductivities) of
electrons and composite fermions are not equivalent.

The quantity measured in drag experiments is the electronic drag
resistivity $\rho_D^{el}$, Eq.~(\ref{rd1}). Given the equality
(\ref{rdcf}), one can calculate either $\rho_D^{cf}$ or
$\rho_D^{el}$. The former approach was developed in \textcite{kim},
while the latter was considered in \textcite{us1}. Both calculations
are based on the standard lowest-order perturbation theory and yield
similar results (albeit with a rather different
interpretation\footnote{The subquadratic temperature dependence
  (\ref{rdn12}) of the drag resistivity at ${\nu=1/2}$ was interpreted
  in \textcite{kim} as a signature of the non-Fermi-Liquid nature of
  composite fermions. In particular it was related to the similar
  power law in the self-energy of the composite fermions leading to
  the ${\omega\sim{q}^3}$ scaling of the typical frequencies.}). The
calculation of \textcite{kim} consists evaluating Eq.~(\ref{td}) for
composite fermions and using the correspondence (\ref{rdcf}).
Alternatively \cite{us1}, one can treat the problem in purely
electronic terms assuming that interlayer interaction is dominated by
the direct Coulomb coupling [the assumption which justifies
  Eq.~(\ref{rdcf})]. At the same time, single-layer electronic
response functions (such as ${\rm Im}\Pi^R$) can be calculated within
the composite-fermion approach of \textcite{hlr}.

Within RPA (including the response of composite fermions to the
external, Coulomb, and Chern-Simons potentials) and in the limit
${q\ll{k}_F}$, ${\omega\ll{v}_Fq}$, the electronic density-density
response function (the polarization operator) is given by \cite{hlr}
\begin{equation}
\label{pcf}
\Pi^R(\bs{q}, \omega) = \frac{dn}{d\mu}
\frac{q^3}{q^3 - 8\pi i \omega k_F (dn/d\mu)},
\end{equation}
where ${dn/d\mu}$ is the thermodynamic compressibility of the
${\nu=1/2}$ state. At large momenta,
${{\rm{I}m}\Pi^{-1}\propto{q}^{-3}}$; consequently, the momentum
integration in Eq.~(\ref{td}) is dominated by the region
${q\approx{k}_F(T/T_0)^{1/3}}$ [i.e., determined by poles of the
  interlayer interaction, rather than Eq.~(\ref{qd})].  As a result,
the temperature dependence of the drag resistivity is weaker than in
the absence of magnetic field \cite{uss,us1}
\begin{equation}
\label{rdn12}
\rho_D = 0.825 (h/e^2)(T/T_0)^{4/3},
\end{equation}
where the characteristic temperature depends on the carrier
density $n$, interlayer spacing $d$, dielectric constant
$\varepsilon$, and thermodynamic compressibility
\[
T_0 = \frac{\pi e^2 n d}{\varepsilon} 
\left[
1+ \frac{\varepsilon}{2\pi e^2 d} \left(\frac{dn}{d\mu}\right)^{-1}
\right].
\]
The same temperature dependence was reported in \textcite{kim,sak}.

For realistic parameter values similar to those of the experiment of
\textcite{lil}, the drag resistivity (\ref{rdn12}) is much larger than
the zero-field result (\ref{rdb}). This fact is associated with the
smallness of the typical momenta involved in the interlayer scattering
processes and slow relaxation of density fluctuations in the
${\nu=1/2}$ state.

\begin{figure}
\begin{center}
\includegraphics[width=0.98\linewidth]{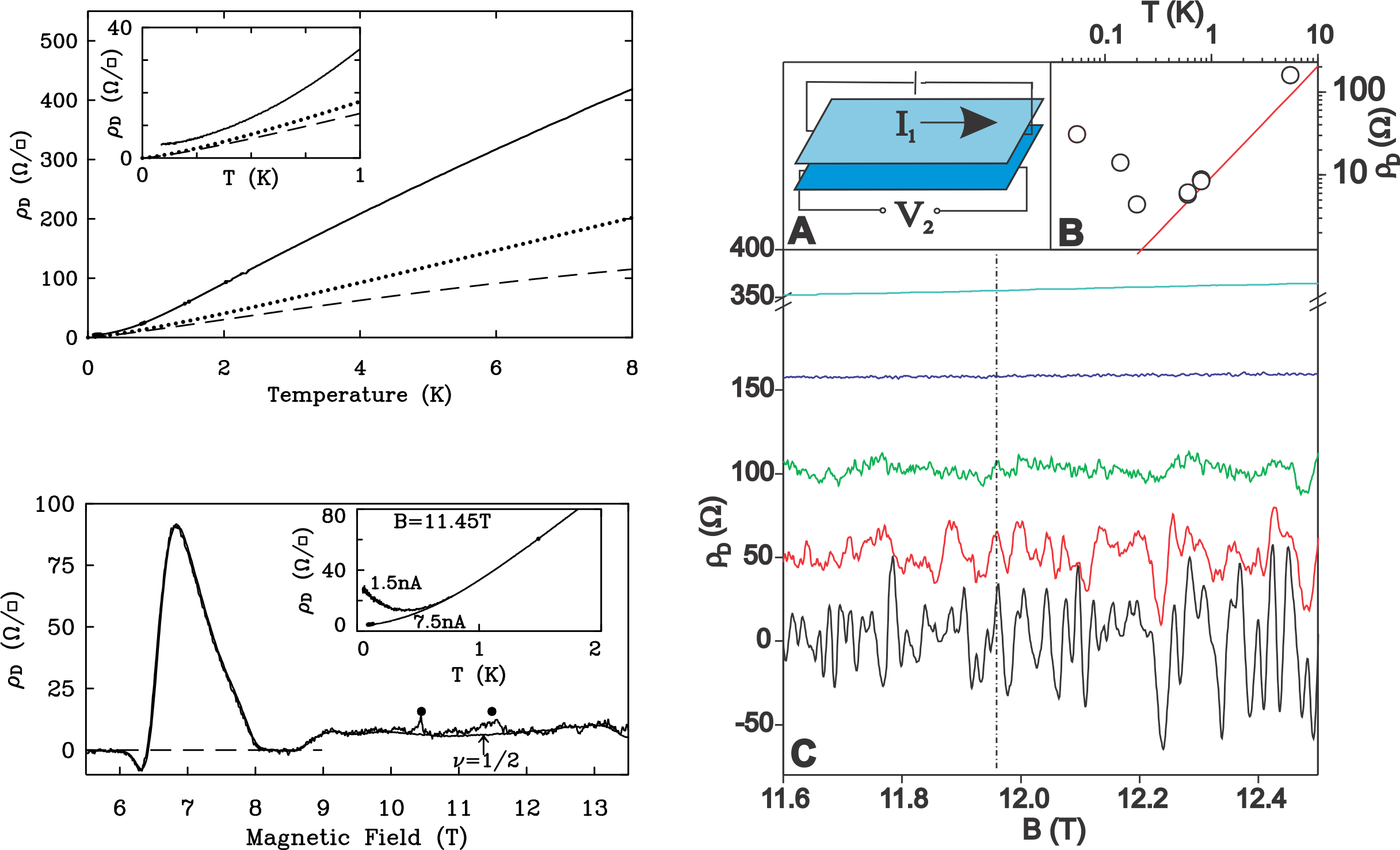}
\end{center}
\caption{(Color online) Coulomb drag measurements at ${\nu=1/2}$. Left
  panel: the top plot shows the experimental ${\rho_D(T)}$ (solid
  line) compared to the theory of \textcite{uss}; the bottom plot
  shows the field dependence. The inset shows ${\rho_D}$ at
  ${B=11.45}$T for two values of the driving current. [Reproduced from
    \textcite{lil}.] Right panel: (A) circuit schematic; (B)
  ${\rho_D(T)}$ (dots) vs Eq.~(\ref{rdn12}); (C) ${\rho_D(B)}$ for
  different temperatures, ${T=0.05-5.6}$K. The vertical line
  corresponds to the $B$ field at which the points plotted in panel
  (B) were measured. [Reproduced from \textcite{prs}.]}
\label{data_cf}
\end{figure}

The role of disorder in the drag effect in the $\nu=1/2$ state was
considered in \textcite{uss}. In the diffusive regime, the
polarization operator is given by the standard form (\ref{prd}) and
hence the drag resistivity is given by Eq.~(\ref{rdd}), albeit with a
different diffusion coefficient than the same system would have in the
absence of magnetic field. The result is much larger than at $B=0$.
In the diffusive regime, this follows from the observation that the
longitudinal conductivity (or the diffusive constant, which encodes
all microscopic details) at ${\nu=1/2}$ is much smaller than at
${B=0}$.

Although the above theory is qualitatively similar to the experimental
observations \cite{lil} (e.g. drag at $\nu=1/2$ is much larger that at
$B=0$; the temperature dependence in clean samples is subquadratic),
theoretical calculations significantly underestimate the overall value
of $\rho_D$ as compared to the experiment of
\textcite{lil}. \textcite{ya3} suggested, that the reason for the
discrepancy is that the interlayer separation in the samples of
\textcite{lil} was close to the critical value, where the system forms
an incompressible interlayer state (for a detailed discussion of
correlated states see Sec.~\ref{qhe}). An alternative suggestion by
\textcite{ust} attributes the unexplained features of the experiment
(including the extrapolated nonvanishing drag at ${T=0}$) to pairing
fluctuations of composite fermions. The two scenarios can be
distinguished by measuring Hall drag, which vanishes in the latter
theory. A later experiment \cite{prs} reported both the magnitude and
temperature dependence of Eq.~(\ref{rdn12}) to be in good agreement
with the measured data.


\section{Mesoscopic fluctuations of Coulomb drag}
\label{mf}


Universal conductance fluctuations \cite{AltLeeWeb1991} is a quantum
interference effect which is a manifestation of the wave nature of
electrons. As the same electrons are responsible for Coulomb drag, it
is natural to expect that the drag resistivity also exhibits
mesoscopic fluctuations. The drag fluctuations were first studied
theoretically in \textcite{me0} and \textcite{mor1,mfj} and then
observed experimentally \cite{mes,tu1,sav,prs}.

In a disordered system, it is impossible to track each individual
impurity and one uses a statistical approach. Impurities are described
by a distribution function and each physical quantity is treated as
being random. Observables correspond to average values of the random
physical quantities with respect to the distribution of impurities. If
a system is large enough, it can be viewed as a combination of smaller
parts, which become statistically independent if they are separated by
distances larger than any relaxation length. Then instead of averaging
over a statistical ensemble, one can average over the volume of one
large system.

In the problem of electronic transport, averaging over the system
volume can be understood as averaging over all possible paths that an
electron can take moving between points A and B
\cite{aar,AltLeeWeb1991,aag}. Such paths can cover all of the system
volume and thus experience all possible local impurity configurations,
making average over the system equivalent to ensemble averaging.

Consider two paths between the points A and B. The total transmission
probability is determined by the absolute value of the sum of the
corresponding quantum amplitudes \cite{AltLeeWeb1991,all}:
\[
W = |A_1+A_2|^2 = |A_1|^2 + |A_2|^2 + 2|A_1||A_2| 
\cos(\varphi_1-\varphi_2),
\]
where $\varphi_{1(2)}$ are the quantum-mechanical phases that an
electron accumulates along the paths. Typically, the phases
$\varphi_{1(2)}$ are random (or incoherent). As a result, the
interference term vanishes upon averaging over all possible paths (or
impurity configurations), leading to the semiclassical sum of
transition probabilities
\footnote{In special cases of coherent paths (for instance,
  time-reversed paths) the phase difference is exactly zero. Then the
  interference term does not vanish and leads to quantum corrections
  to semiclassical transport properties, such as the weak localization
  correction \cite{aar,AltLeeWeb1991}.}
\[
\left\langle\cos(\varphi_1-\varphi_2)\right\rangle = 0 
\,\,\Rightarrow\,\,
W = W_1 + W_2, \, W_{1(2)} = |A_{1(2)}|^2.
\]

Random quantities can be characterized not only by their average
value, but also by higher moments of their statistical distribution,
which may be sensitive to the interference term even if the phases
$\varphi_{1(2)}$ are still random. Indeed, fluctuations of the
transition probability
\[
\langle \left[W-\langle W \rangle \right]^2\rangle =
4 W_1 W_2 \langle \cos^2(\varphi_1-\varphi_2)\rangle
= 2 W_1 W_2.
\]
are completely determined by the interference term.

Fluctuations of the transmission probability result in fluctuations of
transport coefficients. The effect of such fluctuations can be
observed only in small enough samples \cite{AltLeeWeb1991}. Indeed, in
order justify the concept of the phase associated with a given
electronic path, the length of the path should be less than a typical
inelastic relaxation length $L_\varphi$, otherwise coherence would be
lost before the electron reaches point B \cite{anabr}. At the same
time, the path length should be larger than the mean free path in the
system (otherwise electron motion along the path would be
deterministic). Therefore, typical paths (and hence, the sample sizes)
should be characterized by intermediate lengths $L$
\[
\ell \ll L \ll L_\varphi.
\]
Fluctuations observed at such length scales are known as
``mesoscopic fluctuations'' \cite{AltLeeWeb1991}.


\subsection{Drag fluctuations in conventional diffusive samples}
\label{mf0}


Mesoscopic fluctuations of the usual conductance \cite{AltLeeWeb1991}
are known as the ``universal conductance fluctuations'' (UCF). The
universality is manifest when ${T\ll{E}_T}$, where $E_T$ is the
Thouless energy of the sample (i.e. in small samples or at low
temperatures; in the diffusive regime, $E_T = D/L^2=g/(2\pi\nu L^2)$,
with $g$ being the dimensionless conductance). In this case, the
fluctuations are characterized by the universal value
\begin{equation}
\label{ucf}
\delta\sigma \simeq \frac{e^2}{\hbar}, \quad
\left\langle\delta G^2\right\rangle \approx \frac{e^4}{h^2}, 
\quad
\sqrt{\frac{\left\langle\delta G^2\right\rangle}
{\left\langle G \right\rangle^2}}\simeq\frac{1}{g(L)},
\end{equation}
where ${G=ge^2/h}$ is the conductance of the system. The latter
equality emphasizes the fact that the dimensionless conductance is a
function of the system size. 

In larger samples, ${\langle\delta{G}^2\rangle}$ is a function of
temperature and the sample size. Arguments leading to Eq.~(\ref{ucf})
are valid only for coherent samples \cite{AltLeeWeb1991}. At larger
length scales, ${L\gg{L}_\varphi}$ the coherence is lost, and the
disorder averaging should be performed by dividing the sample into
patches of the size $L_\varphi$. Individual self-coherent patches
(\ref{ucf}) can be combined as a network of random conductors. This
yields (in dimension $d$)  
\begin{equation}
\label{rn1}
\left\langle\delta G^2(L)\right\rangle \simeq 
\left\langle\delta G^2(L_\varphi)\right\rangle
\left(L_\varphi/L\right)^d.
\end{equation}

The patches of the size $L_\varphi$ remain self-coherent as long as
${T\ll{E}_T(L_\varphi)}$. At higher temperatures, thermal averaging
should be performed up to energies of order $T$, suppressing the
conductance fluctuations
\[
\left\langle\delta G^2(L_\varphi)\right\rangle \simeq 
(e^2/h)^2 E_T(L_\varphi)/T.
\]
The conductance fluctuations of the sample become
\begin{equation}
\label{ucftl}
\left\langle\delta G^2[L; T\!>\!E_T(L_\varphi)]\right\rangle\!\simeq\!
(e^2/h)^2 (L_\varphi/L)^d\hbar/(T\tau_\varphi),
\end{equation}
where ${\tau_\varphi=\hbar{E}_T^{-1}(L_\varphi)}$ is the dephasing
time \cite{akl}.

The fluctuations (\ref{ucftl}) are only observable in mesoscopic
samples. Assuming the samples to be ``metallic'', ${g\gg{1}}$, the UCF
(\ref{ucf}) yield only a small correction to the average value of
conductance. For example, in the experiment of \textcite{mes} the
single-layer resistance fluctuates by about $200$m$\Omega$ around the
average of about $500\Omega$.

Now, we apply the above arguments to Coulomb drag \cite{me0}. The drag
conductivity depends on (i) the phase space available to electron-hole
excitations; (ii) matrix elements of the interlayer interaction; and
(iii) electron-hole asymmetry, expressed through the energy dependence
of the density of states (or, the density dependence of the
single-layer Drude conductivity). This can be schematically summarized
by
\begin{equation}
\label{sdgr}
\sigma_D \simeq \frac{e^2}{\hbar} 
\left(\frac{\partial}{\partial\mu}\ln g\right)^2
\times
\begin{pmatrix}
{\rm phase} \cr
{\rm volume}
\end{pmatrix}
\times
\begin{pmatrix}
{\rm matrix} \cr
{\rm element}
\end{pmatrix}.
\end{equation}
The average drag conductivity [cf. Eqs.~(\ref{rdd}) and (\ref{sdd})]
can then be understood (up to the logarithmic factor) by estimating
the phase volume by $T^2$, the matrix element by $(\varkappa d)^{-2}$
(coming from static screening), and the factor of the electron-hole
asymmetry by $E_F^{-2}$.

Fluctuations of the drag conductivity can also be estimated with the
help of Eq.~(\ref{sdgr}). Consider first the lowest temperatures
${T\ll{E_T}}$, where the sample is effectively zero-dimensional
(0D). The phase space is then only limited by temperature, yielding
the usual factor of $T^2$. The factor of the electron-hole asymmetry
in Eq.~(\ref{sdgr}) is a random quantity with the typical value
${\sim{E}_T^{-2}}$, since the Thouless energy is the typical scale of
mesoscopic effects. Interaction matrix elements in 0D are independent
of energy \cite{aag}; fluctuations are determined by off-diagonal
elements that contain a small factor of $g^{-2}$. As a result, one
finds the variance of the drag conductivity that strongly exceeds the
average
\begin{equation}
\label{ds0d}
\delta\sigma_D \sim \frac{e^2}{\hbar}\frac{T^2}{g^2E_T^2}, \quad
\frac{\displaystyle\sqrt{\left\langle \delta\sigma_D^2 \right\rangle}}
{\left\langle\sigma_D\right\rangle}
\simeq
\frac{E_F^2}{g^2E_T^2} \simeq \frac{L^4}{\ell^4} \gg 1.
\end{equation}

The quadratic temperature dependence of the variance of the drag
conductivity \cite{me0} for mesoscopic samples (${L\ll{L}_\varphi}$,
${T\ll{E_T}}$) was also obtained in the context of quantum circuits
[see Sec.~\ref{nano} and \textcite{le2}] and within the random
matrix theory \cite{mor1,mor2}.

In order to extend the 0D argument to larger samples,
${L\gg{L}_\varphi}$, we again divide the system into patches of the
size $L_\varphi$. Since the patches are largely uncorrelated (due to
the loss of phase coherence), they can be combined as a network of
random conductors, see Eq.~(\ref{rn1}). Each patch can be analyzed
similarly to the 0D case. However, now the interaction matrix elements
become energy-dependent on the scales larger than $E_T$, decreasing
with the transmitted energy $\omega$ as ${|M|^2\sim\omega^{-2}}$.
Thus the energy transfer is limited by the Thouless energy of the
patch ${\omega\sim E_T(L_\varphi)=\tau_\varphi^{-1}}$, rather than
temperature. As a result, the phase space is limited by
${T\tau_\varphi^{-1}}$, rather than the usual $T^2$. The fluctuations
of the density of states (which determine the factor of electron-hole
asymmetry) should now be calculated on the scale of temperature rather
than the Thouless energy. This suppresses the fluctuations in each
layer by the factor of ${\sqrt{E_T(L_\varphi)/T}}$. Combining the
above estimates, we find
\begin{equation}
\label{ds2d1}
\delta\sigma_D(L_\varphi) \sim \frac{e^2}{\hbar}
\frac{T\tau_\varphi^{-1}}{g^2 E_T^2(L_\varphi)}
\frac{E_T(L_\varphi)}{T} \sim \frac{e^2}{\hbar g^2},
\end{equation}
which is $T$-independent, in contrast to the 0D result
(\ref{ds0d}).

\begin{figure}
\begin{center}
\includegraphics[width=0.97\linewidth]{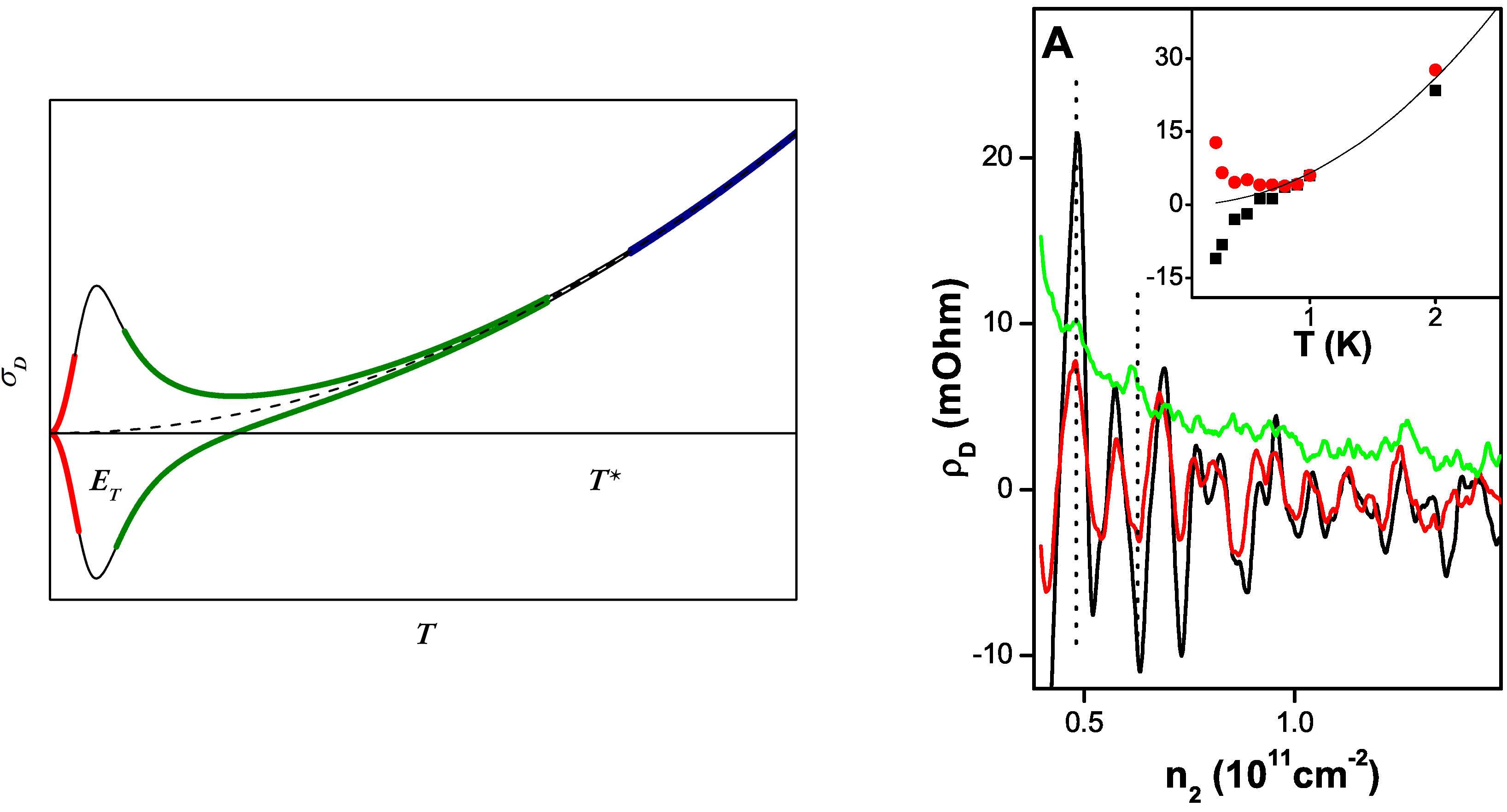}
\end{center}
\caption{(Color online) Left: Qualitative picture of the typical
  measured drag signal. At high enough temperatures ${T>T^*}$ the
  average drag conductivity (\ref{sdd}) is representative,
  ${\sigma_D\propto{T}^2}$, with positive coefficient. Below $T^*$,
  fluctuations dominate and the sign of the measured signal becomes
  random, i.e. dependent on a particular configuration of
  disorder. For ${T<T^*}$ the temperature dependence weakens to
  ${\sigma_D\propto1/\sqrt{T}}$. At very low (most likely,
  experimentally inaccessible) temperatures ${T<E_T\ll{T}^*}$, the
  quadratic temperature dependence is restored, but with a random
  coefficient, as fluctuations in the effectively 0D system are much
  stronger than the average, see Eq.~(\ref{ds0d}).  Right: Measured
  drag resistance as a function of carrier density in the passive
  layer for ${T=1,0.4,0.24}$K. Inset: the temperature dependence of
  the same data for the two values of $n_2$ denoted by vertical dotted
  lines in the main plot. The line indicates the $T^2$
  dependence. [From \textcite{mes}. Reprinted with permission from
    AAAS.]}
\label{ds}
\end{figure}

Final the Coulomb drag fluctuations in 2D samples can be estimated by
combining Eqs.~(\ref{rn1}) and (\ref{ds2d1}):
\begin{eqnarray}
\label{ds2d2}
\left\langle\delta\sigma_D^2(L)\right\rangle \sim 
\frac{e^4}{\hbar^2g^4} \frac{L_\varphi^2}{L^2}
\sim \frac{e^4}{\hbar^2g^4} E_T(L)\tau_\varphi\propto \frac{1}{T}.
\end{eqnarray}
The temperature dependence of the fluctuations (\ref{ds2d2}) is
contained in the dephasing time ${\tau_\varphi\sim{g}/T}$ \cite{aar}.
At high enough temperatures, ${T\gg{T}^*}$, the fluctuations are small
[the average value of $\sigma_D$ (\ref{sdd}) is representative], but
for ${T\ll{T}^*}$ fluctuations dominate, see Fig.~\ref{ds}. The
crossover temperature $T^*$ can be found by setting the relative
fluctuation to unity
\[
T^* \sim E_F(g^2nL^2)^{-1/5}.
\]
The fluctuation-dominated regime is characterized by typical values of
$\sigma_D$ determined by Eq.~(\ref{ds2d2}) rather than the average.
In particular, the temperature dependence of the measured drag
conductivity in this regime appears almost saturating as
${\sigma_D\propto{1}/\sqrt{T}}$. The value of the prefactor in this
expression is sample-dependent and has a random sign. If temperature
is decreased further, then eventually (although probably only in
theory) one may reach the regime where ${T<E_T}$. Then the sample will
become effectively zero-dimensional and the quadratic temperature
dependence ${\sigma_D\propto{T}^2}$ will be restored. In this regime
of lowest temperatures, fluctuations greatly exceed the average [see
  Eq.~(\ref{ds0d})] and therefore the coefficient in the quadratic
temperature dependence will be random (with random sign). The
temperature dependence of a typical drag signal is sketched in the
left panel of Fig.~\ref{ds} (cf. the inset in the right panel of
Fig.~\ref{ds}; see also the inset in the lower left panel of
Fig.~\ref{data_cf}).

The above qualitative picture is in full agreement with microscopic
calculations \cite{me0}. The average square of the drag conductivity
has the form
\begin{subequations}
\label{ds2d}
\begin{equation}
\left\langle \sigma_D^{\alpha\beta}\sigma_D^{\alpha'\beta'} \right\rangle
=
\left(\delta^{\alpha\alpha'}\delta^{\beta\beta'} 
+ \delta^{\alpha\beta'}\delta^{\alpha'\beta}\right)
\left\langle \sigma_D^2 \right\rangle,
\end{equation}
\begin{equation}
\left\langle \sigma_D^2 \right\rangle =\frac{e^4}{\hbar^2}
\frac{\gamma}{18\pi^3}
\left(\frac{32\ln 2 -14}{3} \right)
\frac{E_T\tau_\varphi \ln\varkappa d}{g^4(\varkappa d)^3},
\end{equation}
\end{subequations}
where ${\gamma=1.0086}$. Comparing Eq.~(\ref{ds2d}) with the average
drag conductivity in the diffusive regime (\ref{sdd}), one finds the
cross-over temperature ${T^*=E_F(16\pi g^2 n L^2)^{-1/5}}$.

For heterostructures used in \textcite{ex2,lil}, the value of $T^*$
can be estimated as ${T^*\approx{0.2}}$K, which is below the
temperature range of these experiments. Hence, the average drag
coefficients (\ref{rdd}) and (\ref{sdd}) were sufficient to account
for the observed effect with no trace of the random sign predicted by
Eqs.~(\ref{ds2d}).

The result (\ref{ds2d}) is valid for homogeneous 2D diffusive samples
in the absence of magnetic field. The randomness (i.e. the sample to
sample variation) of the sign of the effect should be contrasted with
the deterministic sign change of the drag resistivity suggested for
bilayer systems with in-plane periodic potential modulation
\cite{alk}. Drag signals of both signs have been observed in
vertically integrated 1D quantum wires \cite{lar}. While the observed
effect has been argued \cite{but} to have a mesoscopic origin
\cite{mor1} dominated by charge fluctuations \cite{le2,sab}, the data
appear to be not random, but reproducible. Very similar data were
obtained in the subsequent experiment \cite{lar2} and interpreted with
the help of the Luttinger Liquid theory \cite{pus} (see
Sec.~\ref{nw}).

\begin{figure}
\begin{center}
\includegraphics[width=0.45\linewidth]{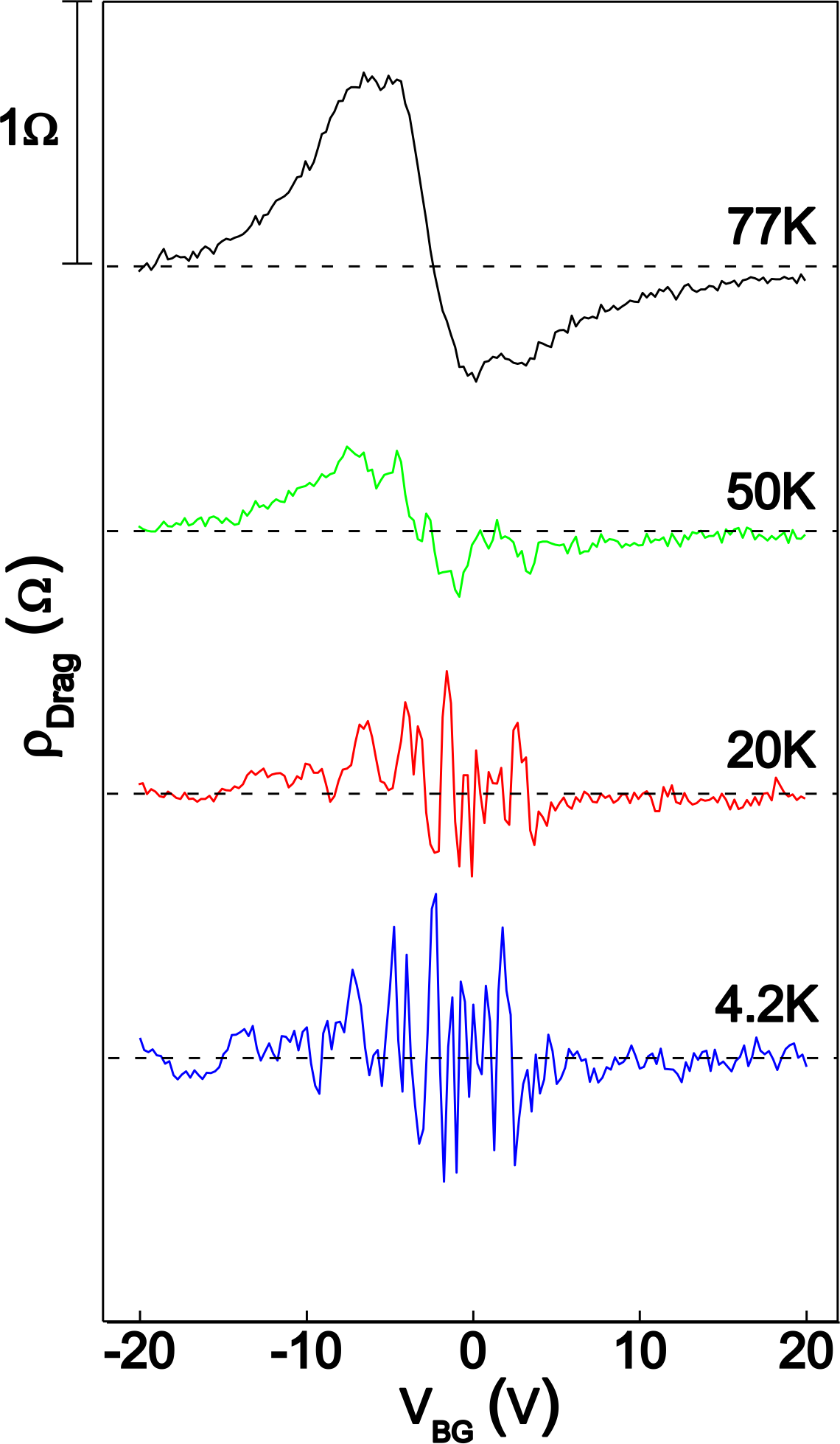}
\end{center}
\caption{(Color online) Mesoscopic fluctuations of Coulomb drag in
  graphene. At low temperatures the fluctuations fully obscure the
  average drag. The curves are shifted for clarity; the horizontal
  dashed lines indicate $0\Omega$ for each curve. [Reproduced from
    \textcite{tu1}.]}
\label{fig_fluc_gr}
\end{figure}

More recently, drag fluctuations were observed in diffusive
graphene-based double-layer samples \cite{tu1,tu2}, see
Fig.~\ref{fig_fluc_gr}. The temperature dependence
${\delta\sigma_D\propto{T}^{-1/2}}$ [following from Eq.~(\ref{ds2d})
  and the assumption that the main phase-breaking mechanism in the
  device is electron-electron scattering \cite{aar}] appears to be in
agreement with the experimental data. Other aspects of this
experiments are specific to graphene. The fluctuations appear to be
more pronounced in the vicinity of the charge neutrality
point. However, at the time of writing, a theory of drag fluctuations
in graphene has not been developed. There is also no explanation for
the most puzzling feature of the data reported in \textcite{tu2}
showing an apparent violation of Onsager reciprocity as the drag
fluctuations depend only on the charge density in the passive layer
and not in the active layer.

Closing this Section, we note that strong fluctuations of Coulomb drag
ultimately follow from strong fluctuations of the nonlinear
susceptibility. The fact that the fluctuations of the drag resistivity
exceed the average is related to the overall smallness of the drag
effect due to electron-hole symmetry. A related phenomenon is the
fluctuations of the electro-acoustic current\cref{fnm}, which is also
determined by the same nonlinear susceptibility.


\subsection{Giant fluctuations of Coulomb drag}


The predictions of the fluctuation theory \cite{me0} were put to the
test in the dedicated experiment \cite{mes}. Both the UCF and drag
fluctuations were measured in the same structure. The observed UCF
have shown the usual behavior \cite{AltLeeWeb1991}. A direct
comparison of the correlation fields for the UCF and drag fluctuations
confirmed that both effects depend on the same coherence length
$L_\varphi$ \cite{aag} and have the same quantum origin.
Surprisingly, the observed giant drag fluctuations \cite{mes} greatly
exceeded the original prediction \cite{me0}. This discrepancy was
attributed to the fact that the experiment was performed in the
ballistic regime \cite{zna,zn2}.

Let us remind the reader (see Sec.~\ref{drag0}) that the in drag
measurements difference between ``diffusive'' and ``ballistic''
samples is in the relation between the mean-free path $\ell$ and
interlayer separation $d$. The latter sets the upper limit for the
interlayer momentum transfer due to the exponential decay of the
Coulomb interaction (\ref{coulomb}). Thus, if the mean free path is
small ${\ell\ll{d}}$, then ${q\ll{d}^{-1}\ll\ell^{-1}}$ and the effect
is dominated by the diffusive motion of charge carriers. In
``cleaner'' samples with $\ell\gg{d}$, both small ${q\ll\ell^{-1}}$
and large ${\ell^{-1}\ll{q}\ll{d}^{-1}}$ momentum transfers are
possible. The conventional statement, that in such samples Coulomb
drag is dominated by ballistically moving carriers \cite{kor}, follows
from observing that processes with large momentum transfers yield a
much larger drag resistivity Eq.~(\ref{rdb}) compared with the
diffusive result (\ref{rdd}), see Eq.~(\ref{baldif}).

Coherence properties of electrons are also sensitive to the nature of
their motion. The dephasing time $\tau_\varphi$ is a manifestation of
inelastic electron-electron scattering \cite{aar,aag}. The
conventional theory of interaction effects in electronic systems
\cite{aar} yields the following estimate for the dephasing time in
diffusive systems
\begin{subequations}
\label{tauf}
\begin{equation}
\label{tfd}
\tau_\varphi^{-1}(T\tau\ll 1) \sim (T\ln g)/g.
\end{equation}
At higher temperatures, transport is dominated by processes with one
or few successive impurity scatterings. In this ``ballistic'' regime
\cite{zna}, the dephasing time exhibits somewhat stronger temperature
dependence \cite{zn2}
\begin{equation}
\label{tfb}
\tau_\varphi^{-1}(T\tau\gg 1) \sim (T^2/E_F) \ln (2E_F/T).
\end{equation}
In Eqs.~(\ref{tfd}) and (\ref{tfb}) the parameter distinguishing the
diffusive and ballistic regimes is ${T\tau}$ which is independent of
the inter-layer separation. This is to be expected since the theory
leading to Eqs.~(\ref{tfd}) and (\ref{tfb}) was devoted to
two-dimensional systems and not bilayers.

The effect of the external magnetic field on the single-layer
conductance fluctuations analyzed in \textcite{mes} demonstrates the
expected crossover between the ballistic and diffusive results :
\begin{equation}
\tau_\varphi^{-1} \propto
\begin{cases}
T, &  T\tau\lesssim 1, \\[1pt]
T^2, & T\tau\gtrsim 1.
\end{cases}
\end{equation}
\end{subequations}
The same sample where Coulomb drag is dominated by the ballistic
motion of electrons with large interlayer momentum transfers,
${\ell^{-1}\ll{q}\ll{d}^{-1}}$, may exhibit {\it both} the diffusive
and ballistic behavior of single-layer transport properties, e.g. of
the dephasing time (\ref{tauf}).

Similar crossover was observed also in the drag fluctuations that
exhibited strikingly different temperature dependence at large and
small $T\tau$ \cite{mes}:
\begin{eqnarray}
\label{dsbex}
\left\langle\delta\sigma_D^2\right\rangle \propto
\begin{cases}
T^{-1}, & T\tau\lesssim 1, \\[1pt]
T^{-4}, & T\tau\gtrsim 1.
\end{cases}
\end{eqnarray}
The crossover temperature in Eq.~(\ref{dsbex}) was found to be about
the same as in Eq.~(\ref{tauf}). This coincidence raised the question
of whether the large magnitude of the observed drag fluctuations and
their unexpected temperature dependence (\ref{dsbex}) had the same
origin that would involve large momentum transfers
${\ell^{-1}\ll{q}\ll{d}^{-1}}$ [given that the small momentum transfers
    lead to Eq.~(\ref{ds2d})].

Interaction processes characterized by large momentum transfers
${q\gg\ell^{-1}}$ involve two electrons separated by a distance that
is smaller than the average impurity separation. Therefore the effect
should be determined by {\it local} electron properties. Local
properties, such as the local density of states, are known to exhibit
mesoscopic fluctuations stronger than those of the global properties
(responsible for drag fluctuations in the diffusive regime). In
particular, fluctuations of the local density of states are given by
\cite{ler}
\begin{eqnarray}
\label{ldos}
\delta\nu^2 \sim \frac{\nu^2}{g} 
\ln\frac{{\rm max}(L_T, L_\varphi)}{\ell},
\end{eqnarray}
where $L_T=\sqrt{D/T}$ is the thermal length.

\begin{figure}
\begin{center}
\includegraphics[width=0.5\linewidth]{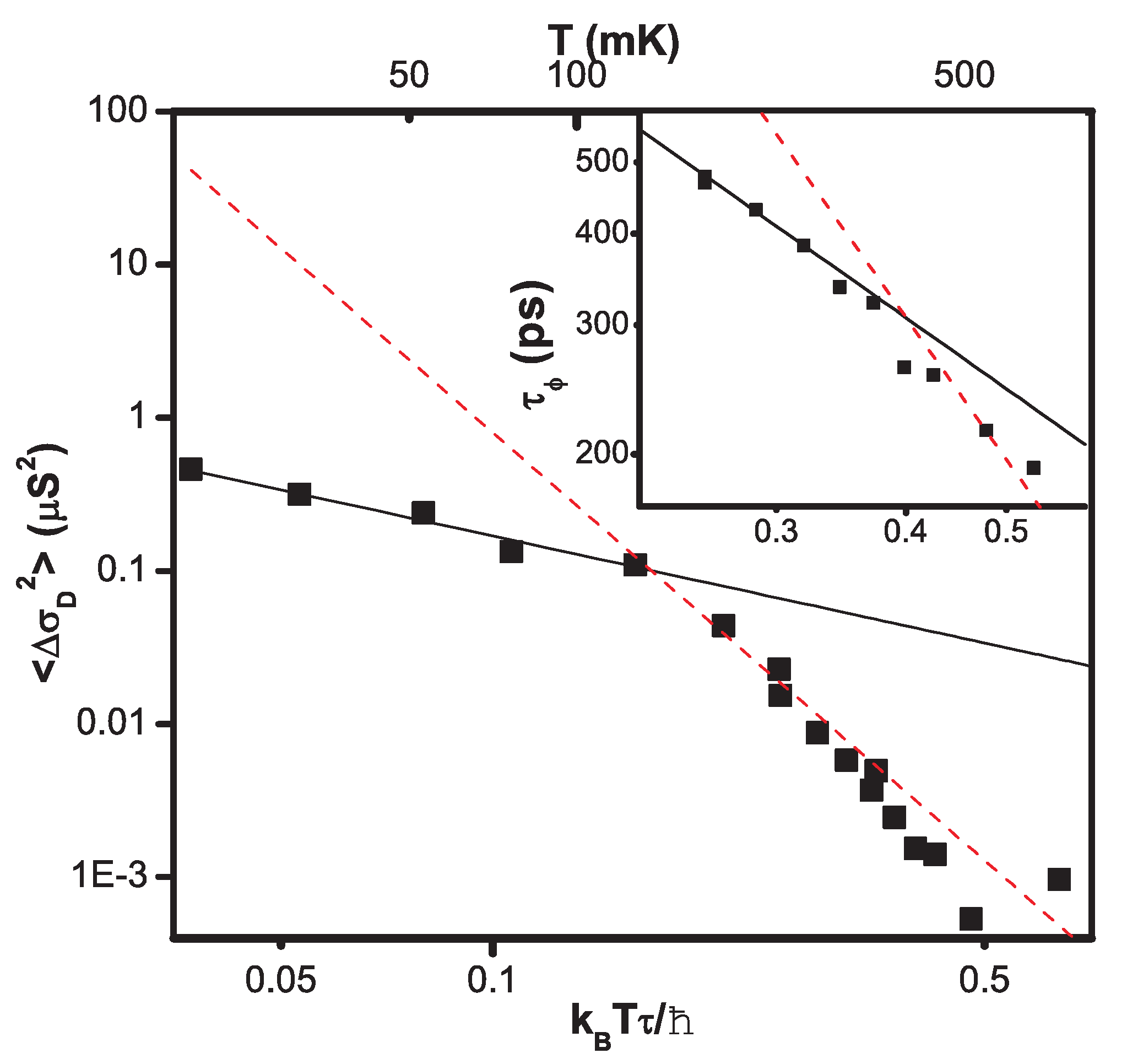}
\end{center}
\caption{Drag fluctuations in ballistic samples. The lines represent
  the asymptotic power laws, see Eq.~(\ref{dsbex}). The inset shows
  the measured dephasing time. The lines in the inset represent the
  power laws from Eq.~(\ref{tauf}).  [Adapted from
    \textcite{mes}. Reprinted with permission from AAAS.]}
\label{fig_dsb}
\end{figure}

Contribution of processes with large momentum transfers to drag
fluctuations can be estimated using Eqs.~(\ref{sdgr}) and
(\ref{rn1}). Electron-electron scattering can be described with the
help of ``ballistic'' expressions discussed in Sec.~\ref{kt}. As small
angle scattering plays the dominant role, the matrix element of the
interlayer interaction is proportional to the ratio of the mean-free
path to the interlayer separation
${|M|^2\sim{g}^{-2}\ell^2/(\varkappa^2d^4)}$. As this interaction is
static, the phase space is only limited by $T$. Assuming that the
fluctuations of the nonlinear susceptibility are dominated by
fluctuations of the local density of states (\ref{ldos}), one finds
(up to a logarithmic factor)
\begin{equation}
\label{dsb1}
\delta\sigma_D(L_\varphi) \sim \frac{e^2}{\hbar} \frac{T}{gE_T(L_\varphi)}
\frac{\ell^2}{\varkappa^2 d^4},
\end{equation}
where the thermal smearing was taken into account similarly to
Eq.~(\ref{ds2d1}). This leads to the estimate for the drag
fluctuations in the whole sample \cite{mes}
\begin{equation}
\label{dsb2}
\left\langle\delta\sigma_D^2\right\rangle \sim \frac{e^4}{\hbar^2}
\frac{\ell^4}{g^2\varkappa^4 d^8}\frac{T^2}{E_T^2(L_\varphi)}
\frac{L_\varphi^2}{L^2}
\propto T^2\tau_\varphi^3.
\end{equation}

The result (\ref{dsb2}) contains two falsifiable predictions: (i) the
magnitude and (ii) the temperature dependence of the drag
fluctuations. In comparison with Eq.~(\ref{ds2d}), the prefactor in
Eq.~(\ref{dsb1}) contains the large factor ${\ell^4/d^4}$ and
moreover, $g^2$ instead of $g^4$ in the denominator. Consequently, the
drag fluctuations (\ref{dsb2}) are much stronger than the diffusive
prediction. At the same time, using the temperature dependence of the
dephasing time (\ref{tauf}), one immediately recovers the measured
temperature dependence of the drag coefficient (\ref{dsbex}). The
crossover between the two temperature regimes in Eqs.~(\ref{dsbex})
and (\ref{dsb2}) is illustrated in Fig.~\ref{fig_dsb}.


\subsection{Drag fluctuations at the half-filled Landau level}


Mesoscopic fluctuations of Coulomb drag of composite fermions were
studied theoretically in \cite{me2} and experimentally in
\cite{prs}. Despite the significant increase in the magnitude of drag
of composite fermions relative to that of normal electrons
\cite{lil,jo1,zel,mur} the fluctuations of the drag resistivity can
still exceed the average, resulting in an alternating sign of the
measured drag resistivity.

Qualitatively, one can estimate the fluctuation effects using
Eq.~(\ref{dia}). Similarly to the ${B=0}$ case, drag fluctuations stem
from the fluctuations of the nonlinear susceptibility. In the
diffusive regime, ${\langle\bs{\Gamma}\rangle}$ is given by
Eq.~(\ref{gd}) with the polarization operator having the standard form
(\ref{prd}), although with a different diffusion constant
\cite{uss}. In contrast to the ${B=0}$ case, the $\nu=1/2$ state is
characterized by a large Hall conductivity. This leads to the
nonlinear susceptibility being approximately orthogonal to the
transferred momentum $\bs{q}$ [unlike Eq.~(\ref{gdr})].

Fluctuations of $\bs{\Gamma}$ (and thus of the drag resistivity)
result from mesoscopic fluctuations of
${\partial\sigma/\partial{n}}$. Other parameters, such as the
compressibility and the diffusion constant can be taken at their
average values (their fluctuations are much smaller than the
averages). To estimate fluctuations of ${\partial\sigma/\partial{n}}$,
one can express the conductivity in terms of the response functions of
composite fermions using Eq.~(\ref{rcf}). On average, the conductivity
matrix of composite fermions is diagonal. Assuming the large
dimensionless conductance of composite fermions, $g_{cf}\gg 1$, the
electronic longitudinal conductivity is inversely proportional to
$g_{cf}$, meaning {\it smallness} of the electronic dimensionless
conductance
\begin{eqnarray}
\label{gcf}
g \approx 1/(4g_{cf}) \ll 1.
\end{eqnarray}
This is the reason one needs to perform calculations in the
composite-fermion basis: the ${B=0}$ theory of Sec.~\ref{mf0} is
justified by the small parameter $1/g$.

Adapting the ${B=0}$ theory to the case of composite fermions,
\textcite{me2} found the fluctuations of the nonlinear
susceptibility (\ref{gd}) of a coherent sample of size $L$ in the
$\nu=1/2$ state to be large
\begin{equation}
\label{dgn12}
\delta\Gamma\sim iq \frac{e}{h} \frac{L^2}{g_{cf}^2} 
{\rm Im}\Pi^R,
\qquad
\frac{\langle\delta\Gamma^2\rangle}
{\langle\Gamma^\gamma\rangle^2}\sim 
\frac{k_F^4 L^4}{g_{cf}^4} \gg 1,
\end{equation}
similarly to Eq.~(\ref{ds0d}). This is already an observable
conclusion: in a fully coherent sample in the diffusive regime,
fluctuations of the acoustoelectric current (determined by the same
nonlinear susceptibility) are much larger that its average. The
result (\ref{dgn12}) is justified as long as the thermal
${L_T^{cf}\equiv\sqrt{\hbar{D}^{cf}/T}}$ and phase breaking
${L_\varphi^{cf}}$ length scales of composite fermions are much larger
than $L$.

For larger samples, the global phase coherence is lost and one has to
employ the averaging procedure described in Sec.~\ref{mf0}. The system
can be divided into ${L^2/(L_\varphi^{cf})^2}$ self-coherent patches
of the size of the phase-breaking length of composite
fermions $L_\varphi^{cf}$. Summing up contributions of all patches
according to Eq.~(\ref{rn1}), one finds
\begin{widetext}
\begin{eqnarray}
\langle\rho_D^2\rangle =\frac{h^2}{e^4}
\frac{1}{g_{cf}^4(\kappa d)^2}
\left(\frac{L_\varphi^{cf}}{L}\right)^2
{\rm min}\left[1, \alpha_1
\left(\frac{g_{cf}^2 T\tau_\varphi^{cf}}{\kappa d \hbar}\right)^2
\right]
{\rm min}\left[\alpha_3, \alpha_2
\left( T\tau_\varphi^{cf}/\hbar\right)^2
\right],
\label{fln12}
\end{eqnarray}
\end{widetext}
where ${\alpha_3\approx0.2(32/9\pi)=0.23}$ and the coefficients
$\alpha_{1,2}$ are of order unity \cite{me2}.

\begin{figure}
\begin{center}
\includegraphics[width=0.5\linewidth]{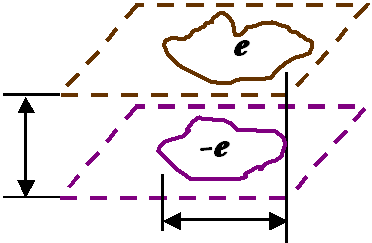}
\end{center}
\caption{(Color online) A cartoon illustration of the phase-breaking
  mechanism at ${\nu=1/2}$. A random flux in the system can be
  generated by charge density fluctuations with the opposite signs of
  excess local charges in the two layers.}
\label{cart}
\end{figure}

The magnitude of the mesoscopic fluctuations depends on the precise
source of phase breaking, but their temperature dependence is robust:
all generic models of phase breaking in two dimensions \cite{aar} lead
to ${1/\tau_{\varphi}\propto{T}}$ in the diffusive regime. In the
${\nu=1/2}$ state phase breaking comes from the {\em quasi-elastic}
scattering of composite fermions off the thermal quasi-static
fluctuations of the Chern-Simons magnetic field. This mechanism can be
illustrated using a cartoon shown in Fig.~\ref{cart}. Consider a
density fluctuation where the excess charges in the two layers have
opposite signs. Such a fluctuation is accompanied by a random flux
that interacts with the composite fermions leading to the loss of
coherence. The energy of this fluctuation is of order $T$. It can also
be estimated as the energy of a simple capacitor,
${2\pi{e}^2d/[\varepsilon(L_\varphi^{cf})^2]\simeq{T}}$,
where $\varepsilon$ is the bulk dielectric constant and
$L_\varphi^{cf}$ is the typical size of the density fluctuation with
the electron number of the order of unity creating the random flux of
approximately $\Phi_0$. As a result,
\begin{equation}
\label{tfcf}
1/\tau_\varphi^{cf} \simeq g_{cf}T/(\varkappa d).
\end{equation}

Substituting the above estimate into Eq.~(\ref{fln12}), one finds
(assuming ${g^{cf}\gg\varkappa{d}}$)
\[
\langle \rho_D^2\rangle \simeq \frac{h^2}{e^4}
\frac{2\pi e^2 d}{T\varepsilon L^2 g_{cf}^6}.
\]
Using realistic parameters \cite{lil} (i.e. $L \simeq 100{\mu}m$;
$d=300$\AA; $T=0.6K$; $R=3k\Omega/\Box$ leading to $g_{cf} \approx 8$;
and $\langle\rho_D\rangle = 15\Omega /\Box$), the magnitude of the
drag fluctuations can be estimated as $\delta \rho_D \approx 0.3
\Omega$. For lower temperatures and smaller samples, the theory
predicts stronger fluctuations (i.a. exceeding the average). Such
strong fluctuations were observed in \textcite{prs}, albeit with a
substantially larger magnitude that follows from the above estimate.

The dephasing time due to above mechanism of quasi-elastic scattering
of composite fermions on thermal fluctuations of the Chern-Simons
field appears to be shorter than the temperature scale
$T\tau_\varphi^{cf}\ll 1$. This does not create any additional
complication since most of the phase breaking results from scattering
off the Chern-Simons field fluctuations whose dynamics (with
characteristic frequency $T/g_{cf}$) is very slow compared to
$\tau_\varphi^{cf}$, but fast compared to the time of the
experiment. Field fluctuations which are static on the scale of the
experiment time affect the mesoscopic fluctuations only by affecting
$g_{cf}$. Field fluctuations that are faster than that scale make the
potential landscape seen by the composite fermions time dependent, and
lead to a suppression of the mesoscopic fluctuations by partial
ensemble averaging.

Consider the correlation function
\[\langle\rho_D(B)\rho_D(B+\delta B)\rangle
-\langle\rho_D(B)\rangle\langle\rho_D(B+\delta B)\rangle = 
F_1\left(\frac{\delta B}{B^*}\right),
\]
with the field $B$ near the ${\nu=1/2}$ value. An experimental study
of the decay of this correlation function is a way to measure
$L_\varphi^{cf}$: the characteristic magnetic field of the decay is
${B^*\sim\left(L_\varphi^{cf}\right)^{-2}\Phi_0}$. The decay of this
correlator as a function of density
\[
\langle\rho_D(n)\rho_D(n+\delta n)\rangle -
\langle\rho_D(n)\rangle\langle\rho_D(n+\delta n)\rangle =
F_2\left(\frac{\delta n}{n^*}\right).
\] 
also yields $L_\varphi^{cf}$: the characteristic density change $n^*$
at which it decays is expected to correspond to half of an electron in
a phase coherent region, i.e.
${n^*=\left(L_\varphi^{cf}\right)^{-2}/2}$.  This statement holds as
long as the composite fermion cyclotron radius is much larger than its
mean free path, i.e., for ${|\nu-1/2|<(2g_{cf})^{-1}}$.

\begin{figure}
\begin{center}
\includegraphics[width=0.6\linewidth]{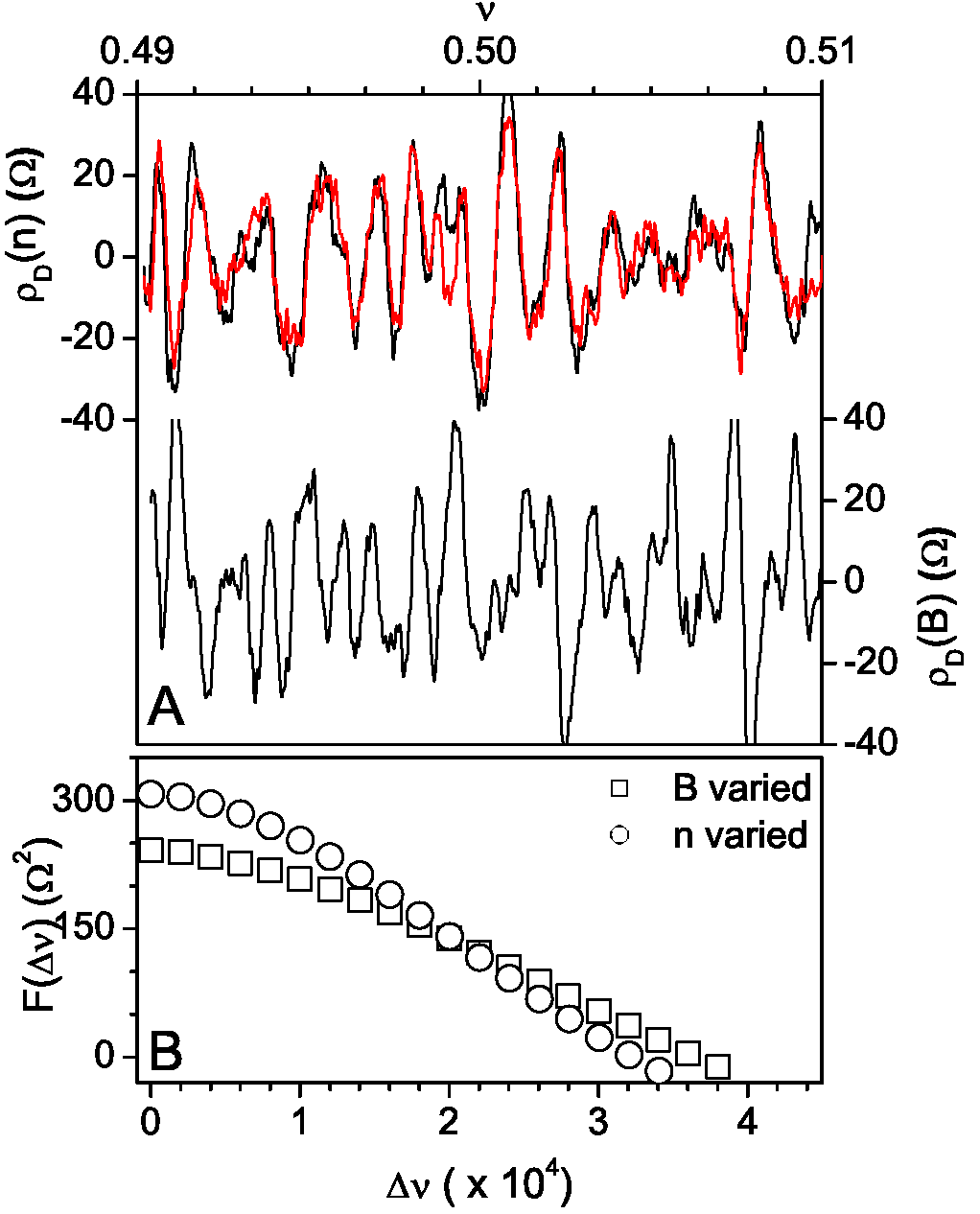}
\end{center}
\caption{(Color online) Mesoscopic fluctuations of Coulomb drag at
  ${\nu=1/2}$, ${T=50}$mK. Upper panel: comparison of ${\delta\rho_D}$
  as a function of the filling fraction $\nu$ obtained by varying
  either the carrier density or magnetic field. The red curve shows a
  different measurement run demonstrating the reproducibility of the
  fluctuations. Similarity of the periods of ${\rho_D(n)}$ and
  ${\rho_D(B)}$ is the proof of composite-fermion drag. Lower panel:
  autocorrelation function of the fluctuations shown in the upper
  panel. Squares represent ${\rho_D(B)}$ and circles -
  ${\rho_D(n)}$. [Reproduced from \textcite{prs}.]}
\label{data6}
\end{figure}

The ratio of the above characteristic field $B^*$ to the characteristic 
density $n^*$ yields two flux quanta
\begin{equation}
\label{fa}
\frac{B^*}{n^*} \simeq 2 \Phi_0.
\end{equation}
This should be contrasted to the zero field case, where
\[
B^*\rightarrow\frac{\Phi_0}{L_\varphi^2}, \quad
n^*\rightarrow\frac{k_F\ell}{L_\varphi^2}
\quad\Rightarrow\quad
\frac{B^*}{n^*} \simeq \frac{\Phi_0}{g}.
\]
At $B=0$ the electrons do not carry any attached flux. Therefore the
characteristic density $n^*$ corresponds to a change in the chemical
potential of order $\tau_\varphi^{-1}$. Consequently, observation of
the ratio (\ref{fa}) in a laboratory experiment serves as a
verification of the concept of the flux attachment and the fact that
charge carriers in the system are indeed composite fermions.

In single-layer measurements of mesoscopic fluctuations, the ratio
(\ref{fa}) has been reported in \textcite{kvo}. In double-layer
systems, mesoscopic fluctuations of Coulomb drag were investigated in
\textcite{prs}, where it was shown, that the fluctuations of drag
resistivity obtained either by varying of the magnetic field (with
$n=const$) or by varying the carrier density (holding $B$ constant)
exhibit the same characteristic scale (or a ``period''), if plotted as
a function of the filling factor $\nu=nh/eB$, see Fig.~\ref{data6}.
The similarity of the two ``periods'' is equivalent to the ratio
(\ref{fa}).


\section{Drag in graphene-based double-layer devices}
\label{dig}


The physical picture of frictional drag outlined in the preceding
Sections is based on the following assumptions: (i) each of the layers
is in a Fermi-liquid state, which at the very least means
${\mu_{1(2)}\gg{T}}$; (ii) electron-electron interaction does not
contribute to the intralayer transport scattering time; (iii) the
interlayer Coulomb interaction is assumed to be weak enough,
${\alpha=e^2/v_F\ll1}$, such that $\rho_D$ is determined by the
lowest-order perturbation theory \cite{kor,jho,fln,fl2,mac} leading to
Eq.~(\ref{td}).

The drag resistivity (\ref{rdb}) depends on the carrier density,
temperature, and interlayer spacing. In semiconductor devices, only
temperature can be varied with relative ease, and even then only
within the limit ${\mu_{1(2)}\gg{T}}$ [with the notable exception of
  \textcite{pi2}].

Lifting one or more of the above assumptions leads to significant
changes in the drag effect. Recently drag measurements were performed
in a system of two parallel graphene sheets
\cite{tu1,exg,tu2,meg}. This system offers much greater flexibility
compared to prior experiments in semiconductor heterostructures. The
graphene-based system allows one to scan a wide range of chemical
potentials (by electrostatically controlling carrier density) from the
Fermi-liquid regime to the charge neutrality (or Dirac) point
${\mu_i=0}$. Moreover, using hexagonal boron nitride as a substrate
\cite{meg,ge1}, one can decrease disorder strength in the system and
reach the regime, where transport properties of the two layers are
dominated by electron-electron interaction, ${\tau\gg\tau_{ee}}$. In
addition, modern technology allows for a controlled growth of boron
nitride yielding devices with a relatively wide range of the
interlayer separations, which can be as low as ${d=1}$nm
(corresponding to only three atomic layers!). While the experiments
\cite{tu1,exg,tu2,meg} were performed at relatively low temperatures
${T<v_g/d}$ ($v_g$ is the quasiparticle velocity in
graphene), the range of temperatures available for these measurements
(typically, ${4-240}$K) is much wider than in earlier studies. In a
parallel development, Coulomb drag measurements in graphene double
ribbon structures were reported in \textcite{chen}.

\begin{figure}
\begin{center}
\includegraphics[width=0.8\linewidth]{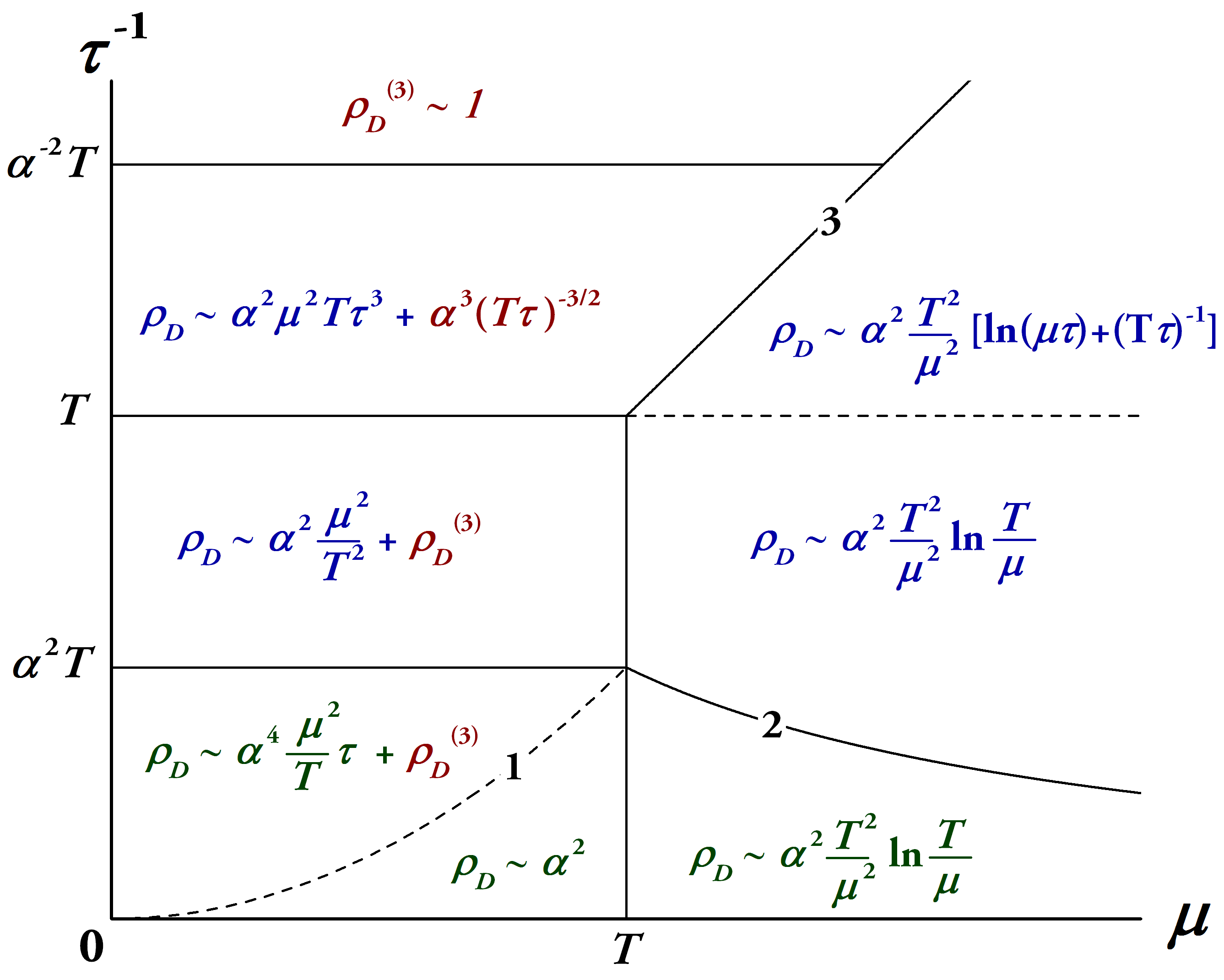}
\end{center}
\caption{(Color online) Summary of the parameter regimes and the
  resulting drag coefficient in graphene for identical layers,
  ${\mu\ll{\rm{min}}(T/\alpha,v/d)}$, and uncorrelated
  disorder. Bottom row (below the curve 2,
  ${\tau^{-1}\ll\alpha^2T^2/\mu}$): solutions to the quantum kinetic
  equation, see Sec.~\ref{hydro}. Curve 1
  (${\tau^{-1}=\alpha^2\mu^2/T}$) separates the two regimes in
  Eq.~(\ref{r1dp}). Middle row (${\alpha^2T\ll\tau^{-1}\ll{T}}$): the
  region where the QKE approach overlaps with the perturbation theory
  \cite{met}. The third-order contribution
  ${\rho_D^{(3)}={\cal{O}}(\alpha^3)}$ yielding nonzero drag at
  ${\mu=0}$ is shown in red. Upper row (${\tau^{-1}\gg{T}}$): the
  diffusive regime, where $\rho_D^{(3)}$ saturates for
  ${\tau^{-1}\gg{T}/\alpha^2}$). [Reproduced from \textcite{mem}.]}
\label{pdg}
\end{figure}

In graphene one can reach parameter regimes, which were inaccessible
in semiconductor samples, see Fig.~\ref{pdg}: (i) near charge
neutrality, the chemical potential may become smaller then
temperature, ${\mu_{1(2)}\ll{T}}$; the electronic system becomes
nondegenerate; (ii) low-energy excitations in graphene are
characterized by the linear Dirac-like dispersion; there is no
Galilean invariance in the system and transport properties are
strongly affected by electron-electron interaction
\cite{kats2012,pol}. Moreover, electrons interact by means of 3D
nonrelativistic Coulomb interaction, which breaks the Lorenz
invariance of the Dirac Hamiltonian.

Nondegenerate systems were considered in thee early work on
frictional drag \cite{pog,pri,jac,bo1} in the context of
semiconductors, where elementary excitations are typically modeled by
quasiparticles with parabolic dispersion. In that case,
electron-electron interaction plays a subleading role in single-layer
transport (due to Galilean invariance). In contrast, in ultra-clean
graphene near the Dirac point single-layer transport is dominated by
electron-electron interaction \cite{pol,met}.

The low-temperature degenerate regime ${T\ll\mu}$ can be achieved in
doped graphene, which is expected to exhibit the same qualitative
behavior as the semiconductor devices. Indeed, in ballistic samples
and under the additional assumption of the small screening length
${\varkappa{d}\gg1}$, one recovers \cite{da1,met} the standard
expression for the drag resistivity (\ref{rdb}), albeit with an extra
factor ${N=4}$ reflecting higher degeneracy of the single-particle
spectrum in graphene. However, this regime might be outside of the
experimentally accessible parameter range of drag measurements in
graphene-based samples \cite{tu1,tu2,exg,meg}.

For weaker doping, the assumption of the small screening length is
invalid and the standard result (\ref{rdb}) has to be modified
\cite{met}. A perturbative treatment can still be developed as long as
the transport properties of both layers are dominated by disorder
(i.e., ${\tau_{ee}\gg\tau}$). If electron-electron interaction is weak
enough
\begin{equation}
\label{pt}
\alpha^2T\tau \, {\rm min}(1, T/\mu_i)\ll 1,
\end{equation}
then the drag conductivity is given by the standard expression
(\ref{dia}). Close to the Dirac point (${\mu_i\ll{T}}$), this yields
${\rho_D\propto\mu_1\mu_2}$. At intermediate densities
(${\mu\sim{T}}$), the drag coefficient reaches a maximum and then
decays towards the asymptotic limit (\ref{rdb}). This decay is
characterized by a long crossover from the logarithmic behavior at
${\mu_i>T}$ to the standard result (\ref{rdb}) that is only achieved
for small screening lengths, ${\varkappa{d}\gg1}$. As a
result, the density dependence of ${\rho_D(\mu_i\gtrsim{v}_g/d)}$
cannot be described by a power law. Partially due to this fact,
several conflicting results for $\rho_D$ have been reported in
literature \cite{da1,kac,net,dss,car,lux,am2,sl1,sl2}.


\subsection{Perturbative regime in ballistic samples}
\label{ptg}


The perturbation theory is valid when transport properties of the
sample are dominated by potential disorder, such that
${\tau\ll\tau_{ee}}$, see Eq.~(\ref{pt}). In ballistic samples the
mean-free path is large compared to the interlayer separation
${\ell\gg{d}}$. For experimentally relevant temperature range
${T<v_g/d}$, the latter condition is compatible with the more standard
condition for ballistic transport in disordered systems
${T\tau\gg{1}}$. The resulting parameter range occupies the middle row
of the ``phase diagram'' shown in Fig.~\ref{pdg} between the line
${T\tau\simeq1}$ and curve 2.

Perturbative calculations in the ballistic regime can be performed
using either the diagrammatic (see Fig.~\ref{fig_fd}), or
kinetic-equation approach (see Sec.~\ref{kt}). In both cases, one
arrives at the expression similar to (\ref{td}), where the nonlinear
susceptibility and screened interlayer interaction (and hence the
polarization operator) have to be specified for Dirac fermions in
graphene.

\subsubsection{Nonlinear susceptibility in graphene}
\hfill

In contrast to the theory reviewed in Sec.~\ref{drag0}, here we are
interested in a wide range of chemical potentials including the Dirac
point ${\mu=0}$. The nonlinear susceptibility and polarization
operator in graphene for arbitrary $\mu$ and $T$ were derived in
\textcite{met}. Assuming the long, energy-independent impurity
scattering time $\tau$ and neglecting intralayer interaction, the
nonlinear susceptibility has the form
\begin{equation}
\label{gg}
\bs{\Gamma}(\omega, \bs{q}) = 
-2\frac{e\tau\bs{q}}{\pi} 
g\left(\frac{\omega}{2T}, \frac{v_gq}{2T}; \frac{\mu}{T}\right),
\end{equation}
where [with ${W=\omega/(2T)}$, ${Q=v_gq/(2T)}$, and ${x=\mu/T}$]
\begin{widetext}
\begin{equation}
\label{gg1}
g\left(W,Q; x\right) = 
\begin{cases}
\sqrt{\frac{W^2}{Q^2}-1}
\displaystyle
\int\limits_0^1\!dz 
\frac{z\sqrt{1-z^2}}{z^2-W^2/Q^2}
\; I(z; W, Q; x), & |W|>Q \cr
-\sqrt{1-\frac{W^2}{Q^2}}
\displaystyle
\int\limits^\infty_1\!dz \;
\displaystyle
\frac{z\sqrt{z^2-1}}{z^2-W^2/Q^2} 
\; I(z; W, Q; x), & |W|<Q
\end{cases}.
\end{equation}
\begin{equation}
\label{i2}
I(z; W, Q; x) =
\tanh\frac{zQ+W+x}{2} -
\tanh\frac{zQ+W-x}{2} +
\tanh\frac{zQ-W-x}{2} -
\tanh\frac{zQ-W+x}{2}.
\end{equation}
Under the same assumptions, the polarization operator is given by
\begin{subequations}
\label{pgt}
\begin{eqnarray}
&&
\Pi^R = \frac{q}{4\pi^2 v_g} \int\limits_0^1\!
\int\limits_0^1\!\frac{dz_1dz_2}{z_1\sqrt{(1-z_1^2)(1-z_2^2)}}
\left[
(z_1^{-2}-1)\left(\frac{Q}{z_2 Q + W + i\eta} + 
\frac{Q}{z_2 Q - W - i\eta} \right) J_1(z_1^{-1}, z_2, x) \right.
\\
&&
\nonumber\\
&&
\quad\quad\quad\quad\quad\quad\quad\quad\quad
+
\left.
(1-z_2^2)\left(\frac{Q}{z_1^{-1} Q + W + i\eta} + 
\frac{Q}{z_1^{-1} Q - W - i\eta} \right) J_2(z_1^{-1}, z_2, x)
\right],
\nonumber
\end{eqnarray}
where
\begin{equation}
J_{1(2)}(z_1, z_2, x) = \tanh\frac{(z_1+z_2)Q+x}{2} +
\tanh\frac{(z_1+z_2)Q-x}{2}
\mp
\tanh\frac{(z_1-z_2)Q+x}{2}
\mp
\tanh\frac{(z_1-z_2)Q-x}{2}.
\quad
\end{equation}
\end{subequations}
\end{widetext}
The perturbative calculation amounts to using the polarization
operator (\ref{pgt}) to determine the effective interlayer interaction
(\ref{d12}) and then evaluating the drag conductivity
[cf. Eqs.~(\ref{td}) and (\ref{dia})] \cite{da1}
\begin{eqnarray}
\label{sd}
\sigma^{\alpha\beta}_D = \frac{1}{16\pi T}
\sum_{\bs{q}}\!
\int\!d\omega\frac{|{\cal D}^R_{12}|^2}{\sinh^2\frac{\omega}{2T}}
\Gamma_1^\beta(\omega, \bs{q}) 
\Gamma_2^\alpha(\omega, \bs{q}) ,
\quad\quad
\end{eqnarray}
using the nonlinear susceptibility (\ref{gg}). The drag resistivity
is then given by Eq.~(\ref{rd1}). For arbitrary $\mu$ and $T$ this
calculation has to be performed numerically \cite{net,lux}. At the
same time, all qualitative features of the drag effect can be
elucidated by using simple limiting values.

The nonlinear susceptibility (\ref{gg}) decays exponentially for
${q\gg{\rm{max}}(\mu,T)}$.
In the vicinity of the Dirac point, $T\gg\mu$, the integral that
determines the function ${g(W,Q,x)}$ cannot be evaluated in terms of
elementary functions. It can be shown, however, that in this case the
nonlinear susceptibility is proportional to $\mu/T$ \cite{met}
\begin{equation}
\label{gdp}
g(x\ll 1) \propto \mu/T,
\end{equation}
which could be expected since drag is supposed to vanish -- or, more
precisely, to change sign -- at the Dirac point.

In the degenerate limit ${T\ll\mu}$, the dimensionless function
${g(W,Q,x)}$ may be approximated by
\begin{equation}
\label{ggr}
g(x\gg 1, |W|<Q) \approx
\frac{4W}{Q} \sqrt{1-\frac{W^2}{Q^2}}
\frac{\sinh x}
{\cosh Q + \cosh x}.
\end{equation}
Furthermore, for ${\mu\gg{v}_Fq\gg\omega\sim{T}}$ (or ${Q\gg{W}}$) the
nonlinear susceptibility becomes similar to the standard \cite{kor}
``Fermi-liquid'' expression for the ballistic regime (\ref{gb})
\begin{equation}
\label{ggr2}
g(x\gg 1, |W|\ll Q) \approx 4\omega/(v_g q),
\end{equation}
where the extra factor of $4$ corresponds to extra degeneracy of Dirac
fermions in graphene \cite{da1,am2}.

The relation (\ref{gip}) between $\bs{\Gamma}$ and ${{\rm{Im}}\Pi}$ is
not satisfied in graphene. This follows from a direct comparison
between their respective integral representations. In particular, 
the nonlinear susceptibility (\ref{gg}) vanishes at the Dirac
point due to exact electron-hole symmetry, 
${\bs{\Gamma}(\mu=0)=0}$
\cite{da1}, while the polarization operator (\ref{pgt}) remains finite,
${{\rm{Im}}\Pi(\mu=0)\ne0}$ \cite{pol}.

Similarly to the usual Lindhard function \cite{Giuliani,lind}, the
polarization operator in doped graphene has the simple static limit
\begin{equation}
\label{stpo}
\Pi^R(\omega=T=0) = 2k_F/(\pi v_g).
\end{equation}
At the Dirac point, the result is somewhat different
\begin{equation}
\label{stpodp}
\Pi^R(\mu=\omega=0) =
\begin{cases}
q/(4v_g), & T\ll v_gq, \cr
4T\ln2/(\pi v_g^2), & T\gg v_gq.
\end{cases}
\end{equation}

\subsubsection{Lowest-order perturbation theory}
\hfill

We now use the above approximations to find the limiting expressions
for the drag resistivity in the perturbative regime \cite{met}.

In the simplest limit ${N\alpha\mu\ll{T}}$, the perturbative approach
is justified automatically. In this case, the single-layer
conductivity is determined by weak impurity scattering and has the
form
\begin{subequations}
\label{s0}
\begin{equation}
\sigma_0 = e^2 T\tau \; h_0\left(\mu/T\right),
\end{equation}
where
\begin{equation}
\label{h0}
h_0(x) = \frac{2}{\pi}\!
\int\limits_{-\infty}^\infty\!
\frac{ dz |z|}{\cosh^2\left(z+\frac{x}{2}\right)}
=\frac{2}{\pi}\left\{
\begin{matrix}
x, & x\gg 1,\cr
2\ln 2, & x\ll 1.
\end{matrix}
\right.
\end{equation}
\end{subequations}
In this limit screening is ineffective and for
${\mu_i,T\ll{v}_g/d}$ the interlayer spacing drops out of the
problem. Then we may use the ``bare'' Coulomb potential
(\ref{coulomb}), while the frequency and momentum integration
in Eq.~(\ref{sd}) are determined by the nonlinear susceptibility
(\ref{gg}).

Close to the double Dirac point, ${\mu_i\ll{T}}$, the nonlinear
susceptibility can be approximated by Eq.~(\ref{gdp}), while the
remaining integration is dominated by frequencies and momenta of order
temperature, ${\omega,v_gq\sim{T}}$, yielding a dimensionless
coefficient. The resulting drag resistivity is given by
\begin{subequations}
\label{zd}
\begin{equation}
\label{rddp}
\rho_D(\mu_i\ll T) \approx 1.41 \alpha^2(\hbar/e^2)
(\mu_1\mu_2/T^2).
\end{equation}

If only one of the layers is tuned close to the Dirac point,
${\mu_1\ll{T}\ll\mu_2}$, the drag conductivity (\ref{sd}) is
independent of the properties of the second layer, as the integration
in Eq.~(\ref{sd}) is still determined by the region
${\omega,v_gq\sim{T}}$. The single-layer conductivity in the second
layer is still determined by $\mu_2$, see Eq.~(\ref{s0}). As a result,
\begin{equation}
\label{rd-12}
\rho_D(\mu_1\ll T\ll \mu_2) \approx 5.8\alpha^2(\hbar/e^2)
(\mu_1/\mu_2).
\end{equation}

In the opposite limit ${\mu\gg{T}}$, the nonlinear susceptibility is
given by Eq.~(\ref{ggr}). Now the momentum integral in Eq.~(\ref{sd})
is logarithmic and is dominated by large values of momentum
${Q\gg{W}}$. The ratio of the hyperbolic functions in
Eq.~(\ref{ggr}) is similar to the step function: it's equal to unity
for ${Q\ll{x}}$ and vanishes at larger values of momentum
${Q\gg{x}}$. Therefore $x$ effectively acts as the upper cut-off and
the momentum integral can be approximated by a logarithm
\begin{equation}
\label{logint}
\int\limits_W^\infty\! \frac{dQ}{Q}
\frac{\sinh^2x}{\left(\cosh Q + \cosh x\right)^2}
\approx \ln \frac{x}{W}.
\end{equation}
Consequently the drag coefficient is similar to the standard results
of Sec.~\ref{drag0} 
\begin{equation}
\label{rdlt}
\rho_D (\mu\gg T)\approx  
\alpha^2\frac{\hbar}{e^2} 
\frac{8\pi^2}{3} 
\frac{T^2}{\mu^2} \ln \frac{\mu}{T}.
\end{equation}
\end{subequations}
This is to be expected, since at low temperatures ${T\ll\mu}$ the
phase-space argument yielding the $T^2$ dependence is justified and
the electron-hole asymmetry determines the dependence on the chemical
potential. The logarithmic factor is beyond such qualitative
estimates (the result (\ref{rdlt}) was calculated with
logarithmic accuracy).

\paragraph{Static screening for vanishing interaction strength}
\hfill

For slightly stronger interaction (i.e. smaller dielectric
permittivity of the insulating substrate) or slightly larger
interlayer spacing the condition $N\alpha\mu_{i} \ll T, \mu_i \ll
v_g/d$ breaks down and one needs to take into account static
screening. Static screening corresponds to the approximation
(\ref{stpo}) to the polarization operator. If the interaction strength
is still small $\alpha\rightarrow 0$, then the interaction can be
described by
\begin{equation}
\label{d12p2}
{\cal D}^R_{12} = - \frac{2\pi\alpha v_g^2}
{v_gq+2N\alpha\mu } \;
e^{-qd},
\end{equation}
where $N=4$ is due to spin and valley degeneracy. The additional
constant in the denominator affects the logarithmic integral
(\ref{logint}). As the chemical potential is being increased away from
the Dirac point, the following regimes may be gradually achieved
[here we discuss these regimes for the case of identical layers;
generalization to the case of two inequivalent layers is
straightforward]:

(i) ${N\alpha\mu\ll{T}\ll\mu\ll{v}_g/d}$. This regime is identical to
the above arguments leading to Eq.~(\ref{rdlt}).

(ii) ${N\alpha\mu\ll{T}\ll{v}_g/d\ll\mu}$. If the chemical potential
is increased beyond the inverse interlayer spacing, then the momentum
integration in Eq.~(\ref{logint}) is cut off by $v_g/d$ instead of
$\mu$.  The logarithmic behavior of the drag conductivity will be
modified and $\sigma_D$ no longer depends on the chemical potential
\begin{subequations}
\label{logsd}
\begin{equation}
\label{s2}
\sigma_D \sim \alpha^2e^2T^2\tau^2\ln[v_g/(Td)].
\end{equation}

(iia) ${T\ll{N}\alpha\mu\ll\mu\ll{v}_g/d}$. In this case one finds
instead of Eq.~(\ref{s2})
\begin{equation}
\label{s2a}
\sigma_D \sim \alpha^2e^2T^2\tau^2\ln[1/(N\alpha)].
\end{equation}

(iii) ${T\ll{N}\alpha\mu\ll{v}_g/d\ll\mu}$. Increasing the chemical
potential further leads to the regime where the static screening can
no longer be neglected. Now the lower integration limit in
Eq.~(\ref{logint}) is effectively given by the inverse screening
length rather than the frequency. The upper limit is still determined
by the interlayer spacing. Therefore the drag conductivity again
depends logarithmically on the chemical potential \cite{kac}
\begin{equation}
\label{s3}
\sigma_D \sim \alpha^2e^2T^2\tau^2\ln[v_g/(N\alpha\mu d)],
\end{equation}
but now this is a {\it decreasing} function, indicating the existence
of the absolute maximum of the drag conductivity as a function of the
chemical potential.

{
\begin{figure}
\begin{center}
\includegraphics[width=0.6\linewidth]{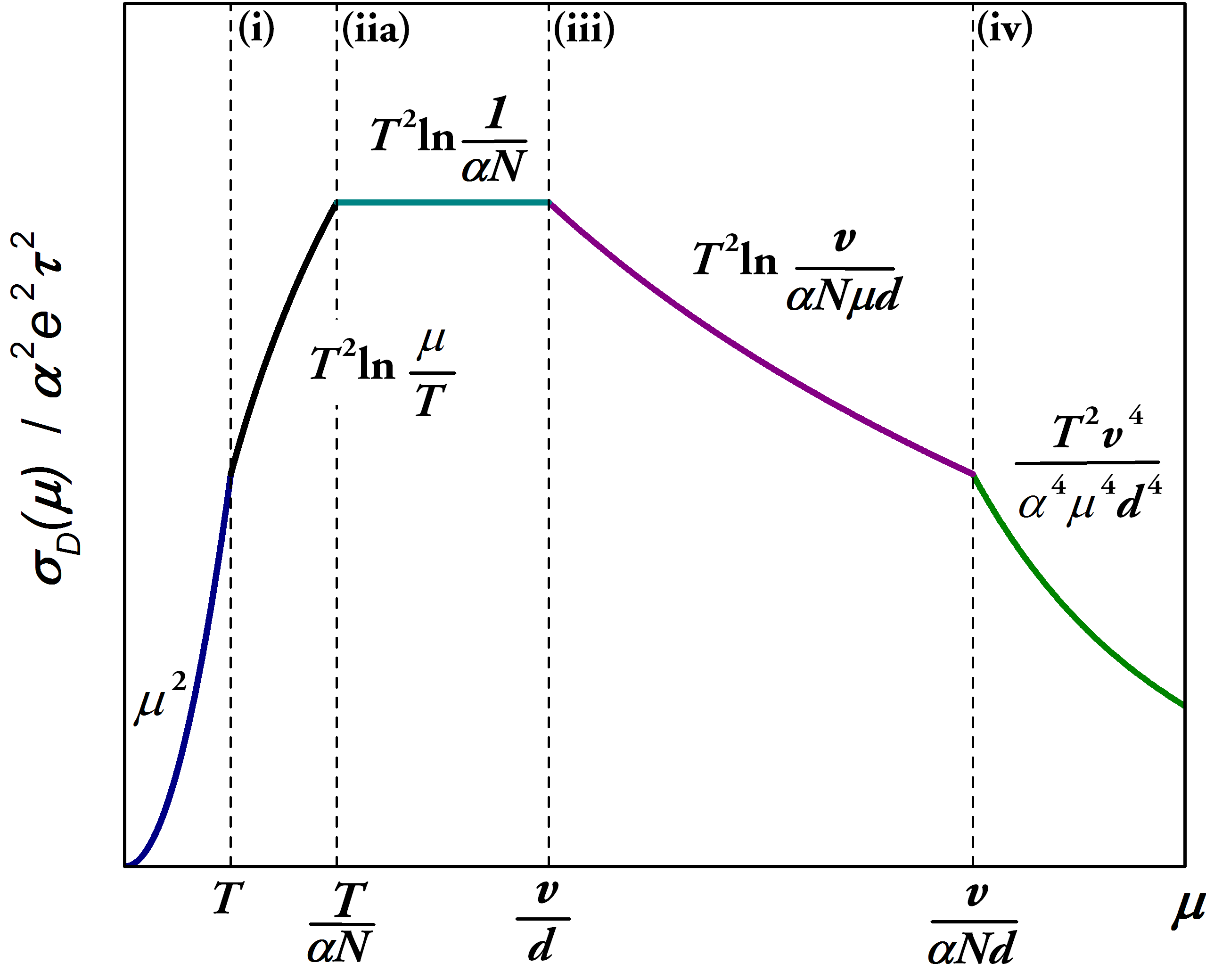}
\end{center}
\caption{(Color online) Sketch of the drag conductivity (in the units
  of ${\alpha^2e^2\tau^2}$) as a function of the chemical potential
  illustrating the results (\ref{logsd}). The blue line shows the
  quadratic dependence (\ref{rddp}) in the vicinity of the Dirac
  point. If ${T\gg{N}\alpha{v}_g/d}$, then the region (iia) should be
  replaced by (ii): the logarithmic dependence ${T^2\ln(1/\alpha)}$
  should be replaced by ${T^2\ln[v_g/(Td)]}$, and the limits $v_g/d$
  and $T/(N \alpha)$ should be exchanged. [Reproduced from
    \textcite{met}.]}
\label{sk1}
\end{figure}
}

(iv) ${T\ll{v}_g/d\ll{N}\alpha\mu\ll\mu}$. Finally, if the chemical
potential is so large that the screening length becomes smaller than
the interlayer spacing the momentum integral in Eq.~(\ref{logint}) is
no longer logarithmic. As the integration is now dominated by momenta
large compared to $T$, the nonlinear susceptibility may be
approximated by Eq.~(\ref{ggr2}), leading to the standard Fermi-liquid
result
\begin{equation}
\label{ft}
\sigma_D = \frac{\zeta(3)}{4}
\frac{e^2\tau^2T^2}{(k_Fd)^2(\varkappa d)^2}, \quad \varkappa = 4\alpha k_F,
\end{equation}
\end{subequations}
which differs from that of \textcite{kor} only by the factor
reflecting valley degeneracy in graphene \cite{da1,kac,am2}. The above
results are illustrated in Fig~\ref{sk1}.

\paragraph{Static screening for intermediate interaction strength}
\hfill

The results (\ref{zd}) and (\ref{logsd}) rely on the interaction
weakness. For stronger interaction, ${N\alpha>1}$, (i) the
approximation (\ref{d12p2}) might be unjustified and the full
expression (\ref{d12}) for the interaction propagator should be used;
(ii) the four regimes (\ref{logsd}) may not exist, since it might
happen that ${T/(N\alpha)\ll{T}<v_g/(N\alpha d)\ll{v}_g/d}$. In this
case, perturbative analysis can still be justified in the degenerate
regime, ${\mu\gg{T}}$, where there are two distinct regimes, (a)
$\mu\ll v_g/d$, and (b) $\mu\gg v_g/d$ \cite{met}; the latter regime
is usually identified with the Fermi-liquid result (\ref{ft}). As the
single-layer conductivity is still large and dominated by disorder,
the condition (\ref{pt}) can be somewhat relaxed:
\begin{equation}
\label{pt2}
\tau_{ee}\gg\tau \;\; \Rightarrow \;\;
\tau^{-1}\!\gg\alpha^2 T^2/\mu
\;\; \Rightarrow \;\;
\alpha^2T\tau\ll\mu/T.
\end{equation}
Proceeding under the assumptions of static screening and the ballistic
regime (i.e., the dominant contribution to the effect comes from large
momenta $v_gq>\omega$), the result of momentum integration is
determined by the upper limit and can be assumed independent of
$\omega$.  The frequency and momentum integrals factorize and 
neglecting ${W/Q}$ under the square root in Eq.~(\ref{ggr}) one
finds
\begin{widetext}
\begin{equation}
\label{r1}
\sigma_D = \alpha^2 e^2 T^2 \tau^2 
f_0\left(\frac{\mu}{T}; \alpha; \frac{Td}{v_g}\right),
\quad
f_0(x; \alpha ;\lambda) \approx \frac{32}{3}
\int\limits_{1}^\infty 
\frac{dQ Q^3 e^{-4\lambda Q}}
{\left[(Q+\tilde\alpha(x))^2 - \tilde\alpha(x)^2e^{-4\lambda Q}\right]^2}
\frac{\sinh^2x}{\left(\cosh Q + \cosh x\right)^2},
\end{equation}
\end{widetext}
where
\begin{equation}
\label{at}
\tilde\alpha(x) = N\alpha x/2.
\end{equation}

The results for weaker interaction, Eqs.~(\ref{logsd}), can be
recovered from Eq.~(\ref{r1}) by neglecting terms proportional to
${\tilde\alpha^2}$ in the denominator [which corresponds to
approximating the interlayer interaction (\ref{d12}) by
Eq.~(\ref{d12p2})]. In the limit ${\mu\gg{v}_g/d}$, 
the function $f_0$ depends on a single parameter
\begin{subequations}
\begin{equation}
f_0(x\lambda\gg 1) \approx \tilde f_0(4\lambda\tilde\alpha),
\end{equation}
where
\begin{equation}
\label{rf0}
\tilde f_0(y) = \frac{32}{3}
\int\limits_{0}^\infty
\frac{dZ Z^3 e^{-Z}}
{\left[(Z+y)^2 - y^2e^{-Z}\right]^2}.
\end{equation}
\end{subequations}
The function (\ref{rf0}) describes the crossover between the regimes
(iii) and (iv) of Eqs.~(\ref{logsd}) (see Fig.~\ref{sk1}).  This can
be seen by evaluating the integral in the two limits (here
$\gamma_0\approx 0.577216$ is the Euler's constant)
\begin{subequations}
\begin{eqnarray}
\label{rlog}
\tilde f_0(y\ll 1) \approx -32/3 \left(\ln y
+\gamma_0+11/6\right),
\end{eqnarray}
\begin{eqnarray}
\label{rfl}
\tilde f_0(y\gg 1) \approx 64\zeta(3)y^{-4}.
\end{eqnarray}
\end{subequations}
Numerically, this crossover spans a large interval of values of the
chemical potential such that the Fermi-liquid result (\ref{ft}) is
practically unattainable in graphene-based drag measurements
\cite{tut}, see Fig.~\ref{cp1}.

In experiment one typically measures carrier density rather than the
chemical potential \cite{tu1,tu2,exg,meg}. In graphene, the electron
density is given by
\begin{subequations}
\label{nm}
\begin{equation}
n = \int\limits_{-\infty}^\infty \frac{d\epsilon |\epsilon|}{\pi v_g^2} 
\left[\tanh\frac{\epsilon}{2T} - \tanh\frac{\epsilon-\mu}{2T}\right].
\end{equation}
Using the asymptotic expressions 
\begin{equation}
\label{nl}
n= \frac{1}{\pi v_g^2}
\begin{cases}
\mu^2, & \mu\gg T, \\[1pt]
(4\ln 2) \mu T, & \mu\ll T, 
\end{cases}
\end{equation}
\end{subequations}
one can obtain the qualitative dependence of $\rho_D$ on $n$, see
Table~\ref{table0}.

\begin{table}
\begin{center}
\begin{tabular}{cc}
parameter region & drag coefficient \\
\hline\noalign{\smallskip}
${\mu \ll T}$ & 
$\rho_D\sim n T^{-2}$ \\\noalign{\smallskip}
$T\ll \mu \ll v_g/d$ & 
$\rho_D \sim T^2 n^{-1} \ln (\alpha N n^{1/2} d/v_g) $ \\\noalign{\smallskip}
$\mu \gg v_g/d$ & 
$\rho_D = \rho_D^{FL} \sim T^2 n^{-3} d^{-4} $ \\
\end{tabular}
\end{center}
\caption{Asymptotic expressions for the drag coefficient to the 
  leading order of perturbation theory assuming ``realistic'' interaction 
  strength ${\alpha{N}\gtrsim1}$, identical layers ${n_1=n_2=n}$, and the
  experimentally relevant situation ${T<v_g/d}$. In the opposite regime 
  ${T\gg{v}_g/d}$ all results for $\rho_D$ should be divided by ${Td/v_g}$ 
  \cite{met,lux}.}
\label{table0}
\end{table}

The strongly doped, Fermi-liquid regime has attracted the most
attention in literature. Most authors report the standard
${\rho_D\sim{T}^2n^{-3}d^{-4}}$ behavior
\cite{da1,am2,kac,dss,car,met} assuming the energy-independent
impurity scattering time.

\subsubsection{Energy-dependent scattering time}
\hfill

In graphene, the impurity scattering time strongly depends on the type
of disorder and on energy \cite{kats2012}. In particular, for Coulomb
scatterers \cite{and,nom1,nom2,fal} or strong short-range impurities
\cite{ost}
\begin{subequations}
\begin{equation}
\label{tauclmb}
\tau(\epsilon)=\tau_0^2|\epsilon|.
\end{equation}
For weak short-ranged disorder \cite{and2}
\begin{equation}
\label{tausr}
\tau(\epsilon)=\gamma/|\epsilon|.
\end{equation}
\end{subequations}
Moreover, quenched disorder in graphene experiences logarithmic
renormalization \cite{alef}.

Drag in the presence of Coulomb impurities was first considered in
\textcite{net,dss}. Both papers reported a stronger dependence of the
drag coefficient on the carrier density and interlayer separation,
${\rho_D\sim{T}^2n^{-4}d^{-6}}$. This result was later disputed in
\textcite{car,met,amo}. These authors showed that the energy (or
momentum) dependence of $\tau$ is qualitatively irrelevant for the
asymptotic behavior of $\rho_D$. In the degenerate limit, microscopic
calculations lead to the same results with ${\tau(\mu)}$ substituted
in place of $\tau$. Close to the neutrality point, the drag
coefficient acquires an additional logarithmic factor
\begin{equation}
\label{rdte}
\rho_D(\mu_i\ll T)\sim\alpha^2 (\hbar /e^{2}) (\mu_1\mu_2/T^2)\ln T\tau(T).
\end{equation}

\subsubsection{Plasmon contribution}
\hfill

The dynamically screened interaction propagator contains plasmon
poles, that may (see Section~\ref{kt}) affect the resulting drag
resistivity. Theoretically, plasmons were studied in graphene
monolayers (at ${T=0}$) in \textcite{plas0,plas1,pol} and in
double-layer graphene systems in \textcite{plas2,plas3,bad}.
Renormalization of the plasmon spectrum due to electron-electron
interaction was considered in \textcite{plas4}. Experimentally,
plasmons were observed in graphene on SiO$_2$ substrate
\cite{plex1,plex11}, graphene-insulator stacks \cite{plex2}, and in
graphene micro-ribbon arrays \cite{plex3}. Bound states of plasmons
with charge carriers, the so-called ``plasmarons'' were observed in
\textcite{plasmaron1,plasmaron2}. Plasmons subjected to high magnetic
field were studied in \textcite{magplas}. For reviews of graphene
plasmonics see \textcite{plarev,plasmarev2}. More recently, plasmonic
excitations in Coulomb coupled $N$-layer graphene structures were
studied in \textcite{zhu}.

{
\begin{figure}
\begin{center}
\includegraphics[width=0.97\linewidth]{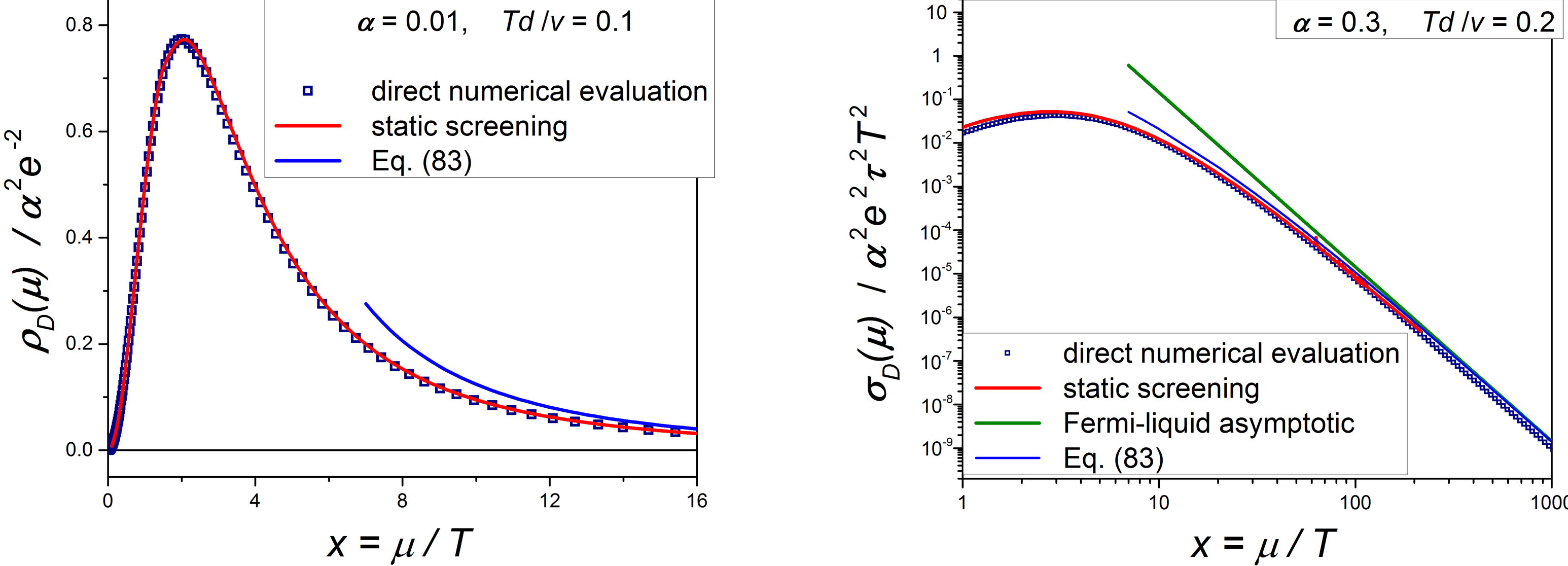}
\end{center}
\caption{(Color online) Results of the numerical evaluation of the
  drag coefficient. The squares represent the calculation of
  Eq.~(\ref{dia}) with the only approximation that the polarization
  operator in the screened interlayer interaction (\ref{d12}) was
  evaluated in the absence of disorder. The red line corresponds to
  the same calculation, with the polarization operator replaced by
  Eq.~(\ref{stpo}). The blue line was calculated with the approximate
  expression (\ref{rf0}), valid for ${\mu\gg{v}_g/d}$.  Left panel:
  ${\alpha=0.01}$, ${Td/v_g=0.1}$. Right panel: ${\alpha=0.3}$ and
  ${Td/v_g=0.2}$; log-log scale. The straight green line represents
  the Fermi-liquid result (\ref{ft}). [Reproduced from
    \textcite{met}.]}
\label{cp1}
\end{figure}
}

Within the above perturbative approach, i.e., Eq.~(\ref{pt}), and for
low enough temperatures, ${\mu,T\ll{v}_g/d}$, the plasmon contribution
to drag is subleading \cite{met}. The plasmon pole appears in the
region ${\omega>v_gq}$. Similarly to the situation in semiconductor
devices (Sec.~\ref{kt}), double-layer graphene systems admit an
acoustic (${\omega\sim{q}}$) and an optical (${\omega\sim\sqrt{q}}$)
plasmon modes \cite{plasds,ppp}. In the case where the plasmon decay
rate is small (as determined be either weak Coulomb interaction or
weak disorder), one can use the $\delta$-function approximation to the
interlayer interaction propagator \cite{fln2}. The corresponding
contribution to the drag conductivity contains a small factor
${g^2}|D|^2\sim\alpha^3$ for small momenta ${v_gq\sim\alpha{T}}$ (or
$\alpha^4$ for ${v_gq\sim{T}}$). If the energy dependence of the
scattering time is taken into account, the small parameter is
${\alpha^2T\tau}$, see Eq.~(\ref{pt}).

The above conclusion is illustrated in Figs.~\ref{cp1} showing a
comparison between the full numerical evaluation of the perturbative
drag coefficient using Eqs.~(\ref{d12}), (\ref{gg}), (\ref{pgt}), and
(\ref{sd}) and the same calculation within the approximation of static
screening (\ref{d12p2}). Numerical modeling of experimental samples,
see Figs.~\ref{fig_neto} and \ref{fig_misha} below, includes the
contribution was automatically by using the dynamically screened
interaction propagator (\ref{d12}).

At the same time, quantitative description of experiments, especially
in devices with wider interlayer spacing, might be significantly
affected by such aspects as inhomogeneous dielectric background
\cite{bad,car} and hybridization between phonon and plasmon modes
\cite{amo}. Plasmon-mediated drag between graphene wave-guides was
suggested in \textcite{shy}.

\subsubsection{Drag between massless and massive fermions}
\hfill

Graphene-based double-layer devices can be used to observe Coulomb
drag between massless and massive particles by coupling Dirac fermions
in monolayer graphene to quasiparticles with parabolic dispersion in
either bilayer graphene \cite{sch} or a usual 2DEG
\cite{ppp,sch}. Experimental realizations were
reported in \textcite{fis,gam}.

Theoretical analysis of \textcite{ppp,sch} is based on the standard
expression (\ref{sd}). Both works focus on the low-temperature,
degenerate regime ${T\ll\mu}$. As expected, in the case of strong
screening ${\varkappa{d}\gg1}$, both works reproduce the standard
result (\ref{rdb}). For ${\varkappa{d}\ll1}$, the resulting drag
coefficient is still quadratic in temperature, but contains also a
logarithmic factor reminiscent of Eqs.~(\ref{logsd}). \textcite{ppp}
report a $d$-independent drag in the special case ${k_{g}=k_{2D}}$
(which implies a density mismatch between the layers due to the
difference in the degeneracies of single-particle states).  In the
low-density limit ${n\rightarrow0}$ this yields
${\rho_D\propto{n}^{-1}}$, similarly to the results of \textcite{car},
see also Table~\ref{table0}. On the other hand, \textcite{sch} report
${\rho_D\propto{n}^{-2}}$ in the limit ${d\rightarrow0}$ and for
${n_g=2n_{2D}}$. Such discrepancies in the asymptotic behavior of
$\rho_D$ may appear due to the complicated structure of the nonlinear
susceptibility in graphene, see Eqs.~(\ref{logsd}) and Fig.~\ref{sk1}.

The predicted $T^2$ dependence is observed in experiment \cite{gam} in
the $10$K${<T<40}$K range, although with the smaller magnitude. At
higher temperatures, a violation of Onsager reciprocity was observed.
This was attributed to the interlayer current. Most interestingly, at
lower temperatures $T<10$K, the measured drag shows a marked upturn
that may indicate a phase transition at $T_c\sim10-100$mK, see
Sec.~\ref{exc1}.

The system of coupled Dirac and Schr\"odinger quasiparticles was also
considered in \textcite{jain}, where it was found that interspecies
interaction plays a significant role in determining collective
(plasmon) modes.

\subsubsection{Numerical evaluation of the drag coefficient}
\hfill

The above discussion demonstrates that already at the perturbative
level, the drag conductivity (\ref{sd}) exhibits multiple asymptotic
dependencies. Consequently, virtually every paper on the subject
presents results of numerical evaluation of Eq.~(\ref{sd}). In
contrast to the earlier work on semiconductor devices (see
Sec.~\ref{drag0}), most authors focus on the density (or chemical
potential) dependence rather than on the $T$-dependence. The overall
shape of ${\rho_D(n)}$ curves is qualitatively the same in all
calculations. At the Dirac point, drag vanishes, ${\rho_D(n=0)=0}$
(this conclusion does not agree with the experiments of
\textcite{exg,meg}, see below). Deep in the degenerate (or
low-temperature) regime, ${T\ll\mu}$, ${\varkappa{d}\gg1}$, the drag
coefficient reaches the standard decaying result
(\ref{rdb}). Therefore, for intermediate densities there has to be
a maximum, roughly at ${\mu\sim{T}}$. The corresponding shape is
shown in Fig.~\ref{cp1}.

{
\begin{figure}
\begin{center}
\includegraphics[width=0.7\linewidth]{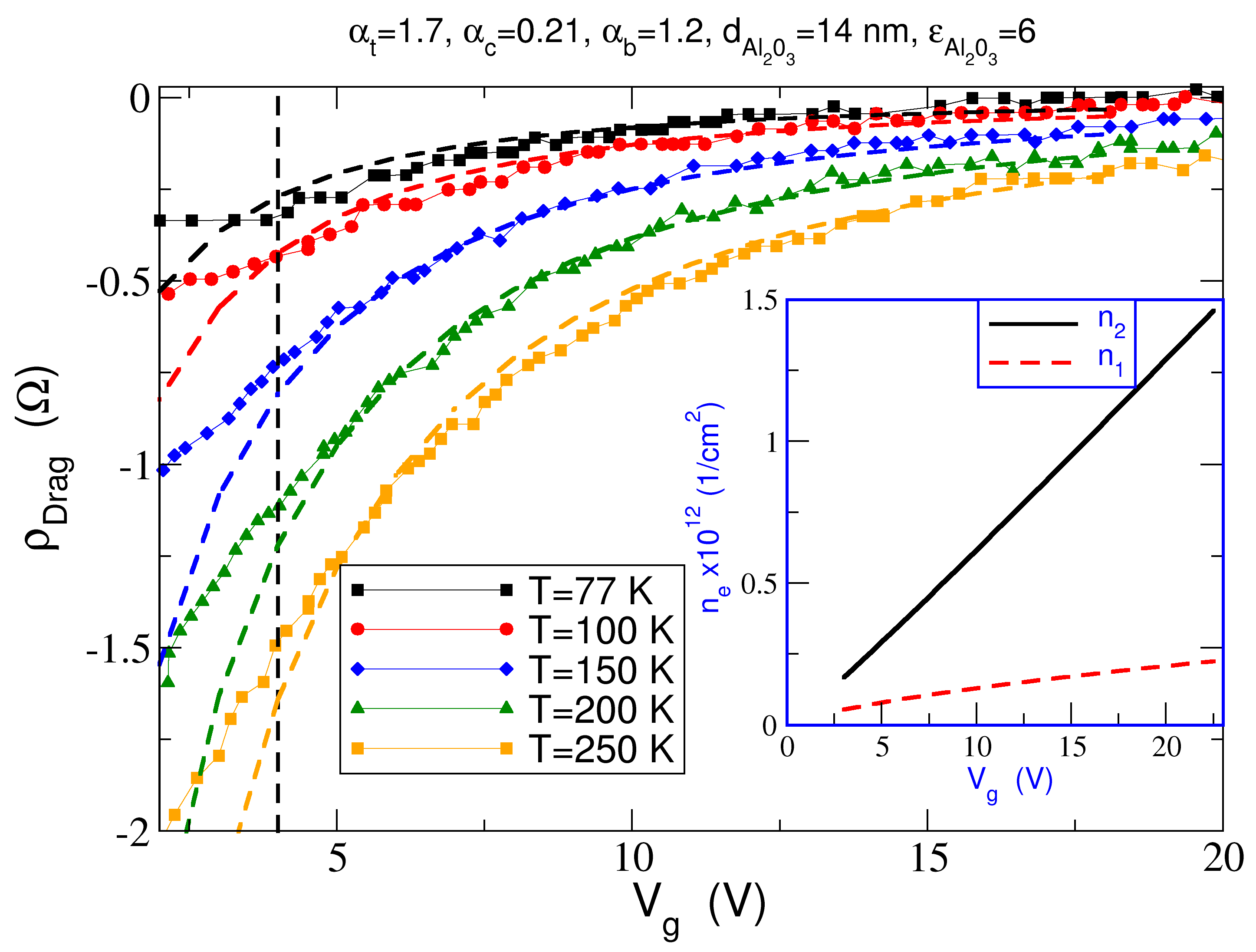}
\end{center}
\caption{(Color online) Results of the numerical evaluation (lines) of
  the drag coefficient and comparison with the data (symbols) of
  \textcite{tu1}. The interlayer spacing ($d=14$nm) and dielectric
  constants of the insulating material were chosen to represent the
  experimental device. Inset: the relation of the carrier densities
  and gate voltage, obtained from the electrostatic model of the
  sample. [Reproduced from \textcite{net}.]}
\label{fig_neto}
\end{figure}
}

\textcite{net} presented detailed numerical calculations aimed at
describing the experimental findings of \textcite{tu1}, see
Fig.~\ref{fig_neto}. This calculation included electrostatic modeling
of the device (which included two insulators, SiO$_2$ and
Al$_2$O$_3$), dynamically screened (within RPA) electron-electron
interaction, and the realistic model of Coulomb impurities. For doped
graphene layers, the results of the calculation show excellent
agreement with the data.

{
\begin{figure}
\begin{center}
\includegraphics[width=0.97\linewidth]{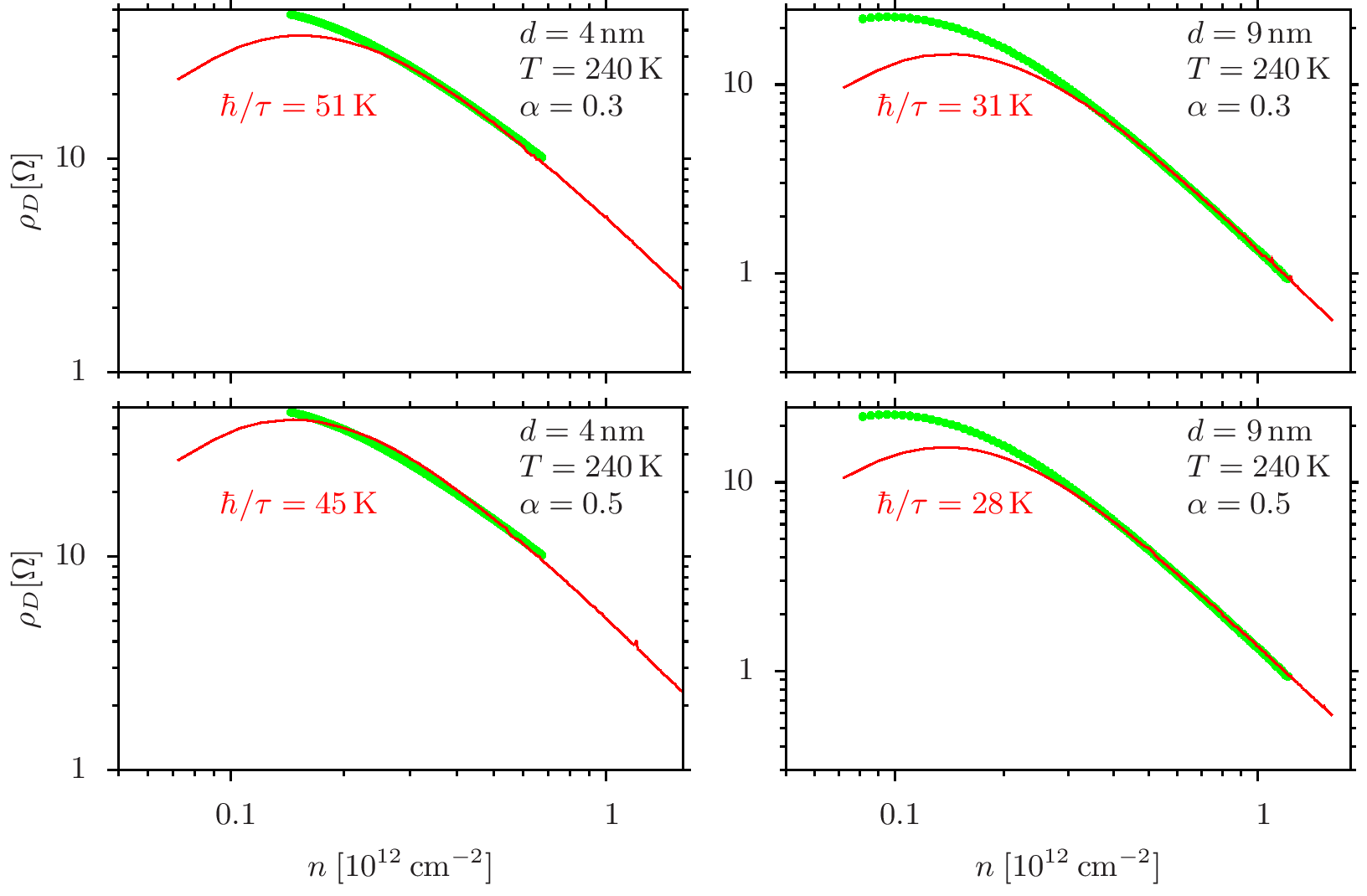}
\end{center}
\caption{(Color online) Results of the numerical evaluation
  \cite{priv2} of the drag coefficient (red line) and the experimental
  data (green dots) \cite{priv1}. The values of ${T=240}$K and $d$ are
  taken from the experiment. The only fitting parameter is the
  energy-independent impurity scattering time (once the value of
  $\alpha_g$ is chosen). The polarization operator was calculated at
  ${T=240}$K and in the presence of disorder (in the ballistic
  regime).}
\label{fig_misha}
\end{figure}
}

Theoretical modeling of ultra-clean graphene double-layers (using
boron nitride as a substrate as well as insulating spacer) based the
theory of \textcite{met} was performed by \textcite{priv2}, see
Fig.~\ref{fig_misha}. In this calculation, the polarization operator
was calculated at the experimental temperature in the presence of
disorder, in contrast to the ${T=0}$, free-electron expression
(\ref{pgt}). The use of full, dynamically screened Coulomb interaction
ensured that all plasmon-related features were taken into account
automatically. Choosing realistic values \cite{koz,net} for the
effective coupling constant, the only fitting parameter in this
calculation was the impurity scattering time $\tau$, which was taken
to be energy-independent similarly to the above discussion. Such
calculation was also able to reproduce the data
\cite{priv1} in the doped regime.

The results shown in Figs.~\ref{fig_neto} and \ref{fig_misha} confirm
the applicability of the perturbative approach to Coulomb drag in
doped graphene. In contrast to similar calculations aimed at
semiconductor devices (see Sec.~\ref{drag0}), these theories are able
to reach quantitative agreement with the experimental data with the
minimum of fitting parameters. This implies that frictional drag in
graphene is dominated by Coulomb interaction, with phonons playing
only a subleading role. The latter conclusion can be expected, given
that electrons in graphene are physically confined to move in a
two-dimensional plane and the rigidity of the crystal lattice
\cite{kats2012}.


\subsection{Hydrodynamic regime}
\label{hydro}


The perturbation theory outlined in Sec.~\ref{ptg} can be justified
either in the case of weak interaction ${\alpha\ll1}$ or in the
degenerate regime ${\mu\gg{T}}$, see Eq.~(\ref{pt}). At the same time,
the applicability condition (\ref{pt}) involves the impurity
scattering time $\tau$: the perturbation theory fails if the system is
``too clean'', or in other words, if electronic transport is dominated
not by disorder, but rather by electron-electron interaction. The
latter affects transport properties of graphene due to the absence of
Galilean invariance: the velocity of Dirac fermions
$\bs{v}=v_g^2\bs{p}/\epsilon$ is independent of the absolute value of
the momentum and therefore total momentum conservation does not
prevent velocity (or current) relaxation. As a result,
electron-electron scattering is characterized by its own transport
relaxation time, which may become smaller than the
scattering time due to potential disorder, ${\tau_{ee}\ll\tau}$.

``Ultra-clean'' graphene double-layers were discussed in
\textcite{meg,mem,mef} within the framework of the quantum kinetic
equation. In principle, solving the kinetic equation in a strongly
interacting system is a formidable problem, that cannot be solved in
general terms using presently available analytic methods. However in
graphene, one can take advantage of the kinematic peculiarity specific
to Dirac fermions. Indeed, scattering of particles with almost
collinear momenta is enhanced since the momentum and energy
conservation laws coincide. This restricts kinematics of the Dirac
fermions \cite{kas,kin,pol} and leads to the singularity in the
collision integral. This singularity leads to the fast thermalization
of particles within a given direction and allows one to derive
macroscopic - or hydrodynamic - equations that generalize
Eq.~(\ref{dt}) for interacting Dirac fermions. In monolayer graphene,
this approach was discussed in \textcite{mu1,foa,kas,kin,ryz}. An
alternative macroscopic approach to Coulomb drag in
graphene\footnote{The theory of \textcite{sl1,sl2,sl3} relies on
  correlations of the disorder potential in the two layers, see
  Sec.~\ref{tod}.} has been suggested in \textcite{sl1,sl2,sl3}.

\subsubsection{Collinear scattering singularity}
\hfill

Singular behavior of the collision integrals in the case of collinear
scattering of the Dirac fermions \cite{kas,kin,kin1,amy,mem} is
central to the hydrodynamic approach to transport in graphene.

The general form of the kinetic equation in layer $i$ is given by
Eq.~(\ref{beq0}) with the addition of the intralayer collision
integral. If the system is weakly perturbed from equilibrium, then the
distribution function can be written in the form (\ref{df}). Weak
deviations from equilibrium are associated with the smallness of the
nonequilibrium correction $h$, allowing one to linearize the
collision integrals \cite{dau10}. The linearized form of the collision
integrals is given by
\begin{widetext}
\begin{subequations}
\label{i0}
\begin{equation}
{\cal I}_{ij} = \sum_{1, 1', 2'} w_{12, 1'2'}  f^{(0)}_{j,1} f^{(0)}_{i,2} 
\left[1-f^{(0)}_{j,1'}\right]
\left[1-f^{(0)}_{i,2'}\right]
\left[ h_{j,1'} + h_{i,2'} - h_{j,1} - h_{i,2} \right],
\end{equation}
where the function
\begin{equation}
\label{grr}
 w_{1,2; 1',2'}
  = \big| \langle 1, 2 | U | 1', 2' \rangle \big|^2
    (2\pi)^3 \delta(\epsilon_1 + \epsilon_2 - \epsilon_{1'} - \epsilon_{2'})\,
    \delta(\bs{k}_1 + \bs{k}_2 - \bs{k}'_1 - \bs{k}'_2),
\end{equation}
\end{subequations}
\end{widetext}
determines the probability of scattering from states ${1',2'}$ into
states ${1,2}$ (within the Fermi Golden Rule approximation). Here
${\langle1,2|U|1',2'\rangle}$ is the interaction matrix element. The
indices ${i,j=1,2}$ denote the two layers\footnote{In the perturbative
  approach of Sec.~\ref{kt}, the kinetic equation (\ref{beq0})
  contained only the interlayer collision integral. Therefore, one
  could associate the states $1$ and $2$ with the active and passive
  layers and avoid extra layer indices.}.

In graphene, the interaction matrix elements are most conveniently
expressed in the basis of the eigenstates of the Dirac Hamiltonian
$|\epsilon, \bs{e}_{\bs{v}}\rangle$ labeled by their energy $\epsilon$
and the unit vector ${\bs{e}_{\bs{v}}=\bs{v}/v_g}$ pointing in the
direction of velocity (for a given spin and valley projection):
\begin{equation}
\label{u12}
 \big| \langle 1, 2 | U | 1', 2' \rangle \big|^2
  = \left|U\left(\bs{q}\right)\right|^2
    \frac{1 + \bs{e}_{\bs{v}}^{(1)}\bs{e}_{\bs{v}}^{(1')} }{2}
    \frac{1 + \bs{e}_{\bs{v}}^{(2)}\bs{e}_{\bs{v}}^{(2')}}{2}.
\end{equation}
Here ${\bs{q}=\bs{k}_1 - \bs{k}'_1}$ is the transferred momentum and
the two fractions are the ``Dirac factors'' \cite{kats2012}. Now one
can separate quantities related to the initial and final states in the
function $w_{12, 1'2'}$ by using the identities
\begin{eqnarray*}
\label{qomr}
&&
\delta(\epsilon_1 + \epsilon_2 - \epsilon'_1 - \epsilon'_2)
= \!\int\!d\omega
\delta(\epsilon_1 - \epsilon'_1 + \omega)
\delta(\epsilon_2 - \epsilon'_2 - \omega),
\\
&&
\nonumber\\
&&
\delta(\bs{k}_1\!+\!\bs{k}_2\!-\!\bs{k}'_1\!-\!\bs{k}'_2)
= \!\int\!d^2q\,
\delta(\bs{k}_1\! -\! \bs{k}'_1\! +\! \bs{q})
\delta(\bs{k}_2\! -\! \bs{k}'_2\! -\! \bs{q}).
\nonumber
\end{eqnarray*}
The $\delta$-functions yield ${\epsilon_1'=v_g|\bs{k}_1+\bs{q}|}$ and
hence allow one to sum over the states $1'$ and $2'$ in the collision
integral (\ref{i0}). Each of these sums result in a
diverging factor\footnote{In Sec.~\ref{ptg}, the nonlinear susceptibility (\ref{gg})
  did not exhibit this divergence due to an accidental cancellation
  that is specific to the particular case of energy-independent
  impurity scattering time. In a more general situation the
  cancellation does not occur and as a result the rate $\tau_D^{-1}$
  contains an extra logarithmic factor, see Eq.~(\ref{rdte}).}.
\begin{equation}
\label{divi}
\sum_{1'}\propto \frac{1}{\sqrt{v_g^2q^2-\omega^2}}.
\end{equation}
One can see that the divergence corresponds to collinear scattering by
examining the angle $\varphi_{\bs{k}_1\bs{q}}$ at the light cone:
\[
\cos\varphi_{\bs{k}_1\bs{q}} (\omega=v_gq) = 1
\quad\Rightarrow\quad
\varphi_{\bs{k}_1\bs{q}} = 0 \,\, ({\rm or} \,\, \pi).
\]
Hence, the argument of one of the above $\delta$-functions
vanishes: ${\epsilon'_1=\epsilon_1+\omega}$. Similar conclusion
follows for the momentum $\bs{k}_2$. Thus, all momenta are
collinear.

Physically, the divergence (\ref{divi}) represents the fact that for
the linear spectrum the energy and momentum conservation laws
coincide. Consequently, any relaxation rate obtained by integrating the
collision integral (\ref{i0}) over the state $2$ will be
logarithmically divergent. In order to regularize this divergence, one
has to go beyond the Golden-Rule approximation and take into account
renormalization of the spectrum \cite{ggv,nls,son}. This
leads \cite{kin,amy} to the appearance of a large factor
${|\ln(\alpha)|\gg1}$ in generic relaxation rates in graphene. In
disordered graphene, this singularity is also cut off by
disorder-induced broadening of the momentum-conservation
delta-function \cite{met}.

\subsubsection{Macroscopic linear-response theory in graphene}
\hfill

The collinear scattering singularity (\ref{divi}) allows for an
approximate, yet nonperturbative solution of the kinetic equation in
graphene \cite{kas,kin,mem,mef}. The idea is to find zero modes of the
collision integral and build macroscopic equations for the
corresponding currents.

The standard perturbative description of Coulomb drag is based on the
energy-independent approximation for the nonequilibrium distribution
function (\ref{df}). This constant solution of the kinetic equation
describes the single zero mode of the intralayer collision integral
corresponding to conservation of the electric charge (or the number of
particles). Macroscopic charge flow is described by the electric
current (\ref{jdef}). Integrating the kinetic equation, one finds the
macroscopic equation for $\bs{j}$ equivalent to the Drude theory, see
Eqs.~(\ref{dt}). Such solution is justified by the condition
(\ref{pt}), which means that the collision integral in the kinetic
equation is dominated by disorder.

In contrast, in ``ultra-clean'' graphene the collision integral is
dominated by Coulomb interaction. Using the collinear scattering
singularity (i.e., for ${|\ln(\alpha)|\gg1}$), one can neglect all but
the zero modes of ${\cal I}$ [treated as an integral operator acting
  on ${h_i(\epsilon)}$]. In practice, this means retaining only those
terms in the power series of the distribution function $h_i$ which
correspond to either zero modes of the collision integral, or to its
eigenmodes with nondivergent eigenvalues. \textcite{kas,kin,mem}
have developed the following {\it two-mode approximation}
\begin{equation}
\label{tm0}
h_i = \left(\bs{a}_0^{(i)} + \bs{a}_1^{(i)} \epsilon\right)\bs{v}.
\end{equation}
The vectors $\bs{a}_i$ can be expressed in terms of the two
macroscopic currents in graphene, the electric current (\ref{jdef})
and the energy current
\begin{equation}
\label{enc}
\bs{Q}_i =  \sum \epsilon \bs{v} \delta f_i.
\end{equation}

The appearance of inequivalent currents is the essential feature of
graphene physics. In general, the collision integral has three
nondecaying eigenmodes; hence \textcite{mef} have used the {\it
  three-mode approximation}:
\begin{equation}
\label{thm}
h_i = 
\left(\bs{a}_0^{(i)} + \bs{a}_s^{(i)} {\rm sign}(\epsilon) + \bs{a}_1^{(i)} \epsilon\right)
\bs{v}.
\end{equation}
The ${{\rm sign}(\epsilon)}$ mode is described by the imbalance
current \cite{foa}
\begin{equation}
\label{imc}
\bs{P}_i =  \sum {\rm sign}(\epsilon) \bs{v} \delta f_i.
\end{equation}

Integrating the kinetic equation with the help of either of the above
approximations for the nonequilibrium distribution function, one
obtains macroscopic equations for the currents $\bs{j}$, $\bs{Q}$, and
$\bs{P}$, that generalize the Ohm's law for graphene
\footnote{The full three-mode equations are too cumbersome to
  reproduce here, the interested reader is referred to
  \textcite{mef}.}. Solutions of these equations yield linear response
transport coefficients. Note, that this approach does not rely on the
Kubo formula. In particular, the drag coefficient can be obtained
without the use of the perturbative expressions (\ref{td}) or
(\ref{sd}).

The simplest macroscopic equation describes the energy current. In
an infinite sample, where all quantities are homogeneous, the
equation reads \cite{mem}
\begin{equation}
\label{qeq1}
e v_g^2 n \bs{E} 
+(v_g^2/c) \left[ \bs{j}\times \bs{B}\right]
= \bs{Q}/\tau,
\end{equation}
where $n$ is the carrier density in graphene (\ref{nm}). The collision
integral does not contribute to Eq.~(\ref{qeq1}) due to energy
conservation. In the limit ${\mu\gg{T}}$, all currents are equivalent,
such that ${\bs{j}(\mu\gg{T})\approx(e/\mu)\bs{Q}(\mu\gg{T})}$, and
Eq.~(\ref{qeq1}) becomes equivalent to the Ohm's law (\ref{o}). In
this limit, the Galilean invariance is restored, all
relaxation rates due to electron-electron interaction vanish, and all
three macroscopic equations become equivalent.

At the charge neutrality point ${n=0}$, the equation (\ref{qeq1})
yields ${(v_g^2/c)\left[\bs{j}\times\bs{B}\right]=\bs{Q}/\tau}$.  This
simple-looking relation illustrates all the essential qualitative
features of linear response transport in graphene. Firstly, in the
absence of disorder, ${\tau\rightarrow\infty}$, the equation becomes
senseless, at least when the system is subjected to external magnetic
field. Physically, this means that in the absence of disorder the
assumption of the steady state in an infinite system becomes invalid:
under external bias, the energy current increases indefinitely.
Secondly, if the system is stabilized by disorder, but ${\bs{B}=0}$,
then one finds ${\bs{Q}=0}$. Finally, if the system is subjected to
external magnetic field, the electric and energy currents are
orthogonal, ${\bs{j}\perp\bs{Q}}$. This leads to appearance of {\it
  classical, positive} magnetoresistance \cite{mef,mu1}
\begin{equation}
\label{pmr}
\delta R(B; \mu=0) \propto (v_g^4\tau/c^2) (B^2/T^3),
\end{equation}
as well as magnetodrag in graphene, see Sec.~\ref{gmd} below. These
results are in sharp contrast with the standard Drude theory, see
Eqs.~(\ref{dres}).

\subsubsection{Coulomb drag in weakly disordered graphene.}
\hfill

\begin{figure}
\centering
\includegraphics[width=0.97\linewidth]{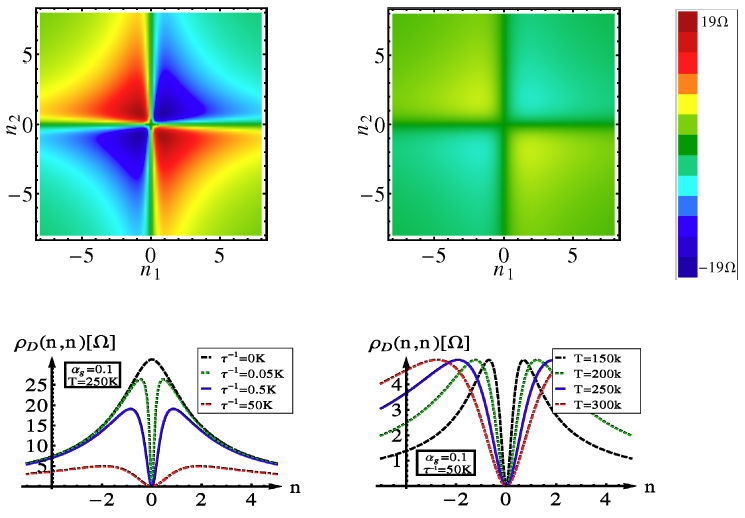}
\caption{(Color online) Leading-order drag coefficient in the
  ballistic regime as a function of carrier densities (in units of
  $10^{11}$cm$^{-2}$) for ${d=9}$nm. Left: $\rho_D$ at ${T=250}$K; the
  upper panel refers to ultra-clean graphene $\tau^{-1}=0.5$K; the
  lower left panel shows the evolution of $\rho_D$ with increasing
  disorder from ${\tau^{-1}=0}$ to ${\tau^{-1}=50}$K. Right: $\rho_D$
  for ${\tau^{-1}=50}$K; the lower panel shows $\rho_D$ for ${T=150}$,
  $200$, $250$, and $300$K. [Reproduced from \textcite{mem}.]}
\label{fr1}
\end{figure}

Close to charge neutrality and in the presence of weak, uncorrelated
disorder ${\alpha^2T\tau\gg1}$ (i.e.  ${\tau^{-1}\ll\tau_{ee}^{-1}}$),
the drag resistivity in the absence of magnetic field was found in
\textcite{mem} and has the form
\begin{equation}
\label{r1dp}
\rho_D(\mu_i\ll T) \approx 2.87
\frac{h}{e^2}
\frac{\alpha^2\mu_1\mu_2}{\mu_1^2+\mu_2^2+0.49T/(\alpha^2\tau)}.
\end{equation}
As long as any (even infinitesimal) disorder is present, $\rho_D$
vanishes at the double Dirac point ${\rho_D(\mu_1=\mu_2=0)=0}$, and
grows sharply in its immediate vicinity, see Fig.~\ref{fr1}.  If only
one of the layers is tuned to the Dirac point (median lines in
Fig.~\ref{fr1}), the drag resistivity always vanishes
\[
\rho_D(\mu_1=0, \mu_2\ne 0) = \rho_D(\mu_1\ne 0, \mu_2=0) =0.
\]

If one varies the carrier density in one of the layers through the
Dirac point, then the drag resistivity changes sign. In the color maps
in Fig.~\ref{fr1} this is represented by the color change between
neighboring quadrants. The same sign pattern of the drag resistivity
(in zero magnetic field) were observed in experiments of
\textcite{tu1,tu2,exg}.

At the double Dirac point in the absence of disorder, one finds (for
${\bs{B}=0}$) ${\rho_D(\mu_1=\mu_2=0)\sim\alpha^2_gh/e^2}$. This
peculiar feature is shown in Fig.~\ref{fr1} by the black curve in the
lower left panel. It is however unlikely that this result is relevant
to the nonzero drag resistivity at the Dirac point observed in
\textcite{exg}. A possible explanation for this observation is
provided by the higher-order effects \cite{mem}.

For intermediate disorder strength, ${\alpha^2T\ll\tau^{-1}\ll{T}}$,
the applicability region of the hydrodynamic approach overlaps with
that of the conventional perturbation theory reviewed in
Sec.~\ref{ptg} and one recovers perturbative results, see
Fig.~\ref{pdg}.

Finally, let us stress the novel qualitative feature of the
hydrodynamic approach: the electron-hole asymmetry does not play a
definitive role in the drag effect. Indeed, the ``drag rate''
$\tau_D^{-1}$ dominates the observable effect only under the standard
assumptions of the Fermi-liquid behavior in the two layers. On the
contrary, in the vicinity of the Dirac point in graphene, another
scattering process, the interplay of fast interlayer energy and
current relaxation which is insensitive to the electron-hole
asymmetry, becomes important. Further examples to such novel behavior
are presented in Section~\ref{gmd}.


\subsection{Diffusive regime}


In strongly disordered graphene samples or, equivalently, at the
lowest temperatures, ${T\tau\ll1}$, the electron motion becomes
diffusive. In this regime, the standard perturbative approach based on
Eq.~(\ref{sd}) is applicable. In particular, the polarization operator
has the standard form (\ref{prd}). The nonlinear susceptibility can
be found using the argument leading to Eq.~(\ref{gd}). In graphene
close to the Dirac point, ${\mu\ll{T}\ll\tau^{-1}}$, the derivative of
the longitudinal conductivity with respect to the carrier density is
independent of the precise nature of disorder and is given by
\cite{mem}
\[
\partial\sigma/\partial n \sim n v^4\tau^4.
\]
In contrast to the theory reviewed in Sec.~\ref{drag0}, in
graphene the Thomas-Fermi screening length is much longer than
the interlayer spacing ${\varkappa{d}\ll1}$; hence one finds
the following expression for the drag resistivity
\begin{eqnarray}
\label{r3dp}
\rho_D\left(\mu_i\ll T \ll\tau^{-1}\right) \sim
(h/e^2) \alpha^2 \mu_1\mu_2 T \tau^3,
\end{eqnarray}
vanishing at $\mu_i=0$ due to the electron-hole symmetry.

In the degenerate regime, ${\mu\gg{T}}$, one recovers the usual
quadratic temperature dependence of the drag resistivity. The behavior
of $\rho_D$ in the diffusive regime is summarized in Fig.~\ref{pdg}
(the upper row). The ``Fermi-liquid'' result (\ref{rdd}) is only
recovered in the academic limit of strong screening $\varkappa d\gg
1$. This regime is not shown in Fig.~\ref{pdg} since in graphene it can
be reached only at the extreme values of the chemical potential, see
Fig.~\ref{cp1}.

Calculations of the lowest-order drag resistivity in the diffusive
regime are essentially the same in any system, see Sec.~\ref{dis}. As
shown in Fig.~\ref{pdg}, the behavior of $\rho_D$ at the lowest
temperatures may be dominated by higher-order drag effects.


\subsection{Giant magneto-drag in graphene}
\label{gmd}


Although the effect of classical magnetoresistance in multi-band
systems is well known in semiconductor physics \cite{seeger2002}, the
equivalent effect in Coulomb drag was only recently observed in
graphene-based devices \cite{exg,meg}. One of the reasons is that the
majority of earlier drag measurements were performed in double-well
semiconductor heterostructures. Then each of the layers is represented
by a two-dimensional electron gas that is formed by electrons
occupying the lowest level in the quantum well at the interface
between two semiconductors in the device. In contrast in graphene, the
conductance and valence bands touch at the Dirac point and as a
result, both electrons and holes participate in transport phenomena at
low doping.

\begin{figure}
\begin{center}
\includegraphics[width=0.97\linewidth]{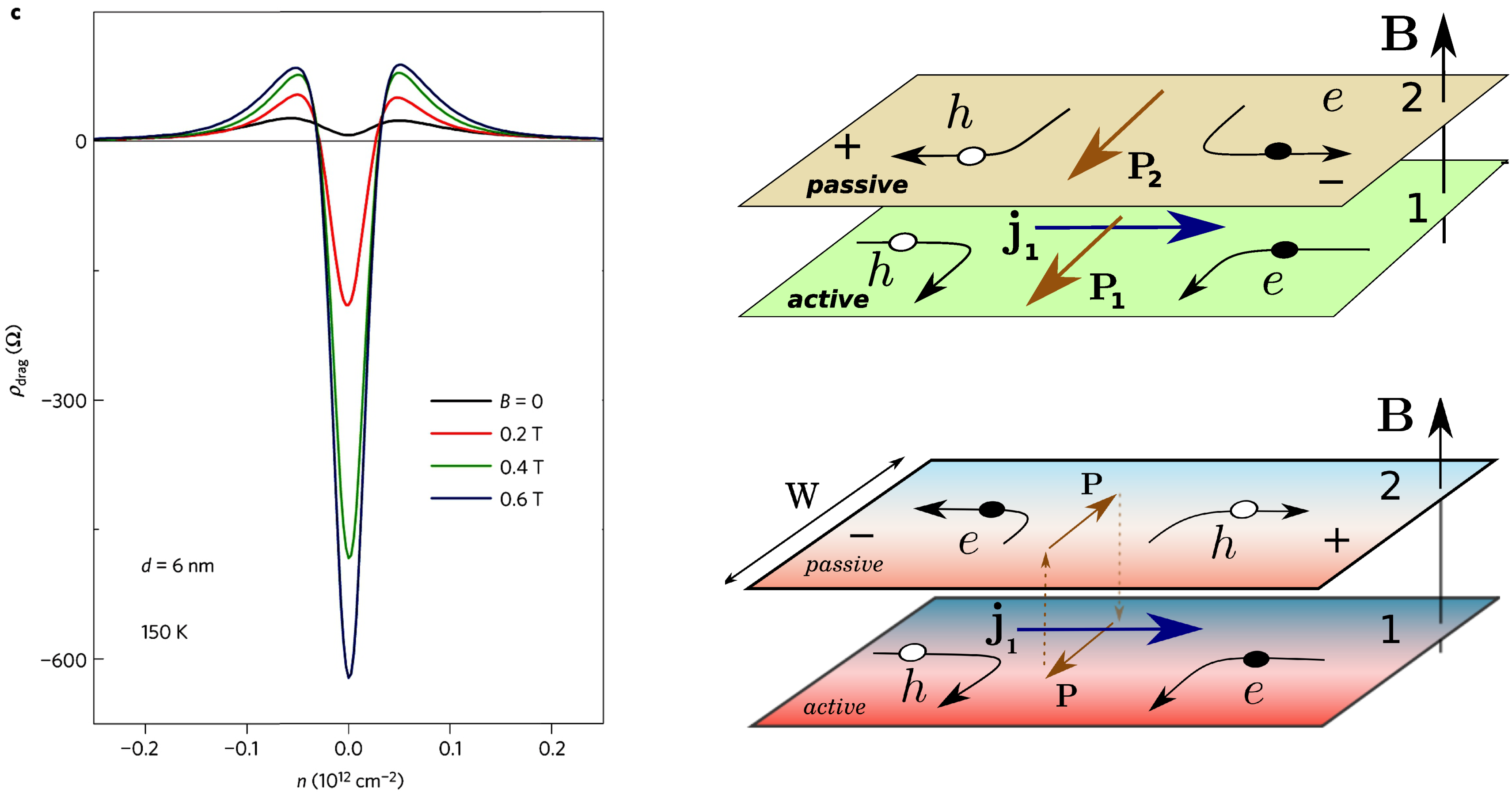}
\end{center}
\caption{(Color online) Left: Off-diagonal resistivity
  ${\rho_{xx}^{12}}$ in magnetic field, measured in a graphene-based
  double-layer device. The two graphene sheets are kept at
  ``opposite'' carrier densities ${n_1=-n_2=n}$ and
  ${T=150}$K. [Reprinted by permission from Macmillan Publishers Ltd:
    Nature Physics, \textcite{exg}] Right: Mechanism of magnetodrag
  at charge neutrality. Upper panel: in an infinite system
  quasi-particle currents in the two layers (denoted by $\bs{P}_i$)
  flow in the same direction, leading to positive
  $\rho_{xx}^{12}$. Lower panel: in a thermally isolated system no net
  quasiparticle flow is possible; the quasiparticle currents in the
  two layers have opposite directions yielding negative
  $\rho_{xx}^{12}$. [Reproduced from \textcite{meg}.]}
\label{fig_gmd}
\end{figure}

The experimental data on magnetodrag in graphene \cite{exg} is shown
in Fig.~\ref{fig_gmd}
\footnote{Notice, that \textcite{exg} adopted an alternative
  definition of the drag resistivity
  ${\rho^D_{xx}=E_{2x}/j_{1x}}$. Therefore in this Section we will
  discuss the off-diagonal resistivity
  ${\rho_{xx}^{12}=E_{2x}/j_{1x}}$ rather that $\rho_D$ that is
  defined in the rest of the paper with the opposite sign.}.  There
are two outstanding features in Fig.~\ref{fig_gmd}. At high carrier
densities (or in the degenerate regime), the effect of the magnetic
field is relatively weak. This observation is consistent with the
expectation, that transport properties of doped graphene are dominated
by one of the two bands (the contribution of the other being
exponentially suppressed). In the vicinity of the Dirac
point both types of carriers contribute to transport. Moreover, the
leading contribution to drag at zero field vanishes right at the
neutrality point due to exact electron-hole symmetry of the Dirac
spectrum. Once the magnetic field is applied, the system develops a
drag signal which is no longer determined by electron-hole
asymmetry. As a result, the drag resistivity near the Dirac point in
the presence of weak magnetic field is much higher than the maximum
value in zero field, see e.g. Fig.~\ref{fig_misha}.

The classical, two-band mechanism of magnetodrag in graphene at charge
neutrality can be readily illustrated in the case, where the system
size is much larger than any characteristic length scale, such that
the two graphene sheets may be considered effectively infinite. In
this case (see Fig.~\ref{fig_gmd}), the driving current
in the active layer, $\bs{j}_1$, corresponds to the
counter-propagating flow of electrons and holes with zero total
momentum (due to the exact electron-hole symmetry). Once the weak
magnetic field is applied, electrons and holes are deflected by the
Lorentz force and drift in the same direction. The resulting
quasi-particle flow, $\bs{P}_1$, carries a nonzero net momentum in the
direction perpendicular to $\bs{j}_1$. This momentum can be
transferred to the passive layer by the interlayer Coulomb interaction
inducing the quasi-particle current, $\bs{P}_2$, in the same direction
as $\bs{P}_1$. The Lorentz forces acting on both types of carriers in
the passive layer drive the charge flow in the direction opposite to
$\bs{j}_1$. If the passive circuit is open, this current is
compensated by a finite drag voltage, yielding a positive drag
resistivity \cite{meg}.

This mechanism of magnetodrag at charge neutrality is closely related
to the anomalous Nernst effect in single-layer graphene
\cite{mu1,zuev,an2}. Indeed, the quasi-particle current is proportional
to the heat current at the Dirac point. The fact that the Lorentz
force in the electron and hole bands has the opposite sign is also the
reason for the vanishing Hall effect at charge neutrality.

\begin{figure}
\begin{center}
\includegraphics[width=0.65\linewidth]{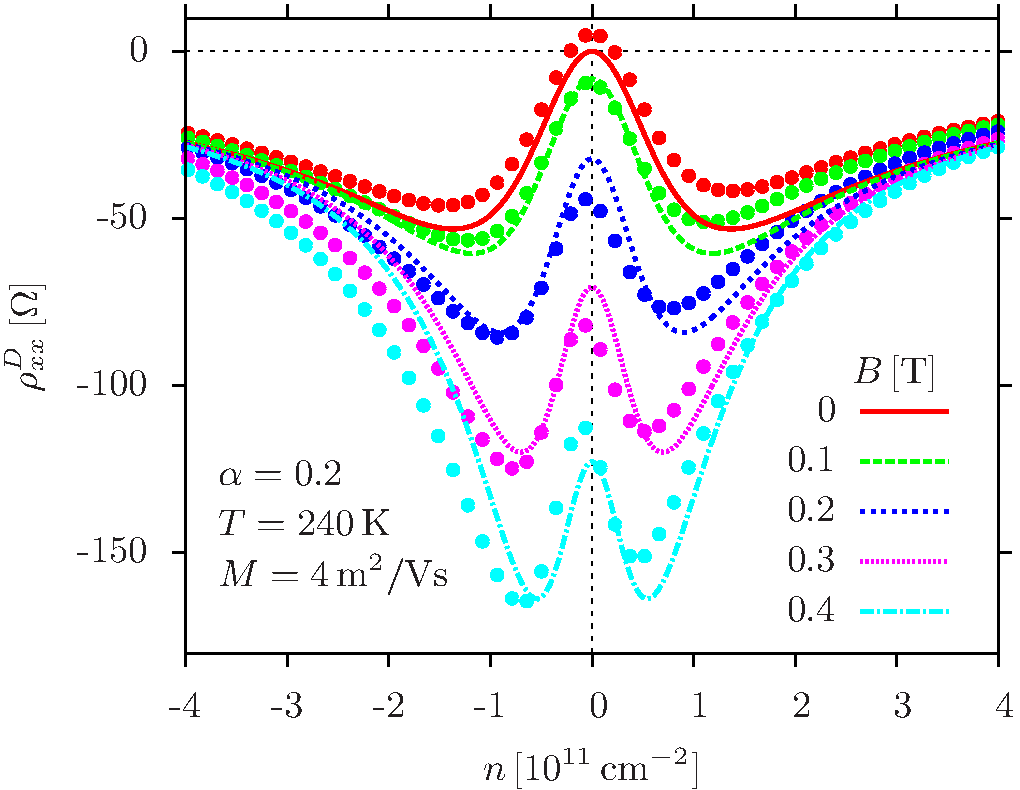}
\end{center}
\caption{(Color online) Off-diagonal resistivity $\rho_{xx}^{12}$ in
  magnetic field, measured in a graphene-based double-layer
  device. Both graphene sheets are kept at the same carrier density
  $n_1=n_2=n$ and at $T=240$K. Solid symbols represent the
  experimental data. [Reproduced from \textcite{meg}.]}
\label{gmd2}
\end{figure}

Despite being qualitatively clear, the above description of
magnetodrag yields the induced drag voltage which has the sign
opposite to that observed in experiment \cite{exg,meg}, see
Fig.~\ref{fig_gmd}. In fact, the {\it negative} drag in
Fig.~\ref{fig_gmd} can only appear if the quasiparticles currents in
the two layers $\bs{P}_1$ and $\bs{P}_2$ have opposite
directions. According to \textcite{meg,mef}, this is what happens in
small, mesoscopic samples used in experiment.

Consider the continuity equation for the quasiparticle current $\bs{P}_1$,
including relaxation by electron-hole recombination \cite{foa,meg}
\begin{equation}
\label{ceq}
\bs{\nabla P}_1 = -(\rho_1-\rho_0)/\tau_{ph}
-(\rho_1-\rho_2)/(2\tau_\textrm{Q}),
\end{equation}
where $\rho_i$ are the quasiparticle densities in the two layers,
${\rho_0=\pi{T}^2/(3v_g^2)}$ is the equilibrium quasiparticle density
at the Dirac point, $\tau_{ph}$ describes the energy loss from the
system dominated by phonon scattering, and $\tau_Q$ characterizes
quasiparticle imbalance relaxation due to interlayer Coulomb
interaction. The equation for the passive layer can be obtained by
interchanging layer indices. In the absence of quasiparticle
recombination, hard-wall boundary conditions at the sample boundaries
allow only for the trivial solution. In contrast, taking into account
inelastic processes, one finds the nontrivial solution illustrated in
Fig.~\ref{fig_gmd}: ${\bs{P}_1=-\bs{P}_2}$.

Combining the continuity equation (\ref{ceq}) with the hydrodynamic
description of linear response transport in graphene (with the
additional gradient terms that account for inhomogeneity of physical
quantities in finite-size systems), one can describe the negative drag
observed in experiment \cite{meg,mef}, see Fig.~\ref{gmd2}. The
exponential collapse of theoretical curves at high carrier density is
an artifact of the two-mode approximation adopted in \textcite{meg}.
The more accurate three-mode approximation \cite{mef} includes
thermoelectric effects formulated in terms of energy currents; the
corresponding hydrodynamic description yields only the power-law decay
of the magnetodrag at ${\mu_i\gg{T}}$, in contrast to the exponential
collapse shown in Fig.~\ref{gmd2}. As compared to lower-temperature
data (see Fig.~\ref{fig_gmd}), the results shown in Fig.~\ref{gmd2}
exhibit qualitatively new features which can be attributed to higher
efficiency of relaxation processes at higher temperature.


\subsection{Hall drag in graphene}
\label{hdg}


Hall drag measurements in graphene were reported in \textcite{meg}.
These experiments were performed at relatively high temperatures
${T=160-240}$K, where macroscopic coherence is not expected to exist.
While disorder effects in graphene are often attributed to Coulomb
scatterers characterized by mean free time that is linear in energy,
the measured Hall drag resistivity is not small as would follow from a
mechanism similar to that suggested in \textcite{hu0,opp}.

\begin{figure}
\begin{center}
\includegraphics[width=0.97\linewidth]{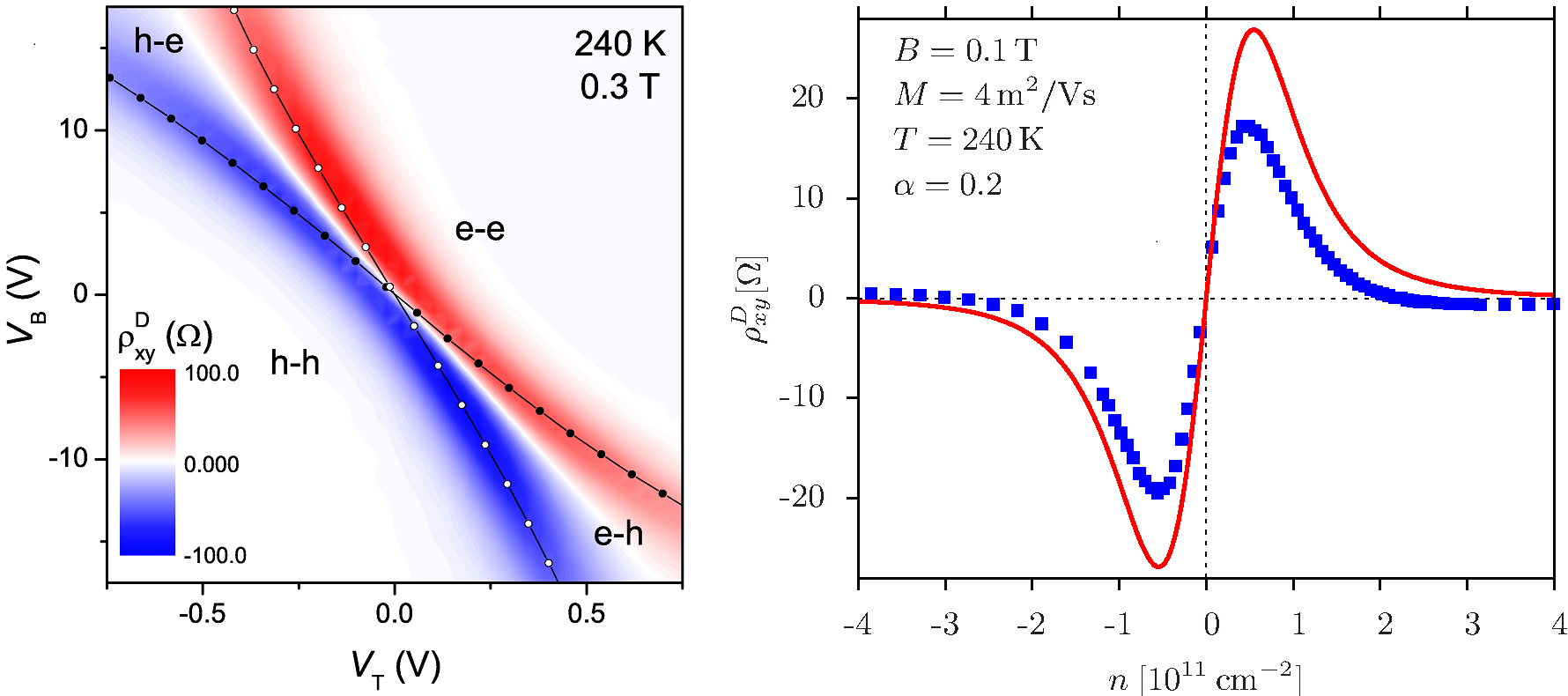}
\end{center}
\caption{(Color online) Left panel: Hall drag resistivity in graphene
  as a function of gate voltages controlling carrier densities in the
  two layers. White diagonal area corresponds to vanishing Hall drag
  for $n_1=-n_2$. Lines track positions of maxima in single-layer
  resistivity in top (open symbols) and bottom (solid symbols)
  layers. Right panel: Hall drag resistivity as a function of carrier
  density for $n_1=n_2=n$. Blue squares represent the experimental
  data. The red curve represents the theoretical
  prediction. [Reproduced from \textcite{meg}.]}
\label{hallexp}
\end{figure}

Instead, double-layer graphene samples demonstrate a much simpler, yet
still strong effect based on the coexistence of electron and hole
liquids in each layer \cite{foa}. Consequently, the observed Hall drag
resistance, Fig.~\ref{hallexp}, is large when one of the layers is
close to the neutrality point and vanishes if two layers have the same
charge densities with opposite signs (a white line running from the
top left to bottom right corner in the left panel Fig.~\ref{hallexp}).

Hall drag effect in graphene can be understood with the help of the
hydrodynamic theory \cite{mef,meg}. Indeed, given the presence of two
noncollinear currents in the model, it is not surprising to see the
nonzero Hall drag away from the Dirac point, where both the
conventional Hall effect and Hall drag change sign together with the
carrier density and thus have to vanish \footnote{A similar two-fluid
  model was used in \textcite{sl3} to explain Hall drag in terms of
  the ``energy-driven drag mechanism''. Indeed, if one omits the
  interlayer frictional force, one would still find nonzero Hall drag
  due to the interlayer energy relaxation.}. Hall drag also has to
vanish in the degenerate regime where only one band contributes to
transport and the standard single-band theory (\ref{dt})
holds. However, this regime lies outside of the parameter range of the
experiment \cite{meg}. Thus, some Hall drag signal is observed at all
densities, but $\rho_{xy}^D$ decays to rather small values as the
density increases beyond ${n\simeq1\times10^{11}}$cm$^{-2}$.
Interestingly enough, the data show a sign change of $\rho_{xy}^D$ at
${n\approx\pm2\times10^{11}}$cm$^{-2}$. This rather weak effect
requires a more accurate consideration.

The right panel of Fig.~\ref{hallexp} shows the results of the
hydrodynamic theory alongside experimental data. This calculation was
performed without any fitting \cite{meg}. The value of impurity
scattering time $\tau$ was determined from the measured single-layer
resistivity. The effective interaction parameter was estimated by the
most plausible value for graphene on hBN, ${\alpha_g\approx0.2}$ [see
  e.g. \textcite{koz,ree} for general considerations and the
  experimental evidence for possible values of $\alpha_g$].

\subsection{Higher-order effects in graphene}
\label{tod}


All theories of Coulomb drag in graphene discussed so far were
concerned with the leading-order contribution of the interlayer
interaction. Indeed, even the nonperturbative results of the
hydrodynamic approach were obtained by solving the kinetic equation
with the collision integral (\ref{i0}), where the transition
probability was estimated using the Fermi Golden Rule.  All such
theories predict vanishing drag at the point of exact electron-hole
symmetry [with the exception of the academic case of pure graphene,
see Eq.~(\ref{r1dp}) and Fig.~\ref{fr1}].

However, measurements \cite{exg,meg} reveal nonzero drag at the double
Dirac point, see Fig.~\ref{fig_gmd}. At the time of writing, there is
no consensus in the community regarding the origin of this effect. At
the same time, higher-order processes (see Sec.~\ref{third}) are known
to be insensitive to the electron-hole symmetry and thus may provide
a plausible explanation \cite{meg}.

\subsubsection{Third-order drag in graphene}
\hfill

\begin{figure}
\begin{center}
\includegraphics[width=0.97\linewidth]{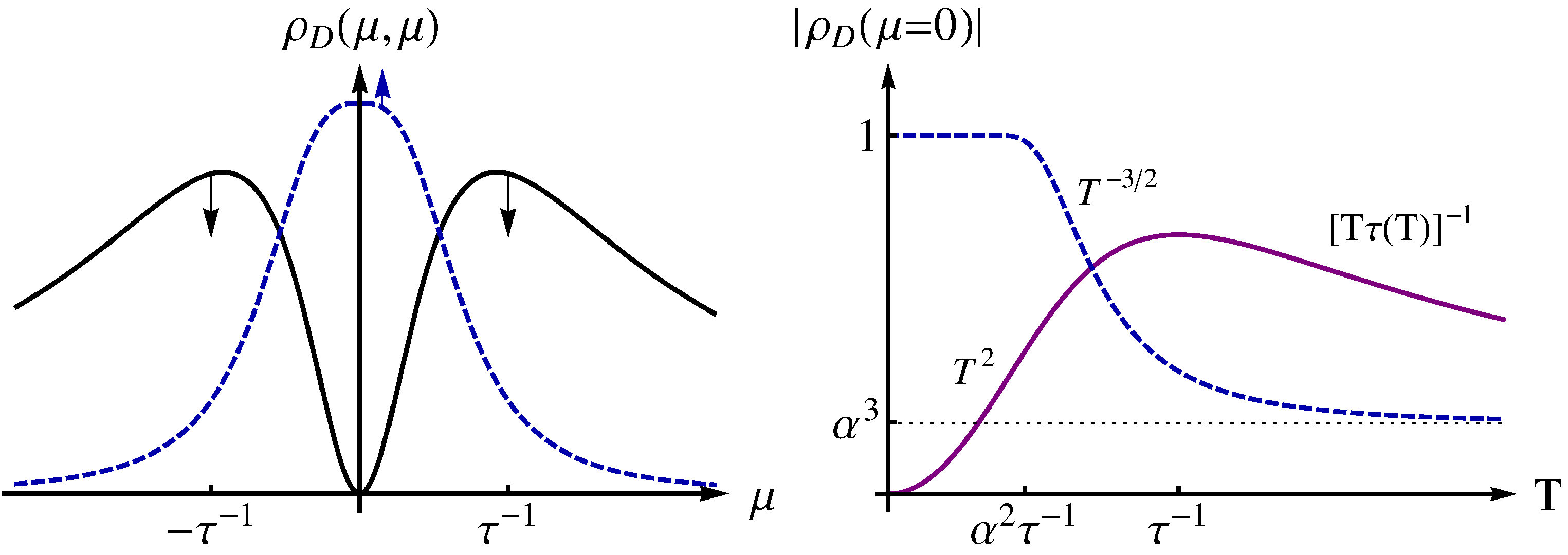}
\end{center}
\caption{(Color online) Schematic view of the drag resistivity at low
  temperatures. The dashed line illustrates the third-order drag
  effect. Left panel: The black solid line represents the lowest order
  contribution to drag. The arrows indicate the tendency of the two
  terms with the decrease of temperature ${T\rightarrow0}$. Right
  panel: The purple solid line represents the contribution of
  correlated disorder. [Reproduced from \textcite{mem}.]}
\label{todg2}
\end{figure}

The third-order drag effect in graphene was considered in
\textcite{mem}. The principle results are shown in Fig.~\ref{pdg} in
red. A schematic illustration of the relative strength of the second-
and third-order contributions is given in the left panel in
Fig.~\ref{todg2}.

The third-order drag resistivity in the diffusive regime can be found
similarly to the conventional case, see Sec.~\ref{third}. All
microscopic details are masked by the diffusive nature of electronic
motion.  However, due to the relatively weak screening and the
possibility to tune the carrier density to the Dirac point, one finds
a richer physical picture with multiple parameter regimes.

The standard ``Fermi-liquid'' regime \cite{le1}
corresponds to the condition
\[
N\varkappa \gg \text{max}\left\{d^{-1},\sqrt{T/D}\right\},
\]
where ${N=4}$ describes spin and valley degeneracy of quasiparticle
states in graphene. Here the temperature-independent result
(\ref{tord1}) is reproduced, although with the extra factors of $N$
\begin{equation}
\rho_D^{(3)}\sim (h/e^2)N^{-5} g^{-3}(\varkappa d)^{-2}.
\end{equation}

At higher temperatures, one can achieve a different, high-temperature
regime with
\[
d^{-1} \ll N\varkappa \ll \sqrt{T/D}.
\]
In this case, the resulting drag resistivity decays rapidly 
\begin{equation}
\rho_D^{(3)}\sim \frac{h}{e^2}\, \frac{1}{g^3}\, 
\frac{1}{(N\varkappa d)^2}  \left(\frac{D \varkappa^2}{T}\right)^{3/2}.
\end{equation}

The experiment of \textcite{exg} was performed on samples with the
small interlayer spacing. In the limit ${\varkappa{d}\ll1}$, one finds
three different temperature regimes.

Close to the Dirac point and at lowest temperatures, the drag
resistivity is temperature-independent:
\begin{eqnarray}
\label{tordp1}
\rho_D^{(3)} (\mu\ll T; T\tau\ll\alpha^2) \sim h/e^2.
\end{eqnarray}
Note, that this result is also independent of the strength of the
Coulomb interaction $\alpha$!

At somewhat higher temperatures (or, equivalently, for slightly weaker
disorder strength), the third-order contribution decays as function of
temperature
\begin{equation}
\label{tordp2}
\rho_D^{(3)} (\mu\ll T\ll\tau^{-1}\!\ll\alpha^{-2}T)
\sim (h/e^2) (\alpha^2T\tau)^{-3/2}.
\end{equation}
These results are illustrated in the right panel in Fig.~\ref{todg2}.

Away from the Dirac point, the third-order contribution decays as a
function of the chemical potential (or equivalently, carrier density)
and quickly becomes subleading, see the left panel in
Fig.~\ref{todg2}:
\begin{equation}
\label{tordp3}
\rho_D^{(3)} 
\left(\mu\tau\gg{\rm max}\left[1, \alpha^{-1}(T\tau)^{1/2}\right]\right)
\sim\frac{h}{e^2} \frac{1}{(\mu\tau)^3}.
\end{equation}
As a result, $\rho_D^{(3)}$ may only be detectable at low temperatures
and in vicinity of the Dirac point.

While estimating ${\rho_D^{(3)}(\mu=0)}$, the single-layer
conductivity was assumed to be of the order of the quantum conductance
${\sigma\sim{e}^2/h}$, i.e. discarding localization effects. Indeed,
single-layer measurements on high-quality samples show
temperature-independent conductivity down to $30$mK \cite{tzs}
[possibly due to the specific character of impurities in graphene
  \cite{mir}].


For weak disorder or higher temperature the diffusive approximation
fails. Drag in vicinity of the Dirac point can then be described by
the quantum kinetic equation approach. The previously reviewed results,
e.g., Eq.~(\ref{r1dp}) were obtained by approximating the collision
integrals with the help of the Fermi Golden Rule, see
Eq.~(\ref{grr}). However, taking into account next-order matrix
elements yields a nonzero contribution, similar to the above
third-order result $\rho_D^{(3)}$.

Taking into account the second-order matrix element, one can
generalize the Golden Rule expression (\ref{u12}) by using the
combination
\begin{equation}
\label{tome}
|U_{12}^{(1)}+U_{12}^{(2)}|^2\simeq |U_{12}^{(1)}|^2
+ 2 \text{Re} \{U_{12}^{(1)} [U_{12}^{(2)}]^*\}.
\end{equation}
Since ${U_{12}^{(1)}\propto\alpha}$ and
${U_{12}^{(2)}\propto\alpha^2}$, all relaxation rates will now get an
additional contribution of the order of $\alpha^3$. In particular, the
``drag rate'' $\tau_D^{-1}$ gets a contribution that is independent of
the carrier density
\begin{equation}
\label{tdto}
\tau_D^{-1} \sim \alpha^2 N (\mu/T)^2 + \alpha^3 N T,
\end{equation}
which dominates near the Dirac point. In this case, one may neglect
the conventional, second-order drag contribution; the result is
\cite{mem}
\[
\rho_D \sim \frac{h}{e^2}
\dfrac{\alpha^3 T+\alpha^4 \mu^2 \tau N}
{T+\alpha^2 \mu^2\tau N}, \qquad \mu\ll\alpha^{1/2} T, \quad T\tau\gg 1.
\]
Exactly at  the Dirac point this yields 
\begin{equation}
\rho_D\sim (h/e^2) \alpha^3.
\label{alpha3}
\end{equation}
This result is illustrated in the right panel of Fig~\ref{todg2} by
the horizontal asymptote at ${T\tau\gg1}$.

\subsubsection{Interlayer disorder correlations}
\hfill

Within the conventional theory, charge carriers in each layer
scatter off an independent disorder potential. This picture is clearly
applicable to the cases where impurities are mostly concentrated in
the substrate insulator sandwiching the double-layer structure. In the
case of the standard double-well heterostructures
\cite{ex1,ex2,eis,hil,lil}, the random potential originates in the
delta-doped layers providing charge carriers. These layers are
typically located on the outer sides of the double-well structure.  In
graphene, disorder potential is often attributed to the insulating
substrate, in particular to SiO$_2$. Indeed, in graphene-based samples
of \textcite{tu1,tu2}, graphene monolayers are exfoliated onto a
thick SiO$_2$ dielectric, while the interlayer spacer consists of
$14$mm-thick Al$_2$O$_3$. In such a structure, the impurity potential
created by the silicon oxide is likely to affect only the nearest
monolayer. 

In contrast, the samples of \textcite{exg} consist of
graphene--hexagonal-boron-nitride heterostructures, where the
interlayer spacer contains only few atomic layers of the same
insulator (boron nitride) that is used as a substrate. In this case,
impurity potential originating from the interlayer spacer would be
equally felt by carriers in both graphene layers. Another scenario for
disorder correlation \cite{exg,sl1} involves interactions between
charge-density inhomogeneities forming due to impurity potential in
the two layers.

The effect of the correlated disorder in the drag measurements is
insensitive to the electron-hole symmetry \cite{go2,hu2}, and thus may
also provide an explanation \cite{mem,sl1} for the observed nonzero
drag in graphene at the Dirac point \cite{exg}.

At high temperatures, $T\tau\gg 1$, the effect of the correlated
disorder can be described by the skeleton diagram similar to the
third-order drag contribution, see the right panel of
Fig.~\ref{todiag}. Interlayer disorder correlations can be
incorporated into the scattering amplitude, but now instead of the
second-order matrix element in Eq.~(\ref{tome}), one has to introduce
an interlayer disorder scattering rate $1/(T\tau_{12})$. The
resulting ``drag rate'' $\tau_D^{-1}$ is given by
\[
1/\tau_D^{\text{corr}} \sim \alpha^2 T /(T \tau_{12})
=\alpha^2/\tau_{12},
\]
corresponding to the drag resistivity
\begin{eqnarray*}
\rho_D^{\text{corr}}\sim \alpha^2/(T \tau_{12}),
\end{eqnarray*}
which overcomes the third-order drag contribution $\rho_D^{(3)} \sim
\alpha^3$ at ${1/\tau_{12}>\alpha{T}}$. This happens in the
perturbative regime (${1/\tau>\alpha^2T}$ for moderately correlated
disorder, ${\tau_{12}\sim\tau}$), where the correlated-disorder
contribution can be calculated diagrammatically.

Macroscopic inhomogeneities can be described in terms of macroscopic
spatial fluctuations $\delta \mu_i$ in chemical potentials of the two
layers \cite{sl1}, characterized by the correlation function
\begin{eqnarray}
F_{ij}^{(\mu)}(\bs{r}-\bs{r}')=
\langle\delta\mu_i(\bs{r})\delta\mu_j(\bs{r}')\rangle\neq 0.
\end{eqnarray}
Assuming the spatial scale of the fluctuations to be much larger than
all characteristic scales related to the particle scattering, one can
solve the hydrodynamic equations locally, yielding the local drag rate
\begin{eqnarray}
1/\tau_D(\bs{r})\sim 
\alpha^2 N \mu_1(\bs{r})\mu_2(\bs{r})/T.
\end{eqnarray}
Averaging over the small fluctuations of the correlated chemical
potentials, one arrives \cite{mem} at the correction to the universal
third-order result (\ref{alpha3}),
\begin{eqnarray}
\Delta \rho_D(\mu=0) \sim 
\frac{h}{e^2}\frac{\alpha^2\, F^{(\mu)}_{12}(0)}{T^2}\, \left(1+\alpha^2 N T\tau\right).
\end{eqnarray}

Finally, in the ultraclean limit 
\begin{eqnarray}
1/\tau\ll \alpha^2 N F_{ii}^{(0)}/T,
\label{cond-delta-mu}
\end{eqnarray}
one can approximate the local drag resistivity by an
analog of Eq.~(\ref{r1dp}): 
\begin{eqnarray}
\Delta \rho_D (\bs{r}; \mu=0) \sim \frac{h}{e^2}\alpha^2 
\frac{\delta\mu_1\delta\mu_2}{\delta\mu_1\delta\mu_1+\delta\mu_2\delta\mu_2}.
\label{DeltaRho}
\end{eqnarray}
In particular, for perfectly correlated chemical potentials,
$\delta\mu_1(\mathbf{r})=\delta\mu_2(\mathbf{r})$, the fluctuations
drops out from Eq.~(\ref{DeltaRho}) and the local resistivity turns
out to be independent of $\bs{r}$.  In a more general case, the
averaging over fluctuations becomes nontrivial, but this can only
affect the numerical prefactor in the final result. Thus, the
correlated large-scale fluctuations of the chemical potentials in the
layers in effect shift the curve 1 in Fig.~\ref{pdg} upwards,
extending the validity of the fully equilibrated result,
\begin{eqnarray}
\rho_D \sim  (h/e^2)\alpha^2,
\end{eqnarray}
to the case of finite disorder, Eq.~(\ref{cond-delta-mu}), at the
Dirac point. This implies that in the case of correlated
inhomogeneities the disorder-induced dip in the lower left panel of
Fig.~\ref{fr1} develops only for sufficiently strong disorder.


\section{Coulomb drag at the nanoscale}
\label{nano}


The effects of Coulomb interaction are especially pronounced at the
nanoscale. In quantum dot devices one can utilize the Coulomb-modified
Fano resonance to detect the electric charge \cite{mar}. Two-level pulse
technique was used to detect individual electron spin
\cite{kou}. Quantum dots were also used as high-frequency noise
detectors \cite{kou2}. Transport measurements on adjacent but
electrically isolated quantum point contacts (QPCs) exhibit a
counterflow of electrons [i.e. detector current flowing in the
  direction opposite to the driving current \cite{khr}]. In nanosize
CdSe-CdS semiconductor tetapods \cite{mau}, Coulomb drag-like effects
lead to photoluminescent emission.

Theoretically, Coulomb drag in a system of two electrically isolated
QPCs was considered in \cite{le2}. Within the linear response the drag
mechanism was found to be similar to that in the bulk 2D electron
systems. Remarkably, already for seemingly modest drive voltages (much
smaller than temperature) the system crosses over to the nonlinear
regime, where the effect is dominated by the excess shot noise of the
drive circuit. Nonlinear transport was also found to be crucial for
drag effects in a system of parallel quantum dots \cite{mol}. An
exciting new development is the proposal to use the drag effects to
study transport properties of polaritons in optical cavities and, in
particular, their superfluidity \cite{loz,loz2}.


\subsection{Quantum dots and quantum point contacts}
\label{dots}


Interactively coupled mesoscopic and nanoscale circuits, such as
quantum wires \cite{de2,deb,mori,ya1,lar}, quantum dots
\cite{agu,kou2} or point contacts \cite{khr2,khr}, provided new
fruitful ways of studying Coulomb drag phenomena and revealed a
plethora of interesting physics. These devices typically have
dimensions smaller than the temperature length ${L_T=v_F/T}$ and
voltage-related length scale ${L_V=v_F/(eV)}$, and differ
substantially form their two-dimensional quantum-well counterparts in
several important ways. (i) The strength of Coulomb interaction is
naturally enhanced by reducing system size that should lead to more
profound dragging effect. (ii) Transmission across the device in the
drag (drive) circuit or both can be efficiently controlled by the gate
voltages that allows to open quantum conduction channels one by
one. (iii) The electron-hole symmetry in such devices is broken much
stronger than in bulk systems. In mesoscopic devices this is due to a
random configurations of impurities, while in the quantum nanocircuits
the effect is due to the energy dependence of transmission
coefficients in Landauer picture of transport
\cite{land1,land2,butt}. (iv) Because of the above reasons, the
quantum circuits may be easily driven out of the linear response
domain and corresponding voltage scale is parametrically smaller than
the temperature. (v) Ultimately, the mere mechanism of drag in the
nonlinear regime is different and governed by the quantum noise
fluctuations.

\begin{figure}
\begin{center}
\includegraphics[width=0.97\linewidth]{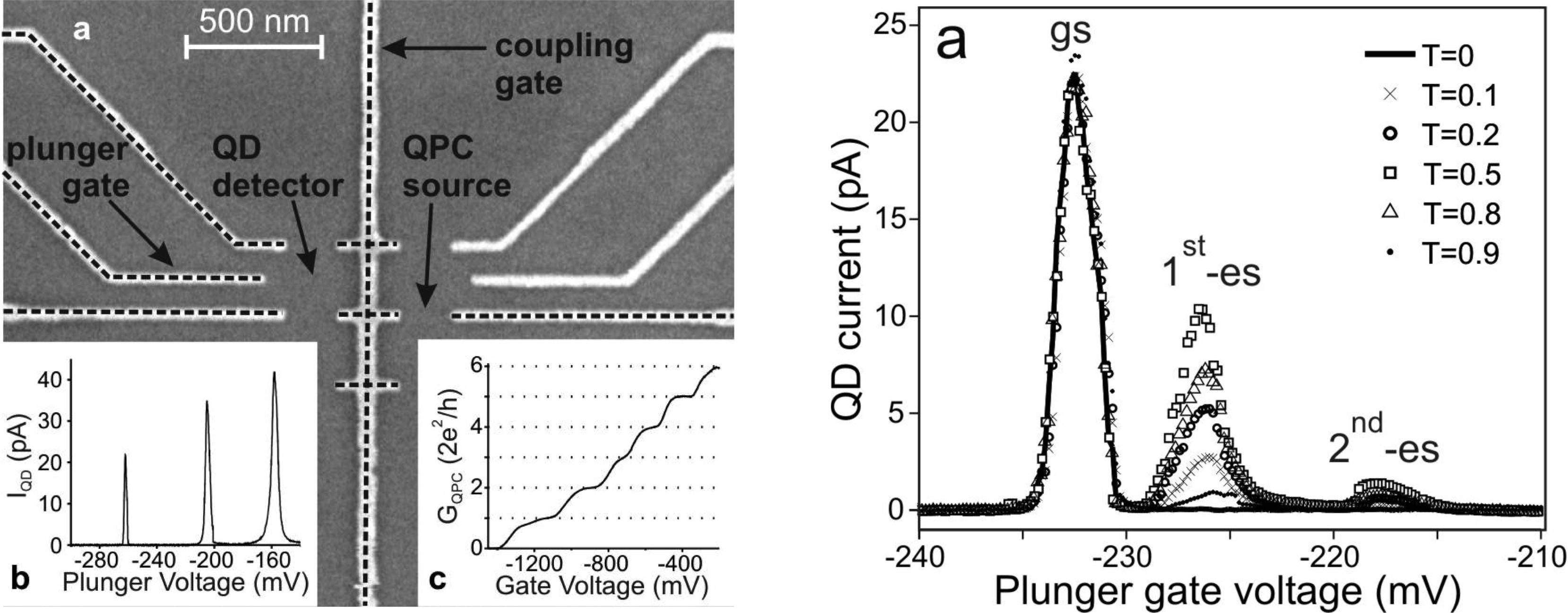}
\end{center}
\caption{Left panel inset-(a) represents scanning electron micrograph
  of the gate structure defined on top of the semiconductor
  heterostructure. The gates highlighted by dashed lines are used to
  define a quantum dot (QD) on the left and a quantum point contact
  (QPC) on the right. Inset-(b) shows current $I_{QD}$ versus plunger
  gate voltage whereas inset-(c) displays QPC conductance $G_{QPC}$ as
  a function of the gate voltage. In such device the QPC is used as a
  noise generator and the QD as a detector. Right panel shows current
  through the QD, as a function of the plunger gate voltage, under the
  influence of shot noise generated by the QPC with characteristic
  peaks. [Reproduced from \textcite{kou2}.]}
\label{Fig-Drag-QPC}
\end{figure}

The most peculiar feature of the observed Coulomb drag in such systems
was that the drag current exhibited maxima for specific values of the
gate voltage, where the drive circuit was tuned to an opening of
another conductance channel, see Fig.~\ref{Fig-Drag-QPC} for the
illustration. This hinted the importance of the electron shot noise in
the drive circuit, which was known to exhibit a qualitatively similar
behavior~\cite{les,res}. Indeed, electron current shot noise power is
proportional to the product of the transmission and reflection
coefficients that is peaked between the conductance plateaus. It was
argued early on that drag may be interpreted as a rectification of
nearly equilibrium classical thermal fluctuations in the drive
circuit~\cite{kor}. The extension of this idea to rectification of the
quantum shot noise was plausible and happened to be correct in a
certain regime. The subtlety of this picture was that such a
rectification is only possible due to electron-hole asymmetry in both
circuits, otherwise drag currents of electrons and holes cancel each
other. The mismatch between transmission probabilities of electron and
hole excitations is maximal at the verge of an opening of the new
conduction channel, which implies that spikes of drag conductance may
originate form the asymmetry alone rather than shot noise.

In order to get an insight into these delicate details consider the
linear response regime when drag conductance $g_D$ can be expressed as
follows \cite{le2}
\begin{equation}
\label{Drag-QPC}
g_D=\int\frac{d\omega}{8\pi T}\frac{|Z_{12}(\omega)|^2}{\omega^2}
\frac{\Gamma_1(\omega)\Gamma_2(\omega)}{\sinh^2(\omega/2T)}.
\end{equation}   
Here ${Z_{12}(\omega)}$ is the interactively-induced trans-impedance
relating local fluctuating currents and voltages between the circuits
\cite{gei}. The corresponding rectification coefficients
are given explicitly by
\begin{equation}
\label{Gamma-QPC}
\Gamma_i(\omega)=\frac{2e}{R_Q}\sum_n\!\!\int\!\! d\epsilon
[f(\epsilon_-)-f(\epsilon_+)]
[|\mathbf{t}_{i n}(\epsilon_+)|^2-|\mathbf{t}_{i n}(\epsilon_-)|^2]
\end{equation} 
where ${R_Q=2\pi/e^2}$ is the quantum of resistance,
$\epsilon_{\pm}=\epsilon\pm\omega/2$, $f(\epsilon)$ is the Fermi
distribution function, and $|\mathbf{t}_{i n}|^2$ is the energy
dependent transmission coefficient in the transversal channel $n$ of
the circuit $i=1,2$. This expression admits a transparent
interpretation: potential fluctuations with frequency $\omega$, say on
the left of the quantum point contact, create electron-hole pairs with
energies $\epsilon_{\pm}$ on the branch of right moving
particles. Consequently the electrons can pass through the quantum
point contact with the probability $|\mathbf{t}_{i
  n}(\epsilon_+)|^2$ , while the holes with the probability
$|\mathbf{t}_{i n}(\epsilon_-)|^2$. The difference between the
two gives the net current flowing across the contact while the Fermi
functions in Eq.~(\ref{Gamma-QPC}) take care of the statistical
occupation of participating scattering states. Notice that unlike in
the Landauer formula for conductance of a single quantum point contact
where transmissions can be treated as being energy independent, the
energy dependence of these probabilities in the drag formula is
crucial in order to have the asymmetry between electrons and holes,
and thus nonzero rectification $\Gamma_i(\omega)$. A particular
functional dependence of $\Gamma$ on frequency depends on a model and
details of device circuitry. It is instructive to focus on a limit of
a single partially open channel in a smooth adiabatic quantum point
contact. One may think then of the potential scattering barrier across
it as being practically parabolic. In such a case its transmission
probability is given by
\begin{equation}\label{T-QPC}
|\mathbf{t}_{i n}(\epsilon)|^2=[\exp[(eV_{gi}-\epsilon)/\Delta_i]+1]^{-1}
\end{equation}
where $\Delta_i$ is an energy scale associated with the curvature
of the parabolic barrier in the point contact $i$, while gate
voltages $V_{gi}$ move the top of the barrier relative to the
Fermi energy within each of the point contact. This form of
transmission was used to explain quantum point contact conductance
quantization \cite{glaz} and it turns out to be useful in
application to the Coulomb drag problem. For the low temperature limit
$T\ll\Delta_i$ using Eq.~(\ref{T-QPC}) in Eq.~(\ref{Gamma-QPC})
and carrying out energy integration yields
\begin{equation}
\Gamma_i(\omega)=\frac{2e\Delta_i}{R_Q}
\ln\left[\frac{\cosh(eV_{gi}/\Delta_i)+\cosh(\omega/\Delta_i)}{\cosh(eV_{gi}/\Delta_i)+1}\right]. 
\end{equation}
In the opposite limit when $T\gg\Delta_i$ one should replace
$\Delta_i\to T$. One should notice that for small frequency
$\Gamma_i\propto \omega^2$ whereas trans-impedance
$Z_{12}(\omega)$ is practically independent of frequency in this limit
since its characteristic scale is typically set by the inverse
$RC$-time of circuits.  Assuming that
$T\ll\mathrm{max}\{\Delta_i,\tau^{-1}_{RC}\}$ one arrives at
\begin{equation}
\frac{g_D}{g_Q}=\frac{\pi^2u^2}{6}\frac{T^2}{\Delta_1\Delta_2}
\frac{1}{\cosh^2(eV_{g1}/\Delta_1)\cosh^2(eV_{g2}/\Delta_2)},
\end{equation}
where $u=Z_{12}(0)/R_Q$.  The resulting expression for the drag
conductance exhibits peak as a function of gate voltage in drag or
drive quantum point contact. Yet at this level it has nothing to do
with the shot noise peaks, but rather reflects rectification of
near-equilibrium thermal fluctuations (hence proportionality to $T^2$)
along with the electron-hole asymmetry (hence a nonmonotonic
dependence on $V_{gi}$). However, one should realize that the
crossover to the nonlinear regime of transport in such devises can
occur at rather low voltages $eV^*\sim T^2/\Delta_i$ such that
Eq.~(\ref{Drag-QPC}) becomes inapplicable already at $V>V^*$. More
general considerations by \cite{le2,chud,sab} revealed that for the
out of equilibrium nonlinear regime the drag current is due to the
rectification of the quantum shot noise and hence proportional to the
Fano factor $\sum_{n}|\mathbf{t}_{ni}|^2[1-|\mathbf{t}_{ni}|^2]$. It
again exhibits a generic nonmonotonic behavior of drag with multiple
peaks but for the entirely different reason independent of asymmetry
factor. Nonlinear transport was also found to be crucial for drag
effects in a system of parallel quantum dots \cite{mol}.
    
Drag phenomena in quantum circuits can be naturally connected to our
earlier discussion of drag in mesoscopic systems in Sec.~\ref{mf}. Indeed,
one or both circuits may be represented by a multichannel
quasi-one-dimensional (or two-dimensional) mesoscopic sample. In this
case $\sum_n|\mathbf{t}_n(\epsilon)|^2=g(\epsilon)$ is a dimensionless
(in units $R^{-1}_Q$) conductance of the sample as a function of its
Fermi energy. As discussed above, such conductance exhibits universal
fluctuations, that is $g(\epsilon)=g+\delta g(\epsilon)$, where
$g\gg1$ is an average conductance and $\delta g(\epsilon)\sim 1$ is a
sample and energy-dependent fluctuating part. Since the characteristic
scale of the energy dependence of the fluctuating part is the Thouless
energy $E_{T}$ one naturally finds from Eq.~(\ref{Gamma-QPC}) that
corresponding mesoscopic fluctuations of the rectification coefficient
are of the order
\begin{equation}
\Gamma(\omega)\sim\pm\frac{e}{R_Q}\frac{\omega^2}{E_T}
\end{equation}   
This result ultimately leads to the an estimate of the variance of
drag in the form of Eq.~(\ref{ds0d})

Quantum Coulomb drag circuits provide rich platform to explore
nanoscale transport far beyond ideas of using them for high-frequency
noise sensing. In particular, a different drag effect may also be
observed in the absence of any drive current if one nanocircuit is
made hotter than the other - the cold circuit is expected to rectify
the thermal charge fluctuations of the hot circuit
\cite{soth}. Furthermore, interactively coupled devices
provide unique tools to test nonlinear fluctuation-dissipation
relations and its closely related Onsager symmetry relations in the
far from equilibrium conditions when detailed balance is explicitly
broken \cite{sab,cue}. 

Other intriguing examples include nanosize CdSe-CdS semiconductor
tetapods \cite{mau} where Coulomb drag-like effects lead to
photoluminescent emission. As an alternative to optical probes,
electrical read-out of a single electron spin becomes possible in a
Coulomb drag-like devices of interactively coupled QPC and QD
\cite{kou}.


\subsection{Optical cavities}
\label{opt}


Coulomb interaction is not exclusive to electrons and can couple any
charges. Moreover, even neutral, composite objects may interact with
charges by means of an effective ``polarization'' or ``charge-dipole''
interaction \cite{marge}, which ultimately stems from the Coulomb
interaction between an external charge and individual charged
constituents of the composite object. In particular, long-ranged
interactions between spatially separated electrons and polaritons may
lead to interesting drag effects \cite{kul,loz,loz2} that can be used,
e.g., for designing electrically controlled optical switches
\cite{ber}.

Two-dimensional excitonic polaritons have been a subject of intensive
research \cite{fran,amop,bali,snoke}. These excitations appear as a
result of resonant exciton-photon interaction in a system consisting
of an optical microcavity and a quantum well embedded within the
cavity. The lower polariton branch is characterized by extremely small
effective mass raising the possibility of achieving the Bose-Einstein
condensation and superfluidity at relatively high temperatures
\cite{bali,lit}.

The optically excited excitons in microcavities should not be confused
with the spontaneously formed excitons in double quantum wells
discussed in Sec.~\ref{qhe1}. In particular, these excitons can be
excited by laser pumping in the single quantum well embedded within
the cavity. Now, if a second quantum well is added to the device
\cite{loz}, then Coulomb interaction binding electrons and holes into
excitons can be screened \cite{guba,gfin} by a 2DEG populating the
second well. As a result, the excitonic binding energy is reduced and
as the density of the 2DEG approaches a certain critical value, the
excitons may disappear altogether. The excitonic collapse manifests
itself through disappearance of the corresponding line in the
photoluminescence spectrum.

Keeping the electron density below the above critical value, one
obtains a system containing coexisting, spatially separated excitons
and electrons. The effective interaction between electrons and
excitons was considered in \textcite{loznik}. This interaction leads
to mutual friction between the two systems that can be observed by
selectively exciting one of them by external probes.

By focusing laser pumping on a particular region within the cavity,
one can generate a gradient of exciton and polariton densities. These
gradients induce a flow of both polaritons and excitons. The
long-range interaction between the excitons (or the exciton component
of the polaritons) may transfer energy and momentum to the electronic
system in the second quantum well, generating an electric current or
inducing voltage, similarly to the standard drag effect discussed in 
Sec.~\ref{drag0}.

\begin{figure}
\begin{center}
\includegraphics[width=0.97\linewidth]{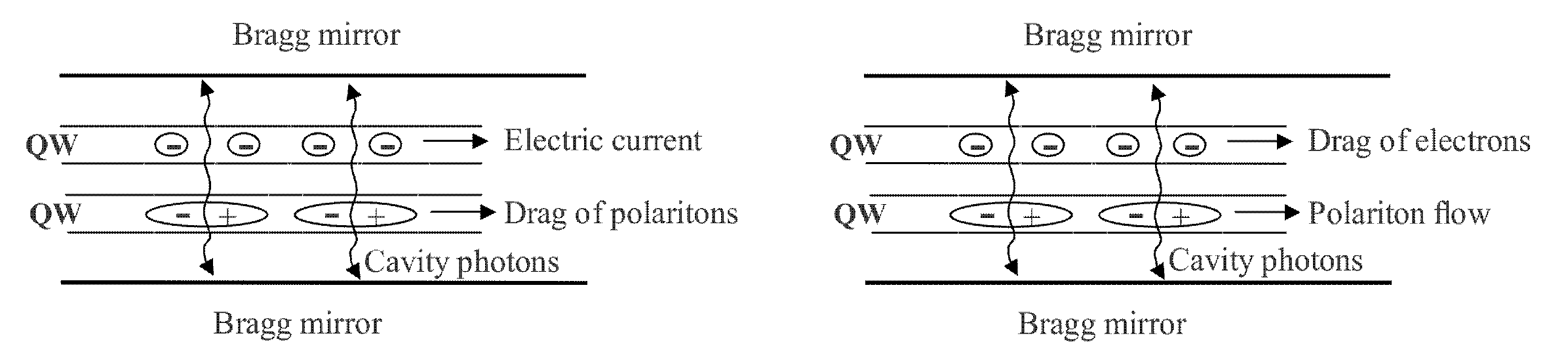}
\end{center}
\caption{Left: quasiparticle flow in the cavity polariton subsystem
  induced by the electric current in the 2DEG at low
  temperatures. Right: electric current in the 2DEG induced by the
  optically excited flow in the polariton subsystem. [Reproduced from
    \textcite{loz2}.] }
\label{optics}
\end{figure}

Alternatively, one can drive a current through the 2DEG. In this case,
the mutual friction will lead to the appearance of the exciton flow.
These excitons are entangled with cavity photons and their flow will
create a flow of polaritons. In other words, the long-ranged
electron-exciton interaction allows one to effectively ``move'' the
cavity photons by applying electric current to the 2DEG
\cite{loz,loz2}. The drag effects in microcavities are schematically
illustrated in Fig.~\ref{optics}.

\begin{figure}
\begin{center}
\includegraphics[width=0.8\linewidth]{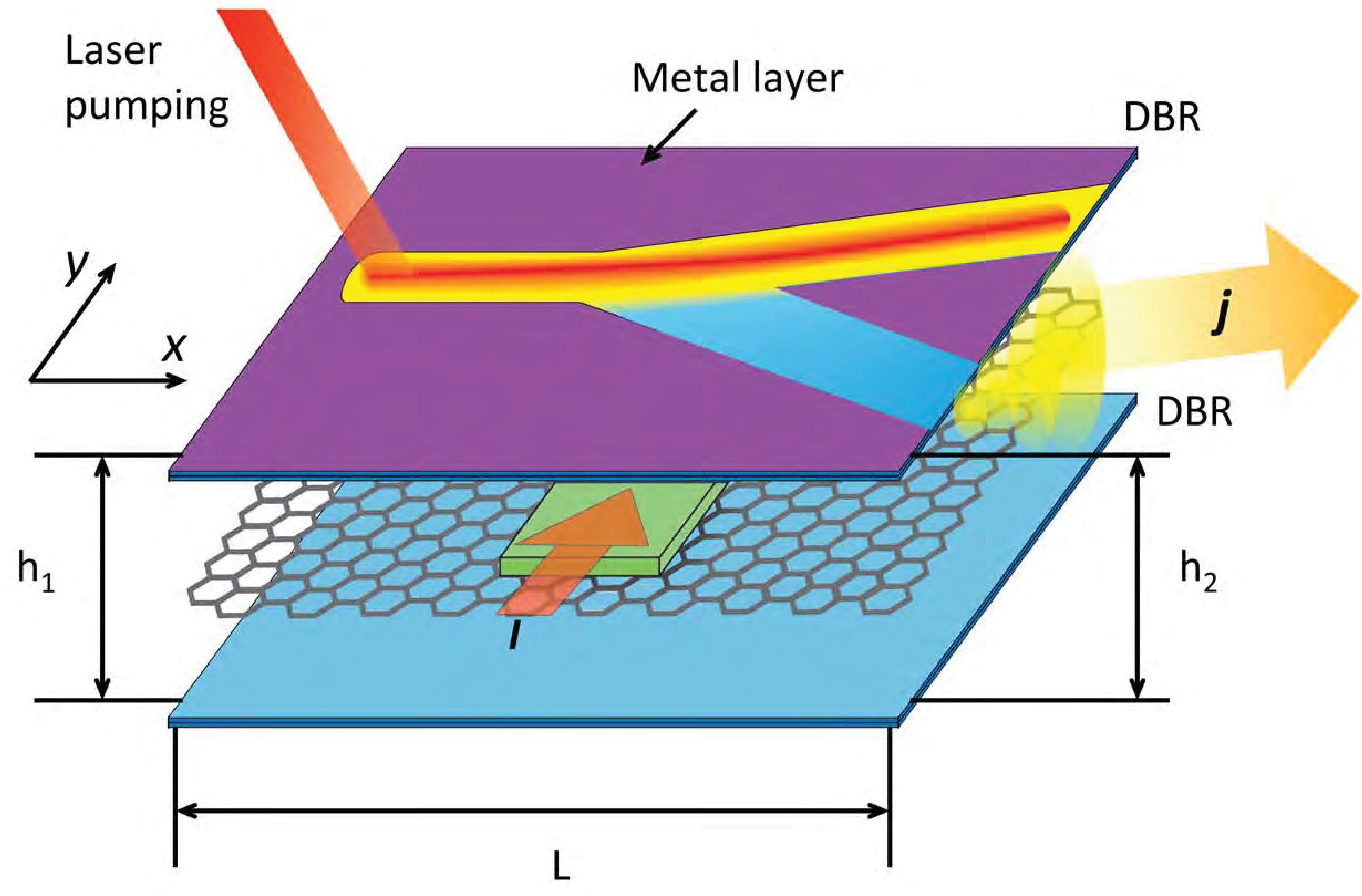}
\end{center}
\caption{(Color online) Schematic of the wedge-shaped microcavity
  formed by two distributed Bragg reflectors (DBR) that encompasses
  the embedded quantum wells The excitons are located in the quantum
  well (gray) between the reflectors. A metal layer deposited on the
  upper DBR creates a Y-shaped potential energy landscape for the
  polaritons. The driving current runs perpendicularly to the stem of
  the channel in the quantum well (green). Reprinted with permission
  from \textcite{ber}. Copyright (2014) American Chemical Society.]}
\label{switch}
\end{figure}

Recently, \textcite{ber} have proposed to use the drag effect in
optical cavities for building an electrically controlled optical
switch, see Fig.~\ref{switch}. The polaritons are assumed to be
created at a constant rate by external laser pumping. The wedge-like
shape of the microcavity is chosen in order the create a force driving
the polaritons along the cavity towards the Y-junction. Without the
drag effect, the polariton flux is distributed equally between the two
branches of the junction. Driving an electric current through a second
quantum well (shown in green in Fig.~\ref{switch}) results in a drag
force in the junction region that effectively redistributes the
polaritons flux between the branches. \textcite{ber} find that for
realistic parameters of the device one can achieve 90\% accuracy of
the switching of the polariton flow.

\section{Coulomb drag between parallel nanowires}
\label{nw}


It is well-known that physics of electrons confined to one spatial
dimension (1D) is dominated by interactions. Coulomb drag between two
closely spaced but electrically isolated quantum wires was used to
observe Wigner crystallization \cite{ya1,yam,ya2} and Luttinger-liquid
effects \cite{deb,la2,lar2}. The effect was also used to study 1D
sub-bands in quasi-1D wires \cite{de2,lar}.

Early theoretical work on drag between 1D systems
\cite{huf,pev,gur,gur2,ra21,ra2} was based on the Fermi-liquid
approach and targeted multiple-channel wires at high enough
temperatures, where electron correlation effects (other than
screening) are not important. \textcite{ta22} considered the role of
disorder. It is however well-known, that the Fermi-liquid theory fails
for purely 1D systems, i.e. single-channel wires \cite{giam}, quasi-1D
wires with single 1D subband occupancy \cite{lar2}, and systems
comprising a small number of coupled 1D channels. Coulomb drag between
two Luttinger liquids with point-like interaction region was discussed
in \textcite{fle,koe}. \textcite{naz} considered two independent
Luttinger liquids coupled by interwire
backscattering. \textcite{slm,slm2} used Bethe-Ansatz methods to solve
the problem of two wires coupled by a particular $\delta$-function
potential. Especially interesting is the prediction of the
Mott-insulator--type state corresponding to formation of two
interlocked charge density waves (CDW) in quantum wires \cite{kle,fks}.

A theory of Coulomb drag based on the Tomonaga-Luttinger liquid (TLL)
theory \cite{tom,lut,hal,hal2} predicts a behavior that qualitatively
deviates from that in higher dimensions. Below a certain crossover
scale $T^*$, the drag resistivity between {\it infinitely long quantum
wires of equal electron density} is predicted to increase exponentially
with decreasing temperature \cite{kle}
\begin{equation}
\label{dr-cdw}
\rho_D\sim\rho_T\exp(\Delta/T). 
\end{equation}  
The energy gap $\Delta$ and crossover temperature $T^*$ are
complicated functions of the interwire distance $d$, width of wires
$w$, effective Bohr radius $a_B$ of the host material, and electron
density $n$. For widely separated wires (${k_Fd\gg1}$) they are
exponentially suppressed
\begin{equation}
\label{Tcdw}
\Delta\sim T^*\sim E_F\exp\left[-k_Fd/(1-K)\right],
\end{equation}
and the drag resistivity exhibits the high-temperature power-law
behavior\footnote{In 1D, ${\rho_D=-\lim_{I_1\to0}(1/L)(dV_2/dI_1)}$,
  is the drag resistivity per unit length \cite{kle,pus}.}
\begin{equation}
\label{rdht}
\rho_D \sim (h/e^2)k_F\lambda^2(T/E_F)^{4K-3},
\end{equation}
for all practically relevant scales. Here $K$ is the TLL interaction
parameter in the relative charge sector determined by the difference
of the small-momentum intra- and interwire couplings and $\lambda$ is
dimensionless interwire backscattering potential strength.

The physical picture behind Eq.~(\ref{dr-cdw}) is that at low
temperatures ${T<T^*}$ the electrons in both wires form zigzag-ordered
interlocked charge density wave (CDW). Then a relative charge
displacement can be created only by overcoming a potential barrier,
which ultimately translates into transport via activation, and
consequently into Arrhenius-like behavior of drag.

For short wires, \textcite{kle} report a qualitatively different
behavior. Here the CDW in one wire may slip as a whole relative to the
CDW in the other wire. These instantaneous slips may stem from either
thermal fluctuations or tunneling events. The latter leads to the drag
resistance that tends to a finite, but exponentially large (in the
wire length $L$) value as ${T\rightarrow0}$. In contrast,
\textcite{poa} find a vanishing drag resistance, ${\rho_D\sim{T}^2}$,
regardless of whether the CDW is formed or not.

For wires with different electron densities, \textcite{fks}
find that the drag resistance (\ref{dr-cdw}) is suppressed by an
additional exponential factor ${\exp(-|\delta\mu|/T)}$, where
${\delta\mu=\mu_1-\mu_2}$ is the difference between the chemical
potentials $\mu_i$ in the two wires. The high-temperature result
(\ref{rdht}) also becomes exponentially suppressed as soon as
$|\delta\mu|$ exceeds the temperature.

Allowing for a spin degree of freedom adds extra complexity to the
problem, since the system might be unstable towards gap opening in the
spin sectors\footnote{For a comprehensive discussion of ground state
  properties of capacitively coupled 1D systems see
  \textcite{giam,carr}.}. If this does not happen (or at temperatures
exceeding the spin gaps), the system shows the same qualitative
behavior as above, but the exponent in Eq.~(\ref{rdht}) changes to
${2K-1}$.  However, if the single wires develop spin gaps, the drag
resistivity vanishes at ${T=0}$ \cite{kle}.

At temperatures above $T^*$, the charge sector is gapless and the
system can be described as two coupled wires in the TLL phase. For
quasiparticles with linear dispersion the only process contributing to
drag is the interwire backscattering characterized by large momentum
transfers ${q\sim2k_F}$. This process can be described by the
usual drag formula (\ref{td}), where one typically assumes the
nonlinear susceptibility to be proportional to the imaginary part of
the polarization operator \cite{fhb,pus}:
\begin{equation}
\label{dr-1D}
\rho_D=\frac{h}{e^2}\int\!\frac{dqd\omega}{4\pi^3}\frac{q^2V^2_q}{n^2T}
\frac{[\mathrm{Im}\Pi(q,\omega)]^2}{\sinh^2(\omega/2T)},
\end{equation}
where $V_q$ describes the interwire interaction. In the limit
${qd\gg1}$, the asymptotic form of $V_q$ is given by
${V_q=(e^2/\epsilon)\sqrt{2\pi/(qd)}\exp(-qd)}$. The polarization
  operator for the TLL model is known \cite{giam}. For spinless
  fermions, the spectral weight of $2k_F$ density fluctuations is
  given by
\begin{eqnarray}
\label{Pi-LL-2kF}
&&
\mathrm{Im}\Pi(q_\pm,\omega)=
-\frac{\sin\pi K}{4\pi^2u}\left(\frac{2\pi\alpha T}{u}\right)^{2K-2}
\\
&&
\nonumber\\
&&
\qquad\qquad\qquad
\times
B\!\left(\!\frac{K}{2}-\frac{i(\omega-uq_\pm)}{4\pi T},1-K\!\right)
\nonumber\\
&&
\nonumber\\
&&
\qquad\qquad\qquad
\times
B\!\left(\!\frac{K}{2}-\frac{i(\omega+uq_\pm)}{4\pi T},1-K\!\right),
\nonumber
\end{eqnarray}  
where ${\alpha\sim k^{-1}_F}$ is the short-distance cut-off of the TLL
theory, ${q_\pm=q\pm2k_F}$, $u$ is the renormalized Fermi velocity,
and $B(x,y)$ is the Euler beta-function. Using Eq.~(\ref{Pi-LL-2kF})
in the expression (\ref{dr-1D}) one recovers Eq.~(\ref{rdht}), which
was obtained by \textcite{kle} by means of a renormalization group
analysis. In the perturbative approach, the interaction parameter
$\lambda$ in Eq.~(\ref{rdht}) is given by ${\lambda=V_{2k_F}/v_F}$.
This leads to the exponential dependence of $\rho_D$ on distance
separating the wires [since ${V_{2k_F}\propto\exp(-2k_Fd)}$]. The
regime of spin-incoherent Luttinger liquid and effect of disorder
modify temperature dependence of Eq.~(\ref{rdht}) \cite{fhb}. In
the weakly interacting limit (${K\simeq1}$) the drag resistivity is
expected to grow linearly with temperature \cite{huf,pev}.

In recent years, a lot of the attention was devoted to 1D liquids with
nonlinear dispersion [for reviews on this topic see
  \textcite{imam,desh,matveev} ]. In the TLL theory, the curvature of
the quasiparticle spectrum is described by an irrelevant operator (in
the renormalization group sense). However, at high enough temperatures
it might lead to important effects and even mask the pure Luttinger
behavior.
In the context of Coulomb drag
\cite{pus,ari,roz,roz2,dgp,sela}, this is particularly important since
nonlinearity of the band kinematically allows drag with small momentum
transfer, ${q\sim{T}/v_F\ll{k}_F}$.

Analytic calculation of the dynamical structure factor
$\mathrm{Im}\Pi(q,\omega)$ for arbitrary interactions and nonlinear
dispersion is a major challenge. However, such calculation is readily
available in the case of weakly interacting electrons. At finite
temperatures, but with the accuracy of the order ${T\ll{m}v^2_F}$,
the one-loop diagram yields
\begin{eqnarray}
\mathrm{Im}\Pi(q,\omega)=\frac{m}{4k}
\frac{\sinh(\omega/2T)}{\cosh(qv_F\xi_+/2T)\cosh(qv_F\xi_-/2T)},
\qquad\,\,
\end{eqnarray}
where ${\xi_\pm=2m(\omega-v_Fq)/q^2\pm1}$. It is now tempting to
follow the conventional path and use this result for $\mathrm{Im}\Pi$
in the expression for the drag (\ref{dr-1D}) to obtain
\begin{equation}
\label{dr-1D-fs}
\rho_D\simeq(h k_F/e^2)(V_0/v_F)^2(T/E_F)^2,
\end{equation}
with the conclusion that curvature effects restore the Fermi-liquid
behavior of drag in 1D wires; furthermore, the contribution
(\ref{dr-1D-fs}) would dominate over the backscattering component
(\ref{rdht}) already at temperatures above ${T>E_Fe^{-4k_Fd}}$ [in
  Eq.~(\ref{dr-1D-fs}) $V_0$ should be understood as
  ${V_{q\sim{T}/v_F}}$]. At even higher temperatures, ${(v/d)<T<E_F}$,
the same approach yields a saturating drag resistivity,
${\rho_D\sim(\hbar{k}_F/e^2)(V_0/v_F)^2(v/dE_F)^2}$, followed by a
falloff ${\rho_D\propto1/T^{3/2}}$ at ${T>E_F}$. For nonidentical
wires there appears an additional energy scale,
${T_\delta=k_F|\delta{v}|}$ describing splitting between symmetric and
antisymmetric plasmons modes in the double-wire system, which is
determined by the difference between Fermi velocities in the wires
${\delta v=v_{F1}-v_{F2}}$. Assuming that ${T_\delta\ll{T}_d}$,
Eq. (\ref{dr-1D-fs}) holds only for ${T_{\delta}<T<T_d}$, whereas
below $T_\delta$ drag resistivity due to forward scattering decreases
as ${\rho_D\propto{T}^5}$ with lowering temperature.

However, as shown by \textcite{dgp} the above conclusions about the
forward scattering contribution to drag may be premature. The reason
is subtle: the expression~(\ref{dr-1D}) was derived under the tacit
assumption that the intralayer relaxation processes due to
electron-electron interaction are faster than the interwire momentum
transfer. Now, in purely 1D systems relaxation is determined by
three-body collisions \cite{gl2,mick,lev3,lev4} as inelastic two-body
interaction is forbidden by energy and momentum conservation. Same
kinematic restrictions require that intrawire backscattering
responsible for equilibration involves states deep at the bottom of
the band. Because of the Pauli statistics, the probability to find
such a state unoccupied is exponentially small. Consequently, the
equilibration rate $\tau^{-1}_{eq}$ in 1D is exponentially suppressed,
${\tau^{-1}_{eq}\propto{e}^{-E_F/T}}$, and the
expressions~(\ref{dr-1D}) and (\ref{dr-1D-fs}) are difficult to justify.

At the same time, interwire backscattering with small momentum
transfer ${q\sim T/V_F\ll{k}_F}$ is also allowed in 1D systems with
nonlinear spectrum. This process involves a pair of scattering states:
one near the Fermi level and another at the bottom of the
band. \textcite{dgp} found a solution of two coupled kinetic equations
[cf. Eqs.~(\ref{beq0})] yielding the drag resistivity in the form
\begin{equation}
\label{rd1d2}
\rho_D\simeq\frac{\hbar k_F}{e^2}
\left(\frac{V_0}{v_F}\right)^2\frac{1}{k_Fd}
\frac{T_d}{T}\sqrt{\frac{E_F}{T}}e^{-2E_F/T}.
\end{equation}     
By comparing the exponential factors in Eqs.~(\ref{rd1d2}) and
(\ref{rdht}), one can see that backscattering-induced drag friction
due to soft collision (namely collisions with small momentum transfer)
dominate over direct backscattering with $2k_F$ momentum transfer at
temperatures $T>T_d$. This is despite the fact that the contribution
of soft collisions being strongly suppressed compared to
Eq.~(\ref{dr-1D-fs}). In addition, depending on the delicate interplay
between the rates of two-particle interwire backscattering with small
momentum transfer and triple-body intrawire chirality changing soft
collisions, one may identify a parameter regime where
${\rho_D\propto{e}^{-E_F/T}}$ is suppressed with only one exponential
factor, namely \cite{dgp}.

\begin{figure}
\includegraphics[width=8cm]{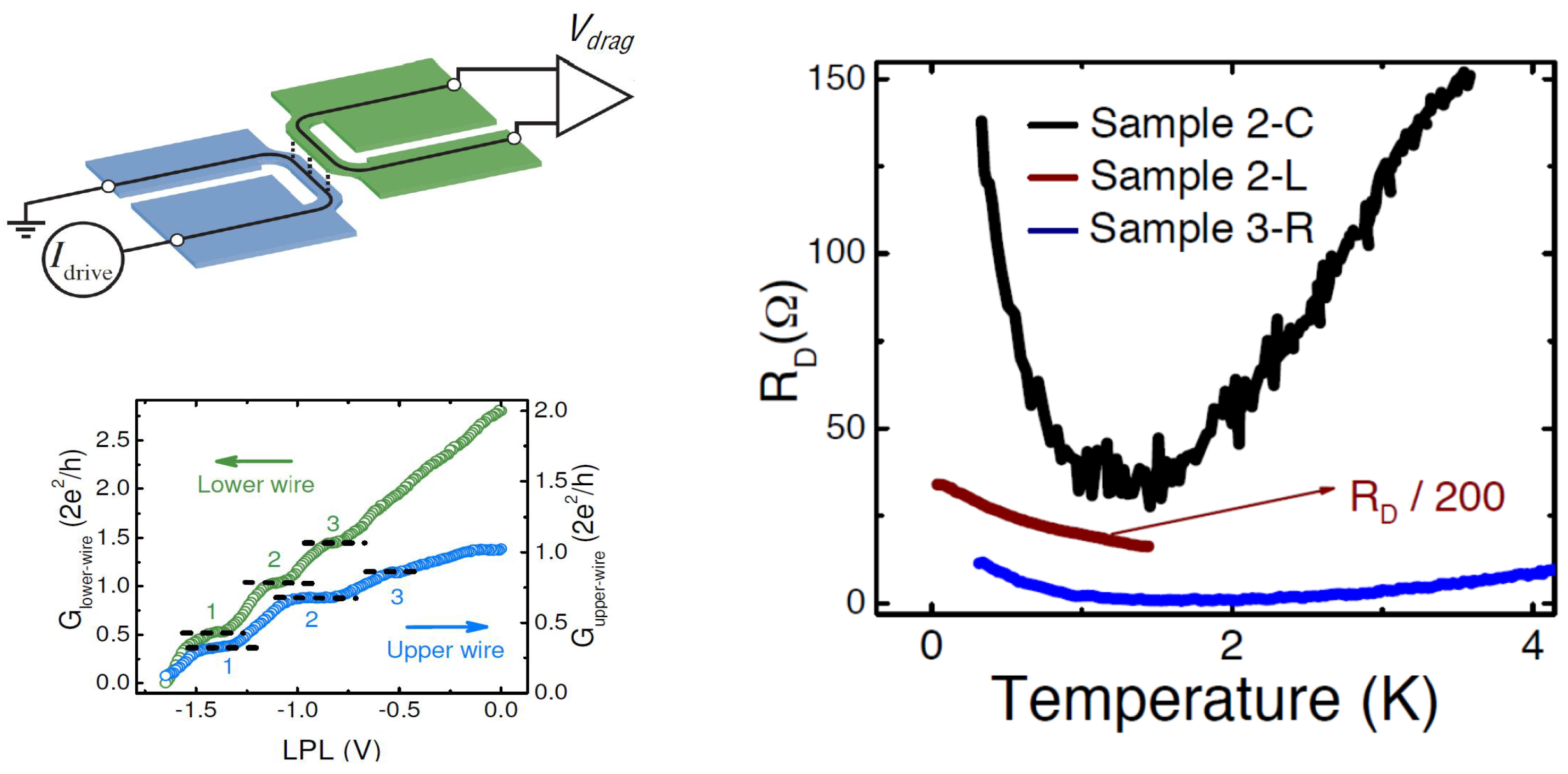}
\caption{(Color online) Left: measurement schematic (top) and
  single-wire conductance quantization (bottom). Right: temperature
  dependence of the drag signal for three different samples (the
  magnitude of $\rho_D$ in sample 2-L is divided by 200 for
  visibility). For samples 2-L and 2-C, the temperature dependence was
  taken with no more than one 1D subband occupancy in each wire,
  whereas the number of 1D subband occupied in sample 3-R is known to
  be bounded by ${0<N_{drive}\leq2}$ and ${0<N_{drag}\leq3}$. [From
    \textcite{lar2}. Reprinted with permission from AAAS.]}
\label{Fig-Drag-1D}
\end{figure}

Coulomb drag between true 1D systems was recently observed by
\textcite{lar2} in a system of vertically integrated quantum wires
where each wire has less than one 1D subband occupied. The most
striking theoretical prediction, i.e., the upturn in the temperature
dependence was revealed below the crossover temperature
${T^*\sim1.6}$K, see Fig.~\ref{Fig-Drag-1D}. However, a quantitative
comparison between the data and above theoretical results proved to be
difficult. Using the experimental estimates for the carrier density
${n_{1D}=\sqrt{n_{2D}}}$ and interwire distance ${d\simeq40}$nm, one
arrives at ${k_Fd\sim2}$. Then, from Eq.~(\ref{Tcdw}) one finds the
values for the Luttinger parameter ${K\simeq0.1-0.2}$ (for samples
3-R and 2-C) corresponding to very strong interaction that is beyond
the applicability of the bosonization theory of \textcite{kle}. On
the other hand, fitting the high-temperature data to the power-law
behavior (\ref{rdht}) yields ${K\simeq1.5}$. This estimate, however,
should be approached with caution, since Eq.~(\ref{rdht}) was
derived for identical wires which is not the case in experiment,
where electronic densities in the parent 2D layers differ by about
20\%. In that case one expects \cite{fks} an exponential suppression
of $\rho_D$. All these issues remain to be clarified both
theoretically and experimentally.

\section{Novel many-body states in double-layer systems}
\label{qhe}

When a double-layer system is subjected to a strong magnetic field,
the standard theoretical description of Coulomb drag
\cite{jho,mac,kor} fails: in contrast to naive expectations, numerous
experiments \cite{hi3,rub,pat,ru2,ru22,fen,hi2,jo1,lok,lo1,pi22} show
significant dependence of the measured drag resistivity $\rho_D$ on
the applied field, especially in the extreme quantum regime
\cite{murph,lil,nan}.

Further experiments revealed the existence of novel quantum Hall
states that are specific to bilayer systems and have no analog in
single-layer samples. Early work in this direction was reviewed in
\textcite{eis,eis1}. Remarkably, the bilayer many-body states
exhibiting the quantum Hall effect \cite{murph} may at the same time
support a condensate of indirect (or interlayer) excitons
\cite{fin,nan,wie}. An interlayer exciton is a bound pair of an
electron from one layer and a hole from another layer of the
device. The exciton carries no electric charge. Nevertheless, exciton
transport (especially in the superfluid state) leads to interesting
electrical effects. The experimental situation in the field is
reviewed in \textcite{eim,eis2}. Here we focus on the manifestations
of this exciting new physics in the drag measurements.

\subsection{Quantum Hall Effect in double-layer systems}
\label{qhe1}

In a seminal paper, \textcite{halp} has suggested a generalization of
the Laughlin wave function for the analysis of multi-component
systems. The simplest example of an extra degree of freedom that can
be accounted for using this approach is the electron spin. A
double-layer system provides another example, which is similar to the
spin-$1/2$ in some respects and is significantly different in other.
The two possible values of the layer index can be represented by the
two orientations of a pseudo-spin \cite{moon,ads}. However, unlike the real
spin, the double-layer system does not possess the $SU(2)$ symmetry
due to the difference between the intra- and interlayer matrix
elements of the Coulomb interaction. Consequently, in the double-layer
system the energy eigenstates do not have to be eigenstates of the
total spin operator $\hat S_T$ \cite{gir}. As a result, states
described by Halperin's wave functions that are not eigenstates of $\hat
S_T$ may be realized in double-layers \cite{suen,eisen}.

\begin{figure}
\begin{center}
\includegraphics[width=0.6\linewidth]{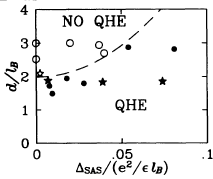}
\end{center}
\caption{Phase diagram of the quantum Hall effect at ${\nu_T=1}$ in
  double-layer systems. $\Delta_{SAS}$ is the tunnel splitting and
  ${e^2/(\epsilon\ell)}$ is the Coulomb energy. Each symbol
  corresponds to a particular double-layer sample. Only the samples
  represented by solid symbols exhibit a quantized Hall plateau at
  ${\nu_T=1}$. The interlayer quantum Hall state exists also in the
  absence of tunneling. [Reproduced from \textcite{murph}.]}
\label{fig_qhe1}
\end{figure}

In this review we are mostly interested in double-layer systems where
tunneling between the two layers is negligible. Such systems support
novel many-body quantum Hall states, that are specific to bilayers and
arise due to interlayer Coulomb interaction \cite{hald,chak}.
\textcite{yosh} investigated a wide class of such states using
Halperin's two-component wave functions \cite{halp}. The ground state
of the system crucially depends on the ratio of the interlayer
separation and magnetic length ${d/\ell_0}$. For a given filling
factor, the magnetic length $\ell_0$ is proportional to the average
separation between electrons in one layer. Therefore, the ratio
${d/\ell_0}$ parametrizes the relative strength of intra- and
interlayer Coulomb interaction. Assuming truly two-dimensional layers
(i.e., setting aside complications that arise due to the finite width
of the quantum wells in GaAs samples), one finds that the interlayer
many-body states are stable for ${d/\ell_0\sim1}$. At large $d$, the
interlayer Coulomb interaction is inefficient and then the system
behaves as if one connects two quantum Hall samples in parallel
\cite{eis1}. This observation can be illustrated with the help of the
typical phase diagram shown in Fig.~\ref{fig_qhe1} for the case of the
total filling factor ${\nu_T=1}$ \cite{murph}. In the opposite limit,
${d/\ell_0\rightarrow0}$, the system approaches the $SU(2)$-symmetric
point, and thus the Halperin states that are not eigenstates of $\hat
S_T$ are expected to collapse \cite{eis1}.

\begin{figure}
\begin{center}
\includegraphics[width=0.97\linewidth]{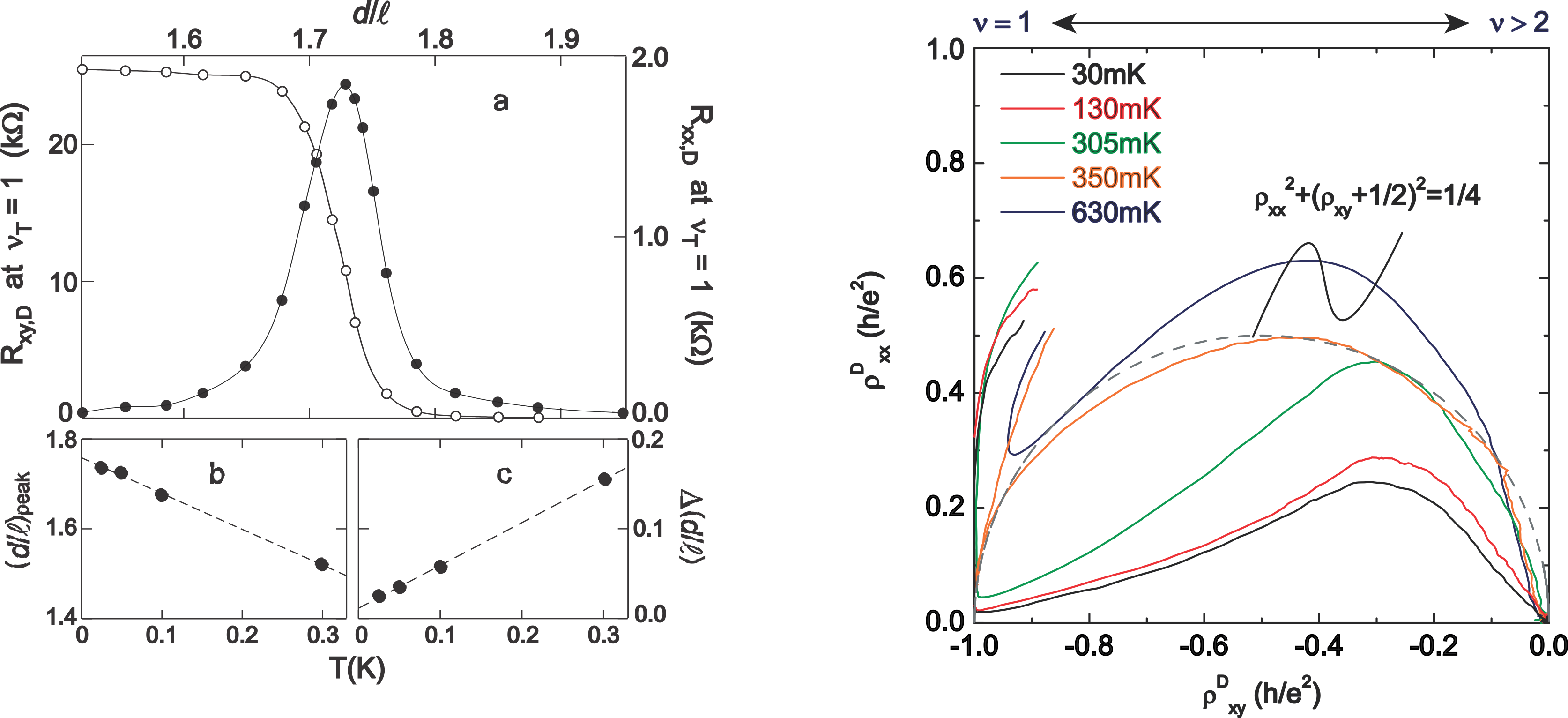}
\end{center}
\caption{(Color online) Left: Hall (open dots) and longitudinal
  (closed dots) drag resistance at ${\nu_T=1}$ and ${T=50}$mK as a
  function of the ratio ${d/\ell_0}$. The two lower panels show the
  temperature dependence of the location and the half-width of the
  peak in $R^D_{xx}$. The lines are guides for the eye. [Reproduced
    from \textcite{kel}.] Right: Drag resistivity and Hall drag
  resistivity in units of ${h/e^2}$ for different temperatures. The
  end points represent the strong- (${\rho^D_{xy}=-1}$) and
  weak-coupling (${\rho^D_{xy}=0}$) regimes. The dashed line
  represents Eq.~(\ref{scl}). [Reproduced from \textcite{tut}.]}
\label{fig_data_tut}
\end{figure}

Double-layers at the total filling factor ${\nu_T=1}$ and with large
interlayer separation (experimentally, ${d/\ell_0\sim2-4}$) behave as
two weakly coupled systems of composite fermions (i.e. each layer is
at ${\nu=1/2}$) while exhibiting strongly enhanced drag as compared to
the zero-field case, see Sec.~\ref{wmf}. As the ratio ${d/\ell_0}$ is
decreased, experiments \cite{ke2,kel} show a gradual development of
the Hall drag signal and a non-monotonic behavior of the longitudinal
drag resistivity $\rho_D$. As ${d/\ell_0}$ approaches the transition
into the strongly correlated, many-body state
\footnote{Since the transition between the weakly and strongly coupled
  quantum Hall states is still poorly understood, one should be
  speaking in terms of the transition region instead of the precise
  critical value of ${d/\ell_0}$.}, $\rho_D$ shows strong enhancement,
followed by a decrease. In the strongly coupled interlayer ${\nu_T=1}$
state $\rho_D$ practically vanishes. At the same time, the Hall drag
resistance develops a quantized plateau, see Fig.~\ref{fig_data_tut}.
Similar behavior was observed in \textcite{tuthuse,wie2,tut}

Early theoretical work on drag in quantum Hall states was focused on
the non-dissipative drag \cite{renn,duan,ya3,ma2}. In contrast to the
case of weak magnetic field (see Sec.~\ref{wmf}), a strong, quantized
Hall drag has been identified as a signature of the interlayer
correlated states. The ${2\times2}$ Hall resistivity matrix (for the
two layers) has been shown \cite{renn,ya3} to be proportional to the
Gram matrix \cite{gram,read} describing topological order in the
quantum Hall state \cite{wen1}:
\begin{equation}
\label{rdhall1}
\rho_{ij}^{xy} = (h/e^2) K_{ij}
\quad\Rightarrow\quad
\rho_{12}^{xy} = n h/e^2, \quad n>0.
\end{equation}
Similar conclusion was reached in \textcite{ma2} on general
topological grounds.

\textcite{dem} have suggested to use the drag resistivity to
distinguish between various quantum Hall states in double-layer
systems at ${\nu_T=1}$. For the compressible (weak-coupling) state at
large interlayer separation, the Hall drag resistivity vanishes, while
the longitudinal drag is determined by gauge-field fluctuations and is
given by Eq.~(\ref{rdn12}). The compressible state exhibits a strong
pairing instability \cite{grei,bone}. If Landau-level mixing is
substantial (as it often is in experimental samples), the paired state
may be described by the ${(3,3,-1)}$ Halperin wave function. This state
resembles a ${p_x+ip_y}$ superconductor of composite fermions. As a
result, it is expected to exhibit the quantized Hall drag resistivity
(\ref{rdhall1}) with ${n=-1}$.

For smaller interlayer separation (${d\simeq\ell_0}$) the system
undergoes a transition into an incompressible, correlated ``quantum
Hall ferromagnet'' state described by the ${(1,1,1)}$ Halperin wave
function. This state possesses a gapless neutral mode and is
characterized by the Hall resistivity (\ref{rdhall1}) with ${n=1}$.

The nature of the transition between the compressible, weak coupling
state at large interlayer separation and the incompressible, strong
coupling state at ${d\simeq\ell_0}$ is not completely understood
\cite{fin,eis2}. Numerical evidence \cite{sgm,bur} suggests a first
order transition at ${T=0}$, which contradicts the experimental
observation of gradual development of the quantized Hall drag
\cite{ke2,kel}, see Fig.~\ref{fig_data_tut}. \textcite{ad2} suggested
a phenomenological description of the drag resistivity in the
transition region. Postulating that in the transition region the
system is split into regions of the strong-coupling ${(1,1,1)}$ phase
and regions of the weak-coupling compressible phase, they describe the
transition as the point where the fraction $f$ of the sample occupied
by the ${(1,1,1)}$ phase reaches the percolation threshold
${f_c=1/2}$.

In a system of identical layers, the linear response theory can be
formulated in terms of symmetric and antisymmetric states. Denoting
the ${2\times2}$ resistivity matrices corresponding to symmetric and
antisymmetric currents by $\rho^s$ and $\rho^a$, one finds the drag
resistivity as ${\rho^D=(\rho^a-\rho^s)/2}$.

In the weak-coupling phase at ${d\gg\ell_0}$, the drag resistivity is
very small, ${\rho^D\ll\rho^{a(s)}}$. Neglecting $\rho^D$, one may
approximate the resistivities as (in units of ${h/e^2}$)
\begin{equation}
\label{rhoas}
\rho^a(d\gg\ell_0)=\rho^s(d\gg\ell_0)=
\begin{pmatrix}
\epsilon & 2 \cr
-2 & \epsilon
\end{pmatrix},
\end{equation}
where ${\epsilon=1/(k_F\ell_{tr})\ll1}$ (within the composite fermion
model), ${k_F=4\pi{n}}$ is the Fermi wave vector, $n$ is the
electronic density, and $\ell_{tr}$ is the transport mean free
path. For ${T<1}$K, the experimentally measured values of $\rho^D$ are
almost two orders of magnitude less than $\epsilon$.

The strong-coupling ${(1,1,1)}$ phase exhibits features of the quantum
Hall state for the symmetric currents
\begin{equation}
\label{rhos0}
\rho^s_0(d\lesssim\ell_0)=
\begin{pmatrix}
0 & 2 \cr
-2 & 0
\end{pmatrix},
\end{equation}
while for the antisymmetric currents it is a superfluid \cite{ad2}
${\rho^a_0(d\lesssim\ell_0)=0}$.

Analyzing the system close to the transition as a composite system
comprising regions of both phases, \textcite{ad2} have found a
phenomenological expression for the drag resistivity
\begin{equation}
\label{rdqh}
\rho_{xx}^D = \frac{8\epsilon f(1-f)(1-2f)}{\epsilon^2+4(1-2f)^2}.
\end{equation}
As $f$ increases from zero, this drag resistivity grows from zero [or
  rather, the very small value in the compressible state that is
  neglected in Eq.~(\ref{rdqh})] reaching a maximum at
${f^*\approx1/2-\epsilon/4}$ (for small $\epsilon$) and again
vanishing at the percolation threshold, in qualitative agreement with
the non-monotonic drag observed in \textcite{kel,ke2}.

Furthermore, using the semicircle law \cite{dyk}, it can be shown
that to the lowest order in $\epsilon$ the drag resistivities satisfy
the relation \cite{ad2}
\begin{equation}
\label{scl}
(\rho^D_{xy}+1/2)^2+(\rho^D_{xx})^2=1/4,
\end{equation}
yielding vanishing Hall drag for the compressible state (${f=0}$) and
the quantized value (\ref{rdhall1}) with ${n=-1}$ for the ${(1,1,1)}$
state at ${f\geqslant1/2}$. Between the two extremes the negative
$\rho^D_{xy}$ varies monotonously. The apparent discrepancy in the
sign of $\rho^D_{xy}$ obtained by \textcite{ad2} and \textcite{dem}
seem to stem from the alternative definition of drag resistivities.
Similar predictions for transport coefficients, in particular
Eq.~(\ref{scl}), but without the explicit phase separation were
obtained by \textcite{sim}. An alternative model invoking the
coexistence of the two phases was suggested by \textcite{spk}.

The semicircle relation (\ref{scl}) was experimentally tested in
\textcite{tut}, see Fig.~\ref{fig_data_tut}. Instead of comparing a
number of double-well devices with different interlayer separations
\cite{kel}, \textcite{tut} varied the electron density and observed
the transition between the strong-coupling state at ${\nu_T=1}$ and
the weakly coupled state at ${\nu_T=2}$. The data at intermediate
temperatures ${T\approx300}$K are in a good quantitative agreement
with the theory. At the same time, Eq.~(\ref{scl}) is only approximate
and is expected to hold if the drag resistivity is much larger than
the symmetric bilayer resistivity at all fillings. Drag resistivity in
the weak-coupling state is also neglected. Given these approximations,
the agreement between the data and the phenomenological theory of
\textcite{ad2} is satisfactory.

\subsection{Interlayer exciton formation}
\label{exc1}

Further experiments revealed the most intriguing feature of the
strong-coupling quantum Hall state at ${\nu_T=1}$: the presence of the
exciton condensate capable of neutral superfluid transport
\cite{eim,eis2}. Originally envisioned for optically generated
excitons in bulk semiconductors \cite{mosk,bla,keld,kelkop}, the
phenomenon has been predicted also for indirect excitons in
double-layer systems \cite{lozyud,shev}.

The quantized Hall effect along with the vanishing longitudinal
resistivity at ${\nu_T=1}$ indicate a gapped spectrum of charged
excitations. In these measurements \cite{suen,eisen,eis2}, electrical
currents in the two layers flow in the same direction. In contrast,
the condensate couples to antiparallel or counterflowing currents
\cite{tuthuse,ke4,wie2} and manifests itself through vanishing Hall
voltage. The simplest explanation for this observation is based on
charge neutrality of excitons: as neutral objects, excitons do not
experience the Lorentz force and hence no Hall voltage develops when
equal, counter-propagating currents are flowing through the two
layers.

Another spectacular manifestation of the exciton condensate is the
Josephson-like tunneling anomaly \cite{spi2,wie,wie3,fin3,tie,yoo}
that theoretically was predicted in \textcite{zee,park} and later 
discussed in \textcite{dolci}.

Finally, the latest experiments revealing the existence of the exciton
condensate utilized the multiple connected Corbino geometry
\cite{nan,tie,tie2,fin2}. For a theoretical discussion of the
superfluid flow in the Corbino geometry see \textcite{sumac}. The
advantage of the Corbino samples is that they support the exciton flow
through the bulk (in contrast to the Hall bar samples where transport
is dominated by the edges).

Coulomb drag has played an important role in discovering the
interlayer correlated state \cite{eis2}. Quantized Hall drag measured
in the simply connected square geometry \cite{ke2} was one of the
first indications of anomalous in-plane transport in double-layer
systems at ${\nu_T=1}$. Remarkably, the quantized Hall voltage has
been found to be the same in both layers. At first glance, this
contradicts the boundary conditions of the drag measurement: drag
experiments involve passing current through one of the layers and
measuring the induced voltage in the other, where no current is
allowed to flow. The absence of the current seems to yield the absence
of the Lorentz force and hence lead to the standard conclusion that no
Hall voltage should be induced in the passive layer, see
Sec.~\ref{wmf}. However, this argument does not take into account
collective effects. In the presence of the condensate, the driving
current can be decomposed into the symmetric and antisymmetric parts
\cite{ad2}. While the symmetric current carries the electric charge,
the antisymmetric -- or counterpropagating -- current is equivalent to
the condensate flow. In the passive layer the two currents cancel each
other thus satisfying the boundary condition. At the same time, it is
the symmetric, charge-carrying current that can couple to the magnetic
field. This current is shared between the layers, yielding the
identical quantized Hall voltage across both layers.

\begin{figure}
\begin{center}
\includegraphics[width=0.7\linewidth]{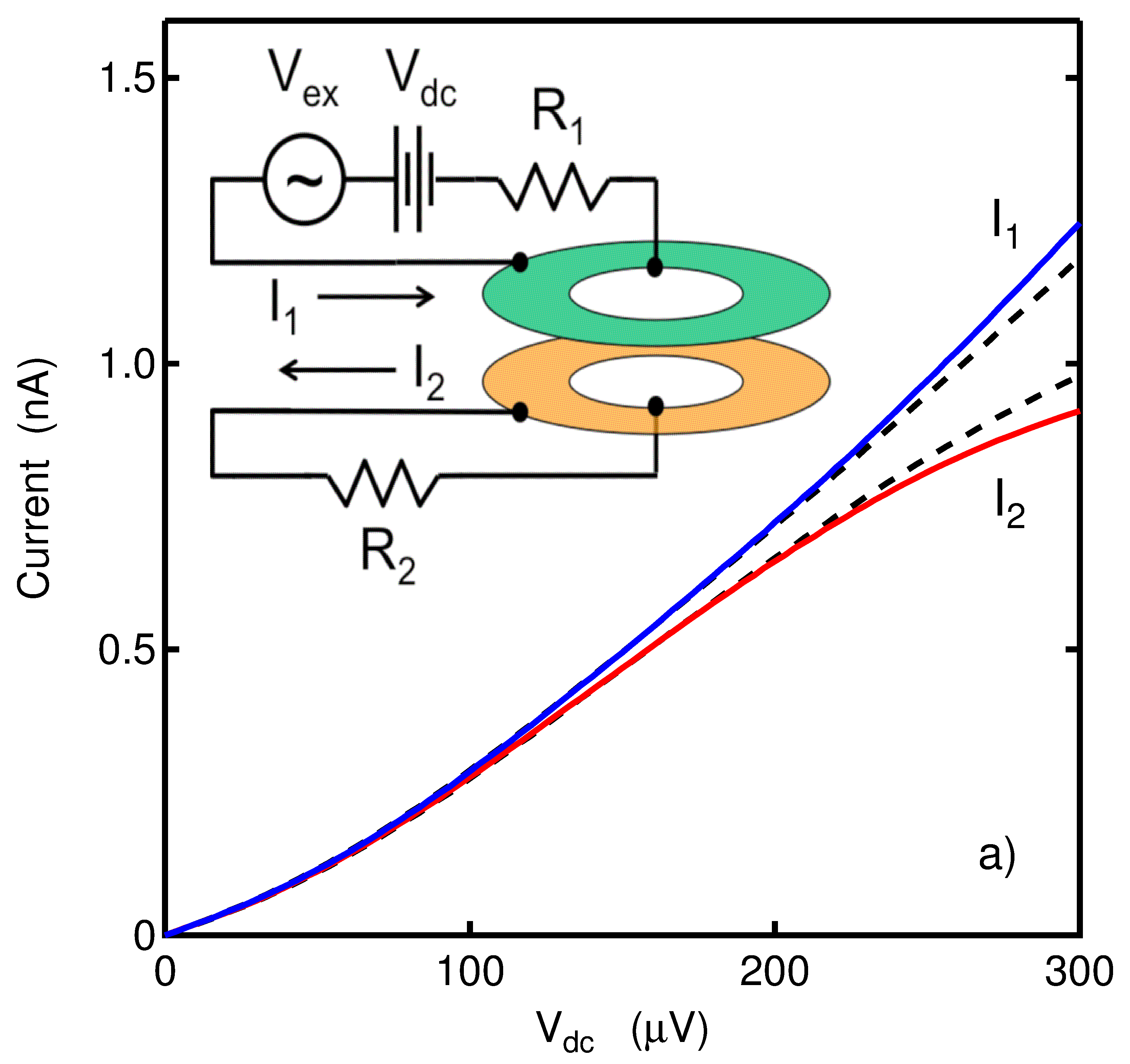}
\end{center}
\caption{(Color online) Corbino Coulomb drag. Solid lines show the
  drive and drag currents. The measurement was performed at $\nu_T=1$,
  ${T=17}$mK, and ${d/\ell_0=1.5}$. Dashed lines represent the results
  of simulations incorporating estimated series resistances and
  measured Corbino conductivity. The inset shows the measurement
  schematic. The resistances $R_i$ comprise both external circuit
  resistors and the resistances intrinsic to the device. [Reprinted by
    permission from Macmillan Publishers Ltd: Nature,
    \textcite{nan}.]}
\label{fig_data_nan}
\end{figure}

Similar arguments lead to the expectation of ``perfect'' longitudinal
drag \cite{sumac}: the symmetric current shared between the layers
should be responsible not only for the Hall, but also for the
longitudinal voltage in the passive layer. This prediction was tested
in a dedicated experiment by \textcite{nan} using Corbino
samples. Deviating from the standard setup, \textcite{nan} have closed
the electric circuit in the passive layer and measured the induced
{\it current}, rather than the voltage. In this case, perfect drag
means that the induced current should be same in magnitude as the
driving current passed through the active layer while flowing in the
opposite direction. This is exactly what has been observed by
\textcite{nan}, at least for small driving currents, see
Fig.~\ref{fig_data_nan}.

The above arguments neglect the impact of disorder that might affect
the presumed dissipationless excitonic transport \cite{sumac} across
the bulk of the device \cite{huse,fert,fil,coop}. Assuming a
phenomenological resistance $R_s$ of the excitonic system, one still
finds [neglecting the Corbino conductance \cite{nan}] perfect drag
${I_1=I_2=V/(R_1+R_2+R_s)}$, where $R_i$ represent the net resistances
in series with the Corbino sample, see the inset in
Fig.~\ref{fig_data_nan}. As the magnitude of ${R_1+R_2}$ is expected
to always exceed ${2h/e^2}$ \cite{sumac,pesin}, the ability of the
experiment to detect small values of $R_s$ is limited. The issue of
dissipation in the excitonic system might be clarified by future
multi-terminal measurements.

So far we have discussed experiments on the exciton physics in
double-layer systems comprising similar electronic layers
\cite{eim}. It is also possible to create devices with oppositely
doped layers, the so-called electron-hole bilayers
\cite{keo,gup}. Coulomb drag measurements in these systems
\cite{cro,sea} do not provide a direct evidence of interlayer
coherence, but nevertheless demonstrate an upturn in $\rho_D$ as the
temperature is lowered below $1$K. The upturn is seen only in devices
with smaller ($20$nm) interlayer separation suggesting exciton
formation.

\begin{figure}
\begin{center}
\includegraphics[width=0.97\linewidth]{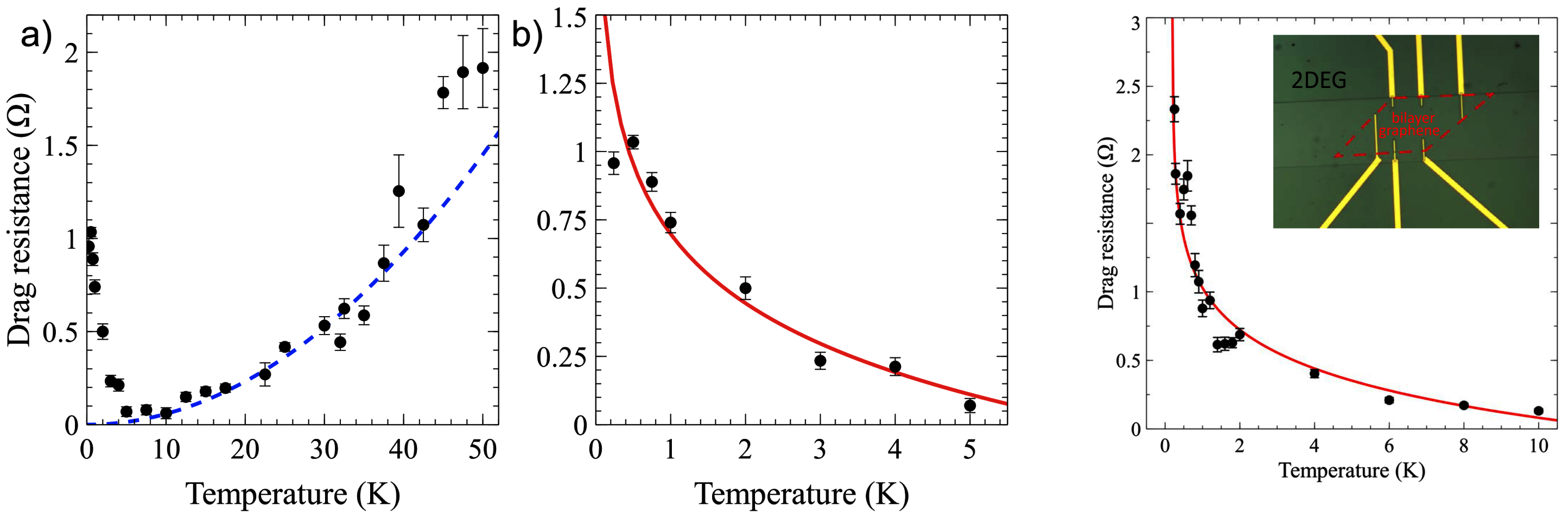}
\end{center}
\caption{(Color online) Coulomb drag in a graphene-2DEG vertical
  heterostructure. Left: measured drag resistivity. The dashed line
  represents the best fit for the standard temperature dependence
  ${R_D=aT^2}$, ${a=(5.8\pm0.3)\times10^{-4}\Omega}$K$^{-2}$. Middle:
  a fit of the low-$T$ upturn based on Eq.~(\ref{rdlog}). The critical
  temperature found from the fit is ${T_c\sim10-100}$mK. Right: the
  low-T upturn in a bilayer graphene-2DEG heterostructure. The fit
  based on Eq.~(\ref{rdlog}) yields ${T_c\sim190}$mK. [Reprinted by
    permission from Macmillan Publishers Ltd: Nature Communications,
    \textcite{gam}.]}
\label{fig_data_gam}
\end{figure}

A microscopic theory of Coulomb drag in proximity to a phase
transition was suggested by \textcite{hu5} and \textcite{min,min2}.
As the system approaches the transition temperature $T_c$ from above,
the drag resistivity was found to exhibit a logarithmic divergence
\begin{equation}
\label{rdlog}
\rho_D=\rho_0 + A\ln[T_c/(T-T_c)],
\end{equation}
where $\rho_0$ and $A$ are two fitting parameters \cite{gam}. While
qualitatively resembling the upturn observed in electron-hole bilayers
\cite{cro,sea}, the theory accounts neither for a subsequent downturn
at the lowest temperatures, nor the apparent violation of Onsager
reciprocity \cite{cro} [although the latter might be related to
  heating effects \cite{sea}]. The theory also does not make
falsifiable predictions regarding the dependence of $\rho_D$ on
carrier densities in the two layers \cite{mor} [at higher
  temperatures, where the data show the standard $T^2$ dependence, the
  density dependence of $\rho_D$ is stronger than expected on the
  basis of the Fermi-liquid many-body calculations \cite{hwa2}].

The logarithmic temperature dependence (\ref{rdlog}) fits well with
the upturn in the drag resistivity observed in \textcite{gam} in
hybrid devices comprising either a monolayer or bilayer graphene sheet
and a GaAs quantum well, see Fig.~\ref{fig_data_gam}. In fact, the
search for exciton physics was one of the main motivations for
experimental studies of Coulomb drag in double-layer graphene-based
structures \cite{exg,tu1}.

Exciton condensation in graphene has attracted considerable
theoretical attention
\cite{bis,jog,loz3,khar,khar2,ale,loze3,fal2,loz4,fil2,fil3,sode,abe}.
Several contradicting values of the transition temperature in
double-layer graphene systems have been reported. The initial estimate
\cite{min,jog} of $T_c$ close to room temperature appeared to be too
optimistic. Screening effects \cite{khar,khar2} were shown to lead to
extremely low values under $1$mk (${T_c\sim10^{-7}E_F}$). More recent
investigations involving detail analysis of screened Coulomb
interaction \cite{loz4,sode,abe}, multiband pairing \cite{min,loz4},
and pairing with nonzero momentum \cite{loze3} suggest somewhat higher
values of $T_c $, making the transition experimentally accessible.

High-temperature coherence and superfluidity has also been suggested
in thin films of topological insulators \cite{moo,loze2,min,min2}.

The effect of exciton condensation on Coulomb drag has been
investigated in graphene numerically by \textcite{zhajin} and in
topological insulator films analytically in \textcite{loze}. The
latter work focused on the drag effect at temperatures exceeding $T_c$,
where the pairing fluctuations are expected to play an important
role. In addition to the Maki-Thompson-type contribution
\cite{hu5,min,min2} to the drag resistivity, \textcite{loze} have 
analyzed the Aslamazov-Larkin-type contribution and found
\begin{equation}
\label{rdal}
\delta\rho_D^{AL} \propto [\ln(T/T_c)]^{-1}.
\end{equation}
Far away from the transition, the result (\ref{rdal}) decays
logarithmically, similarly to Eq.~(\ref{rdlog}), but close to the
transition exhibits a stronger divergence
${\delta\rho_D^{AL}\propto(T-T_c)^{-1}}$.


\section{Open questions and perspectives}
\label{open}


The physics of the Coulomb drag in double-layer systems is well
understood if both layers are in the Fermi-liquid state
\cite{kor,fl2}. The current in the passive layer is created by
exciting electron-hole pairs (each pair consisting of an occupied
state above the Fermi surface and an empty state below) in a state
characterized by finite momentum. The momentum comes from the
electron-hole excitations in the active layer created by the driving
current. The momentum transfer is due to the interlayer Coulomb
interaction. Therefore it follows from the usual phase-space
considerations that the drag coefficient is proportional to the square
of the temperature $\rho_D\propto T^2$. Remarkably, this simple
argument is sufficient to describe the observed low temperature dependence
of $\rho_D$. Deviations from the quadratic dependence at higher temperatures 
are primarily due to the effect of phonons and plasmons \cite{roj}.

The universality of the Landau Fermi-liquid theory \cite{dau10,aar}
can be traced to the linearization of the quasi-particle
spectrum. Within this approximation all details of the microscopic
structure of the system are contained in a limited number of
parameters, such as the Fermi velocity and density of states (DoS) at
the Fermi level. Many observable quantities (e.g. the electronic
contribution to the specific heat, spin susceptibility, period of the
De Haas-van Alphen oscillations, etc.) can be expressed in terms of
these parameters and thus exhibit the ``universal'' behavior (as a
function of temperature or external fields). Same arguments can be
applied to elementary excitations in strongly doped graphene ($\mu\gg
T$), where the Fermi-liquid theory is expected to be applicable.

Coulomb drag belongs to a different class of observables. In
conventional semiconductor devices, it reflects the degree of
electron-hole asymmetry in the system vanishing in the approximation
of linearized spectrum \cite{kor}. The drag coefficient is determined
by the subleading contribution taking into account the curvature of
the quasi-particle spectrum. Indeed, in the passive layer the momentum
is transferred equally to electrons and holes so that the resulting
state can carry current only in the case of electron-hole
asymmetry. Likewise, this asymmetry is necessary for the
current-carrying state in the active layer to be characterized by
nonzero total momentum. The electron-hole asymmetry manifests itself
\cite{me2,opp} in the energy (or chemical potential) dependence of
such quantities as the density of states, single-layer conductivity,
and diffusion coefficient. Within the Fermi-liquid theory \cite{kor},
the asymmetry is weak,
$\partial\sigma_i/\partial\mu_i\approx\sigma_i/\mu_i$, leading to the
drag effect, that is much weaker than the single-layer conductivity.

Coulomb drag in non-Fermi-liquid systems is much more interesting. In
particular, it has been used to study novel strongly-correlated,
many-body states in double quantum wells \cite{eis2}, graphene
\cite{gam}, quantum wires \cite{lar2}, and optical cavities
\cite{ber}, where practical applications in optical switches have been
suggested. In these systems, drag measurements have proved to be an
invaluable tool to study the microscopic structure of complex,
interacting many-body systems.

At the same time, our understanding of many of these systems is
incomplete. In contrast to the Fermi-liquid theory, many aspects of
the strongly-correlated many-body states lack a detailed theoretical
description. Consequently, their transport properties, including
Coulomb drag, can be evaluated only with the help of heuristic or
phenomenological models. One can only hope that a proper microscopic
theory of these effects will eventually be developed.

This brings us to the list of unresolved questions related to the
theory reviewed in this paper and possible direction of the field in
the near future.

(i) At low enough temperatures and especially in strong magnetic
fields, double-layer systems may host excitonic condensates
\cite{eis2}. In monolayer graphene, such condensation is also
possible, but for reasonably weak interactions the condensation
temperature appears to be rather low
\cite{ale,khar,min}. Nevertheless, a possibility of interlayer
correlated states in graphene-based systems [and possibly in hybrid
  devices involving other materials, \cite{geim}] is rather exciting
and certainly requires theoretical attention.

(ii) The hydrodynamic approach of Sec.~\ref{hydro} should be extended
to include thermoelectric effects in graphene-based double-structures
as well as in monolayer and bilayer graphene. As pointed out in
\textcite{foa,mef}, the quasiparticle imbalance in graphene may play a
decisive role in thermal transport. Another promising direction may be
opened by generalization of the macroscopic linear-response equations
to a true, nonlinear hydrodynamics. The relation between the quantum
kinetic equation of \textcite{zna} and the hydrodynamic approach [both
  in graphene \cite{mef} and in 2DEG \cite{apos,aks}] is also of
certain theoretical interest.

(iii) Dirac fermions can be found as low-energy excitations not only
in graphene, but also in topological insulators
\cite{bernevig,shen}. An extension of the present theory of Coulomb
drag to various possible system configurations involving topological
insulators and/or hybrid devices involving topological insulators,
graphene, etc. appears to be very promising \cite{min}.

(iv) Novel aspects of Luttinger liquid physics and role of
equilibration processes on drag can be further explored with the edge
states of quantum Hall systems or topological edge liquids of quantum
spin Hall effect. Some theoretical predictions have already being made
\cite{zuz} and recent experimental advances
\cite{molen,rrdu,gold,alti} bring these exciting perspectives within
reach.

(v) Mesoscopic fluctuations of Coulomb drag in ballistic samples
should be further analyzed on the basis of the microscopic theory. The
theory should be further extended to the cases of Dirac fermions in
graphene and composite fermions at the half-filled Landau
level. Experimental work in this direction has been already initiated
in \textcite{prs,tu1}.

(vi) The third-order drag effect (see Sec.~\ref{tod}) bears a certain
resemblance to the well-known Altshuler-Aronov corrections to
single-layer conductivity \cite{aar,zna}. In \textcite{zna} it was
shown that the dominant contribution to conductivity at low (diffusive
regime) and high (ballistic regime) temperatures technically comes
from different diagrams describing conceptually similar, but at the
same time distinct interference processes. Similarly, we expect that
the third-order drag contribution in ballistic regime might be
governed by scattering processes which are distinct from those
considered in \textcite{le2}.

We would like to close this review by pointing out the surprising
richness of the Coulomb drag problem. The original suggestion of a way
to observe interwell interactions in semiconductor heterostructures
has developed into a vibrant field of research where technological
advances go hand in hand with theoretical developments. New
experiments with novel materials keep being devised and stimulate new
avenues for theoretical thinking. We should be expecting to see
further intriguing discoveries being made related to frictional drag
in the foreseeable future.

\section*{Acknowledgments}

We would like to thank all our collaborators for sharing their
insights over the years and especially I.L. Aleiner, A.V. Andreev,
S. Apostolov, W. Chen, I.V. Gornyi, A. Kamenev, M.I. Katsnelson,
A.D. Mirlin, M. Norman, L. Ponomarenko, P.M. Ostrovsky, M. Sch\"utt,
A. Stern, M. Titov, and the late A. Savchenko. While writing this
review we have greatly benefited from discussions with M. Foster,
G. Gervais, D. Polyakov, and J. Schmalian.

The work of A.L. was supported by NSF grants DMR-1401908 and
ECCS-1407875, and in part by DAAD grant from German Academic Exchange
Services, SPP 1459 and SPP 1666 of the Deutsche
Forschungsgemeinschaft. BN acknowledges support from the EU Network
Grant InterNoM and the Deutsche Forschungsgemeinschaft under grants
SCHO 287/7-1 and SH 81/2-1.

\bibliography{refs-books,references}

\begin{thebibliography}{464}%
\makeatletter
\providecommand \@ifxundefined [1]{%
 \@ifx{#1\undefined}
}%
\providecommand \@ifnum [1]{%
 \ifnum #1\expandafter \@firstoftwo
 \else \expandafter \@secondoftwo
 \fi
}%
\providecommand \@ifx [1]{%
 \ifx #1\expandafter \@firstoftwo
 \else \expandafter \@secondoftwo
 \fi
}%
\providecommand \natexlab [1]{#1}%
\providecommand \enquote  [1]{``#1''}%
\providecommand \bibnamefont  [1]{#1}%
\providecommand \bibfnamefont [1]{#1}%
\providecommand \citenamefont [1]{#1}%
\providecommand \href@noop [0]{\@secondoftwo}%
\providecommand \href [0]{\begingroup \@sanitize@url \@href}%
\providecommand \@href[1]{\@@startlink{#1}\@@href}%
\providecommand \@@href[1]{\endgroup#1\@@endlink}%
\providecommand \@sanitize@url [0]{\catcode `\\12\catcode `\$12\catcode
  `\&12\catcode `\#12\catcode `\^12\catcode `\_12\catcode `\%12\relax}%
\providecommand \@@startlink[1]{}%
\providecommand \@@endlink[0]{}%
\providecommand \url  [0]{\begingroup\@sanitize@url \@url }%
\providecommand \@url [1]{\endgroup\@href {#1}{\urlprefix }}%
\providecommand \urlprefix  [0]{URL }%
\providecommand \Eprint [0]{\href }%
\providecommand \doibase [0]{http://dx.doi.org/}%
\providecommand \selectlanguage [0]{\@gobble}%
\providecommand \bibinfo  [0]{\@secondoftwo}%
\providecommand \bibfield  [0]{\@secondoftwo}%
\providecommand \translation [1]{[#1]}%
\providecommand \BibitemOpen [0]{}%
\providecommand \bibitemStop [0]{}%
\providecommand \bibitemNoStop [0]{.\EOS\space}%
\providecommand \EOS [0]{\spacefactor3000\relax}%
\providecommand \BibitemShut  [1]{\csname bibitem#1\endcsname}%
\let\auto@bib@innerbib\@empty
\bibitem [{\citenamefont {Abedinpour}\ \emph {et~al.}(2011)\citenamefont
  {Abedinpour}, \citenamefont {Vignale}, \citenamefont {Principi},
  \citenamefont {Polini}, \citenamefont {Tse},\ and\ \citenamefont
  {MacDonald}}]{plas4}%
  \BibitemOpen
  \bibfield  {author} {\bibinfo {author} {\bibnamefont {Abedinpour},
  \bibfnamefont {S.~H.}}, \bibinfo {author} {\bibfnamefont {G.}~\bibnamefont
  {Vignale}}, \bibinfo {author} {\bibfnamefont {A.}~\bibnamefont {Principi}},
  \bibinfo {author} {\bibfnamefont {M.}~\bibnamefont {Polini}}, \bibinfo
  {author} {\bibfnamefont {W.-K.}\ \bibnamefont {Tse}}, \ and\ \bibinfo
  {author} {\bibfnamefont {A.~H.}\ \bibnamefont {MacDonald}}} (\bibinfo {year}
  {2011}),\ \href@noop {} {\bibfield  {journal} {\bibinfo  {journal} {Phys.
  Rev. B}\ }\textbf {\bibinfo {volume} {84}},\ \bibinfo {pages}
  {045429}}\BibitemShut {NoStop}%
\bibitem [{\citenamefont {Abergel}\ \emph {et~al.}(2013)\citenamefont
  {Abergel}, \citenamefont {Rodriguez-Vega}, \citenamefont {Rossi},\ and\
  \citenamefont {Das~Sarma}}]{abe}%
  \BibitemOpen
  \bibfield  {author} {\bibinfo {author} {\bibnamefont {Abergel}, \bibfnamefont
  {D.~S.~L.}}, \bibinfo {author} {\bibfnamefont {M.}~\bibnamefont
  {Rodriguez-Vega}}, \bibinfo {author} {\bibfnamefont {E.}~\bibnamefont
  {Rossi}}, \ and\ \bibinfo {author} {\bibfnamefont {S.}~\bibnamefont
  {Das~Sarma}}} (\bibinfo {year} {2013}),\ \href {\doibase
  10.1103/PhysRevB.88.235402} {\bibfield  {journal} {\bibinfo  {journal} {Phys.
  Rev. B}\ }\textbf {\bibinfo {volume} {88}},\ \bibinfo {pages}
  {235402}}\BibitemShut {NoStop}%
\bibitem [{\citenamefont {Abrahams}\ \emph {et~al.}(1979)\citenamefont
  {Abrahams}, \citenamefont {Anderson}, \citenamefont {Licciardello},\ and\
  \citenamefont {Ramakrishnan}}]{gang4}%
  \BibitemOpen
  \bibfield  {author} {\bibinfo {author} {\bibnamefont {Abrahams},
  \bibfnamefont {E.}}, \bibinfo {author} {\bibfnamefont {P.~W.}\ \bibnamefont
  {Anderson}}, \bibinfo {author} {\bibfnamefont {D.~C.}\ \bibnamefont
  {Licciardello}}, \ and\ \bibinfo {author} {\bibfnamefont {T.~V.}\
  \bibnamefont {Ramakrishnan}}} (\bibinfo {year} {1979}),\ \href@noop {}
  {\bibfield  {journal} {\bibinfo  {journal} {Phys. Rev. Lett.}\ }\textbf
  {\bibinfo {volume} {42}},\ \bibinfo {pages} {673}}\BibitemShut {NoStop}%
\bibitem [{\citenamefont {Abrikosov}\ and\ \citenamefont
  {Beneslavskii}(1971)}]{nls}%
  \BibitemOpen
  \bibfield  {author} {\bibinfo {author} {\bibnamefont {Abrikosov},
  \bibfnamefont {A.}}, \ and\ \bibinfo {author} {\bibfnamefont
  {S.}~\bibnamefont {Beneslavskii}}} (\bibinfo {year} {1971}),\ \href@noop {}
  {\bibfield  {journal} {\bibinfo  {journal} {Zh. Eksp. Teor. Fiz.}\ }\textbf
  {\bibinfo {volume} {59}},\ \bibinfo {pages} {1280}},\ \bibinfo {note} {[Sov.
  Phys. JETP {\bf 32}, 699 (1971)]}\BibitemShut {NoStop}%
\bibitem [{\citenamefont {Aguado}\ and\ \citenamefont
  {Kouwenhoven}(2000)}]{agu}%
  \BibitemOpen
  \bibfield  {author} {\bibinfo {author} {\bibnamefont {Aguado}, \bibfnamefont
  {R.}}, \ and\ \bibinfo {author} {\bibfnamefont {L.~P.}\ \bibnamefont
  {Kouwenhoven}}} (\bibinfo {year} {2000}),\ \href {\doibase
  10.1103/PhysRevLett.84.1986} {\bibfield  {journal} {\bibinfo  {journal}
  {Phys. Rev. Lett.}\ }\textbf {\bibinfo {volume} {84}},\ \bibinfo {pages}
  {1986}}\BibitemShut {NoStop}%
\bibitem [{\citenamefont {Aleiner}\ \emph {et~al.}(1999)\citenamefont
  {Aleiner}, \citenamefont {Altshuler},\ and\ \citenamefont
  {Gershenson}}]{aag}%
  \BibitemOpen
  \bibfield  {author} {\bibinfo {author} {\bibnamefont {Aleiner}, \bibfnamefont
  {I.~L.}}, \bibinfo {author} {\bibfnamefont {B.~L.}\ \bibnamefont
  {Altshuler}}, \ and\ \bibinfo {author} {\bibfnamefont {M.~E.}\ \bibnamefont
  {Gershenson}}} (\bibinfo {year} {1999}),\ \href@noop {} {\bibfield  {journal}
  {\bibinfo  {journal} {Waves Random Media}\ }\textbf {\bibinfo {volume} {9}},\
  \bibinfo {pages} {201}}\BibitemShut {NoStop}%
\bibitem [{\citenamefont {Aleiner}\ and\ \citenamefont {Efetov}(2006)}]{alef}%
  \BibitemOpen
  \bibfield  {author} {\bibinfo {author} {\bibnamefont {Aleiner}, \bibfnamefont
  {I.~L.}}, \ and\ \bibinfo {author} {\bibfnamefont {K.~B.}\ \bibnamefont
  {Efetov}}} (\bibinfo {year} {2006}),\ \href@noop {} {\bibfield  {journal}
  {\bibinfo  {journal} {Phys. Rev. Lett.}\ }\textbf {\bibinfo {volume} {97}},\
  \bibinfo {pages} {236801}}\BibitemShut {NoStop}%
\bibitem [{\citenamefont {Aleiner}\ \emph {et~al.}(2007)\citenamefont
  {Aleiner}, \citenamefont {Kharzeev},\ and\ \citenamefont {Tsvelik}}]{ale}%
  \BibitemOpen
  \bibfield  {author} {\bibinfo {author} {\bibnamefont {Aleiner}, \bibfnamefont
  {I.~L.}}, \bibinfo {author} {\bibfnamefont {D.~E.}\ \bibnamefont {Kharzeev}},
  \ and\ \bibinfo {author} {\bibfnamefont {A.~M.}\ \bibnamefont {Tsvelik}}}
  (\bibinfo {year} {2007}),\ \href {\doibase 10.1103/PhysRevB.76.195415}
  {\bibfield  {journal} {\bibinfo  {journal} {Phys. Rev. B}\ }\textbf {\bibinfo
  {volume} {76}},\ \bibinfo {pages} {195415}}\BibitemShut {NoStop}%
\bibitem [{\citenamefont {Alkauskas}\ \emph {et~al.}(2002)\citenamefont
  {Alkauskas}, \citenamefont {Flensberg}, \citenamefont {Hu},\ and\
  \citenamefont {Jauho}}]{alk}%
  \BibitemOpen
  \bibfield  {author} {\bibinfo {author} {\bibnamefont {Alkauskas},
  \bibfnamefont {A.}}, \bibinfo {author} {\bibfnamefont {K.}~\bibnamefont
  {Flensberg}}, \bibinfo {author} {\bibfnamefont {B.~Y.-K.}\ \bibnamefont
  {Hu}}, \ and\ \bibinfo {author} {\bibfnamefont {A.-P.}\ \bibnamefont
  {Jauho}}} (\bibinfo {year} {2002}),\ \href@noop {} {\bibfield  {journal}
  {\bibinfo  {journal} {Phys. Rev. B}\ }\textbf {\bibinfo {volume} {66}},\
  \bibinfo {pages} {201304}}\BibitemShut {NoStop}%
\bibitem [{\citenamefont {Altimiras}\ \emph {et~al.}(2010)\citenamefont
  {Altimiras}, \citenamefont {le~Sueur}, \citenamefont {Gennser}, \citenamefont
  {Cavanna}, \citenamefont {Mailly},\ and\ \citenamefont {Pierre}}]{alti}%
  \BibitemOpen
  \bibfield  {author} {\bibinfo {author} {\bibnamefont {Altimiras},
  \bibfnamefont {C.}}, \bibinfo {author} {\bibfnamefont {H.}~\bibnamefont
  {le~Sueur}}, \bibinfo {author} {\bibfnamefont {U.}~\bibnamefont {Gennser}},
  \bibinfo {author} {\bibfnamefont {A.}~\bibnamefont {Cavanna}}, \bibinfo
  {author} {\bibfnamefont {D.}~\bibnamefont {Mailly}}, \ and\ \bibinfo {author}
  {\bibfnamefont {F.}~\bibnamefont {Pierre}}} (\bibinfo {year} {2010}),\
  \href@noop {} {\bibfield  {journal} {\bibinfo  {journal} {Nature Physics}\
  }\textbf {\bibinfo {volume} {6}},\ \bibinfo {pages} {34}}\BibitemShut
  {NoStop}%
\bibitem [{\citenamefont {Altshuler}\ and\ \citenamefont {Aronov}(1979)}]{aa1}%
  \BibitemOpen
  \bibfield  {author} {\bibinfo {author} {\bibnamefont {Altshuler},
  \bibfnamefont {B.}}, \ and\ \bibinfo {author} {\bibfnamefont
  {A.}~\bibnamefont {Aronov}}} (\bibinfo {year} {1979}),\ \href@noop {}
  {\bibfield  {journal} {\bibinfo  {journal} {Zh. Eksp. Teor. Fiz.}\ }\textbf
  {\bibinfo {volume} {77}},\ \bibinfo {pages} {2028}},\ \bibinfo {note} {[Sov.
  Phys. JETP {\bf 50}, 968 (1979)]}\BibitemShut {NoStop}%
\bibitem [{\citenamefont {Altshuler}(1985)}]{alts}%
  \BibitemOpen
  \bibfield  {author} {\bibinfo {author} {\bibnamefont {Altshuler},
  \bibfnamefont {B.~L.}}} (\bibinfo {year} {1985}),\ \href@noop {} {\bibfield
  {journal} {\bibinfo  {journal} {Pis'ma Zh. Eksp. Teor. Fiz.}\ }\textbf
  {\bibinfo {volume} {41}},\ \bibinfo {pages} {530}},\ \bibinfo {note} {[JETP
  Lett. {\bf 41}, 648 (1985)]}\BibitemShut {NoStop}%
\bibitem [{\citenamefont {Altshuler}\ and\ \citenamefont {Aronov}(1985)}]{aar}%
  \BibitemOpen
  \bibfield  {author} {\bibinfo {author} {\bibnamefont {Altshuler},
  \bibfnamefont {B.~L.}}, \ and\ \bibinfo {author} {\bibfnamefont {A.~G.}\
  \bibnamefont {Aronov}}} (\bibinfo {year} {1985}),\ in\ \href@noop {} {\emph
  {\bibinfo {booktitle} {Electron-Electron Interactions in Disordered
  Systems}}},\ \bibinfo {editor} {edited by\ \bibinfo {editor} {\bibfnamefont
  {A.~L.}\ \bibnamefont {Efros}}\ and\ \bibinfo {editor} {\bibfnamefont
  {M.}~\bibnamefont {Pollak}}}\ (\bibinfo  {publisher} {North-Holland,
  Amsterdam})\BibitemShut {NoStop}%
\bibitem [{\citenamefont {Altshuler}\ \emph {et~al.}(1980)\citenamefont
  {Altshuler}, \citenamefont {Khmel'nitzkii}, \citenamefont {Larkin},\ and\
  \citenamefont {Lee}}]{akl}%
  \BibitemOpen
  \bibfield  {author} {\bibinfo {author} {\bibnamefont {Altshuler},
  \bibfnamefont {B.~L.}}, \bibinfo {author} {\bibfnamefont {D.~E.}\
  \bibnamefont {Khmel'nitzkii}}, \bibinfo {author} {\bibfnamefont {A.~I.}\
  \bibnamefont {Larkin}}, \ and\ \bibinfo {author} {\bibfnamefont {P.~A.}\
  \bibnamefont {Lee}}} (\bibinfo {year} {1980}),\ \href@noop {} {\bibfield
  {journal} {\bibinfo  {journal} {Phys. Rev. B}\ }\textbf {\bibinfo {volume}
  {22}},\ \bibinfo {pages} {5142}}\BibitemShut {NoStop}%
\bibitem [{\citenamefont {Altshuler}\ and\ \citenamefont {Lee}(1988)}]{all}%
  \BibitemOpen
  \bibfield  {author} {\bibinfo {author} {\bibnamefont {Altshuler},
  \bibfnamefont {B.~L.}}, \ and\ \bibinfo {author} {\bibfnamefont {P.~A.}\
  \bibnamefont {Lee}}} (\bibinfo {year} {1988}),\ \href@noop {} {\bibfield
  {journal} {\bibinfo  {journal} {Physics Today}\ }\textbf {\bibinfo {volume}
  {41}},\ \bibinfo {pages} {36}}\BibitemShut {NoStop}%
\bibitem [{\citenamefont {Altshuler}\ \emph {et~al.}(1991)\citenamefont
  {Altshuler}, \citenamefont {Lee},\ and\ \citenamefont
  {Webb}}]{AltLeeWeb1991}%
  \BibitemOpen
  \bibinfo {editor} {\bibnamefont {Altshuler}, \bibfnamefont {B.~L.}}, \bibinfo
  {editor} {\bibfnamefont {P.~A.}\ \bibnamefont {Lee}}, \ and\ \bibinfo
  {editor} {\bibfnamefont {R.~A.}\ \bibnamefont {Webb}},\ Eds. (\bibinfo {year}
  {1991}),\ \href@noop {} {\emph {\bibinfo {title} {Mesoscopic Phenomena in
  Solids}}}\ (\bibinfo  {publisher} {North-Holland, New York})\BibitemShut
  {NoStop}%
\bibitem [{\citenamefont {Amo}\ \emph {et~al.}(2009)\citenamefont {Amo},
  \citenamefont {Sanvitto}, \citenamefont {Laussy}, \citenamefont {Ballarini},
  \citenamefont {{del Valle}}, \citenamefont {Martin}, \citenamefont
  {Lemaître}, \citenamefont {Bloch}, \citenamefont {Krizhanovskii},
  \citenamefont {Skolnick}, \citenamefont {Tejedor},\ and\ \citenamefont
  {Vina}}]{amop}%
  \BibitemOpen
  \bibfield  {author} {\bibinfo {author} {\bibnamefont {Amo}, \bibfnamefont
  {A.}}, \bibinfo {author} {\bibfnamefont {D.}~\bibnamefont {Sanvitto}},
  \bibinfo {author} {\bibfnamefont {F.~P.}\ \bibnamefont {Laussy}}, \bibinfo
  {author} {\bibfnamefont {D.}~\bibnamefont {Ballarini}}, \bibinfo {author}
  {\bibfnamefont {E.}~\bibnamefont {{del Valle}}}, \bibinfo {author}
  {\bibfnamefont {M.~D.}\ \bibnamefont {Martin}}, \bibinfo {author}
  {\bibfnamefont {A.}~\bibnamefont {Lemaître}}, \bibinfo {author}
  {\bibfnamefont {J.}~\bibnamefont {Bloch}}, \bibinfo {author} {\bibfnamefont
  {D.~N.}\ \bibnamefont {Krizhanovskii}}, \bibinfo {author} {\bibfnamefont
  {M.~S.}\ \bibnamefont {Skolnick}}, \bibinfo {author} {\bibfnamefont
  {C.}~\bibnamefont {Tejedor}}, \ and\ \bibinfo {author} {\bibfnamefont
  {L.}~\bibnamefont {Vina}}} (\bibinfo {year} {2009}),\ \href@noop {}
  {\bibfield  {journal} {\bibinfo  {journal} {Nature}\ }\textbf {\bibinfo
  {volume} {457}},\ \bibinfo {pages} {291}}\BibitemShut {NoStop}%
\bibitem [{\citenamefont {Amorim}\ and\ \citenamefont {Peres}(2012)}]{am2}%
  \BibitemOpen
  \bibfield  {author} {\bibinfo {author} {\bibnamefont {Amorim}, \bibfnamefont
  {B.}}, \ and\ \bibinfo {author} {\bibfnamefont {N.~M.~R.}\ \bibnamefont
  {Peres}}} (\bibinfo {year} {2012}),\ \href@noop {} {\bibfield  {journal}
  {\bibinfo  {journal} {J. Phys.: Condens. Matter}\ }\textbf {\bibinfo {volume}
  {24}},\ \bibinfo {pages} {335602}}\BibitemShut {NoStop}%
\bibitem [{\citenamefont {Amorim}\ \emph {et~al.}(2012)\citenamefont {Amorim},
  \citenamefont {Schiefele}, \citenamefont {Sols},\ and\ \citenamefont
  {Guinea}}]{amo}%
  \BibitemOpen
  \bibfield  {author} {\bibinfo {author} {\bibnamefont {Amorim}, \bibfnamefont
  {B.}}, \bibinfo {author} {\bibfnamefont {J.}~\bibnamefont {Schiefele}},
  \bibinfo {author} {\bibfnamefont {F.}~\bibnamefont {Sols}}, \ and\ \bibinfo
  {author} {\bibfnamefont {F.}~\bibnamefont {Guinea}}} (\bibinfo {year}
  {2012}),\ \href@noop {} {\bibfield  {journal} {\bibinfo  {journal} {Phys.
  Rev. B}\ }\textbf {\bibinfo {volume} {86}},\ \bibinfo {pages}
  {125448}}\BibitemShut {NoStop}%
\bibitem [{\citenamefont {An}\ \emph {et~al.}(2006)\citenamefont {An},
  \citenamefont {Gopalakrishnan}, \citenamefont {Shiroyanagi}, \citenamefont
  {Parks}, \citenamefont {Gramila}, \citenamefont {Pfeiffer},\ and\
  \citenamefont {West}}]{san}%
  \BibitemOpen
  \bibfield  {author} {\bibinfo {author} {\bibnamefont {An}, \bibfnamefont
  {S.}}, \bibinfo {author} {\bibfnamefont {G.}~\bibnamefont {Gopalakrishnan}},
  \bibinfo {author} {\bibfnamefont {Y.}~\bibnamefont {Shiroyanagi}}, \bibinfo
  {author} {\bibfnamefont {S.~C.}\ \bibnamefont {Parks}}, \bibinfo {author}
  {\bibfnamefont {T.~J.}\ \bibnamefont {Gramila}}, \bibinfo {author}
  {\bibfnamefont {L.~N.}\ \bibnamefont {Pfeiffer}}, \ and\ \bibinfo {author}
  {\bibfnamefont {K.~W.}\ \bibnamefont {West}}} (\bibinfo {year} {2006}),\
  \href@noop {} {\bibfield  {journal} {\bibinfo  {journal} {Physica E}\
  }\textbf {\bibinfo {volume} {34}},\ \bibinfo {pages} {214}}\BibitemShut
  {NoStop}%
\bibitem [{\citenamefont {Anderson}\ \emph {et~al.}(1979)\citenamefont
  {Anderson}, \citenamefont {Abrahams},\ and\ \citenamefont
  {Ramakrishnan}}]{anabr}%
  \BibitemOpen
  \bibfield  {author} {\bibinfo {author} {\bibnamefont {Anderson},
  \bibfnamefont {P.~W.}}, \bibinfo {author} {\bibfnamefont {E.}~\bibnamefont
  {Abrahams}}, \ and\ \bibinfo {author} {\bibfnamefont {T.~V.}\ \bibnamefont
  {Ramakrishnan}}} (\bibinfo {year} {1979}),\ \href@noop {} {\bibfield
  {journal} {\bibinfo  {journal} {Phys. Rev. Lett.}\ }\textbf {\bibinfo
  {volume} {43}},\ \bibinfo {pages} {718}}\BibitemShut {NoStop}%
\bibitem [{\citenamefont {Ando}(2006)}]{and}%
  \BibitemOpen
  \bibfield  {author} {\bibinfo {author} {\bibnamefont {Ando}, \bibfnamefont
  {T.}}} (\bibinfo {year} {2006}),\ \href@noop {} {\bibfield  {journal}
  {\bibinfo  {journal} {Journal of the Physical Society of Japan}\ }\textbf
  {\bibinfo {volume} {75}}~(\bibinfo {number} {7}),\ \bibinfo {pages}
  {074716}}\BibitemShut {NoStop}%
\bibitem [{\citenamefont {Andreev}\ \emph {et~al.}(2011)\citenamefont
  {Andreev}, \citenamefont {Kivelson},\ and\ \citenamefont {Spivak}}]{aks}%
  \BibitemOpen
  \bibfield  {author} {\bibinfo {author} {\bibnamefont {Andreev}, \bibfnamefont
  {A.~V.}}, \bibinfo {author} {\bibfnamefont {S.~A.}\ \bibnamefont {Kivelson}},
  \ and\ \bibinfo {author} {\bibfnamefont {B.}~\bibnamefont {Spivak}}}
  (\bibinfo {year} {2011}),\ \href {\doibase 10.1103/PhysRevLett.106.256804}
  {\bibfield  {journal} {\bibinfo  {journal} {Phys. Rev. Lett.}\ }\textbf
  {\bibinfo {volume} {106}},\ \bibinfo {pages} {256804}}\BibitemShut {NoStop}%
\bibitem [{\citenamefont {Apalkov}\ and\ \citenamefont {Raikh}(2005)}]{apa}%
  \BibitemOpen
  \bibfield  {author} {\bibinfo {author} {\bibnamefont {Apalkov}, \bibfnamefont
  {V.~M.}}, \ and\ \bibinfo {author} {\bibfnamefont {M.~E.}\ \bibnamefont
  {Raikh}}} (\bibinfo {year} {2005}),\ \href@noop {} {\bibfield  {journal}
  {\bibinfo  {journal} {Phys. Rev. B}\ }\textbf {\bibinfo {volume} {71}},\
  \bibinfo {pages} {245109}}\BibitemShut {NoStop}%
\bibitem [{\citenamefont {Apostolov}\ \emph {et~al.}(2014)\citenamefont
  {Apostolov}, \citenamefont {Levchenko},\ and\ \citenamefont
  {Andreev}}]{apos}%
  \BibitemOpen
  \bibfield  {author} {\bibinfo {author} {\bibnamefont {Apostolov},
  \bibfnamefont {S.~S.}}, \bibinfo {author} {\bibfnamefont {A.}~\bibnamefont
  {Levchenko}}, \ and\ \bibinfo {author} {\bibfnamefont {A.~V.}\ \bibnamefont
  {Andreev}}} (\bibinfo {year} {2014}),\ \href {\doibase
  10.1103/PhysRevB.89.121104} {\bibfield  {journal} {\bibinfo  {journal} {Phys.
  Rev. B}\ }\textbf {\bibinfo {volume} {89}},\ \bibinfo {pages}
  {121104}}\BibitemShut {NoStop}%
\bibitem [{\citenamefont {Aristov}(2007)}]{ari}%
  \BibitemOpen
  \bibfield  {author} {\bibinfo {author} {\bibnamefont {Aristov}, \bibfnamefont
  {D.~N.}}} (\bibinfo {year} {2007}),\ \href@noop {} {\bibfield  {journal}
  {\bibinfo  {journal} {Phys. Rev. B}\ }\textbf {\bibinfo {volume} {76}},\
  \bibinfo {pages} {085327}}\BibitemShut {NoStop}%
\bibitem [{\citenamefont {Arnold}\ \emph {et~al.}(2000)\citenamefont {Arnold},
  \citenamefont {Moore},\ and\ \citenamefont {Yaffe}}]{amy}%
  \BibitemOpen
  \bibfield  {author} {\bibinfo {author} {\bibnamefont {Arnold}, \bibfnamefont
  {P.}}, \bibinfo {author} {\bibfnamefont {G.~D.}\ \bibnamefont {Moore}}, \
  and\ \bibinfo {author} {\bibfnamefont {L.~G.}\ \bibnamefont {Yaffe}}}
  (\bibinfo {year} {2000}),\ \href@noop {} {\bibfield  {journal} {\bibinfo
  {journal} {Journal of High Energy Physics}\ }\textbf {\bibinfo {volume}
  {2000}}~(\bibinfo {number} {11}),\ \bibinfo {pages} {001}}\BibitemShut
  {NoStop}%
\bibitem [{\citenamefont {Aslamazov}\ and\ \citenamefont {Larkin}(1968)}]{asl}%
  \BibitemOpen
  \bibfield  {author} {\bibinfo {author} {\bibnamefont {Aslamazov},
  \bibfnamefont {L.~G.}}, \ and\ \bibinfo {author} {\bibfnamefont {A.~I.}\
  \bibnamefont {Larkin}}} (\bibinfo {year} {1968}),\ \href@noop {} {\bibfield
  {journal} {\bibinfo  {journal} {Fiz. Tverd. Tela.}\ }\textbf {\bibinfo
  {volume} {10}},\ \bibinfo {pages} {1104}},\ \bibinfo {note} {[Sov. Phys.
  Solid State. {\bf 10}, 875 (1968)]}\BibitemShut {NoStop}%
\bibitem [{\citenamefont {Badalyan}\ and\ \citenamefont {Kim}(2003)}]{bkm}%
  \BibitemOpen
  \bibfield  {author} {\bibinfo {author} {\bibnamefont {Badalyan},
  \bibfnamefont {S.~M.}}, \ and\ \bibinfo {author} {\bibfnamefont {C.~S.}\
  \bibnamefont {Kim}}} (\bibinfo {year} {2003}),\ \href@noop {} {\bibfield
  {journal} {\bibinfo  {journal} {Solid State Commun.}\ }\textbf {\bibinfo
  {volume} {127}},\ \bibinfo {pages} {521}}\BibitemShut {NoStop}%
\bibitem [{\citenamefont {Badalyan}\ \emph {et~al.}(2007)\citenamefont
  {Badalyan}, \citenamefont {Kim}, \citenamefont {Vignale},\ and\ \citenamefont
  {Senatore}}]{ba3}%
  \BibitemOpen
  \bibfield  {author} {\bibinfo {author} {\bibnamefont {Badalyan},
  \bibfnamefont {S.~M.}}, \bibinfo {author} {\bibfnamefont {C.~S.}\
  \bibnamefont {Kim}}, \bibinfo {author} {\bibfnamefont {G.}~\bibnamefont
  {Vignale}}, \ and\ \bibinfo {author} {\bibfnamefont {G.}~\bibnamefont
  {Senatore}}} (\bibinfo {year} {2007}),\ \href@noop {} {\bibfield  {journal}
  {\bibinfo  {journal} {Phys. Rev. B}\ }\textbf {\bibinfo {volume} {75}},\
  \bibinfo {pages} {125321}}\BibitemShut {NoStop}%
\bibitem [{\citenamefont {Badalyan}\ and\ \citenamefont {Peeters}(2012)}]{bad}%
  \BibitemOpen
  \bibfield  {author} {\bibinfo {author} {\bibnamefont {Badalyan},
  \bibfnamefont {S.~M.}}, \ and\ \bibinfo {author} {\bibfnamefont {F.~M.}\
  \bibnamefont {Peeters}}} (\bibinfo {year} {2012}),\ \href@noop {} {\bibfield
  {journal} {\bibinfo  {journal} {Phys. Rev. B}\ }\textbf {\bibinfo {volume}
  {86}},\ \bibinfo {pages} {121405}}\BibitemShut {NoStop}%
\bibitem [{\citenamefont {Badalyan}\ and\ \citenamefont
  {R{\"o}ssler}(1999)}]{ba2}%
  \BibitemOpen
  \bibfield  {author} {\bibinfo {author} {\bibnamefont {Badalyan},
  \bibfnamefont {S.~M.}}, \ and\ \bibinfo {author} {\bibfnamefont
  {U.}~\bibnamefont {R{\"o}ssler}}} (\bibinfo {year} {1999}),\ \href@noop {}
  {\bibfield  {journal} {\bibinfo  {journal} {Phys. Rev. B}\ }\textbf {\bibinfo
  {volume} {59}},\ \bibinfo {pages} {5643}}\BibitemShut {NoStop}%
\bibitem [{\citenamefont {Badalyan}\ and\ \citenamefont {Vignale}(2009)}]{bav}%
  \BibitemOpen
  \bibfield  {author} {\bibinfo {author} {\bibnamefont {Badalyan},
  \bibfnamefont {S.~M.}}, \ and\ \bibinfo {author} {\bibfnamefont
  {G.}~\bibnamefont {Vignale}}} (\bibinfo {year} {2009}),\ \href@noop {}
  {\bibfield  {journal} {\bibinfo  {journal} {Phys. Rev. Lett.}\ }\textbf
  {\bibinfo {volume} {103}},\ \bibinfo {pages} {196601}}\BibitemShut {NoStop}%
\bibitem [{\citenamefont {Baker}\ and\ \citenamefont {Rojo}(2001)}]{bak}%
  \BibitemOpen
  \bibfield  {author} {\bibinfo {author} {\bibnamefont {Baker}, \bibfnamefont
  {J.}}, \ and\ \bibinfo {author} {\bibfnamefont {A.~G.}\ \bibnamefont {Rojo}}}
  (\bibinfo {year} {2001}),\ \href@noop {} {\bibfield  {journal} {\bibinfo
  {journal} {J. Phys.: Condens. Matter}\ }\textbf {\bibinfo {volume} {13}},\
  \bibinfo {pages} {3389}}\BibitemShut {NoStop}%
\bibitem [{\citenamefont {Baker}\ \emph {et~al.}(1999)\citenamefont {Baker},
  \citenamefont {Vignale},\ and\ \citenamefont {Rojo}}]{ba4}%
  \BibitemOpen
  \bibfield  {author} {\bibinfo {author} {\bibnamefont {Baker}, \bibfnamefont
  {J.}}, \bibinfo {author} {\bibfnamefont {G.}~\bibnamefont {Vignale}}, \ and\
  \bibinfo {author} {\bibfnamefont {A.~G.}\ \bibnamefont {Rojo}}} (\bibinfo
  {year} {1999}),\ \href@noop {} {\bibfield  {journal} {\bibinfo  {journal}
  {Phys. Rev. B}\ }\textbf {\bibinfo {volume} {60}},\ \bibinfo {pages}
  {8804}}\BibitemShut {NoStop}%
\bibitem [{\citenamefont {Balili}\ \emph {et~al.}(2007)\citenamefont {Balili},
  \citenamefont {Hartwell}, \citenamefont {Snoke}, \citenamefont {Pfeiffer},\
  and\ \citenamefont {West}}]{bali}%
  \BibitemOpen
  \bibfield  {author} {\bibinfo {author} {\bibnamefont {Balili}, \bibfnamefont
  {R.}}, \bibinfo {author} {\bibfnamefont {V.}~\bibnamefont {Hartwell}},
  \bibinfo {author} {\bibfnamefont {D.}~\bibnamefont {Snoke}}, \bibinfo
  {author} {\bibfnamefont {L.}~\bibnamefont {Pfeiffer}}, \ and\ \bibinfo
  {author} {\bibfnamefont {K.}~\bibnamefont {West}}} (\bibinfo {year} {2007}),\
  \href {\doibase 10.1126/science.1140990} {\bibfield  {journal} {\bibinfo
  {journal} {Science}\ }\textbf {\bibinfo {volume} {316}}~(\bibinfo {number}
  {5827}),\ \bibinfo {pages} {1007}}\BibitemShut {NoStop}%
\bibitem [{\citenamefont {Balram}\ \emph {et~al.}(2014)\citenamefont {Balram},
  \citenamefont {Hutasoit},\ and\ \citenamefont {Jain}}]{jain}%
  \BibitemOpen
  \bibfield  {author} {\bibinfo {author} {\bibnamefont {Balram}, \bibfnamefont
  {A.~C.}}, \bibinfo {author} {\bibfnamefont {J.~A.}\ \bibnamefont {Hutasoit}},
  \ and\ \bibinfo {author} {\bibfnamefont {J.~K.}\ \bibnamefont {Jain}}}
  (\bibinfo {year} {2014}),\ \href@noop {} {\bibfield  {journal} {\bibinfo
  {journal} {Phys. Rev. B}\ }\textbf {\bibinfo {volume} {90}},\ \bibinfo
  {pages} {045103}}\BibitemShut {NoStop}%
\bibitem [{\citenamefont {Berk}\ \emph {et~al.}(1995)\citenamefont {Berk},
  \citenamefont {Kamenev}, \citenamefont {Palevski}, \citenamefont {Pfeiffer},\
  and\ \citenamefont {West}}]{berk}%
  \BibitemOpen
  \bibfield  {author} {\bibinfo {author} {\bibnamefont {Berk}, \bibfnamefont
  {Y.}}, \bibinfo {author} {\bibfnamefont {A.}~\bibnamefont {Kamenev}},
  \bibinfo {author} {\bibfnamefont {A.}~\bibnamefont {Palevski}}, \bibinfo
  {author} {\bibfnamefont {L.~N.}\ \bibnamefont {Pfeiffer}}, \ and\ \bibinfo
  {author} {\bibfnamefont {K.~W.}\ \bibnamefont {West}}} (\bibinfo {year}
  {1995}),\ \href {\doibase 10.1103/PhysRevB.51.2604} {\bibfield  {journal}
  {\bibinfo  {journal} {Phys. Rev. B}\ }\textbf {\bibinfo {volume} {51}},\
  \bibinfo {pages} {2604}}\BibitemShut {NoStop}%
\bibitem [{\citenamefont {Berman}\ \emph
  {et~al.}(2010{\natexlab{a}})\citenamefont {Berman}, \citenamefont
  {Kezerashvili},\ and\ \citenamefont {Lozovik}}]{loz}%
  \BibitemOpen
  \bibfield  {author} {\bibinfo {author} {\bibnamefont {Berman}, \bibfnamefont
  {O.~L.}}, \bibinfo {author} {\bibfnamefont {R.}~\bibnamefont {Kezerashvili}},
  \ and\ \bibinfo {author} {\bibfnamefont {Y.~E.}\ \bibnamefont {Lozovik}}}
  (\bibinfo {year} {2010}{\natexlab{a}}),\ \href@noop {} {\bibfield  {journal}
  {\bibinfo  {journal} {Physics Letters A}\ }\textbf {\bibinfo {volume}
  {374}},\ \bibinfo {pages} {3681}}\BibitemShut {NoStop}%
\bibitem [{\citenamefont {Berman}\ \emph
  {et~al.}(2010{\natexlab{b}})\citenamefont {Berman}, \citenamefont
  {Kezerashvili},\ and\ \citenamefont {Lozovik}}]{loz2}%
  \BibitemOpen
  \bibfield  {author} {\bibinfo {author} {\bibnamefont {Berman}, \bibfnamefont
  {O.~L.}}, \bibinfo {author} {\bibfnamefont {R.}~\bibnamefont {Kezerashvili}},
  \ and\ \bibinfo {author} {\bibfnamefont {Y.~E.}\ \bibnamefont {Lozovik}}}
  (\bibinfo {year} {2010}{\natexlab{b}}),\ \href@noop {} {\bibfield  {journal}
  {\bibinfo  {journal} {Phys. Rev. B}\ }\textbf {\bibinfo {volume} {82}},\
  \bibinfo {pages} {125307}}\BibitemShut {NoStop}%
\bibitem [{\citenamefont {Berman}\ \emph {et~al.}(2014)\citenamefont {Berman},
  \citenamefont {Kezerashvili},\ and\ \citenamefont {Kolmakov}}]{ber}%
  \BibitemOpen
  \bibfield  {author} {\bibinfo {author} {\bibnamefont {Berman}, \bibfnamefont
  {O.~L.}}, \bibinfo {author} {\bibfnamefont {R.~Y.}\ \bibnamefont
  {Kezerashvili}}, \ and\ \bibinfo {author} {\bibfnamefont {G.~V.}\
  \bibnamefont {Kolmakov}}} (\bibinfo {year} {2014}),\ \href {\doibase
  10.1021/nn503787q} {\bibfield  {journal} {\bibinfo  {journal} {ACS Nano}\
  }\textbf {\bibinfo {volume} {8}}~(\bibinfo {number} {10}),\ \bibinfo {pages}
  {10437}}\BibitemShut {NoStop}%
\bibitem [{\citenamefont {Bernevig}\ and\ \citenamefont
  {Hughes}(2013)}]{bernevig}%
  \BibitemOpen
  \bibfield  {author} {\bibinfo {author} {\bibnamefont {Bernevig},
  \bibfnamefont {A.}}, \ and\ \bibinfo {author} {\bibfnamefont
  {T.}~\bibnamefont {Hughes}}} (\bibinfo {year} {2013}),\ \href@noop {} {\emph
  {\bibinfo {title} {Topological Insulators and Topological Superconductors}}}\
  (\bibinfo  {publisher} {Princeton University Press})\BibitemShut {NoStop}%
\bibitem [{\citenamefont {Blatt}\ \emph {et~al.}(1962)\citenamefont {Blatt},
  \citenamefont {B\"oer},\ and\ \citenamefont {Brandt}}]{bla}%
  \BibitemOpen
  \bibfield  {author} {\bibinfo {author} {\bibnamefont {Blatt}, \bibfnamefont
  {J.~M.}}, \bibinfo {author} {\bibfnamefont {K.~W.}\ \bibnamefont {B\"oer}}, \
  and\ \bibinfo {author} {\bibfnamefont {W.}~\bibnamefont {Brandt}}} (\bibinfo
  {year} {1962}),\ \href {\doibase 10.1103/PhysRev.126.1691} {\bibfield
  {journal} {\bibinfo  {journal} {Phys. Rev.}\ }\textbf {\bibinfo {volume}
  {126}},\ \bibinfo {pages} {1691}}\BibitemShut {NoStop}%
\bibitem [{\citenamefont {Bloch}(1930)}]{bloch}%
  \BibitemOpen
  \bibfield  {author} {\bibinfo {author} {\bibnamefont {Bloch}, \bibfnamefont
  {F.}}} (\bibinfo {year} {1930}),\ \href@noop {} {\bibfield  {journal}
  {\bibinfo  {journal} {Zeitschrift f{\"u}r Physik}\ }\textbf {\bibinfo
  {volume} {59}},\ \bibinfo {pages} {208}}\BibitemShut {NoStop}%
\bibitem [{\citenamefont {Boebinger}\ \emph {et~al.}(1991)\citenamefont
  {Boebinger}, \citenamefont {Passner}, \citenamefont {Pfeiffer},\ and\
  \citenamefont {West}}]{boeb}%
  \BibitemOpen
  \bibfield  {author} {\bibinfo {author} {\bibnamefont {Boebinger},
  \bibfnamefont {G.~S.}}, \bibinfo {author} {\bibfnamefont {A.}~\bibnamefont
  {Passner}}, \bibinfo {author} {\bibfnamefont {L.~N.}\ \bibnamefont
  {Pfeiffer}}, \ and\ \bibinfo {author} {\bibfnamefont {K.~W.}\ \bibnamefont
  {West}}} (\bibinfo {year} {1991}),\ \href {\doibase
  10.1103/PhysRevB.43.12673} {\bibfield  {journal} {\bibinfo  {journal} {Phys.
  Rev. B}\ }\textbf {\bibinfo {volume} {43}},\ \bibinfo {pages}
  {12673}}\BibitemShut {NoStop}%
\bibitem [{\citenamefont {Boiko}\ and\ \citenamefont {Sirenko}(1988)}]{bo1}%
  \BibitemOpen
  \bibfield  {author} {\bibinfo {author} {\bibnamefont {Boiko}, \bibfnamefont
  {I.~I.}}, \ and\ \bibinfo {author} {\bibfnamefont {Y.~M.}\ \bibnamefont
  {Sirenko}}} (\bibinfo {year} {1988}),\ \href@noop {} {\bibfield  {journal}
  {\bibinfo  {journal} {Zh. Tekh. Fiz.}\ }\textbf {\bibinfo {volume} {58}},\
  \bibinfo {pages} {967}},\ \bibinfo {note} {[Sov. Phys. Tech. Phys. {\bf 33},
  586 (1988)]}\BibitemShut {NoStop}%
\bibitem [{\citenamefont {Boiko}\ and\ \citenamefont {Sirenko}(1990)}]{bo12}%
  \BibitemOpen
  \bibfield  {author} {\bibinfo {author} {\bibnamefont {Boiko}, \bibfnamefont
  {I.~I.}}, \ and\ \bibinfo {author} {\bibfnamefont {Y.~M.}\ \bibnamefont
  {Sirenko}}} (\bibinfo {year} {1990}),\ \href@noop {} {\bibfield  {journal}
  {\bibinfo  {journal} {Phys. Status Solidi B}\ }\textbf {\bibinfo {volume}
  {159}},\ \bibinfo {pages} {805}}\BibitemShut {NoStop}%
\bibitem [{\citenamefont {Boiko}\ \emph {et~al.}(1992)\citenamefont {Boiko},
  \citenamefont {Vasilopoulos},\ and\ \citenamefont {Sirenko}}]{boi}%
  \BibitemOpen
  \bibfield  {author} {\bibinfo {author} {\bibnamefont {Boiko}, \bibfnamefont
  {I.~I.}}, \bibinfo {author} {\bibfnamefont {P.}~\bibnamefont {Vasilopoulos}},
  \ and\ \bibinfo {author} {\bibfnamefont {Y.~M.}\ \bibnamefont {Sirenko}}}
  (\bibinfo {year} {1992}),\ \href@noop {} {\bibfield  {journal} {\bibinfo
  {journal} {Phys. Rev. B}\ }\textbf {\bibinfo {volume} {45}},\ \bibinfo
  {pages} {13538}}\BibitemShut {NoStop}%
\bibitem [{\citenamefont {Bonesteel}(1993)}]{bone}%
  \BibitemOpen
  \bibfield  {author} {\bibinfo {author} {\bibnamefont {Bonesteel},
  \bibfnamefont {N.~E.}}} (\bibinfo {year} {1993}),\ \href {\doibase
  10.1103/PhysRevB.48.11484} {\bibfield  {journal} {\bibinfo  {journal} {Phys.
  Rev. B}\ }\textbf {\bibinfo {volume} {48}},\ \bibinfo {pages}
  {11484}}\BibitemShut {NoStop}%
\bibitem [{\citenamefont {B{\o}nsager}\ \emph {et~al.}(1996)\citenamefont
  {B{\o}nsager}, \citenamefont {Flensberg}, \citenamefont {Hu},\ and\
  \citenamefont {Jauho}}]{bo3}%
  \BibitemOpen
  \bibfield  {author} {\bibinfo {author} {\bibnamefont {B{\o}nsager},
  \bibfnamefont {M.~C.}}, \bibinfo {author} {\bibfnamefont {K.}~\bibnamefont
  {Flensberg}}, \bibinfo {author} {\bibfnamefont {B.~Y.-K.}\ \bibnamefont
  {Hu}}, \ and\ \bibinfo {author} {\bibfnamefont {A.~P.}\ \bibnamefont
  {Jauho}}} (\bibinfo {year} {1996}),\ \href@noop {} {\bibfield  {journal}
  {\bibinfo  {journal} {Phys. Rev. Lett.}\ }\textbf {\bibinfo {volume} {77}},\
  \bibinfo {pages} {1366}}\BibitemShut {NoStop}%
\bibitem [{\citenamefont {B{\o}nsager}\ \emph {et~al.}(1997)\citenamefont
  {B{\o}nsager}, \citenamefont {Flensberg}, \citenamefont {Hu},\ and\
  \citenamefont {Jauho}}]{bo32}%
  \BibitemOpen
  \bibfield  {author} {\bibinfo {author} {\bibnamefont {B{\o}nsager},
  \bibfnamefont {M.~C.}}, \bibinfo {author} {\bibfnamefont {K.}~\bibnamefont
  {Flensberg}}, \bibinfo {author} {\bibfnamefont {B.~Y.-K.}\ \bibnamefont
  {Hu}}, \ and\ \bibinfo {author} {\bibfnamefont {A.~P.}\ \bibnamefont
  {Jauho}}} (\bibinfo {year} {1997}),\ \href@noop {} {\bibfield  {journal}
  {\bibinfo  {journal} {Phys. Rev. B}\ }\textbf {\bibinfo {volume} {56}},\
  \bibinfo {pages} {10314}}\BibitemShut {NoStop}%
\bibitem [{\citenamefont {B{\o}nsager}\ \emph
  {et~al.}(1998{\natexlab{a}})\citenamefont {B{\o}nsager}, \citenamefont
  {Flensberg}, \citenamefont {Hu},\ and\ \citenamefont {MacDonald}}]{bo2}%
  \BibitemOpen
  \bibfield  {author} {\bibinfo {author} {\bibnamefont {B{\o}nsager},
  \bibfnamefont {M.~C.}}, \bibinfo {author} {\bibfnamefont {K.}~\bibnamefont
  {Flensberg}}, \bibinfo {author} {\bibfnamefont {B.~Y.-K.}\ \bibnamefont
  {Hu}}, \ and\ \bibinfo {author} {\bibfnamefont {A.}~\bibnamefont
  {MacDonald}}} (\bibinfo {year} {1998}{\natexlab{a}}),\ \href@noop {}
  {\bibfield  {journal} {\bibinfo  {journal} {Phys. Rev. B}\ }\textbf {\bibinfo
  {volume} {57}},\ \bibinfo {pages} {7085}}\BibitemShut {NoStop}%
\bibitem [{\citenamefont {B{\o}nsager}\ \emph
  {et~al.}(1998{\natexlab{b}})\citenamefont {B{\o}nsager}, \citenamefont
  {Flensberg}, \citenamefont {Hu},\ and\ \citenamefont {MacDonald}}]{bo22}%
  \BibitemOpen
  \bibfield  {author} {\bibinfo {author} {\bibnamefont {B{\o}nsager},
  \bibfnamefont {M.~C.}}, \bibinfo {author} {\bibfnamefont {K.}~\bibnamefont
  {Flensberg}}, \bibinfo {author} {\bibfnamefont {B.~Y.-K.}\ \bibnamefont
  {Hu}}, \ and\ \bibinfo {author} {\bibfnamefont {A.}~\bibnamefont
  {MacDonald}}} (\bibinfo {year} {1998}{\natexlab{b}}),\ \href@noop {}
  {\bibfield  {journal} {\bibinfo  {journal} {Physica B}\ }\textbf {\bibinfo
  {volume} {249-251}},\ \bibinfo {pages} {864}}\BibitemShut {NoStop}%
\bibitem [{\citenamefont {B{\o}nsager}\ \emph {et~al.}(2000)\citenamefont
  {B{\o}nsager}, \citenamefont {Kim},\ and\ \citenamefont {MacDonald}}]{bon}%
  \BibitemOpen
  \bibfield  {author} {\bibinfo {author} {\bibnamefont {B{\o}nsager},
  \bibfnamefont {M.~C.}}, \bibinfo {author} {\bibfnamefont {Y.~B.}\
  \bibnamefont {Kim}}, \ and\ \bibinfo {author} {\bibfnamefont {A.~H.}\
  \bibnamefont {MacDonald}}} (\bibinfo {year} {2000}),\ \href@noop {}
  {\bibfield  {journal} {\bibinfo  {journal} {Phys. Rev. B}\ }\textbf {\bibinfo
  {volume} {62}},\ \bibinfo {pages} {10940}}\BibitemShut {NoStop}%
\bibitem [{\citenamefont {Bostwick}\ \emph {et~al.}(2010)\citenamefont
  {Bostwick}, \citenamefont {Speck}, \citenamefont {Seyller}, \citenamefont
  {Horn}, \citenamefont {Polini}, \citenamefont {Asgari}, \citenamefont
  {MacDonald},\ and\ \citenamefont {Rotenberg}}]{plasmaron1}%
  \BibitemOpen
  \bibfield  {author} {\bibinfo {author} {\bibnamefont {Bostwick},
  \bibfnamefont {A.}}, \bibinfo {author} {\bibfnamefont {F.}~\bibnamefont
  {Speck}}, \bibinfo {author} {\bibfnamefont {T.}~\bibnamefont {Seyller}},
  \bibinfo {author} {\bibfnamefont {K.}~\bibnamefont {Horn}}, \bibinfo {author}
  {\bibfnamefont {M.}~\bibnamefont {Polini}}, \bibinfo {author} {\bibfnamefont
  {R.}~\bibnamefont {Asgari}}, \bibinfo {author} {\bibfnamefont {A.~H.}\
  \bibnamefont {MacDonald}}, \ and\ \bibinfo {author} {\bibfnamefont
  {E.}~\bibnamefont {Rotenberg}}} (\bibinfo {year} {2010}),\ \href {\doibase
  10.1126/science.1186489} {\bibfield  {journal} {\bibinfo  {journal}
  {Science}\ }\textbf {\bibinfo {volume} {328}}~(\bibinfo {number} {5981}),\
  \bibinfo {pages} {999}}\BibitemShut {NoStop}%
\bibitem [{\citenamefont {Braude}\ and\ \citenamefont {Stern}(2001)}]{bra}%
  \BibitemOpen
  \bibfield  {author} {\bibinfo {author} {\bibnamefont {Braude}, \bibfnamefont
  {V.}}, \ and\ \bibinfo {author} {\bibfnamefont {A.}~\bibnamefont {Stern}}}
  (\bibinfo {year} {2001}),\ \href@noop {} {\bibfield  {journal} {\bibinfo
  {journal} {Phys. Rev. B}\ }\textbf {\bibinfo {volume} {64}},\ \bibinfo
  {pages} {115431}}\BibitemShut {NoStop}%
\bibitem [{\citenamefont {Brener}\ and\ \citenamefont {Metzner}(2005)}]{mez}%
  \BibitemOpen
  \bibfield  {author} {\bibinfo {author} {\bibnamefont {Brener}, \bibfnamefont
  {S.}}, \ and\ \bibinfo {author} {\bibfnamefont {W.}~\bibnamefont {Metzner}}}
  (\bibinfo {year} {2005}),\ \href@noop {} {\bibfield  {journal} {\bibinfo
  {journal} {Pis'ma Zh. Eksp. Teor. Fiz.}\ }\textbf {\bibinfo {volume} {81}},\
  \bibinfo {pages} {648}},\ \bibinfo {note} {[JETP Lett. {\bf 81}, 498
  (2005)]}\BibitemShut {NoStop}%
\bibitem [{\citenamefont {Bulnes~Cuetara}\ \emph {et~al.}(2013)\citenamefont
  {Bulnes~Cuetara}, \citenamefont {Esposito}, \citenamefont {Schaller},\ and\
  \citenamefont {Gaspard}}]{cue}%
  \BibitemOpen
  \bibfield  {author} {\bibinfo {author} {\bibnamefont {Bulnes~Cuetara},
  \bibfnamefont {G.}}, \bibinfo {author} {\bibfnamefont {M.}~\bibnamefont
  {Esposito}}, \bibinfo {author} {\bibfnamefont {G.}~\bibnamefont {Schaller}},
  \ and\ \bibinfo {author} {\bibfnamefont {P.}~\bibnamefont {Gaspard}}}
  (\bibinfo {year} {2013}),\ \href {\doibase 10.1103/PhysRevB.88.115134}
  {\bibfield  {journal} {\bibinfo  {journal} {Phys. Rev. B}\ }\textbf {\bibinfo
  {volume} {88}},\ \bibinfo {pages} {115134}}\BibitemShut {NoStop}%
\bibitem [{\citenamefont {Burkov}\ \emph {et~al.}(2002)\citenamefont {Burkov},
  \citenamefont {Schliemann}, \citenamefont {MacDonald},\ and\ \citenamefont
  {Girvin}}]{bur}%
  \BibitemOpen
  \bibfield  {author} {\bibinfo {author} {\bibnamefont {Burkov}, \bibfnamefont
  {A.}}, \bibinfo {author} {\bibfnamefont {J.}~\bibnamefont {Schliemann}},
  \bibinfo {author} {\bibfnamefont {A.}~\bibnamefont {MacDonald}}, \ and\
  \bibinfo {author} {\bibfnamefont {S.}~\bibnamefont {Girvin}}} (\bibinfo
  {year} {2002}),\ \href {\doibase
  http://dx.doi.org/10.1016/S1386-9477(01)00299-5} {\bibfield  {journal}
  {\bibinfo  {journal} {Physica E}\ }\textbf {\bibinfo {volume} {12}}~(\bibinfo
  {number} {1–4}),\ \bibinfo {pages} {28}}\BibitemShut {NoStop}%
\bibitem [{\citenamefont {B\"uttiker}\ \emph {et~al.}(1985)\citenamefont
  {B\"uttiker}, \citenamefont {Imry}, \citenamefont {Landauer},\ and\
  \citenamefont {Pinhas}}]{butt}%
  \BibitemOpen
  \bibfield  {author} {\bibinfo {author} {\bibnamefont {B\"uttiker},
  \bibfnamefont {M.}}, \bibinfo {author} {\bibfnamefont {Y.}~\bibnamefont
  {Imry}}, \bibinfo {author} {\bibfnamefont {R.}~\bibnamefont {Landauer}}, \
  and\ \bibinfo {author} {\bibfnamefont {S.}~\bibnamefont {Pinhas}}} (\bibinfo
  {year} {1985}),\ \href {\doibase 10.1103/PhysRevB.31.6207} {\bibfield
  {journal} {\bibinfo  {journal} {Phys. Rev. B}\ }\textbf {\bibinfo {volume}
  {31}},\ \bibinfo {pages} {6207}}\BibitemShut {NoStop}%
\bibitem [{\citenamefont {B{\"u}ttiker}\ and\ \citenamefont
  {S{\'a}nchez}(2011)}]{but}%
  \BibitemOpen
  \bibfield  {author} {\bibinfo {author} {\bibnamefont {B{\"u}ttiker},
  \bibfnamefont {M.}}, \ and\ \bibinfo {author} {\bibfnamefont
  {R.}~\bibnamefont {S{\'a}nchez}}} (\bibinfo {year} {2011}),\ \href@noop {}
  {\bibfield  {journal} {\bibinfo  {journal} {Nature Natotechnology}\ }\textbf
  {\bibinfo {volume} {6}},\ \bibinfo {pages} {757}}\BibitemShut {NoStop}%
\bibitem [{\citenamefont {Carr}\ \emph {et~al.}(2013)\citenamefont {Carr},
  \citenamefont {Narozhny},\ and\ \citenamefont {Nersesyan}}]{carr}%
  \BibitemOpen
  \bibfield  {author} {\bibinfo {author} {\bibnamefont {Carr}, \bibfnamefont
  {S.~T.}}, \bibinfo {author} {\bibfnamefont {B.~N.}\ \bibnamefont {Narozhny}},
  \ and\ \bibinfo {author} {\bibfnamefont {A.~A.}\ \bibnamefont {Nersesyan}}}
  (\bibinfo {year} {2013}),\ \href {\doibase
  http://dx.doi.org/10.1016/j.aop.2013.08.007} {\bibfield  {journal} {\bibinfo
  {journal} {Annals of Physics}\ }\textbf {\bibinfo {volume} {339}},\ \bibinfo
  {pages} {22}}\BibitemShut {NoStop}%
\bibitem [{\citenamefont {Carrega}\ \emph {et~al.}(2012)\citenamefont
  {Carrega}, \citenamefont {Tudorovskiy}, \citenamefont {Principi},
  \citenamefont {Katsnelson},\ and\ \citenamefont {Polini}}]{car}%
  \BibitemOpen
  \bibfield  {author} {\bibinfo {author} {\bibnamefont {Carrega}, \bibfnamefont
  {M.}}, \bibinfo {author} {\bibfnamefont {T.}~\bibnamefont {Tudorovskiy}},
  \bibinfo {author} {\bibfnamefont {A.}~\bibnamefont {Principi}}, \bibinfo
  {author} {\bibfnamefont {M.~I.}\ \bibnamefont {Katsnelson}}, \ and\ \bibinfo
  {author} {\bibfnamefont {M.}~\bibnamefont {Polini}}} (\bibinfo {year}
  {2012}),\ \href@noop {} {\bibfield  {journal} {\bibinfo  {journal} {New J.
  Phys.}\ }\textbf {\bibinfo {volume} {14}},\ \bibinfo {pages}
  {063033}}\BibitemShut {NoStop}%
\bibitem [{\citenamefont {Chakraborty}\ and\ \citenamefont
  {Pietil\"ainen}(1987)}]{chak}%
  \BibitemOpen
  \bibfield  {author} {\bibinfo {author} {\bibnamefont {Chakraborty},
  \bibfnamefont {T.}}, \ and\ \bibinfo {author} {\bibfnamefont
  {P.}~\bibnamefont {Pietil\"ainen}}} (\bibinfo {year} {1987}),\ \href
  {\doibase 10.1103/PhysRevLett.59.2784} {\bibfield  {journal} {\bibinfo
  {journal} {Phys. Rev. Lett.}\ }\textbf {\bibinfo {volume} {59}},\ \bibinfo
  {pages} {2784}}\BibitemShut {NoStop}%
\bibitem [{\citenamefont {Cheianov}\ and\ \citenamefont {Fal'ko}(2006)}]{fal}%
  \BibitemOpen
  \bibfield  {author} {\bibinfo {author} {\bibnamefont {Cheianov},
  \bibfnamefont {V.~V.}}, \ and\ \bibinfo {author} {\bibfnamefont {V.~I.}\
  \bibnamefont {Fal'ko}}} (\bibinfo {year} {2006}),\ \href@noop {} {\bibfield
  {journal} {\bibinfo  {journal} {Phys. Rev. Lett.}\ }\textbf {\bibinfo
  {volume} {97}},\ \bibinfo {pages} {226801}}\BibitemShut {NoStop}%
\bibitem [{\citenamefont {Chen}\ and\ \citenamefont
  {Appenzeller}(2013)}]{chen}%
  \BibitemOpen
  \bibfield  {author} {\bibinfo {author} {\bibnamefont {Chen}, \bibfnamefont
  {H.~Y.}}, \ and\ \bibinfo {author} {\bibfnamefont {J.}~\bibnamefont
  {Appenzeller}}} (\bibinfo {year} {2013}),\ \href@noop {} {\bibfield
  {journal} {\bibinfo  {journal} {Nano Research}\ }\textbf {\bibinfo {volume}
  {6}},\ \bibinfo {pages} {897}}\BibitemShut {NoStop}%
\bibitem [{\citenamefont {Chen}\ \emph {et~al.}(2015)\citenamefont {Chen},
  \citenamefont {Andreev},\ and\ \citenamefont {Levchenko}}]{chaa}%
  \BibitemOpen
  \bibfield  {author} {\bibinfo {author} {\bibnamefont {Chen}, \bibfnamefont
  {W.}}, \bibinfo {author} {\bibfnamefont {A.~V.}\ \bibnamefont {Andreev}}, \
  and\ \bibinfo {author} {\bibfnamefont {A.}~\bibnamefont {Levchenko}}}
  (\bibinfo {year} {2015}),\ \href@noop {} {}\bibinfo {note} {ArXiv:1503.05566
  (unpublished)}\BibitemShut {NoStop}%
\bibitem [{\citenamefont {Chudnovskiy}(2009)}]{chud}%
  \BibitemOpen
  \bibfield  {author} {\bibinfo {author} {\bibnamefont {Chudnovskiy},
  \bibfnamefont {A.~L.}}} (\bibinfo {year} {2009}),\ \href {\doibase
  10.1103/PhysRevB.80.081309} {\bibfield  {journal} {\bibinfo  {journal} {Phys.
  Rev. B}\ }\textbf {\bibinfo {volume} {80}},\ \bibinfo {pages}
  {081309}}\BibitemShut {NoStop}%
\bibitem [{\citenamefont {Conway}\ and\ \citenamefont {Sloane}(1988)}]{gram}%
  \BibitemOpen
  \bibfield  {author} {\bibinfo {author} {\bibnamefont {Conway}, \bibfnamefont
  {J.~H.}}, \ and\ \bibinfo {author} {\bibfnamefont {N.~J.~A.}\ \bibnamefont
  {Sloane}}} (\bibinfo {year} {1988}),\ \href@noop {} {\emph {\bibinfo {title}
  {Sphere Packings, Lattices, and Groups}}}\ (\bibinfo  {publisher}
  {Springer-Verlag, New York})\BibitemShut {NoStop}%
\bibitem [{\citenamefont {Croxall}\ \emph {et~al.}(2008)\citenamefont
  {Croxall}, \citenamefont {{Das Gupta}}, \citenamefont {Nicoll}, \citenamefont
  {Thangaraj}, \citenamefont {Beere}, \citenamefont {Farrer}, \citenamefont
  {Ritchie},\ and\ \citenamefont {Pepper}}]{cro}%
  \BibitemOpen
  \bibfield  {author} {\bibinfo {author} {\bibnamefont {Croxall}, \bibfnamefont
  {A.~F.}}, \bibinfo {author} {\bibfnamefont {K.}~\bibnamefont {{Das Gupta}}},
  \bibinfo {author} {\bibfnamefont {C.~A.}\ \bibnamefont {Nicoll}}, \bibinfo
  {author} {\bibfnamefont {M.}~\bibnamefont {Thangaraj}}, \bibinfo {author}
  {\bibfnamefont {H.~E.}\ \bibnamefont {Beere}}, \bibinfo {author}
  {\bibfnamefont {I.}~\bibnamefont {Farrer}}, \bibinfo {author} {\bibfnamefont
  {D.~A.}\ \bibnamefont {Ritchie}}, \ and\ \bibinfo {author} {\bibfnamefont
  {M.}~\bibnamefont {Pepper}}} (\bibinfo {year} {2008}),\ \href@noop {}
  {\bibfield  {journal} {\bibinfo  {journal} {Phys. Rev. Lett.}\ }\textbf
  {\bibinfo {volume} {101}},\ \bibinfo {pages} {246801}}\BibitemShut {NoStop}%
\bibitem [{\citenamefont {Cui}\ \emph {et~al.}(1988)\citenamefont {Cui},
  \citenamefont {Lei},\ and\ \citenamefont {Horing}}]{cui}%
  \BibitemOpen
  \bibfield  {author} {\bibinfo {author} {\bibnamefont {Cui}, \bibfnamefont
  {H.~L.}}, \bibinfo {author} {\bibfnamefont {X.~L.}\ \bibnamefont {Lei}}, \
  and\ \bibinfo {author} {\bibfnamefont {N.~J.~M.}\ \bibnamefont {Horing}}}
  (\bibinfo {year} {1988}),\ \href@noop {} {\bibfield  {journal} {\bibinfo
  {journal} {Phys. Rev. B}\ }\textbf {\bibinfo {volume} {37}},\ \bibinfo
  {pages} {8223}}\BibitemShut {NoStop}%
\bibitem [{\citenamefont {Cuoco}\ \emph {et~al.}(2009)\citenamefont {Cuoco},
  \citenamefont {Saldarriaga}, \citenamefont {Polcari}, \citenamefont
  {Guarino}, \citenamefont {Moran}, \citenamefont {Baca}, \citenamefont
  {Vecchione},\ and\ \citenamefont {Romano}}]{cuo}%
  \BibitemOpen
  \bibfield  {author} {\bibinfo {author} {\bibnamefont {Cuoco}, \bibfnamefont
  {M.}}, \bibinfo {author} {\bibfnamefont {W.}~\bibnamefont {Saldarriaga}},
  \bibinfo {author} {\bibfnamefont {A.}~\bibnamefont {Polcari}}, \bibinfo
  {author} {\bibfnamefont {A.}~\bibnamefont {Guarino}}, \bibinfo {author}
  {\bibfnamefont {O.}~\bibnamefont {Moran}}, \bibinfo {author} {\bibfnamefont
  {E.}~\bibnamefont {Baca}}, \bibinfo {author} {\bibfnamefont {A.}~\bibnamefont
  {Vecchione}}, \ and\ \bibinfo {author} {\bibfnamefont {P.}~\bibnamefont
  {Romano}}} (\bibinfo {year} {2009}),\ \href@noop {} {\bibfield  {journal}
  {\bibinfo  {journal} {Phys. Rev. B}\ }\textbf {\bibinfo {volume} {79}},\
  \bibinfo {pages} {014523}}\BibitemShut {NoStop}%
\bibitem [{\citenamefont {D'Amico}\ and\ \citenamefont {Vignale}(2000)}]{dam}%
  \BibitemOpen
  \bibfield  {author} {\bibinfo {author} {\bibnamefont {D'Amico}, \bibfnamefont
  {I.}}, \ and\ \bibinfo {author} {\bibfnamefont {G.}~\bibnamefont {Vignale}}}
  (\bibinfo {year} {2000}),\ \href@noop {} {\bibfield  {journal} {\bibinfo
  {journal} {Phys. Rev. B}\ }\textbf {\bibinfo {volume} {62}},\ \bibinfo
  {pages} {4853}}\BibitemShut {NoStop}%
\bibitem [{\citenamefont {{Das Gupta}}\ \emph {et~al.}(2011)\citenamefont {{Das
  Gupta}}, \citenamefont {Croxall}, \citenamefont {Waldie}, \citenamefont
  {Nicoll}, \citenamefont {Beere}, \citenamefont {Farrer}, \citenamefont
  {Ritchie},\ and\ \citenamefont {Pepper}}]{gup}%
  \BibitemOpen
  \bibfield  {author} {\bibinfo {author} {\bibnamefont {{Das Gupta}},
  \bibfnamefont {K.}}, \bibinfo {author} {\bibfnamefont {A.~F.}\ \bibnamefont
  {Croxall}}, \bibinfo {author} {\bibfnamefont {J.}~\bibnamefont {Waldie}},
  \bibinfo {author} {\bibfnamefont {C.~A.}\ \bibnamefont {Nicoll}}, \bibinfo
  {author} {\bibfnamefont {H.~E.}\ \bibnamefont {Beere}}, \bibinfo {author}
  {\bibfnamefont {I.}~\bibnamefont {Farrer}}, \bibinfo {author} {\bibfnamefont
  {D.~A.}\ \bibnamefont {Ritchie}}, \ and\ \bibinfo {author} {\bibfnamefont
  {M.}~\bibnamefont {Pepper}}} (\bibinfo {year} {2011}),\ \href@noop {}
  {\bibfield  {journal} {\bibinfo  {journal} {Adv. Cond. Matt. Phys.}\ }\textbf
  {\bibinfo {volume} {2011}},\ \bibinfo {pages} {727958}}\BibitemShut {NoStop}%
\bibitem [{\citenamefont {{Das Gupta}}\ \emph {et~al.}(2008)\citenamefont {{Das
  Gupta}}, \citenamefont {Thangaraj}, \citenamefont {Croxall}, \citenamefont
  {Beere}, \citenamefont {Nicoll}, \citenamefont {Ritchie},\ and\ \citenamefont
  {Pepper}}]{gup2}%
  \BibitemOpen
  \bibfield  {author} {\bibinfo {author} {\bibnamefont {{Das Gupta}},
  \bibfnamefont {K.}}, \bibinfo {author} {\bibfnamefont {M.}~\bibnamefont
  {Thangaraj}}, \bibinfo {author} {\bibfnamefont {A.}~\bibnamefont {Croxall}},
  \bibinfo {author} {\bibfnamefont {H.}~\bibnamefont {Beere}}, \bibinfo
  {author} {\bibfnamefont {C.}~\bibnamefont {Nicoll}}, \bibinfo {author}
  {\bibfnamefont {D.}~\bibnamefont {Ritchie}}, \ and\ \bibinfo {author}
  {\bibfnamefont {M.}~\bibnamefont {Pepper}}} (\bibinfo {year} {2008}),\
  \href@noop {} {\bibfield  {journal} {\bibinfo  {journal} {Physica E}\
  }\textbf {\bibinfo {volume} {40}}~(\bibinfo {number} {5}),\ \bibinfo {pages}
  {1693}}\BibitemShut {NoStop}%
\bibitem [{\citenamefont {Das~Sarma}\ and\ \citenamefont
  {Madhukar}(1981)}]{plasds}%
  \BibitemOpen
  \bibfield  {author} {\bibinfo {author} {\bibnamefont {Das~Sarma},
  \bibfnamefont {S.}}, \ and\ \bibinfo {author} {\bibfnamefont
  {A.}~\bibnamefont {Madhukar}}} (\bibinfo {year} {1981}),\ \href@noop {}
  {\bibfield  {journal} {\bibinfo  {journal} {Phys. Rev. B}\ }\textbf {\bibinfo
  {volume} {23}},\ \bibinfo {pages} {805}}\BibitemShut {NoStop}%
\bibitem [{\citenamefont {Das~Sarma}\ and\ \citenamefont {Mason}(1985)}]{sdas}%
  \BibitemOpen
  \bibfield  {author} {\bibinfo {author} {\bibnamefont {Das~Sarma},
  \bibfnamefont {S.}}, \ and\ \bibinfo {author} {\bibfnamefont {B.~A.}\
  \bibnamefont {Mason}}} (\bibinfo {year} {1985}),\ \href@noop {} {\bibfield
  {journal} {\bibinfo  {journal} {Ann. Phys. (NY)}\ }\textbf {\bibinfo {volume}
  {163}},\ \bibinfo {pages} {78}}\BibitemShut {NoStop}%
\bibitem [{\citenamefont {Debray}\ \emph {et~al.}(2000)\citenamefont {Debray},
  \citenamefont {Vasilopoulos}, \citenamefont {Raichev}, \citenamefont
  {Perrin}, \citenamefont {Rahman},\ and\ \citenamefont {Mitchel}}]{de2}%
  \BibitemOpen
  \bibfield  {author} {\bibinfo {author} {\bibnamefont {Debray}, \bibfnamefont
  {P.}}, \bibinfo {author} {\bibfnamefont {P.}~\bibnamefont {Vasilopoulos}},
  \bibinfo {author} {\bibfnamefont {O.}~\bibnamefont {Raichev}}, \bibinfo
  {author} {\bibfnamefont {R.}~\bibnamefont {Perrin}}, \bibinfo {author}
  {\bibfnamefont {M.}~\bibnamefont {Rahman}}, \ and\ \bibinfo {author}
  {\bibfnamefont {W.~C.}\ \bibnamefont {Mitchel}}} (\bibinfo {year} {2000}),\
  \href@noop {} {\bibfield  {journal} {\bibinfo  {journal} {Physica E}\
  }\textbf {\bibinfo {volume} {6}},\ \bibinfo {pages} {694}}\BibitemShut
  {NoStop}%
\bibitem [{\citenamefont {Debray}\ \emph {et~al.}(2001)\citenamefont {Debray},
  \citenamefont {Zverev}, \citenamefont {Raichev}, \citenamefont {Klesse},
  \citenamefont {Vasilopoulos},\ and\ \citenamefont {Newrock}}]{deb}%
  \BibitemOpen
  \bibfield  {author} {\bibinfo {author} {\bibnamefont {Debray}, \bibfnamefont
  {P.}}, \bibinfo {author} {\bibfnamefont {V.}~\bibnamefont {Zverev}}, \bibinfo
  {author} {\bibfnamefont {O.}~\bibnamefont {Raichev}}, \bibinfo {author}
  {\bibfnamefont {R.}~\bibnamefont {Klesse}}, \bibinfo {author} {\bibfnamefont
  {P.}~\bibnamefont {Vasilopoulos}}, \ and\ \bibinfo {author} {\bibfnamefont
  {R.~S.}\ \bibnamefont {Newrock}}} (\bibinfo {year} {2001}),\ \href@noop {}
  {\bibfield  {journal} {\bibinfo  {journal} {J. Phys.: Condens. Matter}\
  }\textbf {\bibinfo {volume} {13}},\ \bibinfo {pages} {3389}}\BibitemShut
  {NoStop}%
\bibitem [{\citenamefont {Debray}\ \emph {et~al.}(2002)\citenamefont {Debray},
  \citenamefont {Zverev}, \citenamefont {Gurevich}, \citenamefont {Klesse},\
  and\ \citenamefont {Newrock}}]{dzg}%
  \BibitemOpen
  \bibfield  {author} {\bibinfo {author} {\bibnamefont {Debray}, \bibfnamefont
  {P.}}, \bibinfo {author} {\bibfnamefont {V.~N.}\ \bibnamefont {Zverev}},
  \bibinfo {author} {\bibfnamefont {V.}~\bibnamefont {Gurevich}}, \bibinfo
  {author} {\bibfnamefont {R.}~\bibnamefont {Klesse}}, \ and\ \bibinfo {author}
  {\bibfnamefont {R.~S.}\ \bibnamefont {Newrock}}} (\bibinfo {year} {2002}),\
  \href@noop {} {\bibfield  {journal} {\bibinfo  {journal} {Semicond. Sci.
  Technol.}\ }\textbf {\bibinfo {volume} {17}},\ \bibinfo {pages}
  {R21}}\BibitemShut {NoStop}%
\bibitem [{\citenamefont {Deshpande}\ \emph {et~al.}(2010)\citenamefont
  {Deshpande}, \citenamefont {Bockrath}, \citenamefont {Glazman},\ and\
  \citenamefont {Yacoby}}]{desh}%
  \BibitemOpen
  \bibfield  {author} {\bibinfo {author} {\bibnamefont {Deshpande},
  \bibfnamefont {V.~V.}}, \bibinfo {author} {\bibfnamefont {M.}~\bibnamefont
  {Bockrath}}, \bibinfo {author} {\bibfnamefont {L.~I.}\ \bibnamefont
  {Glazman}}, \ and\ \bibinfo {author} {\bibfnamefont {A.}~\bibnamefont
  {Yacoby}}} (\bibinfo {year} {2010}),\ \href@noop {} {\bibfield  {journal}
  {\bibinfo  {journal} {Nature}\ }\textbf {\bibinfo {volume} {464}},\ \bibinfo
  {pages} {209}}\BibitemShut {NoStop}%
\bibitem [{\citenamefont {Dmitriev}\ \emph {et~al.}(2012)\citenamefont
  {Dmitriev}, \citenamefont {Gornyi},\ and\ \citenamefont {Polyakov}}]{dgp}%
  \BibitemOpen
  \bibfield  {author} {\bibinfo {author} {\bibnamefont {Dmitriev},
  \bibfnamefont {A.~P.}}, \bibinfo {author} {\bibfnamefont {I.~V.}\
  \bibnamefont {Gornyi}}, \ and\ \bibinfo {author} {\bibfnamefont {D.~G.}\
  \bibnamefont {Polyakov}}} (\bibinfo {year} {2012}),\ \href@noop {} {\bibfield
   {journal} {\bibinfo  {journal} {Phys. Rev. B}\ }\textbf {\bibinfo {volume}
  {86}},\ \bibinfo {pages} {245402}}\BibitemShut {NoStop}%
\bibitem [{\citenamefont {Dmitriev}\ \emph {et~al.}(2008)\citenamefont
  {Dmitriev}, \citenamefont {Evers}, \citenamefont {Gornyi}, \citenamefont
  {Mirlin}, \citenamefont {Polyakov},\ and\ \citenamefont {W{\"o}lfle}}]{dmi}%
  \BibitemOpen
  \bibfield  {author} {\bibinfo {author} {\bibnamefont {Dmitriev},
  \bibfnamefont {I.~A.}}, \bibinfo {author} {\bibfnamefont {F.}~\bibnamefont
  {Evers}}, \bibinfo {author} {\bibfnamefont {I.~V.}\ \bibnamefont {Gornyi}},
  \bibinfo {author} {\bibfnamefont {A.~D.}\ \bibnamefont {Mirlin}}, \bibinfo
  {author} {\bibfnamefont {D.~G.}\ \bibnamefont {Polyakov}}, \ and\ \bibinfo
  {author} {\bibfnamefont {P.}~\bibnamefont {W{\"o}lfle}}} (\bibinfo {year}
  {2008}),\ \href@noop {} {\bibfield  {journal} {\bibinfo  {journal} {Phys.
  Status Solidi B}\ }\textbf {\bibinfo {volume} {245}},\ \bibinfo {pages}
  {239}}\BibitemShut {NoStop}%
\bibitem [{\citenamefont {Dolcini}\ \emph {et~al.}(2010)\citenamefont
  {Dolcini}, \citenamefont {Rainis}, \citenamefont {Taddei}, \citenamefont
  {Polini}, \citenamefont {Fazio},\ and\ \citenamefont {MacDonald}}]{dolci}%
  \BibitemOpen
  \bibfield  {author} {\bibinfo {author} {\bibnamefont {Dolcini}, \bibfnamefont
  {F.}}, \bibinfo {author} {\bibfnamefont {D.}~\bibnamefont {Rainis}}, \bibinfo
  {author} {\bibfnamefont {F.}~\bibnamefont {Taddei}}, \bibinfo {author}
  {\bibfnamefont {M.}~\bibnamefont {Polini}}, \bibinfo {author} {\bibfnamefont
  {R.}~\bibnamefont {Fazio}}, \ and\ \bibinfo {author} {\bibfnamefont {A.~H.}\
  \bibnamefont {MacDonald}}} (\bibinfo {year} {2010}),\ \href {\doibase
  10.1103/PhysRevLett.104.027004} {\bibfield  {journal} {\bibinfo  {journal}
  {Phys. Rev. Lett.}\ }\textbf {\bibinfo {volume} {104}},\ \bibinfo {pages}
  {027004}}\BibitemShut {NoStop}%
\bibitem [{\citenamefont {Donarini}\ \emph {et~al.}(2003)\citenamefont
  {Donarini}, \citenamefont {Ferrari}, \citenamefont {Jauho},\ and\
  \citenamefont {Molinari}}]{don}%
  \BibitemOpen
  \bibfield  {author} {\bibinfo {author} {\bibnamefont {Donarini},
  \bibfnamefont {A.}}, \bibinfo {author} {\bibfnamefont {R.}~\bibnamefont
  {Ferrari}}, \bibinfo {author} {\bibfnamefont {A.~P.}\ \bibnamefont {Jauho}},
  \ and\ \bibinfo {author} {\bibfnamefont {L.}~\bibnamefont {Molinari}}}
  (\bibinfo {year} {2003}),\ \href@noop {} {\bibfield  {journal} {\bibinfo
  {journal} {Phys. Lett. A}\ }\textbf {\bibinfo {volume} {312}},\ \bibinfo
  {pages} {123}}\BibitemShut {NoStop}%
\bibitem [{\citenamefont {Du}\ \emph {et~al.}(2015)\citenamefont {Du},
  \citenamefont {Knez}, \citenamefont {Sullivan},\ and\ \citenamefont
  {Du}}]{rrdu}%
  \BibitemOpen
  \bibfield  {author} {\bibinfo {author} {\bibnamefont {Du}, \bibfnamefont
  {L.}}, \bibinfo {author} {\bibfnamefont {I.}~\bibnamefont {Knez}}, \bibinfo
  {author} {\bibfnamefont {G.}~\bibnamefont {Sullivan}}, \ and\ \bibinfo
  {author} {\bibfnamefont {R.-R.}\ \bibnamefont {Du}}} (\bibinfo {year}
  {2015}),\ \href {\doibase 10.1103/PhysRevLett.114.096802} {\bibfield
  {journal} {\bibinfo  {journal} {Phys. Rev. Lett.}\ }\textbf {\bibinfo
  {volume} {114}},\ \bibinfo {pages} {096802}}\BibitemShut {NoStop}%
\bibitem [{\citenamefont {Duan}(1995)}]{duan}%
  \BibitemOpen
  \bibfield  {author} {\bibinfo {author} {\bibnamefont {Duan}, \bibfnamefont
  {J.-M.}}} (\bibinfo {year} {1995}),\ \href
  {http://stacks.iop.org/0295-5075/29/i=6/a=010} {\bibfield  {journal}
  {\bibinfo  {journal} {EPL (Europhysics Letters)}\ }\textbf {\bibinfo {volume}
  {29}}~(\bibinfo {number} {6}),\ \bibinfo {pages} {489}}\BibitemShut {NoStop}%
\bibitem [{\citenamefont {Duan}\ and\ \citenamefont {Yip}(1993)}]{dua}%
  \BibitemOpen
  \bibfield  {author} {\bibinfo {author} {\bibnamefont {Duan}, \bibfnamefont
  {J.-M.}}, \ and\ \bibinfo {author} {\bibfnamefont {S.}~\bibnamefont {Yip}}}
  (\bibinfo {year} {1993}),\ \href@noop {} {\bibfield  {journal} {\bibinfo
  {journal} {Phys. Rev. Lett.}\ }\textbf {\bibinfo {volume} {70}},\ \bibinfo
  {pages} {3647}}\BibitemShut {NoStop}%
\bibitem [{\citenamefont {Duine}\ \emph {et~al.}(2011)\citenamefont {Duine},
  \citenamefont {Polini}, \citenamefont {Raoux}, \citenamefont {Stoof},\ and\
  \citenamefont {Vignale}}]{dui}%
  \BibitemOpen
  \bibfield  {author} {\bibinfo {author} {\bibnamefont {Duine}, \bibfnamefont
  {R.~A.}}, \bibinfo {author} {\bibfnamefont {M.}~\bibnamefont {Polini}},
  \bibinfo {author} {\bibfnamefont {A.}~\bibnamefont {Raoux}}, \bibinfo
  {author} {\bibfnamefont {H.}~\bibnamefont {Stoof}}, \ and\ \bibinfo {author}
  {\bibfnamefont {G.}~\bibnamefont {Vignale}}} (\bibinfo {year} {2011}),\
  \href@noop {} {\bibfield  {journal} {\bibinfo  {journal} {New J. Phys.}\
  }\textbf {\bibinfo {volume} {13}},\ \bibinfo {pages} {045010}}\BibitemShut
  {NoStop}%
\bibitem [{\citenamefont {Duine}\ \emph {et~al.}(2010)\citenamefont {Duine},
  \citenamefont {Polini}, \citenamefont {Stoof},\ and\ \citenamefont
  {Vignale}}]{du1}%
  \BibitemOpen
  \bibfield  {author} {\bibinfo {author} {\bibnamefont {Duine}, \bibfnamefont
  {R.~A.}}, \bibinfo {author} {\bibfnamefont {M.}~\bibnamefont {Polini}},
  \bibinfo {author} {\bibfnamefont {H.}~\bibnamefont {Stoof}}, \ and\ \bibinfo
  {author} {\bibfnamefont {G.}~\bibnamefont {Vignale}}} (\bibinfo {year}
  {2010}),\ \href@noop {} {\bibfield  {journal} {\bibinfo  {journal} {Phys.
  Rev. Lett.}\ }\textbf {\bibinfo {volume} {104}},\ \bibinfo {pages}
  {220403}}\BibitemShut {NoStop}%
\bibitem [{\citenamefont {Duine}\ and\ \citenamefont {Stoof}(2009)}]{du2}%
  \BibitemOpen
  \bibfield  {author} {\bibinfo {author} {\bibnamefont {Duine}, \bibfnamefont
  {R.~A.}}, \ and\ \bibinfo {author} {\bibfnamefont {H.~T.~C.}\ \bibnamefont
  {Stoof}}} (\bibinfo {year} {2009}),\ \href@noop {} {\bibfield  {journal}
  {\bibinfo  {journal} {Phys. Rev. Lett.}\ }\textbf {\bibinfo {volume} {103}},\
  \bibinfo {pages} {170401}}\BibitemShut {NoStop}%
\bibitem [{\citenamefont {Dykhne}\ and\ \citenamefont {Ruzin}(1994)}]{dyk}%
  \BibitemOpen
  \bibfield  {author} {\bibinfo {author} {\bibnamefont {Dykhne}, \bibfnamefont
  {A.~M.}}, \ and\ \bibinfo {author} {\bibfnamefont {I.~M.}\ \bibnamefont
  {Ruzin}}} (\bibinfo {year} {1994}),\ \href {\doibase
  10.1103/PhysRevB.50.2369} {\bibfield  {journal} {\bibinfo  {journal} {Phys.
  Rev. B}\ }\textbf {\bibinfo {volume} {50}},\ \bibinfo {pages}
  {2369}}\BibitemShut {NoStop}%
\bibitem [{\citenamefont {Efetov}\ and\ \citenamefont {Kim}(2010)}]{efk}%
  \BibitemOpen
  \bibfield  {author} {\bibinfo {author} {\bibnamefont {Efetov}, \bibfnamefont
  {D.~K.}}, \ and\ \bibinfo {author} {\bibfnamefont {P.}~\bibnamefont {Kim}}}
  (\bibinfo {year} {2010}),\ \href@noop {} {\bibfield  {journal} {\bibinfo
  {journal} {Phys. Rev. Lett.}\ }\textbf {\bibinfo {volume} {105}},\ \bibinfo
  {pages} {256805}}\BibitemShut {NoStop}%
\bibitem [{\citenamefont {Efimkin}\ and\ \citenamefont
  {Lozovik}(2011)}]{loze3}%
  \BibitemOpen
  \bibfield  {author} {\bibinfo {author} {\bibnamefont {Efimkin}, \bibfnamefont
  {D.~K.}}, \ and\ \bibinfo {author} {\bibfnamefont {Y.~E.}\ \bibnamefont
  {Lozovik}}} (\bibinfo {year} {2011}),\ \href@noop {} {\bibfield  {journal}
  {\bibinfo  {journal} {Zh. Eksp. Teor. Fiz.}\ }\textbf {\bibinfo {volume}
  {140}},\ \bibinfo {pages} {1009}},\ \bibinfo {note} {[JETP {\bf 113}, 880
  (2011)]}\BibitemShut {NoStop}%
\bibitem [{\citenamefont {Efimkin}\ and\ \citenamefont {Lozovik}(2013)}]{loze}%
  \BibitemOpen
  \bibfield  {author} {\bibinfo {author} {\bibnamefont {Efimkin}, \bibfnamefont
  {D.~K.}}, \ and\ \bibinfo {author} {\bibfnamefont {Y.~E.}\ \bibnamefont
  {Lozovik}}} (\bibinfo {year} {2013}),\ \href@noop {} {\bibfield  {journal}
  {\bibinfo  {journal} {Phys. Rev. B}\ }\textbf {\bibinfo {volume} {88}},\
  \bibinfo {pages} {235420}}\BibitemShut {NoStop}%
\bibitem [{\citenamefont {Efimkin}\ \emph {et~al.}(2012)\citenamefont
  {Efimkin}, \citenamefont {Lozovik},\ and\ \citenamefont {Sokolik}}]{loze2}%
  \BibitemOpen
  \bibfield  {author} {\bibinfo {author} {\bibnamefont {Efimkin}, \bibfnamefont
  {D.~K.}}, \bibinfo {author} {\bibfnamefont {Y.~E.}\ \bibnamefont {Lozovik}},
  \ and\ \bibinfo {author} {\bibfnamefont {A.~A.}\ \bibnamefont {Sokolik}}}
  (\bibinfo {year} {2012}),\ \href {\doibase 10.1103/PhysRevB.86.115436}
  {\bibfield  {journal} {\bibinfo  {journal} {Phys. Rev. B}\ }\textbf {\bibinfo
  {volume} {86}},\ \bibinfo {pages} {115436}}\BibitemShut {NoStop}%
\bibitem [{\citenamefont {Eisenstein}(2014)}]{eis2}%
  \BibitemOpen
  \bibfield  {author} {\bibinfo {author} {\bibnamefont {Eisenstein},
  \bibfnamefont {J.}}} (\bibinfo {year} {2014}),\ \href {\doibase
  10.1146/annurev-conmatphys-031113-133832} {\bibfield  {journal} {\bibinfo
  {journal} {Annual Review of Condensed Matter Physics}\ }\textbf {\bibinfo
  {volume} {5}}~(\bibinfo {number} {1}),\ \bibinfo {pages} {159}}\BibitemShut
  {NoStop}%
\bibitem [{\citenamefont {Eisenstein}(1992)}]{eis}%
  \BibitemOpen
  \bibfield  {author} {\bibinfo {author} {\bibnamefont {Eisenstein},
  \bibfnamefont {J.~P.}}} (\bibinfo {year} {1992}),\ \href@noop {} {\bibfield
  {journal} {\bibinfo  {journal} {Superlattices Microstruct.}\ }\textbf
  {\bibinfo {volume} {12}},\ \bibinfo {pages} {107}}\BibitemShut {NoStop}%
\bibitem [{\citenamefont {Eisenstein}(1997)}]{eis1}%
  \BibitemOpen
  \bibfield  {author} {\bibinfo {author} {\bibnamefont {Eisenstein},
  \bibfnamefont {J.~P.}}} (\bibinfo {year} {1997}),\ in\ \href@noop {} {\emph
  {\bibinfo {booktitle} {Perspectives in Quantum Hall Effects}}},\ \bibinfo
  {editor} {edited by\ \bibinfo {editor} {\bibfnamefont {S.}~\bibnamefont {{Das
  Sarma}}}\ and\ \bibinfo {editor} {\bibfnamefont {A.}~\bibnamefont
  {Pinczuk}}}\ (\bibinfo  {publisher} {Wiley, New York})\BibitemShut {NoStop}%
\bibitem [{\citenamefont {Eisenstein}\ \emph {et~al.}(1992)\citenamefont
  {Eisenstein}, \citenamefont {Boebinger}, \citenamefont {Pfeiffer},
  \citenamefont {West},\ and\ \citenamefont {He}}]{eisen}%
  \BibitemOpen
  \bibfield  {author} {\bibinfo {author} {\bibnamefont {Eisenstein},
  \bibfnamefont {J.~P.}}, \bibinfo {author} {\bibfnamefont {G.~S.}\
  \bibnamefont {Boebinger}}, \bibinfo {author} {\bibfnamefont {L.~N.}\
  \bibnamefont {Pfeiffer}}, \bibinfo {author} {\bibfnamefont {K.~W.}\
  \bibnamefont {West}}, \ and\ \bibinfo {author} {\bibfnamefont
  {S.}~\bibnamefont {He}}} (\bibinfo {year} {1992}),\ \href {\doibase
  10.1103/PhysRevLett.68.1383} {\bibfield  {journal} {\bibinfo  {journal}
  {Phys. Rev. Lett.}\ }\textbf {\bibinfo {volume} {68}},\ \bibinfo {pages}
  {1383}}\BibitemShut {NoStop}%
\bibitem [{\citenamefont {Eisenstein}\ and\ \citenamefont
  {MacDonald}(2004)}]{eim}%
  \BibitemOpen
  \bibfield  {author} {\bibinfo {author} {\bibnamefont {Eisenstein},
  \bibfnamefont {J.~P.}}, \ and\ \bibinfo {author} {\bibfnamefont {A.~H.}\
  \bibnamefont {MacDonald}}} (\bibinfo {year} {2004}),\ \href@noop {}
  {\bibfield  {journal} {\bibinfo  {journal} {Nature}\ }\textbf {\bibinfo
  {volume} {432}},\ \bibinfo {pages} {691}}\BibitemShut {NoStop}%
\bibitem [{\citenamefont {Elsayad}\ \emph {et~al.}(2008)\citenamefont
  {Elsayad}, \citenamefont {Carini},\ and\ \citenamefont {Baxter}}]{els}%
  \BibitemOpen
  \bibfield  {author} {\bibinfo {author} {\bibnamefont {Elsayad}, \bibfnamefont
  {K.}}, \bibinfo {author} {\bibfnamefont {J.~P.}\ \bibnamefont {Carini}}, \
  and\ \bibinfo {author} {\bibfnamefont {D.}~\bibnamefont {Baxter}}} (\bibinfo
  {year} {2008}),\ \href@noop {} {\bibfield  {journal} {\bibinfo  {journal}
  {Solid State Comm.}\ }\textbf {\bibinfo {volume} {148}},\ \bibinfo {pages}
  {261}}\BibitemShut {NoStop}%
\bibitem [{\citenamefont {Elzerman}\ \emph {et~al.}(2004)\citenamefont
  {Elzerman}, \citenamefont {Hanson}, \citenamefont {van Beveren},
  \citenamefont {Witkamp}, \citenamefont {Vandersypen},\ and\ \citenamefont
  {Kouwenhoven}}]{kou}%
  \BibitemOpen
  \bibfield  {author} {\bibinfo {author} {\bibnamefont {Elzerman},
  \bibfnamefont {J.}}, \bibinfo {author} {\bibfnamefont {R.}~\bibnamefont
  {Hanson}}, \bibinfo {author} {\bibfnamefont {L.~W.}\ \bibnamefont {van
  Beveren}}, \bibinfo {author} {\bibfnamefont {B.}~\bibnamefont {Witkamp}},
  \bibinfo {author} {\bibfnamefont {L.}~\bibnamefont {Vandersypen}}, \ and\
  \bibinfo {author} {\bibfnamefont {L.}~\bibnamefont {Kouwenhoven}}} (\bibinfo
  {year} {2004}),\ \href@noop {} {\bibfield  {journal} {\bibinfo  {journal}
  {Nature}\ }\textbf {\bibinfo {volume} {430}},\ \bibinfo {pages}
  {431}}\BibitemShut {NoStop}%
\bibitem [{\citenamefont {Farina}\ \emph {et~al.}(2004)\citenamefont {Farina},
  \citenamefont {Lewis}, \citenamefont {Kurdak}, \citenamefont {Ghosh},\ and\
  \citenamefont {Bhattacharya}}]{far}%
  \BibitemOpen
  \bibfield  {author} {\bibinfo {author} {\bibnamefont {Farina}, \bibfnamefont
  {L.~A.}}, \bibinfo {author} {\bibfnamefont {K.~M.}\ \bibnamefont {Lewis}},
  \bibinfo {author} {\bibfnamefont {C.}~\bibnamefont {Kurdak}}, \bibinfo
  {author} {\bibfnamefont {S.}~\bibnamefont {Ghosh}}, \ and\ \bibinfo {author}
  {\bibfnamefont {P.}~\bibnamefont {Bhattacharya}}} (\bibinfo {year} {2004}),\
  \href@noop {} {\bibfield  {journal} {\bibinfo  {journal} {Phys. Rev. B}\
  }\textbf {\bibinfo {volume} {70}},\ \bibinfo {pages} {153302}}\BibitemShut
  {NoStop}%
\bibitem [{\citenamefont {Fei}\ \emph {et~al.}(2011)\citenamefont {Fei},
  \citenamefont {Andreev}, \citenamefont {Bao}, \citenamefont {Zhang},
  \citenamefont {S.~McLeod}, \citenamefont {Wang}, \citenamefont {Stewart},
  \citenamefont {Zhao}, \citenamefont {Dominguez}, \citenamefont {Thiemens},
  \citenamefont {Fogler}, \citenamefont {Tauber}, \citenamefont {Castro~Neto},
  \citenamefont {Lau}, \citenamefont {Keilmann},\ and\ \citenamefont
  {Basov}}]{plex1}%
  \BibitemOpen
  \bibfield  {author} {\bibinfo {author} {\bibnamefont {Fei}, \bibfnamefont
  {Z.}}, \bibinfo {author} {\bibfnamefont {G.~O.}\ \bibnamefont {Andreev}},
  \bibinfo {author} {\bibfnamefont {W.}~\bibnamefont {Bao}}, \bibinfo {author}
  {\bibfnamefont {L.~M.}\ \bibnamefont {Zhang}}, \bibinfo {author}
  {\bibfnamefont {A.}~\bibnamefont {S.~McLeod}}, \bibinfo {author}
  {\bibfnamefont {C.}~\bibnamefont {Wang}}, \bibinfo {author} {\bibfnamefont
  {M.~K.}\ \bibnamefont {Stewart}}, \bibinfo {author} {\bibfnamefont
  {Z.}~\bibnamefont {Zhao}}, \bibinfo {author} {\bibfnamefont {G.}~\bibnamefont
  {Dominguez}}, \bibinfo {author} {\bibfnamefont {M.}~\bibnamefont {Thiemens}},
  \bibinfo {author} {\bibfnamefont {M.~M.}\ \bibnamefont {Fogler}}, \bibinfo
  {author} {\bibfnamefont {M.~J.}\ \bibnamefont {Tauber}}, \bibinfo {author}
  {\bibfnamefont {A.~H.}\ \bibnamefont {Castro~Neto}}, \bibinfo {author}
  {\bibfnamefont {C.~N.}\ \bibnamefont {Lau}}, \bibinfo {author} {\bibfnamefont
  {F.}~\bibnamefont {Keilmann}}, \ and\ \bibinfo {author} {\bibfnamefont
  {D.~N.}\ \bibnamefont {Basov}}} (\bibinfo {year} {2011}),\ \href@noop {}
  {\bibfield  {journal} {\bibinfo  {journal} {Nano Letters}\ }\textbf {\bibinfo
  {volume} {11}}~(\bibinfo {number} {11}),\ \bibinfo {pages}
  {4701}}\BibitemShut {NoStop}%
\bibitem [{\citenamefont {Fei}\ \emph {et~al.}(2012)\citenamefont {Fei},
  \citenamefont {Rodin}, \citenamefont {Andreev}, \citenamefont {Bao},
  \citenamefont {McLeod}, \citenamefont {Wagner}, \citenamefont {Zhang},
  \citenamefont {Zhao}, \citenamefont {Thiemens}, \citenamefont {Dominguez},
  \citenamefont {Fogler}, \citenamefont {Castro~Neto}, \citenamefont {Lau},
  \citenamefont {Keilmann},\ and\ \citenamefont {Basov}}]{plex11}%
  \BibitemOpen
  \bibfield  {author} {\bibinfo {author} {\bibnamefont {Fei}, \bibfnamefont
  {Z.}}, \bibinfo {author} {\bibfnamefont {A.~S.}\ \bibnamefont {Rodin}},
  \bibinfo {author} {\bibfnamefont {G.~O.}\ \bibnamefont {Andreev}}, \bibinfo
  {author} {\bibfnamefont {W.}~\bibnamefont {Bao}}, \bibinfo {author}
  {\bibfnamefont {A.~S.}\ \bibnamefont {McLeod}}, \bibinfo {author}
  {\bibfnamefont {M.}~\bibnamefont {Wagner}}, \bibinfo {author} {\bibfnamefont
  {L.~M.}\ \bibnamefont {Zhang}}, \bibinfo {author} {\bibfnamefont
  {Z.}~\bibnamefont {Zhao}}, \bibinfo {author} {\bibfnamefont {M.}~\bibnamefont
  {Thiemens}}, \bibinfo {author} {\bibfnamefont {G.}~\bibnamefont {Dominguez}},
  \bibinfo {author} {\bibfnamefont {M.~M.}\ \bibnamefont {Fogler}}, \bibinfo
  {author} {\bibfnamefont {A.~H.}\ \bibnamefont {Castro~Neto}}, \bibinfo
  {author} {\bibfnamefont {C.~N.}\ \bibnamefont {Lau}}, \bibinfo {author}
  {\bibfnamefont {F.}~\bibnamefont {Keilmann}}, \ and\ \bibinfo {author}
  {\bibfnamefont {D.~N.}\ \bibnamefont {Basov}}} (\bibinfo {year} {2012}),\
  \href@noop {} {\bibfield  {journal} {\bibinfo  {journal} {Nature (London)}\
  }\textbf {\bibinfo {volume} {487}},\ \bibinfo {pages} {82}}\BibitemShut
  {NoStop}%
\bibitem [{\citenamefont {Feng}\ \emph {et~al.}(1998)\citenamefont {Feng},
  \citenamefont {Zelakiewicz}, \citenamefont {Noh}, \citenamefont {Ragucci},
  \citenamefont {Gramila}, \citenamefont {Pfeiffer},\ and\ \citenamefont
  {West}}]{fen}%
  \BibitemOpen
  \bibfield  {author} {\bibinfo {author} {\bibnamefont {Feng}, \bibfnamefont
  {X.~G.}}, \bibinfo {author} {\bibfnamefont {S.}~\bibnamefont {Zelakiewicz}},
  \bibinfo {author} {\bibfnamefont {H.}~\bibnamefont {Noh}}, \bibinfo {author}
  {\bibfnamefont {T.~J.}\ \bibnamefont {Ragucci}}, \bibinfo {author}
  {\bibfnamefont {T.~J.}\ \bibnamefont {Gramila}}, \bibinfo {author}
  {\bibfnamefont {L.~N.}\ \bibnamefont {Pfeiffer}}, \ and\ \bibinfo {author}
  {\bibfnamefont {K.~W.}\ \bibnamefont {West}}} (\bibinfo {year} {1998}),\
  \href@noop {} {\bibfield  {journal} {\bibinfo  {journal} {Phys. Rev. Lett.}\
  }\textbf {\bibinfo {volume} {81}},\ \bibinfo {pages} {3219}}\BibitemShut
  {NoStop}%
\bibitem [{\citenamefont {Fertig}\ and\ \citenamefont {Murthy}(2005)}]{fert}%
  \BibitemOpen
  \bibfield  {author} {\bibinfo {author} {\bibnamefont {Fertig}, \bibfnamefont
  {H.~A.}}, \ and\ \bibinfo {author} {\bibfnamefont {G.}~\bibnamefont
  {Murthy}}} (\bibinfo {year} {2005}),\ \href {\doibase
  10.1103/PhysRevLett.95.156802} {\bibfield  {journal} {\bibinfo  {journal}
  {Phys. Rev. Lett.}\ }\textbf {\bibinfo {volume} {95}},\ \bibinfo {pages}
  {156802}}\BibitemShut {NoStop}%
\bibitem [{\citenamefont {Fiete}\ \emph {et~al.}(2006)\citenamefont {Fiete},
  \citenamefont {Hur},\ and\ \citenamefont {Balents}}]{fhb}%
  \BibitemOpen
  \bibfield  {author} {\bibinfo {author} {\bibnamefont {Fiete}, \bibfnamefont
  {G.~A.}}, \bibinfo {author} {\bibfnamefont {K.~L.}\ \bibnamefont {Hur}}, \
  and\ \bibinfo {author} {\bibfnamefont {L.}~\bibnamefont {Balents}}} (\bibinfo
  {year} {2006}),\ \href@noop {} {\bibfield  {journal} {\bibinfo  {journal}
  {Phys. Rev. B}\ }\textbf {\bibinfo {volume} {73}},\ \bibinfo {pages}
  {165104}}\BibitemShut {NoStop}%
\bibitem [{\citenamefont {Fil'}\ and\ \citenamefont {Kravchenko}(2009)}]{fil2}%
  \BibitemOpen
  \bibfield  {author} {\bibinfo {author} {\bibnamefont {Fil'}, \bibfnamefont
  {D.~V.}}, \ and\ \bibinfo {author} {\bibfnamefont {L.~Y.}\ \bibnamefont
  {Kravchenko}}} (\bibinfo {year} {2009}),\ \href {\doibase
  http://dx.doi.org/10.1063/1.3224730} {\bibfield  {journal} {\bibinfo
  {journal} {Low Temperature Physics}\ }\textbf {\bibinfo {volume}
  {35}}~(\bibinfo {number} {8}),\ \bibinfo {pages} {712}}\BibitemShut {NoStop}%
\bibitem [{\citenamefont {Fil}\ and\ \citenamefont {Shevchenko}(2007)}]{fil}%
  \BibitemOpen
  \bibfield  {author} {\bibinfo {author} {\bibnamefont {Fil}, \bibfnamefont
  {D.~V.}}, \ and\ \bibinfo {author} {\bibfnamefont {S.~I.}\ \bibnamefont
  {Shevchenko}}} (\bibinfo {year} {2007}),\ \href {\doibase
  http://dx.doi.org/10.1063/1.2781504} {\bibfield  {journal} {\bibinfo
  {journal} {Low Temperature Physics}\ }\textbf {\bibinfo {volume}
  {33}}~(\bibinfo {number} {9}),\ \bibinfo {pages} {780}}\BibitemShut {NoStop}%
\bibitem [{\citenamefont {Finck}\ \emph {et~al.}(2008)\citenamefont {Finck},
  \citenamefont {Champagne}, \citenamefont {Eisenstein}, \citenamefont
  {Pfeiffer},\ and\ \citenamefont {West}}]{fin3}%
  \BibitemOpen
  \bibfield  {author} {\bibinfo {author} {\bibnamefont {Finck}, \bibfnamefont
  {A.~D.~K.}}, \bibinfo {author} {\bibfnamefont {A.~R.}\ \bibnamefont
  {Champagne}}, \bibinfo {author} {\bibfnamefont {J.~P.}\ \bibnamefont
  {Eisenstein}}, \bibinfo {author} {\bibfnamefont {L.~N.}\ \bibnamefont
  {Pfeiffer}}, \ and\ \bibinfo {author} {\bibfnamefont {K.~W.}\ \bibnamefont
  {West}}} (\bibinfo {year} {2008}),\ \href {\doibase
  10.1103/PhysRevB.78.075302} {\bibfield  {journal} {\bibinfo  {journal} {Phys.
  Rev. B}\ }\textbf {\bibinfo {volume} {78}},\ \bibinfo {pages}
  {075302}}\BibitemShut {NoStop}%
\bibitem [{\citenamefont {Finck}\ \emph {et~al.}(2010)\citenamefont {Finck},
  \citenamefont {Eisenstein}, \citenamefont {Pfeiffer},\ and\ \citenamefont
  {West}}]{fin}%
  \BibitemOpen
  \bibfield  {author} {\bibinfo {author} {\bibnamefont {Finck}, \bibfnamefont
  {A.~D.~K.}}, \bibinfo {author} {\bibfnamefont {J.~P.}\ \bibnamefont
  {Eisenstein}}, \bibinfo {author} {\bibfnamefont {L.~N.}\ \bibnamefont
  {Pfeiffer}}, \ and\ \bibinfo {author} {\bibfnamefont {K.~W.}\ \bibnamefont
  {West}}} (\bibinfo {year} {2010}),\ \href@noop {} {\bibfield  {journal}
  {\bibinfo  {journal} {Phys. Rev. Lett.}\ }\textbf {\bibinfo {volume} {104}},\
  \bibinfo {pages} {016801}}\BibitemShut {NoStop}%
\bibitem [{\citenamefont {Finck}\ \emph {et~al.}(2011)\citenamefont {Finck},
  \citenamefont {Eisenstein}, \citenamefont {Pfeiffer},\ and\ \citenamefont
  {West}}]{fin2}%
  \BibitemOpen
  \bibfield  {author} {\bibinfo {author} {\bibnamefont {Finck}, \bibfnamefont
  {A.~D.~K.}}, \bibinfo {author} {\bibfnamefont {J.~P.}\ \bibnamefont
  {Eisenstein}}, \bibinfo {author} {\bibfnamefont {L.~N.}\ \bibnamefont
  {Pfeiffer}}, \ and\ \bibinfo {author} {\bibfnamefont {K.~W.}\ \bibnamefont
  {West}}} (\bibinfo {year} {2011}),\ \href@noop {} {\bibfield  {journal}
  {\bibinfo  {journal} {Phys. Rev. Lett.}\ }\textbf {\bibinfo {volume} {106}},\
  \bibinfo {pages} {236807}}\BibitemShut {NoStop}%
\bibitem [{\citenamefont {Finkelstein}(1983)}]{fnk}%
  \BibitemOpen
  \bibfield  {author} {\bibinfo {author} {\bibnamefont {Finkelstein},
  \bibfnamefont {A.~M.}}} (\bibinfo {year} {1983}),\ \href@noop {} {\bibfield
  {journal} {\bibinfo  {journal} {Zh. Eksp. Teor. Fiz.}\ }\textbf {\bibinfo
  {volume} {84}},\ \bibinfo {pages} {168}},\ \bibinfo {note} {[Sov. Phys. JETP
  {\bf 57}, 97 (1983)]}\BibitemShut {NoStop}%
\bibitem [{\citenamefont {Finkelstein}(1984)}]{fnk1}%
  \BibitemOpen
  \bibfield  {author} {\bibinfo {author} {\bibnamefont {Finkelstein},
  \bibfnamefont {A.~M.}}} (\bibinfo {year} {1984}),\ \href@noop {} {\bibfield
  {journal} {\bibinfo  {journal} {Z. Phys. B: Condens. Matter}\ }\textbf
  {\bibinfo {volume} {56}},\ \bibinfo {pages} {189}}\BibitemShut {NoStop}%
\bibitem [{\citenamefont {Finkelstein}\ \emph {et~al.}(1995)\citenamefont
  {Finkelstein}, \citenamefont {Shtrikman},\ and\ \citenamefont
  {Bar-Joseph}}]{gfin}%
  \BibitemOpen
  \bibfield  {author} {\bibinfo {author} {\bibnamefont {Finkelstein},
  \bibfnamefont {G.}}, \bibinfo {author} {\bibfnamefont {H.}~\bibnamefont
  {Shtrikman}}, \ and\ \bibinfo {author} {\bibfnamefont {I.}~\bibnamefont
  {Bar-Joseph}}} (\bibinfo {year} {1995}),\ \href {\doibase
  10.1103/PhysRevLett.74.976} {\bibfield  {journal} {\bibinfo  {journal} {Phys.
  Rev. Lett.}\ }\textbf {\bibinfo {volume} {74}},\ \bibinfo {pages}
  {976}}\BibitemShut {NoStop}%
\bibitem [{\citenamefont {Fisichella}\ \emph {et~al.}(2014)\citenamefont
  {Fisichella}, \citenamefont {Greco}, \citenamefont {Roccaforte},\ and\
  \citenamefont {Giannazzo}}]{fis}%
  \BibitemOpen
  \bibfield  {author} {\bibinfo {author} {\bibnamefont {Fisichella},
  \bibfnamefont {G.}}, \bibinfo {author} {\bibfnamefont {G.}~\bibnamefont
  {Greco}}, \bibinfo {author} {\bibfnamefont {F.}~\bibnamefont {Roccaforte}}, \
  and\ \bibinfo {author} {\bibfnamefont {F.}~\bibnamefont {Giannazzo}}}
  (\bibinfo {year} {2014}),\ \href@noop {} {\bibfield  {journal} {\bibinfo
  {journal} {Nanoscale}\ }\textbf {\bibinfo {volume} {6}},\ \bibinfo {pages}
  {8671}}\BibitemShut {NoStop}%
\bibitem [{\citenamefont {Flensberg}(1998)}]{fle}%
  \BibitemOpen
  \bibfield  {author} {\bibinfo {author} {\bibnamefont {Flensberg},
  \bibfnamefont {K.}}} (\bibinfo {year} {1998}),\ \href@noop {} {\bibfield
  {journal} {\bibinfo  {journal} {Phys. Rev. Lett.}\ }\textbf {\bibinfo
  {volume} {81}},\ \bibinfo {pages} {184}}\BibitemShut {NoStop}%
\bibitem [{\citenamefont {Flensberg}\ and\ \citenamefont {Hu}(1994)}]{fln}%
  \BibitemOpen
  \bibfield  {author} {\bibinfo {author} {\bibnamefont {Flensberg},
  \bibfnamefont {K.}}, \ and\ \bibinfo {author} {\bibfnamefont {B.~Y.-K.}\
  \bibnamefont {Hu}}} (\bibinfo {year} {1994}),\ \href@noop {} {\bibfield
  {journal} {\bibinfo  {journal} {Phys. Rev. Lett.}\ }\textbf {\bibinfo
  {volume} {73}},\ \bibinfo {pages} {3572}}\BibitemShut {NoStop}%
\bibitem [{\citenamefont {Flensberg}\ and\ \citenamefont {Hu}(1995)}]{fln2}%
  \BibitemOpen
  \bibfield  {author} {\bibinfo {author} {\bibnamefont {Flensberg},
  \bibfnamefont {K.}}, \ and\ \bibinfo {author} {\bibfnamefont {B.~Y.-K.}\
  \bibnamefont {Hu}}} (\bibinfo {year} {1995}),\ \href@noop {} {\bibfield
  {journal} {\bibinfo  {journal} {Phys. Rev. B}\ }\textbf {\bibinfo {volume}
  {52}},\ \bibinfo {pages} {14796}}\BibitemShut {NoStop}%
\bibitem [{\citenamefont {Flensberg}\ \emph {et~al.}(1995)\citenamefont
  {Flensberg}, \citenamefont {Hu}, \citenamefont {Jauho},\ and\ \citenamefont
  {Kinaret}}]{fl2}%
  \BibitemOpen
  \bibfield  {author} {\bibinfo {author} {\bibnamefont {Flensberg},
  \bibfnamefont {K.}}, \bibinfo {author} {\bibfnamefont {B.~Y.-K.}\
  \bibnamefont {Hu}}, \bibinfo {author} {\bibfnamefont {A.-P.}\ \bibnamefont
  {Jauho}}, \ and\ \bibinfo {author} {\bibfnamefont {J.~M.}\ \bibnamefont
  {Kinaret}}} (\bibinfo {year} {1995}),\ \href@noop {} {\bibfield  {journal}
  {\bibinfo  {journal} {Phys. Rev. B}\ }\textbf {\bibinfo {volume} {52}},\
  \bibinfo {pages} {14761}}\BibitemShut {NoStop}%
\bibitem [{\citenamefont {Foster}\ and\ \citenamefont {Aleiner}(2009)}]{foa}%
  \BibitemOpen
  \bibfield  {author} {\bibinfo {author} {\bibnamefont {Foster}, \bibfnamefont
  {M.~S.}}, \ and\ \bibinfo {author} {\bibfnamefont {I.~L.}\ \bibnamefont
  {Aleiner}}} (\bibinfo {year} {2009}),\ \href@noop {} {\bibfield  {journal}
  {\bibinfo  {journal} {Phys. Rev. B}\ }\textbf {\bibinfo {volume} {79}},\
  \bibinfo {pages} {085415}}\BibitemShut {NoStop}%
\bibitem [{\citenamefont {Fredrikse}(1953{\natexlab{a}})}]{fre1}%
  \BibitemOpen
  \bibfield  {author} {\bibinfo {author} {\bibnamefont {Fredrikse},
  \bibfnamefont {H.~P.~R.}}} (\bibinfo {year} {1953}{\natexlab{a}}),\
  \href@noop {} {\bibfield  {journal} {\bibinfo  {journal} {Phys. Rev.}\
  }\textbf {\bibinfo {volume} {91}},\ \bibinfo {pages} {491}}\BibitemShut
  {NoStop}%
\bibitem [{\citenamefont {Fredrikse}(1953{\natexlab{b}})}]{fre2}%
  \BibitemOpen
  \bibfield  {author} {\bibinfo {author} {\bibnamefont {Fredrikse},
  \bibfnamefont {H.~P.~R.}}} (\bibinfo {year} {1953}{\natexlab{b}}),\
  \href@noop {} {\bibfield  {journal} {\bibinfo  {journal} {Phys. Rev.}\
  }\textbf {\bibinfo {volume} {92}},\ \bibinfo {pages} {248}}\BibitemShut
  {NoStop}%
\bibitem [{\citenamefont {Fritz}\ \emph {et~al.}(2008)\citenamefont {Fritz},
  \citenamefont {Schmalian}, \citenamefont {M{\"u}ller},\ and\ \citenamefont
  {Sachdev}}]{kin}%
  \BibitemOpen
  \bibfield  {author} {\bibinfo {author} {\bibnamefont {Fritz}, \bibfnamefont
  {L.}}, \bibinfo {author} {\bibfnamefont {J.}~\bibnamefont {Schmalian}},
  \bibinfo {author} {\bibfnamefont {M.}~\bibnamefont {M{\"u}ller}}, \ and\
  \bibinfo {author} {\bibfnamefont {S.}~\bibnamefont {Sachdev}}} (\bibinfo
  {year} {2008}),\ \href@noop {} {\bibfield  {journal} {\bibinfo  {journal}
  {Phys. Rev. B}\ }\textbf {\bibinfo {volume} {78}},\ \bibinfo {pages}
  {085416}}\BibitemShut {NoStop}%
\bibitem [{\citenamefont {Fuchs}\ \emph {et~al.}(2005)\citenamefont {Fuchs},
  \citenamefont {Klesse},\ and\ \citenamefont {Stern}}]{fks}%
  \BibitemOpen
  \bibfield  {author} {\bibinfo {author} {\bibnamefont {Fuchs}, \bibfnamefont
  {T.}}, \bibinfo {author} {\bibfnamefont {R.}~\bibnamefont {Klesse}}, \ and\
  \bibinfo {author} {\bibfnamefont {A.}~\bibnamefont {Stern}}} (\bibinfo {year}
  {2005}),\ \href@noop {} {\bibfield  {journal} {\bibinfo  {journal} {Phys.
  Rev. B}\ }\textbf {\bibinfo {volume} {71}},\ \bibinfo {pages}
  {045321}}\BibitemShut {NoStop}%
\bibitem [{\citenamefont {Gamucci}\ \emph {et~al.}(2014)\citenamefont
  {Gamucci}, \citenamefont {Spirito}, \citenamefont {Carrega}, \citenamefont
  {Karmakar}, \citenamefont {Lombardo}, \citenamefont {Bruna}, \citenamefont
  {Pfeiffer}, \citenamefont {West}, \citenamefont {Ferrari}, \citenamefont
  {Polini},\ and\ \citenamefont {Pellegrini}}]{gam}%
  \BibitemOpen
  \bibfield  {author} {\bibinfo {author} {\bibnamefont {Gamucci}, \bibfnamefont
  {A.}}, \bibinfo {author} {\bibfnamefont {D.}~\bibnamefont {Spirito}},
  \bibinfo {author} {\bibfnamefont {M.}~\bibnamefont {Carrega}}, \bibinfo
  {author} {\bibfnamefont {B.}~\bibnamefont {Karmakar}}, \bibinfo {author}
  {\bibfnamefont {A.}~\bibnamefont {Lombardo}}, \bibinfo {author}
  {\bibfnamefont {M.}~\bibnamefont {Bruna}}, \bibinfo {author} {\bibfnamefont
  {L.~N.}\ \bibnamefont {Pfeiffer}}, \bibinfo {author} {\bibfnamefont {K.~W.}\
  \bibnamefont {West}}, \bibinfo {author} {\bibfnamefont {A.~C.}\ \bibnamefont
  {Ferrari}}, \bibinfo {author} {\bibfnamefont {M.}~\bibnamefont {Polini}}, \
  and\ \bibinfo {author} {\bibfnamefont {V.}~\bibnamefont {Pellegrini}}}
  (\bibinfo {year} {2014}),\ \href@noop {} {\bibfield  {journal} {\bibinfo
  {journal} {Nature Communications}\ }\textbf {\bibinfo {volume} {5}},\
  \bibinfo {pages} {5824}}\BibitemShut {NoStop}%
\bibitem [{\citenamefont {Geballe}(1953)}]{geb1}%
  \BibitemOpen
  \bibfield  {author} {\bibinfo {author} {\bibnamefont {Geballe}, \bibfnamefont
  {T.~H.}}} (\bibinfo {year} {1953}),\ \href@noop {} {\bibfield  {journal}
  {\bibinfo  {journal} {Phys. Rev.}\ }\textbf {\bibinfo {volume} {92}},\
  \bibinfo {pages} {857}}\BibitemShut {NoStop}%
\bibitem [{\citenamefont {Geballe}\ and\ \citenamefont {Hull}(1954)}]{geb2}%
  \BibitemOpen
  \bibfield  {author} {\bibinfo {author} {\bibnamefont {Geballe}, \bibfnamefont
  {T.~H.}}, \ and\ \bibinfo {author} {\bibfnamefont {G.~W.}\ \bibnamefont
  {Hull}}} (\bibinfo {year} {1954}),\ \href@noop {} {\bibfield  {journal}
  {\bibinfo  {journal} {Phys. Rev.}\ }\textbf {\bibinfo {volume} {94}},\
  \bibinfo {pages} {1134}}\BibitemShut {NoStop}%
\bibitem [{\citenamefont {Geigenm\"uller}\ and\ \citenamefont
  {Nazarov}(1991)}]{gei}%
  \BibitemOpen
  \bibfield  {author} {\bibinfo {author} {\bibnamefont {Geigenm\"uller},
  \bibfnamefont {U.}}, \ and\ \bibinfo {author} {\bibfnamefont {Y.~V.}\
  \bibnamefont {Nazarov}}} (\bibinfo {year} {1991}),\ \href {\doibase
  10.1103/PhysRevB.44.10953} {\bibfield  {journal} {\bibinfo  {journal} {Phys.
  Rev. B}\ }\textbf {\bibinfo {volume} {44}},\ \bibinfo {pages}
  {10953}}\BibitemShut {NoStop}%
\bibitem [{\citenamefont {Geim}\ and\ \citenamefont {Grigorieva}(2013)}]{geim}%
  \BibitemOpen
  \bibfield  {author} {\bibinfo {author} {\bibnamefont {Geim}, \bibfnamefont
  {A.~K.}}, \ and\ \bibinfo {author} {\bibfnamefont {I.~V.}\ \bibnamefont
  {Grigorieva}}} (\bibinfo {year} {2013}),\ \href@noop {} {\bibfield  {journal}
  {\bibinfo  {journal} {Nature}\ }\textbf {\bibinfo {volume} {499}},\ \bibinfo
  {pages} {419}}\BibitemShut {NoStop}%
\bibitem [{\citenamefont {Giamarchi}(2004)}]{giam}%
  \BibitemOpen
  \bibfield  {author} {\bibinfo {author} {\bibnamefont {Giamarchi},
  \bibfnamefont {T.}}} (\bibinfo {year} {2004}),\ \href@noop {} {\emph
  {\bibinfo {title} {Quantum Physics in One Dimension}}}\ (\bibinfo
  {publisher} {Oxford University Press})\BibitemShut {NoStop}%
\bibitem [{\citenamefont {Giordano}\ and\ \citenamefont {Monnier}(1994)}]{gio}%
  \BibitemOpen
  \bibfield  {author} {\bibinfo {author} {\bibnamefont {Giordano},
  \bibfnamefont {N.}}, \ and\ \bibinfo {author} {\bibfnamefont {J.~D.}\
  \bibnamefont {Monnier}}} (\bibinfo {year} {1994}),\ \href@noop {} {\bibfield
  {journal} {\bibinfo  {journal} {Phys. Rev. B}\ }\textbf {\bibinfo {volume}
  {50}},\ \bibinfo {pages} {9363}}\BibitemShut {NoStop}%
\bibitem [{\citenamefont {Girvin}\ and\ \citenamefont {MacDonald}(1997)}]{gir}%
  \BibitemOpen
  \bibfield  {author} {\bibinfo {author} {\bibnamefont {Girvin}, \bibfnamefont
  {S.~M.}}, \ and\ \bibinfo {author} {\bibfnamefont {A.~H.}\ \bibnamefont
  {MacDonald}}} (\bibinfo {year} {1997}),\ in\ \href@noop {} {\emph {\bibinfo
  {booktitle} {Perspectives in Quantum Hall Effects}}},\ \bibinfo {editor}
  {edited by\ \bibinfo {editor} {\bibfnamefont {S.}~\bibnamefont {{Das
  Sarma}}}\ and\ \bibinfo {editor} {\bibfnamefont {A.}~\bibnamefont
  {Pinczuk}}}\ (\bibinfo  {publisher} {Wiley, New York})\BibitemShut {NoStop}%
\bibitem [{\citenamefont {Giuliani}\ and\ \citenamefont
  {Vignale}(2005)}]{Giuliani}%
  \BibitemOpen
  \bibfield  {author} {\bibinfo {author} {\bibnamefont {Giuliani},
  \bibfnamefont {G.}}, \ and\ \bibinfo {author} {\bibfnamefont
  {G.}~\bibnamefont {Vignale}}} (\bibinfo {year} {2005}),\ \href@noop {} {\emph
  {\bibinfo {title} {The Theory of the Properties of Metals and Alloys}}}\
  (\bibinfo  {publisher} {Cambridge University Press})\BibitemShut {NoStop}%
\bibitem [{\citenamefont {Giuliani}\ and\ \citenamefont {Quinn}(1982)}]{qui}%
  \BibitemOpen
  \bibfield  {author} {\bibinfo {author} {\bibnamefont {Giuliani},
  \bibfnamefont {G.~F.}}, \ and\ \bibinfo {author} {\bibfnamefont {J.~J.}\
  \bibnamefont {Quinn}}} (\bibinfo {year} {1982}),\ \href {\doibase
  10.1103/PhysRevB.26.4421} {\bibfield  {journal} {\bibinfo  {journal} {Phys.
  Rev. B}\ }\textbf {\bibinfo {volume} {26}},\ \bibinfo {pages}
  {4421}}\BibitemShut {NoStop}%
\bibitem [{\citenamefont {Glazman}\ \emph {et~al.}(1988)\citenamefont
  {Glazman}, \citenamefont {Lesovik}, \citenamefont {Khmel'nitskii},\ and\
  \citenamefont {Shekhter}}]{glaz}%
  \BibitemOpen
  \bibfield  {author} {\bibinfo {author} {\bibnamefont {Glazman}, \bibfnamefont
  {L.~I.}}, \bibinfo {author} {\bibfnamefont {G.~B.}\ \bibnamefont {Lesovik}},
  \bibinfo {author} {\bibfnamefont {D.~E.}\ \bibnamefont {Khmel'nitskii}}, \
  and\ \bibinfo {author} {\bibfnamefont {R.~I.}\ \bibnamefont {Shekhter}}}
  (\bibinfo {year} {1988}),\ \href@noop {} {\bibfield  {journal} {\bibinfo
  {journal} {Pis'ma Zh. Eksp. Teor. Fiz.}\ }\textbf {\bibinfo {volume} {48}},\
  \bibinfo {pages} {218}},\ \bibinfo {note} {[JETP Lett. {\bf 48}, 238
  (1988)]}\BibitemShut {NoStop}%
\bibitem [{\citenamefont {Glazov}\ \emph {et~al.}(2011)\citenamefont {Glazov},
  \citenamefont {Semina}, \citenamefont {Badalyan},\ and\ \citenamefont
  {Vignale}}]{gla}%
  \BibitemOpen
  \bibfield  {author} {\bibinfo {author} {\bibnamefont {Glazov}, \bibfnamefont
  {M.~M.}}, \bibinfo {author} {\bibfnamefont {M.~A.}\ \bibnamefont {Semina}},
  \bibinfo {author} {\bibfnamefont {S.~M.}\ \bibnamefont {Badalyan}}, \ and\
  \bibinfo {author} {\bibfnamefont {G.}~\bibnamefont {Vignale}}} (\bibinfo
  {year} {2011}),\ \href@noop {} {\bibfield  {journal} {\bibinfo  {journal}
  {Phys. Rev. B}\ }\textbf {\bibinfo {volume} {84}},\ \bibinfo {pages}
  {033305}}\BibitemShut {NoStop}%
\bibitem [{\citenamefont {Gonz\'alez}\ \emph {et~al.}(1999)\citenamefont
  {Gonz\'alez}, \citenamefont {Guinea},\ and\ \citenamefont
  {Vozmediano}}]{ggv}%
  \BibitemOpen
  \bibfield  {author} {\bibinfo {author} {\bibnamefont {Gonz\'alez},
  \bibfnamefont {J.}}, \bibinfo {author} {\bibfnamefont {F.}~\bibnamefont
  {Guinea}}, \ and\ \bibinfo {author} {\bibfnamefont {M.~A.~H.}\ \bibnamefont
  {Vozmediano}}} (\bibinfo {year} {1999}),\ \href@noop {} {\bibfield  {journal}
  {\bibinfo  {journal} {Phys. Rev. B}\ }\textbf {\bibinfo {volume} {59}},\
  \bibinfo {pages} {R2474}}\BibitemShut {NoStop}%
\bibitem [{\citenamefont {Gorbachev}\ \emph {et~al.}(2012)\citenamefont
  {Gorbachev}, \citenamefont {Geim}, \citenamefont {Katsnelson}, \citenamefont
  {Novoselov}, \citenamefont {Tudorovskiy}, \citenamefont {Grigorieva},
  \citenamefont {MacDonald}, \citenamefont {Morozov}, \citenamefont {Watanabe},
  \citenamefont {Taniguchi},\ and\ \citenamefont {Ponomarenko}}]{exg}%
  \BibitemOpen
  \bibfield  {author} {\bibinfo {author} {\bibnamefont {Gorbachev},
  \bibfnamefont {R.~V.}}, \bibinfo {author} {\bibfnamefont {A.~K.}\
  \bibnamefont {Geim}}, \bibinfo {author} {\bibfnamefont {M.~I.}\ \bibnamefont
  {Katsnelson}}, \bibinfo {author} {\bibfnamefont {K.~S.}\ \bibnamefont
  {Novoselov}}, \bibinfo {author} {\bibfnamefont {T.}~\bibnamefont
  {Tudorovskiy}}, \bibinfo {author} {\bibfnamefont {I.~V.}\ \bibnamefont
  {Grigorieva}}, \bibinfo {author} {\bibfnamefont {A.~H.}\ \bibnamefont
  {MacDonald}}, \bibinfo {author} {\bibfnamefont {S.~V.}\ \bibnamefont
  {Morozov}}, \bibinfo {author} {\bibfnamefont {K.}~\bibnamefont {Watanabe}},
  \bibinfo {author} {\bibfnamefont {T.}~\bibnamefont {Taniguchi}}, \ and\
  \bibinfo {author} {\bibfnamefont {L.~A.}\ \bibnamefont {Ponomarenko}}}
  (\bibinfo {year} {2012}),\ \href@noop {} {\bibfield  {journal} {\bibinfo
  {journal} {Nature Physics}\ }\textbf {\bibinfo {volume} {8}},\ \bibinfo
  {pages} {896}}\BibitemShut {NoStop}%
\bibitem [{\citenamefont {Gorkov}\ \emph {et~al.}(1979)\citenamefont {Gorkov},
  \citenamefont {Larkin},\ and\ \citenamefont {Khmel'nitzkii}}]{glk}%
  \BibitemOpen
  \bibfield  {author} {\bibinfo {author} {\bibnamefont {Gorkov}, \bibfnamefont
  {L.~P.}}, \bibinfo {author} {\bibfnamefont {A.~I.}\ \bibnamefont {Larkin}}, \
  and\ \bibinfo {author} {\bibfnamefont {D.~E.}\ \bibnamefont {Khmel'nitzkii}}}
  (\bibinfo {year} {1979}),\ \href@noop {} {\bibfield  {journal} {\bibinfo
  {journal} {Pis'ma Zh. Eksp. Teor. Fiz.}\ }\textbf {\bibinfo {volume} {30}},\
  \bibinfo {pages} {248}},\ \bibinfo {note} {[JETP Lett. {\bf 30}, 228
  (1979)]}\BibitemShut {NoStop}%
\bibitem [{\citenamefont {Gornyi}\ \emph {et~al.}(2000)\citenamefont {Gornyi},
  \citenamefont {Yashenkin},\ and\ \citenamefont {Khveshchenko}}]{go22}%
  \BibitemOpen
  \bibfield  {author} {\bibinfo {author} {\bibnamefont {Gornyi}, \bibfnamefont
  {I.}}, \bibinfo {author} {\bibfnamefont {A.}~\bibnamefont {Yashenkin}}, \
  and\ \bibinfo {author} {\bibfnamefont {D.}~\bibnamefont {Khveshchenko}}}
  (\bibinfo {year} {2000}),\ \href@noop {} {\bibfield  {journal} {\bibinfo
  {journal} {Physica B: Condensed Matter}\ }\textbf {\bibinfo {volume} {284 -
  288, Part 2}},\ \bibinfo {pages} {1930}}\BibitemShut {NoStop}%
\bibitem [{\citenamefont {Gornyi}\ \emph {et~al.}(2004)\citenamefont {Gornyi},
  \citenamefont {Mirlin},\ and\ \citenamefont {von Oppen}}]{gor}%
  \BibitemOpen
  \bibfield  {author} {\bibinfo {author} {\bibnamefont {Gornyi}, \bibfnamefont
  {I.~V.}}, \bibinfo {author} {\bibfnamefont {A.~D.}\ \bibnamefont {Mirlin}}, \
  and\ \bibinfo {author} {\bibfnamefont {F.}~\bibnamefont {von Oppen}}}
  (\bibinfo {year} {2004}),\ \href@noop {} {\bibfield  {journal} {\bibinfo
  {journal} {Phys. Rev. B}\ }\textbf {\bibinfo {volume} {70}},\ \bibinfo
  {pages} {245302}}\BibitemShut {NoStop}%
\bibitem [{\citenamefont {Gornyi}\ and\ \citenamefont {Narozhny}(2014)}]{gorn}%
  \BibitemOpen
  \bibfield  {author} {\bibinfo {author} {\bibnamefont {Gornyi}, \bibfnamefont
  {I.~V.}}, \ and\ \bibinfo {author} {\bibfnamefont {B.~N.}\ \bibnamefont
  {Narozhny}}} (\bibinfo {year} {2014}),\ \href@noop {} {}\bibinfo {note}
  {Unpublished}\BibitemShut {NoStop}%
\bibitem [{\citenamefont {Gornyi}\ \emph {et~al.}(1999)\citenamefont {Gornyi},
  \citenamefont {Yashenkin},\ and\ \citenamefont {Khveshchenko}}]{go2}%
  \BibitemOpen
  \bibfield  {author} {\bibinfo {author} {\bibnamefont {Gornyi}, \bibfnamefont
  {I.~V.}}, \bibinfo {author} {\bibfnamefont {A.~G.}\ \bibnamefont
  {Yashenkin}}, \ and\ \bibinfo {author} {\bibfnamefont {D.~V.}\ \bibnamefont
  {Khveshchenko}}} (\bibinfo {year} {1999}),\ \href@noop {} {\bibfield
  {journal} {\bibinfo  {journal} {Phys. Rev. Lett.}\ }\textbf {\bibinfo
  {volume} {83}},\ \bibinfo {pages} {152}}\BibitemShut {NoStop}%
\bibitem [{\citenamefont {Gramila}\ \emph {et~al.}(1991)\citenamefont
  {Gramila}, \citenamefont {Eisenstein}, \citenamefont {MacDonald},
  \citenamefont {Pfeiffer},\ and\ \citenamefont {West}}]{ex2}%
  \BibitemOpen
  \bibfield  {author} {\bibinfo {author} {\bibnamefont {Gramila}, \bibfnamefont
  {T.~J.}}, \bibinfo {author} {\bibfnamefont {J.~P.}\ \bibnamefont
  {Eisenstein}}, \bibinfo {author} {\bibfnamefont {A.~H.}\ \bibnamefont
  {MacDonald}}, \bibinfo {author} {\bibfnamefont {L.~N.}\ \bibnamefont
  {Pfeiffer}}, \ and\ \bibinfo {author} {\bibfnamefont {K.~W.}\ \bibnamefont
  {West}}} (\bibinfo {year} {1991}),\ \href@noop {} {\bibfield  {journal}
  {\bibinfo  {journal} {Phys. Rev. Lett.}\ }\textbf {\bibinfo {volume} {66}},\
  \bibinfo {pages} {1216}}\BibitemShut {NoStop}%
\bibitem [{\citenamefont {Gramila}\ \emph {et~al.}(1992)\citenamefont
  {Gramila}, \citenamefont {Eisenstein}, \citenamefont {MacDonald},
  \citenamefont {Pfeiffer},\ and\ \citenamefont {West}}]{ex22}%
  \BibitemOpen
  \bibfield  {author} {\bibinfo {author} {\bibnamefont {Gramila}, \bibfnamefont
  {T.~J.}}, \bibinfo {author} {\bibfnamefont {J.~P.}\ \bibnamefont
  {Eisenstein}}, \bibinfo {author} {\bibfnamefont {A.~H.}\ \bibnamefont
  {MacDonald}}, \bibinfo {author} {\bibfnamefont {L.~N.}\ \bibnamefont
  {Pfeiffer}}, \ and\ \bibinfo {author} {\bibfnamefont {K.~W.}\ \bibnamefont
  {West}}} (\bibinfo {year} {1992}),\ \href@noop {} {\bibfield  {journal}
  {\bibinfo  {journal} {Surf. Sci.}\ }\textbf {\bibinfo {volume} {263}},\
  \bibinfo {pages} {446}}\BibitemShut {NoStop}%
\bibitem [{\citenamefont {Gramila}\ \emph {et~al.}(1993)\citenamefont
  {Gramila}, \citenamefont {Eisenstein}, \citenamefont {MacDonald},
  \citenamefont {Pfeiffer},\ and\ \citenamefont {West}}]{ex23}%
  \BibitemOpen
  \bibfield  {author} {\bibinfo {author} {\bibnamefont {Gramila}, \bibfnamefont
  {T.~J.}}, \bibinfo {author} {\bibfnamefont {J.~P.}\ \bibnamefont
  {Eisenstein}}, \bibinfo {author} {\bibfnamefont {A.~H.}\ \bibnamefont
  {MacDonald}}, \bibinfo {author} {\bibfnamefont {L.~N.}\ \bibnamefont
  {Pfeiffer}}, \ and\ \bibinfo {author} {\bibfnamefont {K.~W.}\ \bibnamefont
  {West}}} (\bibinfo {year} {1993}),\ \href@noop {} {\bibfield  {journal}
  {\bibinfo  {journal} {Phys. Rev. B}\ }\textbf {\bibinfo {volume} {47}},\
  \bibinfo {pages} {12957}}\BibitemShut {NoStop}%
\bibitem [{\citenamefont {Gramila}\ \emph {et~al.}(1994)\citenamefont
  {Gramila}, \citenamefont {Eisenstein}, \citenamefont {MacDonald},
  \citenamefont {Pfeiffer},\ and\ \citenamefont {West}}]{ex24}%
  \BibitemOpen
  \bibfield  {author} {\bibinfo {author} {\bibnamefont {Gramila}, \bibfnamefont
  {T.~J.}}, \bibinfo {author} {\bibfnamefont {J.~P.}\ \bibnamefont
  {Eisenstein}}, \bibinfo {author} {\bibfnamefont {A.~H.}\ \bibnamefont
  {MacDonald}}, \bibinfo {author} {\bibfnamefont {L.~N.}\ \bibnamefont
  {Pfeiffer}}, \ and\ \bibinfo {author} {\bibfnamefont {K.~W.}\ \bibnamefont
  {West}}} (\bibinfo {year} {1994}),\ \href@noop {} {\bibfield  {journal}
  {\bibinfo  {journal} {Physica B}\ }\textbf {\bibinfo {volume} {197}},\
  \bibinfo {pages} {442}}\BibitemShut {NoStop}%
\bibitem [{\citenamefont {Greiter}\ \emph {et~al.}(1991)\citenamefont
  {Greiter}, \citenamefont {Wen},\ and\ \citenamefont {Wilczek}}]{grei}%
  \BibitemOpen
  \bibfield  {author} {\bibinfo {author} {\bibnamefont {Greiter}, \bibfnamefont
  {M.}}, \bibinfo {author} {\bibfnamefont {X.-G.}\ \bibnamefont {Wen}}, \ and\
  \bibinfo {author} {\bibfnamefont {F.}~\bibnamefont {Wilczek}}} (\bibinfo
  {year} {1991}),\ \href {\doibase 10.1103/PhysRevLett.66.3205} {\bibfield
  {journal} {\bibinfo  {journal} {Phys. Rev. Lett.}\ }\textbf {\bibinfo
  {volume} {66}},\ \bibinfo {pages} {3205}}\BibitemShut {NoStop}%
\bibitem [{\citenamefont {Grigorenko}\ \emph {et~al.}(2012)\citenamefont
  {Grigorenko}, \citenamefont {Polini},\ and\ \citenamefont
  {Novoselov}}]{plarev}%
  \BibitemOpen
  \bibfield  {author} {\bibinfo {author} {\bibnamefont {Grigorenko},
  \bibfnamefont {A.~N.}}, \bibinfo {author} {\bibfnamefont {M.}~\bibnamefont
  {Polini}}, \ and\ \bibinfo {author} {\bibfnamefont {K.~S.}\ \bibnamefont
  {Novoselov}}} (\bibinfo {year} {2012}),\ \href@noop {} {\bibfield  {journal}
  {\bibinfo  {journal} {Nature Photonics}\ }\textbf {\bibinfo {volume} {6}},\
  \bibinfo {pages} {749}}\BibitemShut {NoStop}%
\bibitem [{\citenamefont {Gr{\"u}neisen}(1933)}]{gru}%
  \BibitemOpen
  \bibfield  {author} {\bibinfo {author} {\bibnamefont {Gr{\"u}neisen},
  \bibfnamefont {E.}}} (\bibinfo {year} {1933}),\ \href@noop {} {\bibfield
  {journal} {\bibinfo  {journal} {Ann. Phys. (Leipzig)}\ }\textbf {\bibinfo
  {volume} {16}},\ \bibinfo {pages} {530}}\BibitemShut {NoStop}%
\bibitem [{\citenamefont {Gubarev}\ \emph {et~al.}(2000)\citenamefont
  {Gubarev}, \citenamefont {Kukushkin}, \citenamefont {Tovstonog},
  \citenamefont {Akimov}, \citenamefont {Smet}, \citenamefont {{von
  Klitzing}},\ and\ \citenamefont {Wegscheider}}]{guba}%
  \BibitemOpen
  \bibfield  {author} {\bibinfo {author} {\bibnamefont {Gubarev}, \bibfnamefont
  {S.~I.}}, \bibinfo {author} {\bibfnamefont {I.~V.}\ \bibnamefont
  {Kukushkin}}, \bibinfo {author} {\bibfnamefont {S.~V.}\ \bibnamefont
  {Tovstonog}}, \bibinfo {author} {\bibfnamefont {M.~Y.}\ \bibnamefont
  {Akimov}}, \bibinfo {author} {\bibfnamefont {J.}~\bibnamefont {Smet}},
  \bibinfo {author} {\bibfnamefont {K.}~\bibnamefont {{von Klitzing}}}, \ and\
  \bibinfo {author} {\bibfnamefont {W.}~\bibnamefont {Wegscheider}}} (\bibinfo
  {year} {2000}),\ \href@noop {} {\bibfield  {journal} {\bibinfo  {journal}
  {Pis'ma Zh. Eksp. Teor. Fiz.}\ }\textbf {\bibinfo {volume} {72}},\ \bibinfo
  {pages} {469}},\ \bibinfo {note} {[JETP Lett. {\bf 72}, 324
  (2000)]}\BibitemShut {NoStop}%
\bibitem [{\citenamefont {Gurevich}(1946{\natexlab{a}})}]{lgu1}%
  \BibitemOpen
  \bibfield  {author} {\bibinfo {author} {\bibnamefont {Gurevich},
  \bibfnamefont {L.~E.}}} (\bibinfo {year} {1946}{\natexlab{a}}),\ \href@noop
  {} {\bibfield  {journal} {\bibinfo  {journal} {Zh. Eksp. Teor. Fiz.}\
  }\textbf {\bibinfo {volume} {16}},\ \bibinfo {pages} {193}},\ \bibinfo {note}
  {[J. Phys. (USSR) {\bf 9}, 857 (1945)]}\BibitemShut {NoStop}%
\bibitem [{\citenamefont {Gurevich}(1946{\natexlab{b}})}]{lgu2}%
  \BibitemOpen
  \bibfield  {author} {\bibinfo {author} {\bibnamefont {Gurevich},
  \bibfnamefont {L.~E.}}} (\bibinfo {year} {1946}{\natexlab{b}}),\ \href@noop
  {} {\bibfield  {journal} {\bibinfo  {journal} {Zh. Eksp. Teor. Fiz.}\
  }\textbf {\bibinfo {volume} {16}},\ \bibinfo {pages} {416}},\ \bibinfo {note}
  {[J. Phys. (USSR) {\bf 10}, 67 (1946)]}\BibitemShut {NoStop}%
\bibitem [{\citenamefont {Gurevich}\ and\ \citenamefont
  {Muradov}(2000)}]{gur2}%
  \BibitemOpen
  \bibfield  {author} {\bibinfo {author} {\bibnamefont {Gurevich},
  \bibfnamefont {V.~L.}}, \ and\ \bibinfo {author} {\bibfnamefont {M.~I.}\
  \bibnamefont {Muradov}}} (\bibinfo {year} {2000}),\ \href@noop {} {\bibfield
  {journal} {\bibinfo  {journal} {Pis'ma Zh. Eksp. Teor. Fiz.}\ }\textbf
  {\bibinfo {volume} {71}},\ \bibinfo {pages} {164}},\ \bibinfo {note} {[JETP
  Lett. {\bf 71}, 111 (2000)]}\BibitemShut {NoStop}%
\bibitem [{\citenamefont {Gurevich}\ and\ \citenamefont {Muradov}(2005)}]{gur}%
  \BibitemOpen
  \bibfield  {author} {\bibinfo {author} {\bibnamefont {Gurevich},
  \bibfnamefont {V.~L.}}, \ and\ \bibinfo {author} {\bibfnamefont {M.~I.}\
  \bibnamefont {Muradov}}} (\bibinfo {year} {2005}),\ \href@noop {} {\bibfield
  {journal} {\bibinfo  {journal} {J. Phys.: Condens. Matter}\ }\textbf
  {\bibinfo {volume} {17}},\ \bibinfo {pages} {87}}\BibitemShut {NoStop}%
\bibitem [{\citenamefont {Gurevich}\ \emph {et~al.}(1998)\citenamefont
  {Gurevich}, \citenamefont {Pevzner},\ and\ \citenamefont {Fenton}}]{pev}%
  \BibitemOpen
  \bibfield  {author} {\bibinfo {author} {\bibnamefont {Gurevich},
  \bibfnamefont {V.~L.}}, \bibinfo {author} {\bibfnamefont {V.~B.}\
  \bibnamefont {Pevzner}}, \ and\ \bibinfo {author} {\bibfnamefont {E.~W.}\
  \bibnamefont {Fenton}}} (\bibinfo {year} {1998}),\ \href
  {http://stacks.iop.org/0953-8984/10/i=11/a=018} {\bibfield  {journal}
  {\bibinfo  {journal} {Journal of Physics: Condensed Matter}\ }\textbf
  {\bibinfo {volume} {10}}~(\bibinfo {number} {11}),\ \bibinfo {pages}
  {2551}}\BibitemShut {NoStop}%
\bibitem [{\citenamefont {Gurevich}\ and\ \citenamefont
  {Mashkevich}(1989)}]{gum}%
  \BibitemOpen
  \bibfield  {author} {\bibinfo {author} {\bibnamefont {Gurevich},
  \bibfnamefont {Y.~G.}}, \ and\ \bibinfo {author} {\bibfnamefont {O.~L.}\
  \bibnamefont {Mashkevich}}} (\bibinfo {year} {1989}),\ \href@noop {}
  {\bibfield  {journal} {\bibinfo  {journal} {Phys. Rep.}\ }\textbf {\bibinfo
  {volume} {181}},\ \bibinfo {pages} {327}}\BibitemShut {NoStop}%
\bibitem [{\citenamefont {G{\"u}ven}\ and\ \citenamefont
  {Tanatar}(1997{\natexlab{a}})}]{tan}%
  \BibitemOpen
  \bibfield  {author} {\bibinfo {author} {\bibnamefont {G{\"u}ven},
  \bibfnamefont {K.}}, \ and\ \bibinfo {author} {\bibfnamefont
  {B.}~\bibnamefont {Tanatar}}} (\bibinfo {year} {1997}{\natexlab{a}}),\
  \href@noop {} {\bibfield  {journal} {\bibinfo  {journal} {Phys. Rev. B}\
  }\textbf {\bibinfo {volume} {56}},\ \bibinfo {pages} {7535}}\BibitemShut
  {NoStop}%
\bibitem [{\citenamefont {G{\"u}ven}\ and\ \citenamefont
  {Tanatar}(1997{\natexlab{b}})}]{tan2}%
  \BibitemOpen
  \bibfield  {author} {\bibinfo {author} {\bibnamefont {G{\"u}ven},
  \bibfnamefont {K.}}, \ and\ \bibinfo {author} {\bibfnamefont
  {B.}~\bibnamefont {Tanatar}}} (\bibinfo {year} {1997}{\natexlab{b}}),\
  \href@noop {} {\bibfield  {journal} {\bibinfo  {journal} {Solid State Comm.}\
  }\textbf {\bibinfo {volume} {104}},\ \bibinfo {pages} {439}}\BibitemShut
  {NoStop}%
\bibitem [{\citenamefont {Haldane}\ and\ \citenamefont {Rezayi}(1987)}]{hald}%
  \BibitemOpen
  \bibfield  {author} {\bibinfo {author} {\bibnamefont {Haldane}, \bibfnamefont
  {F.}}, \ and\ \bibinfo {author} {\bibfnamefont {E.}~\bibnamefont {Rezayi}}}
  (\bibinfo {year} {1987}),\ \href@noop {} {\bibfield  {journal} {\bibinfo
  {journal} {Bull. Am. Phys. Soc.}\ }\textbf {\bibinfo {volume} {32}},\
  \bibinfo {pages} {892}}\BibitemShut {NoStop}%
\bibitem [{\citenamefont {Haldane}(1981{\natexlab{a}})}]{hal}%
  \BibitemOpen
  \bibfield  {author} {\bibinfo {author} {\bibnamefont {Haldane}, \bibfnamefont
  {F.~D.~M.}}} (\bibinfo {year} {1981}{\natexlab{a}}),\ \href@noop {}
  {\bibfield  {journal} {\bibinfo  {journal} {J. Phys. C}\ }\textbf {\bibinfo
  {volume} {14}},\ \bibinfo {pages} {2585}}\BibitemShut {NoStop}%
\bibitem [{\citenamefont {Haldane}(1981{\natexlab{b}})}]{hal2}%
  \BibitemOpen
  \bibfield  {author} {\bibinfo {author} {\bibnamefont {Haldane}, \bibfnamefont
  {F.~D.~M.}}} (\bibinfo {year} {1981}{\natexlab{b}}),\ \href@noop {}
  {\bibfield  {journal} {\bibinfo  {journal} {Phys. Rev. Lett.}\ }\textbf
  {\bibinfo {volume} {47}},\ \bibinfo {pages} {1840}}\BibitemShut {NoStop}%
\bibitem [{\citenamefont {Halperin}(1983)}]{halp}%
  \BibitemOpen
  \bibfield  {author} {\bibinfo {author} {\bibnamefont {Halperin},
  \bibfnamefont {B.}}} (\bibinfo {year} {1983}),\ \href@noop {} {\bibfield
  {journal} {\bibinfo  {journal} {Helv. Phys. Acta}\ }\textbf {\bibinfo
  {volume} {56}},\ \bibinfo {pages} {75}}\BibitemShut {NoStop}%
\bibitem [{\citenamefont {Halperin}\ \emph {et~al.}(1993)\citenamefont
  {Halperin}, \citenamefont {Lee},\ and\ \citenamefont {Read}}]{hlr}%
  \BibitemOpen
  \bibfield  {author} {\bibinfo {author} {\bibnamefont {Halperin},
  \bibfnamefont {B.~I.}}, \bibinfo {author} {\bibfnamefont {P.~A.}\
  \bibnamefont {Lee}}, \ and\ \bibinfo {author} {\bibfnamefont
  {N.}~\bibnamefont {Read}}} (\bibinfo {year} {1993}),\ \href@noop {}
  {\bibfield  {journal} {\bibinfo  {journal} {Phys. Rev. B}\ }\textbf {\bibinfo
  {volume} {47}},\ \bibinfo {pages} {7312}}\BibitemShut {NoStop}%
\bibitem [{\citenamefont {H{\"a}nsch}\ and\ \citenamefont {Mahan}(1983)}]{mah}%
  \BibitemOpen
  \bibfield  {author} {\bibinfo {author} {\bibnamefont {H{\"a}nsch},
  \bibfnamefont {W.}}, \ and\ \bibinfo {author} {\bibfnamefont {G.~D.}\
  \bibnamefont {Mahan}}} (\bibinfo {year} {1983}),\ \href@noop {} {\bibfield
  {journal} {\bibinfo  {journal} {J. Phys. Chem. Solids}\ }\textbf {\bibinfo
  {volume} {44}},\ \bibinfo {pages} {663}}\BibitemShut {NoStop}%
\bibitem [{\citenamefont {Herring}(1954)}]{her}%
  \BibitemOpen
  \bibfield  {author} {\bibinfo {author} {\bibnamefont {Herring}, \bibfnamefont
  {C.}}} (\bibinfo {year} {1954}),\ \href@noop {} {\bibfield  {journal}
  {\bibinfo  {journal} {Phys. Rev.}\ }\textbf {\bibinfo {volume} {96}},\
  \bibinfo {pages} {1163}}\BibitemShut {NoStop}%
\bibitem [{\citenamefont {Hill}\ \emph {et~al.}(1996)\citenamefont {Hill},
  \citenamefont {Nicholls}, \citenamefont {Linfield}, \citenamefont {Pepper},
  \citenamefont {Ritchie}, \citenamefont {Hamilton},\ and\ \citenamefont
  {Jones}}]{hi3}%
  \BibitemOpen
  \bibfield  {author} {\bibinfo {author} {\bibnamefont {Hill}, \bibfnamefont
  {N.~P.~R.}}, \bibinfo {author} {\bibfnamefont {J.~T.}\ \bibnamefont
  {Nicholls}}, \bibinfo {author} {\bibfnamefont {E.~H.}\ \bibnamefont
  {Linfield}}, \bibinfo {author} {\bibfnamefont {M.}~\bibnamefont {Pepper}},
  \bibinfo {author} {\bibfnamefont {D.~A.}\ \bibnamefont {Ritchie}}, \bibinfo
  {author} {\bibfnamefont {A.~R.}\ \bibnamefont {Hamilton}}, \ and\ \bibinfo
  {author} {\bibfnamefont {G.~A.~C.}\ \bibnamefont {Jones}}} (\bibinfo {year}
  {1996}),\ \href@noop {} {\bibfield  {journal} {\bibinfo  {journal} {J. Phys.:
  Condens. Matter}\ }\textbf {\bibinfo {volume} {8}},\ \bibinfo {pages}
  {L557}}\BibitemShut {NoStop}%
\bibitem [{\citenamefont {Hill}\ \emph {et~al.}(1998)\citenamefont {Hill},
  \citenamefont {Nicholls}, \citenamefont {Linfield}, \citenamefont {Pepper},
  \citenamefont {Ritchie}, \citenamefont {Hu},\ and\ \citenamefont
  {Flensberg}}]{hi2}%
  \BibitemOpen
  \bibfield  {author} {\bibinfo {author} {\bibnamefont {Hill}, \bibfnamefont
  {N.~P.~R.}}, \bibinfo {author} {\bibfnamefont {J.~T.}\ \bibnamefont
  {Nicholls}}, \bibinfo {author} {\bibfnamefont {E.~H.}\ \bibnamefont
  {Linfield}}, \bibinfo {author} {\bibfnamefont {M.}~\bibnamefont {Pepper}},
  \bibinfo {author} {\bibfnamefont {D.~A.}\ \bibnamefont {Ritchie}}, \bibinfo
  {author} {\bibfnamefont {B.~Y.-K.}\ \bibnamefont {Hu}}, \ and\ \bibinfo
  {author} {\bibfnamefont {K.}~\bibnamefont {Flensberg}}} (\bibinfo {year}
  {1998}),\ \href@noop {} {\bibfield  {journal} {\bibinfo  {journal} {Physica
  B}\ }\textbf {\bibinfo {volume} {249-251}},\ \bibinfo {pages}
  {868}}\BibitemShut {NoStop}%
\bibitem [{\citenamefont {Hill}\ \emph {et~al.}(1997)\citenamefont {Hill},
  \citenamefont {Nicholls}, \citenamefont {Linfield}, \citenamefont {Pepper},
  \citenamefont {Ritchie}, \citenamefont {Jones}, \citenamefont {Hu},\ and\
  \citenamefont {Flensberg}}]{hil}%
  \BibitemOpen
  \bibfield  {author} {\bibinfo {author} {\bibnamefont {Hill}, \bibfnamefont
  {N.~P.~R.}}, \bibinfo {author} {\bibfnamefont {J.~T.}\ \bibnamefont
  {Nicholls}}, \bibinfo {author} {\bibfnamefont {E.~H.}\ \bibnamefont
  {Linfield}}, \bibinfo {author} {\bibfnamefont {M.}~\bibnamefont {Pepper}},
  \bibinfo {author} {\bibfnamefont {D.~A.}\ \bibnamefont {Ritchie}}, \bibinfo
  {author} {\bibfnamefont {G.~A.~C.}\ \bibnamefont {Jones}}, \bibinfo {author}
  {\bibfnamefont {B.~Y.-K.}\ \bibnamefont {Hu}}, \ and\ \bibinfo {author}
  {\bibfnamefont {K.}~\bibnamefont {Flensberg}}} (\bibinfo {year} {1997}),\
  \href@noop {} {\bibfield  {journal} {\bibinfo  {journal} {Phys. Rev. Lett.}\
  }\textbf {\bibinfo {volume} {78}},\ \bibinfo {pages} {2204}}\BibitemShut
  {NoStop}%
\bibitem [{\citenamefont {H{\"o}pfel}\ and\ \citenamefont {Shah}(1988)}]{ex01}%
  \BibitemOpen
  \bibfield  {author} {\bibinfo {author} {\bibnamefont {H{\"o}pfel},
  \bibfnamefont {R.}}, \ and\ \bibinfo {author} {\bibfnamefont
  {J.}~\bibnamefont {Shah}}} (\bibinfo {year} {1988}),\ \href@noop {}
  {\bibfield  {journal} {\bibinfo  {journal} {Solid-State Electronics}\
  }\textbf {\bibinfo {volume} {31}},\ \bibinfo {pages} {643}}\BibitemShut
  {NoStop}%
\bibitem [{\citenamefont {H{\"o}pfel}\ \emph {et~al.}(1986)\citenamefont
  {H{\"o}pfel}, \citenamefont {Shah}, \citenamefont {Wolff},\ and\
  \citenamefont {Gossard}}]{ex0}%
  \BibitemOpen
  \bibfield  {author} {\bibinfo {author} {\bibnamefont {H{\"o}pfel},
  \bibfnamefont {R.~A.}}, \bibinfo {author} {\bibfnamefont {J.}~\bibnamefont
  {Shah}}, \bibinfo {author} {\bibfnamefont {P.~A.}\ \bibnamefont {Wolff}}, \
  and\ \bibinfo {author} {\bibfnamefont {A.~C.}\ \bibnamefont {Gossard}}}
  (\bibinfo {year} {1986}),\ \href@noop {} {\bibfield  {journal} {\bibinfo
  {journal} {Phys. Rev. Lett.}\ }\textbf {\bibinfo {volume} {56}},\ \bibinfo
  {pages} {2736}}\BibitemShut {NoStop}%
\bibitem [{\citenamefont {Hruska}\ and\ \citenamefont {Spivak}(2002)}]{hru}%
  \BibitemOpen
  \bibfield  {author} {\bibinfo {author} {\bibnamefont {Hruska}, \bibfnamefont
  {M.}}, \ and\ \bibinfo {author} {\bibfnamefont {B.}~\bibnamefont {Spivak}}}
  (\bibinfo {year} {2002}),\ \href {\doibase 10.1103/PhysRevB.65.033315}
  {\bibfield  {journal} {\bibinfo  {journal} {Phys. Rev. B}\ }\textbf {\bibinfo
  {volume} {65}},\ \bibinfo {pages} {033315}}\BibitemShut {NoStop}%
\bibitem [{\citenamefont {Hu}(1997)}]{hu0}%
  \BibitemOpen
  \bibfield  {author} {\bibinfo {author} {\bibnamefont {Hu}, \bibfnamefont
  {B.~Y.-K.}}} (\bibinfo {year} {1997}),\ \href@noop {} {\bibfield  {journal}
  {\bibinfo  {journal} {Physica Scripta}\ }\textbf {\bibinfo {volume} {T69}},\
  \bibinfo {pages} {170}}\BibitemShut {NoStop}%
\bibitem [{\citenamefont {Hu}(1998)}]{hu3}%
  \BibitemOpen
  \bibfield  {author} {\bibinfo {author} {\bibnamefont {Hu}, \bibfnamefont
  {B.~Y.-K.}}} (\bibinfo {year} {1998}),\ \href@noop {} {\bibfield  {journal}
  {\bibinfo  {journal} {Phys. Rev. B}\ }\textbf {\bibinfo {volume} {57}},\
  \bibinfo {pages} {12345}}\BibitemShut {NoStop}%
\bibitem [{\citenamefont {Hu}(2000{\natexlab{a}})}]{hu2}%
  \BibitemOpen
  \bibfield  {author} {\bibinfo {author} {\bibnamefont {Hu}, \bibfnamefont
  {B.~Y.-K.}}} (\bibinfo {year} {2000}{\natexlab{a}}),\ \href@noop {}
  {\bibfield  {journal} {\bibinfo  {journal} {Physica E}\ }\textbf {\bibinfo
  {volume} {6}},\ \bibinfo {pages} {611}}\BibitemShut {NoStop}%
\bibitem [{\citenamefont {Hu}(2000{\natexlab{b}})}]{hu5}%
  \BibitemOpen
  \bibfield  {author} {\bibinfo {author} {\bibnamefont {Hu}, \bibfnamefont
  {B.~Y.-K.}}} (\bibinfo {year} {2000}{\natexlab{b}}),\ \href@noop {}
  {\bibfield  {journal} {\bibinfo  {journal} {Phys. Rev. Lett.}\ }\textbf
  {\bibinfo {volume} {85}},\ \bibinfo {pages} {820}}\BibitemShut {NoStop}%
\bibitem [{\citenamefont {Hu}\ and\ \citenamefont {Flensberg}(1996)}]{huf}%
  \BibitemOpen
  \bibfield  {author} {\bibinfo {author} {\bibnamefont {Hu}, \bibfnamefont
  {B.~Y.-K.}}, \ and\ \bibinfo {author} {\bibfnamefont {K.}~\bibnamefont
  {Flensberg}}} (\bibinfo {year} {1996}),\ in\ \href@noop {} {\emph {\bibinfo
  {booktitle} {Hot Carriers in Semiconductors}}},\ \bibinfo {editor} {edited
  by\ \bibinfo {editor} {\bibfnamefont {K.}~\bibnamefont {Hess}}}\ (\bibinfo
  {publisher} {Plenum Press, New York})\BibitemShut {NoStop}%
\bibitem [{\citenamefont {Huang}\ \emph {et~al.}(1995)\citenamefont {Huang},
  \citenamefont {Bazan},\ and\ \citenamefont {Bernstein}}]{hua}%
  \BibitemOpen
  \bibfield  {author} {\bibinfo {author} {\bibnamefont {Huang}, \bibfnamefont
  {X.}}, \bibinfo {author} {\bibfnamefont {G.}~\bibnamefont {Bazan}}, \ and\
  \bibinfo {author} {\bibfnamefont {G.}~\bibnamefont {Bernstein}}} (\bibinfo
  {year} {1995}),\ \href@noop {} {\bibfield  {journal} {\bibinfo  {journal}
  {Phys. Rev. Lett.}\ }\textbf {\bibinfo {volume} {74}},\ \bibinfo {pages}
  {4051}}\BibitemShut {NoStop}%
\bibitem [{\citenamefont {Huse}(2005)}]{huse}%
  \BibitemOpen
  \bibfield  {author} {\bibinfo {author} {\bibnamefont {Huse}, \bibfnamefont
  {D.~A.}}} (\bibinfo {year} {2005}),\ \href {\doibase
  10.1103/PhysRevB.72.064514} {\bibfield  {journal} {\bibinfo  {journal} {Phys.
  Rev. B}\ }\textbf {\bibinfo {volume} {72}},\ \bibinfo {pages}
  {064514}}\BibitemShut {NoStop}%
\bibitem [{\citenamefont {Hwang}\ and\ \citenamefont
  {Das~Sarma}(2007)}]{plas1}%
  \BibitemOpen
  \bibfield  {author} {\bibinfo {author} {\bibnamefont {Hwang}, \bibfnamefont
  {E.~H.}}, \ and\ \bibinfo {author} {\bibfnamefont {S.}~\bibnamefont
  {Das~Sarma}}} (\bibinfo {year} {2007}),\ \href@noop {} {\bibfield  {journal}
  {\bibinfo  {journal} {Phys. Rev. B}\ }\textbf {\bibinfo {volume} {75}},\
  \bibinfo {pages} {205418}}\BibitemShut {NoStop}%
\bibitem [{\citenamefont {Hwang}\ and\ \citenamefont
  {Das~Sarma}(2008{\natexlab{a}})}]{hwang}%
  \BibitemOpen
  \bibfield  {author} {\bibinfo {author} {\bibnamefont {Hwang}, \bibfnamefont
  {E.~H.}}, \ and\ \bibinfo {author} {\bibfnamefont {S.}~\bibnamefont
  {Das~Sarma}}} (\bibinfo {year} {2008}{\natexlab{a}}),\ \href@noop {}
  {\bibfield  {journal} {\bibinfo  {journal} {Phys. Rev. B}\ }\textbf {\bibinfo
  {volume} {77}},\ \bibinfo {pages} {115449}}\BibitemShut {NoStop}%
\bibitem [{\citenamefont {Hwang}\ and\ \citenamefont
  {Das~Sarma}(2008{\natexlab{b}})}]{hwa2}%
  \BibitemOpen
  \bibfield  {author} {\bibinfo {author} {\bibnamefont {Hwang}, \bibfnamefont
  {E.~H.}}, \ and\ \bibinfo {author} {\bibfnamefont {S.}~\bibnamefont
  {Das~Sarma}}} (\bibinfo {year} {2008}{\natexlab{b}}),\ \href@noop {}
  {\bibfield  {journal} {\bibinfo  {journal} {Phys. Rev. B}\ }\textbf {\bibinfo
  {volume} {78}},\ \bibinfo {pages} {075430}}\BibitemShut {NoStop}%
\bibitem [{\citenamefont {Hwang}\ \emph {et~al.}(2003)\citenamefont {Hwang},
  \citenamefont {Sarma}, \citenamefont {Braude},\ and\ \citenamefont
  {Stern}}]{hbs}%
  \BibitemOpen
  \bibfield  {author} {\bibinfo {author} {\bibnamefont {Hwang}, \bibfnamefont
  {E.~H.}}, \bibinfo {author} {\bibfnamefont {S.~D.}\ \bibnamefont {Sarma}},
  \bibinfo {author} {\bibfnamefont {V.}~\bibnamefont {Braude}}, \ and\ \bibinfo
  {author} {\bibfnamefont {A.}~\bibnamefont {Stern}}} (\bibinfo {year}
  {2003}),\ \href@noop {} {\bibfield  {journal} {\bibinfo  {journal} {Phys.
  Rev. Lett.}\ }\textbf {\bibinfo {volume} {90}},\ \bibinfo {pages}
  {086801}}\BibitemShut {NoStop}%
\bibitem [{\citenamefont {Hwang}\ \emph {et~al.}(2011)\citenamefont {Hwang},
  \citenamefont {Sensarma},\ and\ \citenamefont {Sarma}}]{dss}%
  \BibitemOpen
  \bibfield  {author} {\bibinfo {author} {\bibnamefont {Hwang}, \bibfnamefont
  {E.~H.}}, \bibinfo {author} {\bibfnamefont {R.}~\bibnamefont {Sensarma}}, \
  and\ \bibinfo {author} {\bibfnamefont {S.~D.}\ \bibnamefont {Sarma}}}
  (\bibinfo {year} {2011}),\ \href@noop {} {\bibfield  {journal} {\bibinfo
  {journal} {Phys. Rev. B}\ }\textbf {\bibinfo {volume} {84}},\ \bibinfo
  {pages} {245441}}\BibitemShut {NoStop}%
\bibitem [{\citenamefont {Imambekov}\ \emph {et~al.}(2012)\citenamefont
  {Imambekov}, \citenamefont {Schmidt},\ and\ \citenamefont {Glazman}}]{imam}%
  \BibitemOpen
  \bibfield  {author} {\bibinfo {author} {\bibnamefont {Imambekov},
  \bibfnamefont {A.}}, \bibinfo {author} {\bibfnamefont {T.~L.}\ \bibnamefont
  {Schmidt}}, \ and\ \bibinfo {author} {\bibfnamefont {L.~I.}\ \bibnamefont
  {Glazman}}} (\bibinfo {year} {2012}),\ \href {\doibase
  10.1103/RevModPhys.84.1253} {\bibfield  {journal} {\bibinfo  {journal} {Rev.
  Mod. Phys.}\ }\textbf {\bibinfo {volume} {84}},\ \bibinfo {pages}
  {1253}}\BibitemShut {NoStop}%
\bibitem [{\citenamefont {Jacoboni}\ and\ \citenamefont {Price}(1988)}]{jac}%
  \BibitemOpen
  \bibfield  {author} {\bibinfo {author} {\bibnamefont {Jacoboni},
  \bibfnamefont {C.}}, \ and\ \bibinfo {author} {\bibfnamefont {P.~J.}\
  \bibnamefont {Price}}} (\bibinfo {year} {1988}),\ \href@noop {} {\bibfield
  {journal} {\bibinfo  {journal} {Solid-State Electron.}\ }\textbf {\bibinfo
  {volume} {31}},\ \bibinfo {pages} {649}}\BibitemShut {NoStop}%
\bibitem [{\citenamefont {Jain}(1989)}]{jai}%
  \BibitemOpen
  \bibfield  {author} {\bibinfo {author} {\bibnamefont {Jain}, \bibfnamefont
  {J.~K.}}} (\bibinfo {year} {1989}),\ \href@noop {} {\bibfield  {journal}
  {\bibinfo  {journal} {Phys. Rev. Lett.}\ }\textbf {\bibinfo {volume} {63}},\
  \bibinfo {pages} {199}}\BibitemShut {NoStop}%
\bibitem [{\citenamefont {Jalabert}\ and\ \citenamefont
  {Das~Sarma}(1989)}]{sdas2}%
  \BibitemOpen
  \bibfield  {author} {\bibinfo {author} {\bibnamefont {Jalabert},
  \bibfnamefont {R.}}, \ and\ \bibinfo {author} {\bibfnamefont
  {S.}~\bibnamefont {Das~Sarma}}} (\bibinfo {year} {1989}),\ \href@noop {}
  {\bibfield  {journal} {\bibinfo  {journal} {Phys. Rev. B}\ }\textbf {\bibinfo
  {volume} {40}},\ \bibinfo {pages} {9723}}\BibitemShut {NoStop}%
\bibitem [{\citenamefont {Jauho}\ and\ \citenamefont {Smith}(1993)}]{jho}%
  \BibitemOpen
  \bibfield  {author} {\bibinfo {author} {\bibnamefont {Jauho}, \bibfnamefont
  {A.-P.}}, \ and\ \bibinfo {author} {\bibfnamefont {H.}~\bibnamefont {Smith}}}
  (\bibinfo {year} {1993}),\ \href@noop {} {\bibfield  {journal} {\bibinfo
  {journal} {Phys. Rev. B}\ }\textbf {\bibinfo {volume} {47}},\ \bibinfo
  {pages} {4420}}\BibitemShut {NoStop}%
\bibitem [{\citenamefont {Johnson}\ \emph {et~al.}(2004)\citenamefont
  {Johnson}, \citenamefont {Marcus}, \citenamefont {Hanson},\ and\
  \citenamefont {Gossard}}]{mar}%
  \BibitemOpen
  \bibfield  {author} {\bibinfo {author} {\bibnamefont {Johnson}, \bibfnamefont
  {A.~C.}}, \bibinfo {author} {\bibfnamefont {C.~M.}\ \bibnamefont {Marcus}},
  \bibinfo {author} {\bibfnamefont {M.~P.}\ \bibnamefont {Hanson}}, \ and\
  \bibinfo {author} {\bibfnamefont {A.~C.}\ \bibnamefont {Gossard}}} (\bibinfo
  {year} {2004}),\ \href@noop {} {\bibfield  {journal} {\bibinfo  {journal}
  {Phys. Rev. Lett.}\ }\textbf {\bibinfo {volume} {93}},\ \bibinfo {pages}
  {106803}}\BibitemShut {NoStop}%
\bibitem [{\citenamefont {J{\"o}rger}\ \emph
  {et~al.}(2000{\natexlab{a}})\citenamefont {J{\"o}rger}, \citenamefont
  {Cheng}, \citenamefont {Dietsche}, \citenamefont {Gerhardts}, \citenamefont
  {Specht}, \citenamefont {Eberl},\ and\ \citenamefont {von Klitzing}}]{jo2}%
  \BibitemOpen
  \bibfield  {author} {\bibinfo {author} {\bibnamefont {J{\"o}rger},
  \bibfnamefont {C.}}, \bibinfo {author} {\bibfnamefont {S.~J.}\ \bibnamefont
  {Cheng}}, \bibinfo {author} {\bibfnamefont {W.}~\bibnamefont {Dietsche}},
  \bibinfo {author} {\bibfnamefont {R.}~\bibnamefont {Gerhardts}}, \bibinfo
  {author} {\bibfnamefont {P.}~\bibnamefont {Specht}}, \bibinfo {author}
  {\bibfnamefont {K.}~\bibnamefont {Eberl}}, \ and\ \bibinfo {author}
  {\bibfnamefont {K.}~\bibnamefont {von Klitzing}}} (\bibinfo {year}
  {2000}{\natexlab{a}}),\ \href@noop {} {\bibfield  {journal} {\bibinfo
  {journal} {Physica E}\ }\textbf {\bibinfo {volume} {6}},\ \bibinfo {pages}
  {598}}\BibitemShut {NoStop}%
\bibitem [{\citenamefont {J{\"o}rger}\ \emph
  {et~al.}(2000{\natexlab{b}})\citenamefont {J{\"o}rger}, \citenamefont
  {Cheng}, \citenamefont {Rubel}, \citenamefont {Dietsche}, \citenamefont
  {Gerhardts}, \citenamefont {Specht}, \citenamefont {Eberl},\ and\
  \citenamefont {von Klitzing}}]{jor}%
  \BibitemOpen
  \bibfield  {author} {\bibinfo {author} {\bibnamefont {J{\"o}rger},
  \bibfnamefont {C.}}, \bibinfo {author} {\bibfnamefont {S.~J.}\ \bibnamefont
  {Cheng}}, \bibinfo {author} {\bibfnamefont {H.}~\bibnamefont {Rubel}},
  \bibinfo {author} {\bibfnamefont {W.}~\bibnamefont {Dietsche}}, \bibinfo
  {author} {\bibfnamefont {R.}~\bibnamefont {Gerhardts}}, \bibinfo {author}
  {\bibfnamefont {P.}~\bibnamefont {Specht}}, \bibinfo {author} {\bibfnamefont
  {K.}~\bibnamefont {Eberl}}, \ and\ \bibinfo {author} {\bibfnamefont
  {K.}~\bibnamefont {von Klitzing}}} (\bibinfo {year} {2000}{\natexlab{b}}),\
  \href@noop {} {\bibfield  {journal} {\bibinfo  {journal} {Phys. Rev. B}\
  }\textbf {\bibinfo {volume} {62}},\ \bibinfo {pages} {1572}}\BibitemShut
  {NoStop}%
\bibitem [{\citenamefont {J{\"o}rger}\ \emph
  {et~al.}(2000{\natexlab{c}})\citenamefont {J{\"o}rger}, \citenamefont
  {Dietsche}, \citenamefont {Wegschneider},\ and\ \citenamefont
  {Klitzing}}]{jo1}%
  \BibitemOpen
  \bibfield  {author} {\bibinfo {author} {\bibnamefont {J{\"o}rger},
  \bibfnamefont {C.}}, \bibinfo {author} {\bibfnamefont {W.}~\bibnamefont
  {Dietsche}}, \bibinfo {author} {\bibfnamefont {W.}~\bibnamefont
  {Wegschneider}}, \ and\ \bibinfo {author} {\bibfnamefont {K.}~\bibnamefont
  {Klitzing}}} (\bibinfo {year} {2000}{\natexlab{c}}),\ \href@noop {}
  {\bibfield  {journal} {\bibinfo  {journal} {Physca E}\ }\textbf {\bibinfo
  {volume} {6}},\ \bibinfo {pages} {586}}\BibitemShut {NoStop}%
\bibitem [{\citenamefont {Ju}\ \emph {et~al.}(2011)\citenamefont {Ju},
  \citenamefont {Geng}, \citenamefont {Horng}, \citenamefont {Girit},
  \citenamefont {Martin}, \citenamefont {Hao}, \citenamefont {Bechtel},
  \citenamefont {Liang}, \citenamefont {Zettl}, \citenamefont {Ron~Shen},\ and\
  \citenamefont {Wang}}]{plex3}%
  \BibitemOpen
  \bibfield  {author} {\bibinfo {author} {\bibnamefont {Ju}, \bibfnamefont
  {L.}}, \bibinfo {author} {\bibfnamefont {B.}~\bibnamefont {Geng}}, \bibinfo
  {author} {\bibfnamefont {J.}~\bibnamefont {Horng}}, \bibinfo {author}
  {\bibfnamefont {C.}~\bibnamefont {Girit}}, \bibinfo {author} {\bibfnamefont
  {M.}~\bibnamefont {Martin}}, \bibinfo {author} {\bibfnamefont
  {Z.}~\bibnamefont {Hao}}, \bibinfo {author} {\bibfnamefont {H.~A.}\
  \bibnamefont {Bechtel}}, \bibinfo {author} {\bibfnamefont {X.}~\bibnamefont
  {Liang}}, \bibinfo {author} {\bibfnamefont {A.}~\bibnamefont {Zettl}},
  \bibinfo {author} {\bibfnamefont {Y.}~\bibnamefont {Ron~Shen}}, \ and\
  \bibinfo {author} {\bibfnamefont {F.}~\bibnamefont {Wang}}} (\bibinfo {year}
  {2011}),\ \href@noop {} {\bibfield  {journal} {\bibinfo  {journal} {Nature
  Nanotech.}\ }\textbf {\bibinfo {volume} {6}},\ \bibinfo {pages}
  {630}}\BibitemShut {NoStop}%
\bibitem [{\citenamefont {Kamenev}\ and\ \citenamefont {Oreg}(1995)}]{kor}%
  \BibitemOpen
  \bibfield  {author} {\bibinfo {author} {\bibnamefont {Kamenev}, \bibfnamefont
  {A.}}, \ and\ \bibinfo {author} {\bibfnamefont {Y.}~\bibnamefont {Oreg}}}
  (\bibinfo {year} {1995}),\ \href@noop {} {\bibfield  {journal} {\bibinfo
  {journal} {Phys. Rev. B}\ }\textbf {\bibinfo {volume} {52}},\ \bibinfo
  {pages} {7516}}\BibitemShut {NoStop}%
\bibitem [{\citenamefont {Kashuba}(2008)}]{kas}%
  \BibitemOpen
  \bibfield  {author} {\bibinfo {author} {\bibnamefont {Kashuba}, \bibfnamefont
  {A.~B.}}} (\bibinfo {year} {2008}),\ \href@noop {} {\bibfield  {journal}
  {\bibinfo  {journal} {Phys. Rev. B}\ }\textbf {\bibinfo {volume} {78}},\
  \bibinfo {pages} {085415}}\BibitemShut {NoStop}%
\bibitem [{\citenamefont {Kasprzak}\ \emph {et~al.}(2006)\citenamefont
  {Kasprzak}, \citenamefont {Richard}, \citenamefont {Kundermann},
  \citenamefont {Baas}, \citenamefont {Jeambrun}, \citenamefont {Keeling},
  \citenamefont {Marchetti}, \citenamefont {Szymanska}, \citenamefont {Andre},
  \citenamefont {Staehli}, \citenamefont {Savona}, \citenamefont {Littlewood},
  \citenamefont {Deveaud},\ and\ \citenamefont {Dang}}]{fran}%
  \BibitemOpen
  \bibfield  {author} {\bibinfo {author} {\bibnamefont {Kasprzak},
  \bibfnamefont {J.}}, \bibinfo {author} {\bibfnamefont {M.}~\bibnamefont
  {Richard}}, \bibinfo {author} {\bibfnamefont {S.}~\bibnamefont {Kundermann}},
  \bibinfo {author} {\bibfnamefont {A.}~\bibnamefont {Baas}}, \bibinfo {author}
  {\bibfnamefont {P.}~\bibnamefont {Jeambrun}}, \bibinfo {author}
  {\bibfnamefont {J.~M.~J.}\ \bibnamefont {Keeling}}, \bibinfo {author}
  {\bibfnamefont {F.~M.}\ \bibnamefont {Marchetti}}, \bibinfo {author}
  {\bibfnamefont {M.~H.}\ \bibnamefont {Szymanska}}, \bibinfo {author}
  {\bibfnamefont {R.}~\bibnamefont {Andre}}, \bibinfo {author} {\bibfnamefont
  {J.~L.}\ \bibnamefont {Staehli}}, \bibinfo {author} {\bibfnamefont
  {V.}~\bibnamefont {Savona}}, \bibinfo {author} {\bibfnamefont {P.~B.}\
  \bibnamefont {Littlewood}}, \bibinfo {author} {\bibfnamefont
  {B.}~\bibnamefont {Deveaud}}, \ and\ \bibinfo {author} {\bibfnamefont
  {L.~S.}\ \bibnamefont {Dang}}} (\bibinfo {year} {2006}),\ \href@noop {}
  {\bibfield  {journal} {\bibinfo  {journal} {Nature}\ }\textbf {\bibinfo
  {volume} {443}},\ \bibinfo {pages} {409}}\BibitemShut {NoStop}%
\bibitem [{\citenamefont {Katsnelson}(2011)}]{kac}%
  \BibitemOpen
  \bibfield  {author} {\bibinfo {author} {\bibnamefont {Katsnelson},
  \bibfnamefont {M.~I.}}} (\bibinfo {year} {2011}),\ \href@noop {} {\bibfield
  {journal} {\bibinfo  {journal} {Phys. Rev. B}\ }\textbf {\bibinfo {volume}
  {84}},\ \bibinfo {pages} {041407}}\BibitemShut {NoStop}%
\bibitem [{\citenamefont {Katsnelson}(2012)}]{kats2012}%
  \BibitemOpen
  \bibfield  {author} {\bibinfo {author} {\bibnamefont {Katsnelson},
  \bibfnamefont {M.~I.}}} (\bibinfo {year} {2012}),\ \href@noop {} {\emph
  {\bibinfo {title} {Graphene}}}\ (\bibinfo  {publisher} {Cambridge University
  Press})\BibitemShut {NoStop}%
\bibitem [{\citenamefont {Keldysh}\ and\ \citenamefont
  {Kopaev}(1964)}]{kelkop}%
  \BibitemOpen
  \bibfield  {author} {\bibinfo {author} {\bibnamefont {Keldysh}, \bibfnamefont
  {L.~V.}}, \ and\ \bibinfo {author} {\bibfnamefont {Y.~V.}\ \bibnamefont
  {Kopaev}}} (\bibinfo {year} {1964}),\ \href@noop {} {\bibfield  {journal}
  {\bibinfo  {journal} {Fiz. Tverd. Tela.}\ }\textbf {\bibinfo {volume} {6}},\
  \bibinfo {pages} {2791}},\ \bibinfo {note} {[Sov. Phys. Solid State. {\bf 6},
  2219 (1965)]}\BibitemShut {NoStop}%
\bibitem [{\citenamefont {Keldysh}\ and\ \citenamefont {Kozlov}(1968)}]{keld}%
  \BibitemOpen
  \bibfield  {author} {\bibinfo {author} {\bibnamefont {Keldysh}, \bibfnamefont
  {L.~V.}}, \ and\ \bibinfo {author} {\bibfnamefont {A.~N.}\ \bibnamefont
  {Kozlov}}} (\bibinfo {year} {1968}),\ \href@noop {} {\bibfield  {journal}
  {\bibinfo  {journal} {Zh. Eksp. Teor. Fiz.}\ }\textbf {\bibinfo {volume}
  {54}},\ \bibinfo {pages} {978}},\ \bibinfo {note} {[Sov. Phys. JETP {\bf 27},
  521 (1968)]}\BibitemShut {NoStop}%
\bibitem [{\citenamefont {Kellogg}\ \emph {et~al.}(2003)\citenamefont
  {Kellogg}, \citenamefont {Eisenstein}, \citenamefont {Pfeiffer},\ and\
  \citenamefont {K.W.West}}]{kel}%
  \BibitemOpen
  \bibfield  {author} {\bibinfo {author} {\bibnamefont {Kellogg}, \bibfnamefont
  {M.}}, \bibinfo {author} {\bibfnamefont {J.~P.}\ \bibnamefont {Eisenstein}},
  \bibinfo {author} {\bibfnamefont {L.~N.}\ \bibnamefont {Pfeiffer}}, \ and\
  \bibinfo {author} {\bibnamefont {K.W.West}}} (\bibinfo {year} {2003}),\
  \href@noop {} {\bibfield  {journal} {\bibinfo  {journal} {Phys. Rev. Lett.}\
  }\textbf {\bibinfo {volume} {90}},\ \bibinfo {pages} {246801}}\BibitemShut
  {NoStop}%
\bibitem [{\citenamefont {Kellogg}\ \emph {et~al.}(2004)\citenamefont
  {Kellogg}, \citenamefont {Eisenstein}, \citenamefont {Pfeiffer},\ and\
  \citenamefont {West}}]{ke4}%
  \BibitemOpen
  \bibfield  {author} {\bibinfo {author} {\bibnamefont {Kellogg}, \bibfnamefont
  {M.}}, \bibinfo {author} {\bibfnamefont {J.~P.}\ \bibnamefont {Eisenstein}},
  \bibinfo {author} {\bibfnamefont {L.~N.}\ \bibnamefont {Pfeiffer}}, \ and\
  \bibinfo {author} {\bibfnamefont {K.~W.}\ \bibnamefont {West}}} (\bibinfo
  {year} {2004}),\ \href {\doibase 10.1103/PhysRevLett.93.036801} {\bibfield
  {journal} {\bibinfo  {journal} {Phys. Rev. Lett.}\ }\textbf {\bibinfo
  {volume} {93}},\ \bibinfo {pages} {036801}}\BibitemShut {NoStop}%
\bibitem [{\citenamefont {Kellogg}\ \emph
  {et~al.}(2002{\natexlab{a}})\citenamefont {Kellogg}, \citenamefont
  {Eisenstein}, \citenamefont {Pfeiffer},\ and\ \citenamefont {W.West}}]{ke1}%
  \BibitemOpen
  \bibfield  {author} {\bibinfo {author} {\bibnamefont {Kellogg}, \bibfnamefont
  {M.}}, \bibinfo {author} {\bibfnamefont {J.~P.}\ \bibnamefont {Eisenstein}},
  \bibinfo {author} {\bibfnamefont {L.~N.}\ \bibnamefont {Pfeiffer}}, \ and\
  \bibinfo {author} {\bibfnamefont {K.}~\bibnamefont {W.West}}} (\bibinfo
  {year} {2002}{\natexlab{a}}),\ \href@noop {} {\bibfield  {journal} {\bibinfo
  {journal} {Solid State Comm.}\ }\textbf {\bibinfo {volume} {123}},\ \bibinfo
  {pages} {515}}\BibitemShut {NoStop}%
\bibitem [{\citenamefont {Kellogg}\ \emph
  {et~al.}(2002{\natexlab{b}})\citenamefont {Kellogg}, \citenamefont
  {Spielman}, \citenamefont {Eisenstein}, \citenamefont {Pfeiffer},\ and\
  \citenamefont {W.West}}]{ke2}%
  \BibitemOpen
  \bibfield  {author} {\bibinfo {author} {\bibnamefont {Kellogg}, \bibfnamefont
  {M.}}, \bibinfo {author} {\bibfnamefont {I.~B.}\ \bibnamefont {Spielman}},
  \bibinfo {author} {\bibfnamefont {J.~P.}\ \bibnamefont {Eisenstein}},
  \bibinfo {author} {\bibfnamefont {L.~N.}\ \bibnamefont {Pfeiffer}}, \ and\
  \bibinfo {author} {\bibfnamefont {K.}~\bibnamefont {W.West}}} (\bibinfo
  {year} {2002}{\natexlab{b}}),\ \href@noop {} {\bibfield  {journal} {\bibinfo
  {journal} {Phys. Rev. Lett.}\ }\textbf {\bibinfo {volume} {88}},\ \bibinfo
  {pages} {126804}}\BibitemShut {NoStop}%
\bibitem [{\citenamefont {Keogh}\ \emph {et~al.}(2005)\citenamefont {Keogh},
  \citenamefont {{Das Gupta}}, \citenamefont {Beere}, \citenamefont {Ritchie},\
  and\ \citenamefont {Pepper}}]{keo}%
  \BibitemOpen
  \bibfield  {author} {\bibinfo {author} {\bibnamefont {Keogh}, \bibfnamefont
  {J.~A.}}, \bibinfo {author} {\bibfnamefont {K.}~\bibnamefont {{Das Gupta}}},
  \bibinfo {author} {\bibfnamefont {H.~E.}\ \bibnamefont {Beere}}, \bibinfo
  {author} {\bibfnamefont {D.~A.}\ \bibnamefont {Ritchie}}, \ and\ \bibinfo
  {author} {\bibfnamefont {M.}~\bibnamefont {Pepper}}} (\bibinfo {year}
  {2005}),\ \href@noop {} {\bibfield  {journal} {\bibinfo  {journal} {Appl.
  Phys. Lett.}\ }\textbf {\bibinfo {volume} {87}},\ \bibinfo {pages}
  {202104}}\BibitemShut {NoStop}%
\bibitem [{\citenamefont {Khaetskii}\ and\ \citenamefont
  {Nazarov}(1999)}]{kha}%
  \BibitemOpen
  \bibfield  {author} {\bibinfo {author} {\bibnamefont {Khaetskii},
  \bibfnamefont {A.~V.}}, \ and\ \bibinfo {author} {\bibfnamefont {Y.~V.}\
  \bibnamefont {Nazarov}}} (\bibinfo {year} {1999}),\ \href@noop {} {\bibfield
  {journal} {\bibinfo  {journal} {Phys. Rev. B}\ }\textbf {\bibinfo {volume}
  {59}},\ \bibinfo {pages} {7551}}\BibitemShut {NoStop}%
\bibitem [{\citenamefont {Kharitonov}\ and\ \citenamefont
  {Efetov}(2008)}]{khar}%
  \BibitemOpen
  \bibfield  {author} {\bibinfo {author} {\bibnamefont {Kharitonov},
  \bibfnamefont {M.~Y.}}, \ and\ \bibinfo {author} {\bibfnamefont {K.~B.}\
  \bibnamefont {Efetov}}} (\bibinfo {year} {2008}),\ \href {\doibase
  10.1103/PhysRevB.78.241401} {\bibfield  {journal} {\bibinfo  {journal} {Phys.
  Rev. B}\ }\textbf {\bibinfo {volume} {78}},\ \bibinfo {pages}
  {241401}}\BibitemShut {NoStop}%
\bibitem [{\citenamefont {Kharitonov}\ and\ \citenamefont
  {Efetov}(2010)}]{khar2}%
  \BibitemOpen
  \bibfield  {author} {\bibinfo {author} {\bibnamefont {Kharitonov},
  \bibfnamefont {M.~Y.}}, \ and\ \bibinfo {author} {\bibfnamefont {K.~B.}\
  \bibnamefont {Efetov}}} (\bibinfo {year} {2010}),\ \href
  {http://stacks.iop.org/0268-1242/25/i=3/a=034004} {\bibfield  {journal}
  {\bibinfo  {journal} {Semiconductor Science and Technology}\ }\textbf
  {\bibinfo {volume} {25}}~(\bibinfo {number} {3}),\ \bibinfo {pages}
  {034004}}\BibitemShut {NoStop}%
\bibitem [{\citenamefont {Khrapai}\ \emph {et~al.}(2006)\citenamefont
  {Khrapai}, \citenamefont {Ludwig}, \citenamefont {Kotthaus}, \citenamefont
  {Tranitz},\ and\ \citenamefont {Wegscheider}}]{khr2}%
  \BibitemOpen
  \bibfield  {author} {\bibinfo {author} {\bibnamefont {Khrapai}, \bibfnamefont
  {V.~S.}}, \bibinfo {author} {\bibfnamefont {S.}~\bibnamefont {Ludwig}},
  \bibinfo {author} {\bibfnamefont {J.~P.}\ \bibnamefont {Kotthaus}}, \bibinfo
  {author} {\bibfnamefont {H.~P.}\ \bibnamefont {Tranitz}}, \ and\ \bibinfo
  {author} {\bibfnamefont {W.}~\bibnamefont {Wegscheider}}} (\bibinfo {year}
  {2006}),\ \href {\doibase 10.1103/PhysRevLett.97.176803} {\bibfield
  {journal} {\bibinfo  {journal} {Phys. Rev. Lett.}\ }\textbf {\bibinfo
  {volume} {97}},\ \bibinfo {pages} {176803}}\BibitemShut {NoStop}%
\bibitem [{\citenamefont {Khrapai}\ \emph {et~al.}(2007)\citenamefont
  {Khrapai}, \citenamefont {Ludwig}, \citenamefont {Kotthaus}, \citenamefont
  {Tranitz},\ and\ \citenamefont {Wegscheider}}]{khr}%
  \BibitemOpen
  \bibfield  {author} {\bibinfo {author} {\bibnamefont {Khrapai}, \bibfnamefont
  {V.~S.}}, \bibinfo {author} {\bibfnamefont {S.}~\bibnamefont {Ludwig}},
  \bibinfo {author} {\bibfnamefont {J.~P.}\ \bibnamefont {Kotthaus}}, \bibinfo
  {author} {\bibfnamefont {H.~P.}\ \bibnamefont {Tranitz}}, \ and\ \bibinfo
  {author} {\bibfnamefont {W.}~\bibnamefont {Wegscheider}}} (\bibinfo {year}
  {2007}),\ \href@noop {} {\bibfield  {journal} {\bibinfo  {journal} {Phys.
  Rev. Lett.}\ }\textbf {\bibinfo {volume} {99}},\ \bibinfo {pages}
  {096803}}\BibitemShut {NoStop}%
\bibitem [{\citenamefont {Khveshchenko}(2000)}]{khv}%
  \BibitemOpen
  \bibfield  {author} {\bibinfo {author} {\bibnamefont {Khveshchenko},
  \bibfnamefont {D.~V.}}} (\bibinfo {year} {2000}),\ \href@noop {} {\bibfield
  {journal} {\bibinfo  {journal} {Phys. Rev. B}\ }\textbf {\bibinfo {volume}
  {61}},\ \bibinfo {pages} {7227}}\BibitemShut {NoStop}%
\bibitem [{\citenamefont {Kim}\ \emph {et~al.}(2011)\citenamefont {Kim},
  \citenamefont {Jo}, \citenamefont {Nah}, \citenamefont {Yao}, \citenamefont
  {Banerjee},\ and\ \citenamefont {Tutuc}}]{tu1}%
  \BibitemOpen
  \bibfield  {author} {\bibinfo {author} {\bibnamefont {Kim}, \bibfnamefont
  {S.}}, \bibinfo {author} {\bibfnamefont {I.}~\bibnamefont {Jo}}, \bibinfo
  {author} {\bibfnamefont {J.}~\bibnamefont {Nah}}, \bibinfo {author}
  {\bibfnamefont {Z.}~\bibnamefont {Yao}}, \bibinfo {author} {\bibfnamefont
  {S.~K.}\ \bibnamefont {Banerjee}}, \ and\ \bibinfo {author} {\bibfnamefont
  {E.}~\bibnamefont {Tutuc}}} (\bibinfo {year} {2011}),\ \href@noop {}
  {\bibfield  {journal} {\bibinfo  {journal} {Phys. Rev. B}\ }\textbf {\bibinfo
  {volume} {83}},\ \bibinfo {pages} {161401(R)}}\BibitemShut {NoStop}%
\bibitem [{\citenamefont {Kim}\ and\ \citenamefont {Tutuc}(2012)}]{tu2}%
  \BibitemOpen
  \bibfield  {author} {\bibinfo {author} {\bibnamefont {Kim}, \bibfnamefont
  {S.}}, \ and\ \bibinfo {author} {\bibfnamefont {E.}~\bibnamefont {Tutuc}}}
  (\bibinfo {year} {2012}),\ \href@noop {} {\bibfield  {journal} {\bibinfo
  {journal} {Solid State Commun.}\ }\textbf {\bibinfo {volume} {152}},\
  \bibinfo {pages} {1283}}\BibitemShut {NoStop}%
\bibitem [{\citenamefont {Kim}\ and\ \citenamefont {Millis}(1999)}]{kim}%
  \BibitemOpen
  \bibfield  {author} {\bibinfo {author} {\bibnamefont {Kim}, \bibfnamefont
  {Y.~B.}}, \ and\ \bibinfo {author} {\bibfnamefont {A.~J.}\ \bibnamefont
  {Millis}}} (\bibinfo {year} {1999}),\ \href@noop {} {\bibfield  {journal}
  {\bibinfo  {journal} {Physica E}\ }\textbf {\bibinfo {volume} {4}},\ \bibinfo
  {pages} {171}}\BibitemShut {NoStop}%
\bibitem [{\citenamefont {Kim}\ \emph {et~al.}(2001)\citenamefont {Kim},
  \citenamefont {Nayak}, \citenamefont {Demler}, \citenamefont {Read},\ and\
  \citenamefont {Sarma}}]{dem}%
  \BibitemOpen
  \bibfield  {author} {\bibinfo {author} {\bibnamefont {Kim}, \bibfnamefont
  {Y.~B.}}, \bibinfo {author} {\bibfnamefont {C.}~\bibnamefont {Nayak}},
  \bibinfo {author} {\bibfnamefont {E.}~\bibnamefont {Demler}}, \bibinfo
  {author} {\bibfnamefont {N.}~\bibnamefont {Read}}, \ and\ \bibinfo {author}
  {\bibfnamefont {S.~D.}\ \bibnamefont {Sarma}}} (\bibinfo {year} {2001}),\
  \href@noop {} {\bibfield  {journal} {\bibinfo  {journal} {Phys. Rev. B}\
  }\textbf {\bibinfo {volume} {63}},\ \bibinfo {pages} {205315}}\BibitemShut
  {NoStop}%
\bibitem [{\citenamefont {Klesse}\ and\ \citenamefont {Stern}(2000)}]{kle}%
  \BibitemOpen
  \bibfield  {author} {\bibinfo {author} {\bibnamefont {Klesse}, \bibfnamefont
  {R.}}, \ and\ \bibinfo {author} {\bibfnamefont {A.}~\bibnamefont {Stern}}}
  (\bibinfo {year} {2000}),\ \href@noop {} {\bibfield  {journal} {\bibinfo
  {journal} {Phys. Rev. B}\ }\textbf {\bibinfo {volume} {62}},\ \bibinfo
  {pages} {16912}}\BibitemShut {NoStop}%
\bibitem [{\citenamefont {Komnik}\ and\ \citenamefont {Egger}(2001)}]{koe}%
  \BibitemOpen
  \bibfield  {author} {\bibinfo {author} {\bibnamefont {Komnik}, \bibfnamefont
  {A.}}, \ and\ \bibinfo {author} {\bibfnamefont {R.}~\bibnamefont {Egger}}}
  (\bibinfo {year} {2001}),\ \href@noop {} {\bibfield  {journal} {\bibinfo
  {journal} {Eur. Phys. J. B}\ }\textbf {\bibinfo {volume} {19}},\ \bibinfo
  {pages} {271}}\BibitemShut {NoStop}%
\bibitem [{\citenamefont {K\"onig}\ \emph {et~al.}(2013)\citenamefont
  {K\"onig}, \citenamefont {Baenninger}, \citenamefont {Garcia}, \citenamefont
  {Harjee}, \citenamefont {Pruitt}, \citenamefont {Ames}, \citenamefont
  {Leubner}, \citenamefont {Br\"une}, \citenamefont {Buhmann}, \citenamefont
  {Molenkamp},\ and\ \citenamefont {Goldhaber-Gordon}}]{gold}%
  \BibitemOpen
  \bibfield  {author} {\bibinfo {author} {\bibnamefont {K\"onig}, \bibfnamefont
  {M.}}, \bibinfo {author} {\bibfnamefont {M.}~\bibnamefont {Baenninger}},
  \bibinfo {author} {\bibfnamefont {A.~G.~F.}\ \bibnamefont {Garcia}}, \bibinfo
  {author} {\bibfnamefont {N.}~\bibnamefont {Harjee}}, \bibinfo {author}
  {\bibfnamefont {B.~L.}\ \bibnamefont {Pruitt}}, \bibinfo {author}
  {\bibfnamefont {C.}~\bibnamefont {Ames}}, \bibinfo {author} {\bibfnamefont
  {P.}~\bibnamefont {Leubner}}, \bibinfo {author} {\bibfnamefont
  {C.}~\bibnamefont {Br\"une}}, \bibinfo {author} {\bibfnamefont
  {H.}~\bibnamefont {Buhmann}}, \bibinfo {author} {\bibfnamefont {L.~W.}\
  \bibnamefont {Molenkamp}}, \ and\ \bibinfo {author} {\bibfnamefont
  {D.}~\bibnamefont {Goldhaber-Gordon}}} (\bibinfo {year} {2013}),\ \href
  {\doibase 10.1103/PhysRevX.3.021003} {\bibfield  {journal} {\bibinfo
  {journal} {Phys. Rev. X}\ }\textbf {\bibinfo {volume} {3}},\ \bibinfo {pages}
  {021003}}\BibitemShut {NoStop}%
\bibitem [{\citenamefont {Kozikov}\ \emph {et~al.}(2010)\citenamefont
  {Kozikov}, \citenamefont {Savchenko}, \citenamefont {Narozhny},\ and\
  \citenamefont {Shytov}}]{koz}%
  \BibitemOpen
  \bibfield  {author} {\bibinfo {author} {\bibnamefont {Kozikov}, \bibfnamefont
  {A.~A.}}, \bibinfo {author} {\bibfnamefont {A.~K.}\ \bibnamefont
  {Savchenko}}, \bibinfo {author} {\bibfnamefont {B.~N.}\ \bibnamefont
  {Narozhny}}, \ and\ \bibinfo {author} {\bibfnamefont {A.~V.}\ \bibnamefont
  {Shytov}}} (\bibinfo {year} {2010}),\ \href@noop {} {\bibfield  {journal}
  {\bibinfo  {journal} {Phys. Rev. B}\ }\textbf {\bibinfo {volume} {82}},\
  \bibinfo {pages} {075424}}\BibitemShut {NoStop}%
\bibitem [{\citenamefont {Krishnaswamy}\ \emph {et~al.}(1999)\citenamefont
  {Krishnaswamy}, \citenamefont {Goodnick},\ and\ \citenamefont {Bird}}]{kri}%
  \BibitemOpen
  \bibfield  {author} {\bibinfo {author} {\bibnamefont {Krishnaswamy},
  \bibfnamefont {A.~E.}}, \bibinfo {author} {\bibfnamefont {S.~M.}\
  \bibnamefont {Goodnick}}, \ and\ \bibinfo {author} {\bibfnamefont
  {J.}~\bibnamefont {Bird}}} (\bibinfo {year} {1999}),\ \href@noop {}
  {\bibfield  {journal} {\bibinfo  {journal} {Microelectronic Engineering}\
  }\textbf {\bibinfo {volume} {47}},\ \bibinfo {pages} {81}}\BibitemShut
  {NoStop}%
\bibitem [{\citenamefont {Kulakovskii}\ and\ \citenamefont
  {Lozovik}(2004)}]{kul}%
  \BibitemOpen
  \bibfield  {author} {\bibinfo {author} {\bibnamefont {Kulakovskii},
  \bibfnamefont {D.~V.}}, \ and\ \bibinfo {author} {\bibfnamefont {Y.~E.}\
  \bibnamefont {Lozovik}}} (\bibinfo {year} {2004}),\ \href@noop {} {\bibfield
  {journal} {\bibinfo  {journal} {Zh. Eksp. Teor. Fiz.}\ }\textbf {\bibinfo
  {volume} {125}},\ \bibinfo {pages} {1375}},\ \bibinfo {note} {[Sov. Phys.
  JETP {\bf 98}, 1205 (2004)]}\BibitemShut {NoStop}%
\bibitem [{\citenamefont {Kvon}\ \emph {et~al.}(1997)\citenamefont {Kvon},
  \citenamefont {Olshanetsky}, \citenamefont {Gusev}, \citenamefont {Portal},\
  and\ \citenamefont {Maude}}]{kvo}%
  \BibitemOpen
  \bibfield  {author} {\bibinfo {author} {\bibnamefont {Kvon}, \bibfnamefont
  {Z.~D.}}, \bibinfo {author} {\bibfnamefont {E.~B.}\ \bibnamefont
  {Olshanetsky}}, \bibinfo {author} {\bibfnamefont {G.~M.}\ \bibnamefont
  {Gusev}}, \bibinfo {author} {\bibfnamefont {J.~C.}\ \bibnamefont {Portal}}, \
  and\ \bibinfo {author} {\bibfnamefont {D.~K.}\ \bibnamefont {Maude}}}
  (\bibinfo {year} {1997}),\ \href@noop {} {\bibfield  {journal} {\bibinfo
  {journal} {Phys. Rev. B}\ }\textbf {\bibinfo {volume} {56}},\ \bibinfo
  {pages} {12112}}\BibitemShut {NoStop}%
\bibitem [{\citenamefont {Laikhtman}\ and\ \citenamefont
  {Solomon}(1990)}]{las}%
  \BibitemOpen
  \bibfield  {author} {\bibinfo {author} {\bibnamefont {Laikhtman},
  \bibfnamefont {B.}}, \ and\ \bibinfo {author} {\bibfnamefont {P.~M.}\
  \bibnamefont {Solomon}}} (\bibinfo {year} {1990}),\ \href@noop {} {\bibfield
  {journal} {\bibinfo  {journal} {Phys. Rev. B}\ }\textbf {\bibinfo {volume}
  {41}},\ \bibinfo {pages} {9921}}\BibitemShut {NoStop}%
\bibitem [{\citenamefont {Laikhtman}\ and\ \citenamefont
  {Solomon}(2005)}]{la3}%
  \BibitemOpen
  \bibfield  {author} {\bibinfo {author} {\bibnamefont {Laikhtman},
  \bibfnamefont {B.}}, \ and\ \bibinfo {author} {\bibfnamefont {P.~M.}\
  \bibnamefont {Solomon}}} (\bibinfo {year} {2005}),\ \href@noop {} {\bibfield
  {journal} {\bibinfo  {journal} {Phys. Rev. B}\ }\textbf {\bibinfo {volume}
  {72}},\ \bibinfo {pages} {125338}}\BibitemShut {NoStop}%
\bibitem [{\citenamefont {Laikhtman}\ and\ \citenamefont
  {Solomon}(2006)}]{lai}%
  \BibitemOpen
  \bibfield  {author} {\bibinfo {author} {\bibnamefont {Laikhtman},
  \bibfnamefont {B.}}, \ and\ \bibinfo {author} {\bibfnamefont {P.~M.}\
  \bibnamefont {Solomon}}} (\bibinfo {year} {2006}),\ \href@noop {} {\bibfield
  {journal} {\bibinfo  {journal} {Solid State Commun.}\ }\textbf {\bibinfo
  {volume} {138}},\ \bibinfo {pages} {143}}\BibitemShut {NoStop}%
\bibitem [{\citenamefont {Landau}\ \emph {et~al.}(1984)\citenamefont {Landau},
  \citenamefont {Lifshitz},\ and\ \citenamefont {Pitaevskii}}]{dau8}%
  \BibitemOpen
  \bibfield  {author} {\bibinfo {author} {\bibnamefont {Landau}, \bibfnamefont
  {L.~D.}}, \bibinfo {author} {\bibfnamefont {E.~M.}\ \bibnamefont {Lifshitz}},
  \ and\ \bibinfo {author} {\bibfnamefont {L.~P.}\ \bibnamefont {Pitaevskii}}}
  (\bibinfo {year} {1984}),\ \href@noop {} {\emph {\bibinfo {title}
  {Electrodynamics of Continuous Media}}}\ (\bibinfo  {publisher} {Pergamon
  Press})\BibitemShut {NoStop}%
\bibitem [{\citenamefont {Landauer}(1957)}]{land1}%
  \BibitemOpen
  \bibfield  {author} {\bibinfo {author} {\bibnamefont {Landauer},
  \bibfnamefont {R.}}} (\bibinfo {year} {1957}),\ \href@noop {} {\bibfield
  {journal} {\bibinfo  {journal} {IBM J. Res. Dev.}\ }\textbf {\bibinfo
  {volume} {1}},\ \bibinfo {pages} {223}}\BibitemShut {NoStop}%
\bibitem [{\citenamefont {Landauer}(1970)}]{land2}%
  \BibitemOpen
  \bibfield  {author} {\bibinfo {author} {\bibnamefont {Landauer},
  \bibfnamefont {R.}}} (\bibinfo {year} {1970}),\ \href@noop {} {\bibfield
  {journal} {\bibinfo  {journal} {Philos. Mag.}\ }\textbf {\bibinfo {volume}
  {21}},\ \bibinfo {pages} {863}}\BibitemShut {NoStop}%
\bibitem [{\citenamefont {Laroche}\ \emph {et~al.}(2008)\citenamefont
  {Laroche}, \citenamefont {Bielejec}, \citenamefont {Reno}, \citenamefont
  {Gervais},\ and\ \citenamefont {Lilly}}]{la2}%
  \BibitemOpen
  \bibfield  {author} {\bibinfo {author} {\bibnamefont {Laroche}, \bibfnamefont
  {D.}}, \bibinfo {author} {\bibfnamefont {E.~S.}\ \bibnamefont {Bielejec}},
  \bibinfo {author} {\bibfnamefont {J.~L.}\ \bibnamefont {Reno}}, \bibinfo
  {author} {\bibfnamefont {G.}~\bibnamefont {Gervais}}, \ and\ \bibinfo
  {author} {\bibfnamefont {M.}~\bibnamefont {Lilly}}} (\bibinfo {year}
  {2008}),\ \href@noop {} {\bibfield  {journal} {\bibinfo  {journal} {Physica
  E}\ }\textbf {\bibinfo {volume} {40}},\ \bibinfo {pages} {1569}}\BibitemShut
  {NoStop}%
\bibitem [{\citenamefont {Laroche}\ \emph {et~al.}(2011)\citenamefont
  {Laroche}, \citenamefont {Gervais}, \citenamefont {Lilly},\ and\
  \citenamefont {Reno}}]{lar}%
  \BibitemOpen
  \bibfield  {author} {\bibinfo {author} {\bibnamefont {Laroche}, \bibfnamefont
  {D.}}, \bibinfo {author} {\bibfnamefont {G.}~\bibnamefont {Gervais}},
  \bibinfo {author} {\bibfnamefont {M.~P.}\ \bibnamefont {Lilly}}, \ and\
  \bibinfo {author} {\bibfnamefont {J.~L.}\ \bibnamefont {Reno}}} (\bibinfo
  {year} {2011}),\ \href@noop {} {\bibfield  {journal} {\bibinfo  {journal}
  {Nature Natotechnology}\ }\textbf {\bibinfo {volume} {6}},\ \bibinfo {pages}
  {793}}\BibitemShut {NoStop}%
\bibitem [{\citenamefont {Laroche}\ \emph {et~al.}(2014)\citenamefont
  {Laroche}, \citenamefont {Gervais}, \citenamefont {Lilly},\ and\
  \citenamefont {Reno}}]{lar2}%
  \BibitemOpen
  \bibfield  {author} {\bibinfo {author} {\bibnamefont {Laroche}, \bibfnamefont
  {D.}}, \bibinfo {author} {\bibfnamefont {G.}~\bibnamefont {Gervais}},
  \bibinfo {author} {\bibfnamefont {M.~P.}\ \bibnamefont {Lilly}}, \ and\
  \bibinfo {author} {\bibfnamefont {J.~L.}\ \bibnamefont {Reno}}} (\bibinfo
  {year} {2014}),\ \href@noop {} {\bibfield  {journal} {\bibinfo  {journal}
  {Science}\ }\textbf {\bibinfo {volume} {343}},\ \bibinfo {pages}
  {631}}\BibitemShut {NoStop}%
\bibitem [{\citenamefont {Lee}\ \emph {et~al.}(2011)\citenamefont {Lee},
  \citenamefont {Eastham},\ and\ \citenamefont {Cooper}}]{coop}%
  \BibitemOpen
  \bibfield  {author} {\bibinfo {author} {\bibnamefont {Lee}, \bibfnamefont
  {D.~K.~K.}}, \bibinfo {author} {\bibfnamefont {P.~R.}\ \bibnamefont
  {Eastham}}, \ and\ \bibinfo {author} {\bibfnamefont {N.~R.}\ \bibnamefont
  {Cooper}}} (\bibinfo {year} {2011}),\ \href@noop {} {\bibfield  {journal}
  {\bibinfo  {journal} {Adv. Cond. Matt. Phys.}\ }\textbf {\bibinfo {volume}
  {2011}},\ \bibinfo {pages} {792125}}\BibitemShut {NoStop}%
\bibitem [{\citenamefont {Lee}\ and\ \citenamefont {Stone}(1985)}]{lees}%
  \BibitemOpen
  \bibfield  {author} {\bibinfo {author} {\bibnamefont {Lee}, \bibfnamefont
  {P.~A.}}, \ and\ \bibinfo {author} {\bibfnamefont {A.~D.}\ \bibnamefont
  {Stone}}} (\bibinfo {year} {1985}),\ \href@noop {} {\bibfield  {journal}
  {\bibinfo  {journal} {Phys. Rev. Lett.}\ }\textbf {\bibinfo {volume} {55}},\
  \bibinfo {pages} {1622}}\BibitemShut {NoStop}%
\bibitem [{\citenamefont {Lerner}(1988)}]{ler}%
  \BibitemOpen
  \bibfield  {author} {\bibinfo {author} {\bibnamefont {Lerner}, \bibfnamefont
  {I.~V.}}} (\bibinfo {year} {1988}),\ \href@noop {} {\bibfield  {journal}
  {\bibinfo  {journal} {Phys. Lett. A}\ }\textbf {\bibinfo {volume} {133}},\
  \bibinfo {pages} {253}}\BibitemShut {NoStop}%
\bibitem [{\citenamefont {Lesovik}(1989)}]{les}%
  \BibitemOpen
  \bibfield  {author} {\bibinfo {author} {\bibnamefont {Lesovik}, \bibfnamefont
  {G.~B.}}} (\bibinfo {year} {1989}),\ \href@noop {} {\bibfield  {journal}
  {\bibinfo  {journal} {Pis'ma Zh. Eksp. Teor. Fiz.}\ }\textbf {\bibinfo
  {volume} {49}},\ \bibinfo {pages} {513}},\ \bibinfo {note} {[JETP Lett. {\bf
  49}, 592 (1989)]}\BibitemShut {NoStop}%
\bibitem [{\citenamefont {Levchenko}\ and\ \citenamefont
  {Kamenev}(2008{\natexlab{a}})}]{le2}%
  \BibitemOpen
  \bibfield  {author} {\bibinfo {author} {\bibnamefont {Levchenko},
  \bibfnamefont {A.}}, \ and\ \bibinfo {author} {\bibfnamefont
  {A.}~\bibnamefont {Kamenev}}} (\bibinfo {year} {2008}{\natexlab{a}}),\
  \href@noop {} {\bibfield  {journal} {\bibinfo  {journal} {Phys. Rev. Lett.}\
  }\textbf {\bibinfo {volume} {101}},\ \bibinfo {pages} {216806}}\BibitemShut
  {NoStop}%
\bibitem [{\citenamefont {Levchenko}\ and\ \citenamefont
  {Kamenev}(2008{\natexlab{b}})}]{le1}%
  \BibitemOpen
  \bibfield  {author} {\bibinfo {author} {\bibnamefont {Levchenko},
  \bibfnamefont {A.}}, \ and\ \bibinfo {author} {\bibfnamefont
  {A.}~\bibnamefont {Kamenev}}} (\bibinfo {year} {2008}{\natexlab{b}}),\
  \href@noop {} {\bibfield  {journal} {\bibinfo  {journal} {Phys. Rev. Lett.}\
  }\textbf {\bibinfo {volume} {100}},\ \bibinfo {pages} {026805}}\BibitemShut
  {NoStop}%
\bibitem [{\citenamefont {Levchenko}\ and\ \citenamefont {Norman}(2011)}]{lev}%
  \BibitemOpen
  \bibfield  {author} {\bibinfo {author} {\bibnamefont {Levchenko},
  \bibfnamefont {A.}}, \ and\ \bibinfo {author} {\bibfnamefont {M.~R.}\
  \bibnamefont {Norman}}} (\bibinfo {year} {2011}),\ \href@noop {} {\bibfield
  {journal} {\bibinfo  {journal} {Phys. Rev. B}\ }\textbf {\bibinfo {volume}
  {83}},\ \bibinfo {pages} {100506}}\BibitemShut {NoStop}%
\bibitem [{\citenamefont {Levchenko}\ \emph {et~al.}(2011)\citenamefont
  {Levchenko}, \citenamefont {Ristivojevic},\ and\ \citenamefont
  {Micklitz}}]{lev3}%
  \BibitemOpen
  \bibfield  {author} {\bibinfo {author} {\bibnamefont {Levchenko},
  \bibfnamefont {A.}}, \bibinfo {author} {\bibfnamefont {Z.}~\bibnamefont
  {Ristivojevic}}, \ and\ \bibinfo {author} {\bibfnamefont {T.}~\bibnamefont
  {Micklitz}}} (\bibinfo {year} {2011}),\ \href {\doibase
  10.1103/PhysRevB.83.041303} {\bibfield  {journal} {\bibinfo  {journal} {Phys.
  Rev. B}\ }\textbf {\bibinfo {volume} {83}},\ \bibinfo {pages}
  {041303}}\BibitemShut {NoStop}%
\bibitem [{\citenamefont {Lifshitz}\ and\ \citenamefont
  {Pitaevskii}(1981)}]{dau10}%
  \BibitemOpen
  \bibfield  {author} {\bibinfo {author} {\bibnamefont {Lifshitz},
  \bibfnamefont {E.~M.}}, \ and\ \bibinfo {author} {\bibfnamefont {L.~P.}\
  \bibnamefont {Pitaevskii}}} (\bibinfo {year} {1981}),\ \href@noop {} {\emph
  {\bibinfo {title} {Physical Kinetics}}}\ (\bibinfo  {publisher} {Pergamon
  Press})\BibitemShut {NoStop}%
\bibitem [{\citenamefont {Lilly}\ \emph {et~al.}(1998)\citenamefont {Lilly},
  \citenamefont {Eisenstein}, \citenamefont {Pfeiffer},\ and\ \citenamefont
  {West}}]{lil}%
  \BibitemOpen
  \bibfield  {author} {\bibinfo {author} {\bibnamefont {Lilly}, \bibfnamefont
  {M.~P.}}, \bibinfo {author} {\bibfnamefont {J.~P.}\ \bibnamefont
  {Eisenstein}}, \bibinfo {author} {\bibfnamefont {L.~N.}\ \bibnamefont
  {Pfeiffer}}, \ and\ \bibinfo {author} {\bibfnamefont {K.~W.}\ \bibnamefont
  {West}}} (\bibinfo {year} {1998}),\ \href@noop {} {\bibfield  {journal}
  {\bibinfo  {journal} {Phys. Rev. Lett.}\ }\textbf {\bibinfo {volume} {80}},\
  \bibinfo {pages} {1714}}\BibitemShut {NoStop}%
\bibitem [{\citenamefont {Lindhard}(1954)}]{lind}%
  \BibitemOpen
  \bibfield  {author} {\bibinfo {author} {\bibnamefont {Lindhard},
  \bibfnamefont {J.}}} (\bibinfo {year} {1954}),\ \href@noop {} {\bibfield
  {journal} {\bibinfo  {journal} {Dan. Mat . Fys. Medd.}\ }\textbf {\bibinfo
  {volume} {28}}~(\bibinfo {number} {8})}\BibitemShut {NoStop}%
\bibitem [{\citenamefont {Littlewood}(2007)}]{lit}%
  \BibitemOpen
  \bibfield  {author} {\bibinfo {author} {\bibnamefont {Littlewood},
  \bibfnamefont {P.}}} (\bibinfo {year} {2007}),\ \href {\doibase
  10.1126/science.1142671} {\bibfield  {journal} {\bibinfo  {journal}
  {Science}\ }\textbf {\bibinfo {volume} {316}}~(\bibinfo {number} {5827}),\
  \bibinfo {pages} {989}}\BibitemShut {NoStop}%
\bibitem [{\citenamefont {Lok}\ \emph {et~al.}(2002)\citenamefont {Lok},
  \citenamefont {Kraus}, \citenamefont {Dietsche}, \citenamefont {von
  Klitzing}, \citenamefont {Schwerdt}, \citenamefont {Hauser}, \citenamefont
  {Wegscheider},\ and\ \citenamefont {Bichler}}]{lkd}%
  \BibitemOpen
  \bibfield  {author} {\bibinfo {author} {\bibnamefont {Lok}, \bibfnamefont
  {J.~G.~S.}}, \bibinfo {author} {\bibfnamefont {S.}~\bibnamefont {Kraus}},
  \bibinfo {author} {\bibfnamefont {W.}~\bibnamefont {Dietsche}}, \bibinfo
  {author} {\bibfnamefont {K.}~\bibnamefont {von Klitzing}}, \bibinfo {author}
  {\bibfnamefont {F.}~\bibnamefont {Schwerdt}}, \bibinfo {author}
  {\bibfnamefont {M.}~\bibnamefont {Hauser}}, \bibinfo {author} {\bibfnamefont
  {W.}~\bibnamefont {Wegscheider}}, \ and\ \bibinfo {author} {\bibfnamefont
  {M.}~\bibnamefont {Bichler}}} (\bibinfo {year} {2002}),\ \href@noop {}
  {\bibfield  {journal} {\bibinfo  {journal} {Physica E}\ }\textbf {\bibinfo
  {volume} {12}},\ \bibinfo {pages} {119}}\BibitemShut {NoStop}%
\bibitem [{\citenamefont {Lok}\ \emph {et~al.}(2001{\natexlab{a}})\citenamefont
  {Lok}, \citenamefont {Kraus}, \citenamefont {Pohlt}, \citenamefont
  {Dietsche}, \citenamefont {von Klitzing}, \citenamefont {Wegschneider},\ and\
  \citenamefont {Bichler}}]{lo1}%
  \BibitemOpen
  \bibfield  {author} {\bibinfo {author} {\bibnamefont {Lok}, \bibfnamefont
  {J.~G.~S.}}, \bibinfo {author} {\bibfnamefont {S.}~\bibnamefont {Kraus}},
  \bibinfo {author} {\bibfnamefont {M.}~\bibnamefont {Pohlt}}, \bibinfo
  {author} {\bibfnamefont {W.}~\bibnamefont {Dietsche}}, \bibinfo {author}
  {\bibfnamefont {K.}~\bibnamefont {von Klitzing}}, \bibinfo {author}
  {\bibfnamefont {W.}~\bibnamefont {Wegschneider}}, \ and\ \bibinfo {author}
  {\bibfnamefont {M.}~\bibnamefont {Bichler}}} (\bibinfo {year}
  {2001}{\natexlab{a}}),\ \href@noop {} {\bibfield  {journal} {\bibinfo
  {journal} {Phys. Rev. B}\ }\textbf {\bibinfo {volume} {63}},\ \bibinfo
  {pages} {041305(R)}}\BibitemShut {NoStop}%
\bibitem [{\citenamefont {Lok}\ \emph {et~al.}(2001{\natexlab{b}})\citenamefont
  {Lok}, \citenamefont {Kraus}, \citenamefont {Pohlt}, \citenamefont
  {G{\"u}ven}, \citenamefont {Dietsche}, \citenamefont {von Klitzing},
  \citenamefont {Wegschneider},\ and\ \citenamefont {Bichler}}]{lok}%
  \BibitemOpen
  \bibfield  {author} {\bibinfo {author} {\bibnamefont {Lok}, \bibfnamefont
  {J.~G.~S.}}, \bibinfo {author} {\bibfnamefont {S.}~\bibnamefont {Kraus}},
  \bibinfo {author} {\bibfnamefont {M.}~\bibnamefont {Pohlt}}, \bibinfo
  {author} {\bibfnamefont {K.}~\bibnamefont {G{\"u}ven}}, \bibinfo {author}
  {\bibfnamefont {W.}~\bibnamefont {Dietsche}}, \bibinfo {author}
  {\bibfnamefont {K.}~\bibnamefont {von Klitzing}}, \bibinfo {author}
  {\bibfnamefont {W.}~\bibnamefont {Wegschneider}}, \ and\ \bibinfo {author}
  {\bibfnamefont {M.}~\bibnamefont {Bichler}}} (\bibinfo {year}
  {2001}{\natexlab{b}}),\ \href@noop {} {\bibfield  {journal} {\bibinfo
  {journal} {Physica B}\ }\textbf {\bibinfo {volume} {298}},\ \bibinfo {pages}
  {135}}\BibitemShut {NoStop}%
\bibitem [{\citenamefont {Lozovik}\ and\ \citenamefont
  {Nikitkov}(1999)}]{loznik}%
  \BibitemOpen
  \bibfield  {author} {\bibinfo {author} {\bibnamefont {Lozovik}, \bibfnamefont
  {Y.~E.}}, \ and\ \bibinfo {author} {\bibfnamefont {M.~V.}\ \bibnamefont
  {Nikitkov}}} (\bibinfo {year} {1999}),\ \href@noop {} {\bibfield  {journal}
  {\bibinfo  {journal} {Zh. Eksp. Teor. Fiz.}\ }\textbf {\bibinfo {volume}
  {116}},\ \bibinfo {pages} {1440}},\ \bibinfo {note} {[JETP {\bf 89}, 775
  (1999)]}\BibitemShut {NoStop}%
\bibitem [{\citenamefont {Lozovik}\ \emph {et~al.}(2012)\citenamefont
  {Lozovik}, \citenamefont {Ogarkov},\ and\ \citenamefont {Sokolik}}]{loz4}%
  \BibitemOpen
  \bibfield  {author} {\bibinfo {author} {\bibnamefont {Lozovik}, \bibfnamefont
  {Y.~E.}}, \bibinfo {author} {\bibfnamefont {S.~L.}\ \bibnamefont {Ogarkov}},
  \ and\ \bibinfo {author} {\bibfnamefont {A.~A.}\ \bibnamefont {Sokolik}}}
  (\bibinfo {year} {2012}),\ \href {\doibase 10.1103/PhysRevB.86.045429}
  {\bibfield  {journal} {\bibinfo  {journal} {Phys. Rev. B}\ }\textbf {\bibinfo
  {volume} {86}},\ \bibinfo {pages} {045429}}\BibitemShut {NoStop}%
\bibitem [{\citenamefont {Lozovik}\ and\ \citenamefont {Sokolik}(2008)}]{loz3}%
  \BibitemOpen
  \bibfield  {author} {\bibinfo {author} {\bibnamefont {Lozovik}, \bibfnamefont
  {Y.~E.}}, \ and\ \bibinfo {author} {\bibfnamefont {A.~A.}\ \bibnamefont
  {Sokolik}}} (\bibinfo {year} {2008}),\ \href@noop {} {\bibfield  {journal}
  {\bibinfo  {journal} {Pis'ma Zh. Eksp. Teor. Fiz.}\ }\textbf {\bibinfo
  {volume} {87}},\ \bibinfo {pages} {61}},\ \bibinfo {note} {[JETP Lett. {\bf
  87}, 55 (2008)]}\BibitemShut {NoStop}%
\bibitem [{\citenamefont {Lozovik}\ and\ \citenamefont
  {Yudson}(1976)}]{lozyud}%
  \BibitemOpen
  \bibfield  {author} {\bibinfo {author} {\bibnamefont {Lozovik}, \bibfnamefont
  {Y.~E.}}, \ and\ \bibinfo {author} {\bibfnamefont {V.~I.}\ \bibnamefont
  {Yudson}}} (\bibinfo {year} {1976}),\ \href@noop {} {\bibfield  {journal}
  {\bibinfo  {journal} {Zh. Eksp. Teor. Fiz.}\ }\textbf {\bibinfo {volume}
  {71}},\ \bibinfo {pages} {738}},\ \bibinfo {note} {[Sov. Phys. JETP {\bf 44},
  389 (1976)]}\BibitemShut {NoStop}%
\bibitem [{\citenamefont {Lunde}\ \emph {et~al.}(2006)\citenamefont {Lunde},
  \citenamefont {Flensberg},\ and\ \citenamefont {Glazman}}]{gl1}%
  \BibitemOpen
  \bibfield  {author} {\bibinfo {author} {\bibnamefont {Lunde}, \bibfnamefont
  {A.~M.}}, \bibinfo {author} {\bibfnamefont {K.}~\bibnamefont {Flensberg}}, \
  and\ \bibinfo {author} {\bibfnamefont {L.~I.}\ \bibnamefont {Glazman}}}
  (\bibinfo {year} {2006}),\ \href {\doibase 10.1103/PhysRevLett.97.256802}
  {\bibfield  {journal} {\bibinfo  {journal} {Phys. Rev. Lett.}\ }\textbf
  {\bibinfo {volume} {97}},\ \bibinfo {pages} {256802}}\BibitemShut {NoStop}%
\bibitem [{\citenamefont {Lunde}\ \emph {et~al.}(2007)\citenamefont {Lunde},
  \citenamefont {Flensberg},\ and\ \citenamefont {Glazman}}]{gl2}%
  \BibitemOpen
  \bibfield  {author} {\bibinfo {author} {\bibnamefont {Lunde}, \bibfnamefont
  {A.~M.}}, \bibinfo {author} {\bibfnamefont {K.}~\bibnamefont {Flensberg}}, \
  and\ \bibinfo {author} {\bibfnamefont {L.~I.}\ \bibnamefont {Glazman}}}
  (\bibinfo {year} {2007}),\ \href {\doibase 10.1103/PhysRevB.75.245418}
  {\bibfield  {journal} {\bibinfo  {journal} {Phys. Rev. B}\ }\textbf {\bibinfo
  {volume} {75}},\ \bibinfo {pages} {245418}}\BibitemShut {NoStop}%
\bibitem [{\citenamefont {Lunde}\ \emph {et~al.}(2005)\citenamefont {Lunde},
  \citenamefont {Flensberg},\ and\ \citenamefont {Jauho}}]{lfj}%
  \BibitemOpen
  \bibfield  {author} {\bibinfo {author} {\bibnamefont {Lunde}, \bibfnamefont
  {A.~M.}}, \bibinfo {author} {\bibfnamefont {K.}~\bibnamefont {Flensberg}}, \
  and\ \bibinfo {author} {\bibfnamefont {A.-P.}\ \bibnamefont {Jauho}}}
  (\bibinfo {year} {2005}),\ \href@noop {} {\bibfield  {journal} {\bibinfo
  {journal} {Phys. Rev. B}\ }\textbf {\bibinfo {volume} {71}},\ \bibinfo
  {pages} {125408}}\BibitemShut {NoStop}%
\bibitem [{\citenamefont {Lunde}\ and\ \citenamefont {Jauho}(2004)}]{luj}%
  \BibitemOpen
  \bibfield  {author} {\bibinfo {author} {\bibnamefont {Lunde}, \bibfnamefont
  {A.~M.}}, \ and\ \bibinfo {author} {\bibfnamefont {A.-P.}\ \bibnamefont
  {Jauho}}} (\bibinfo {year} {2004}),\ \href@noop {} {\bibfield  {journal}
  {\bibinfo  {journal} {Semicond. Sci. Technol.}\ }\textbf {\bibinfo {volume}
  {19}},\ \bibinfo {pages} {S433}}\BibitemShut {NoStop}%
\bibitem [{\citenamefont {Lung}\ and\ \citenamefont {Marinescu}(2011)}]{lun}%
  \BibitemOpen
  \bibfield  {author} {\bibinfo {author} {\bibnamefont {Lung}, \bibfnamefont
  {F.}}, \ and\ \bibinfo {author} {\bibfnamefont {D.~C.}\ \bibnamefont
  {Marinescu}}} (\bibinfo {year} {2011}),\ \href@noop {} {\bibfield  {journal}
  {\bibinfo  {journal} {Physica E}\ }\textbf {\bibinfo {volume} {43}},\
  \bibinfo {pages} {1769}}\BibitemShut {NoStop}%
\bibitem [{\citenamefont {Luo}\ \emph {et~al.}(2013)\citenamefont {Luo},
  \citenamefont {Qiu}, \citenamefont {Lu},\ and\ \citenamefont
  {Ni}}]{plasmarev2}%
  \BibitemOpen
  \bibfield  {author} {\bibinfo {author} {\bibnamefont {Luo}, \bibfnamefont
  {X.}}, \bibinfo {author} {\bibfnamefont {T.}~\bibnamefont {Qiu}}, \bibinfo
  {author} {\bibfnamefont {W.}~\bibnamefont {Lu}}, \ and\ \bibinfo {author}
  {\bibfnamefont {Z.}~\bibnamefont {Ni}}} (\bibinfo {year} {2013}),\ \href@noop
  {} {\bibfield  {journal} {\bibinfo  {journal} {Materials Science and
  Engineering: R: Reports}\ }\textbf {\bibinfo {volume} {74}}~(\bibinfo
  {number} {11}),\ \bibinfo {pages} {351}}\BibitemShut {NoStop}%
\bibitem [{\citenamefont {Luttinger}(1963)}]{lut}%
  \BibitemOpen
  \bibfield  {author} {\bibinfo {author} {\bibnamefont {Luttinger},
  \bibfnamefont {J.~M.}}} (\bibinfo {year} {1963}),\ \href@noop {} {\bibfield
  {journal} {\bibinfo  {journal} {J. Math. Phys.}\ }\textbf {\bibinfo {volume}
  {4}},\ \bibinfo {pages} {1154}}\BibitemShut {NoStop}%
\bibitem [{\citenamefont {Lux}\ and\ \citenamefont {Fritz}(2012)}]{lux}%
  \BibitemOpen
  \bibfield  {author} {\bibinfo {author} {\bibnamefont {Lux}, \bibfnamefont
  {J.}}, \ and\ \bibinfo {author} {\bibfnamefont {L.}~\bibnamefont {Fritz}}}
  (\bibinfo {year} {2012}),\ \href@noop {} {\bibfield  {journal} {\bibinfo
  {journal} {Phys. Rev. B}\ }\textbf {\bibinfo {volume} {86}},\ \bibinfo
  {pages} {165446}}\BibitemShut {NoStop}%
\bibitem [{\citenamefont {Lyo}(2003)}]{lyo}%
  \BibitemOpen
  \bibfield  {author} {\bibinfo {author} {\bibnamefont {Lyo}, \bibfnamefont
  {S.~K.}}} (\bibinfo {year} {2003}),\ \href@noop {} {\bibfield  {journal}
  {\bibinfo  {journal} {Phys. Rev. B}\ }\textbf {\bibinfo {volume} {68}},\
  \bibinfo {pages} {045310}}\BibitemShut {NoStop}%
\bibitem [{\citenamefont {Manolescu}\ and\ \citenamefont
  {Tanatar}(2002)}]{mat}%
  \BibitemOpen
  \bibfield  {author} {\bibinfo {author} {\bibnamefont {Manolescu},
  \bibfnamefont {A.}}, \ and\ \bibinfo {author} {\bibfnamefont
  {B.}~\bibnamefont {Tanatar}}} (\bibinfo {year} {2002}),\ \href@noop {}
  {\bibfield  {journal} {\bibinfo  {journal} {Physica E}\ }\textbf {\bibinfo
  {volume} {13}},\ \bibinfo {pages} {80}}\BibitemShut {NoStop}%
\bibitem [{\citenamefont {Margenau}\ and\ \citenamefont
  {Kestner}(1969)}]{marge}%
  \BibitemOpen
  \bibfield  {author} {\bibinfo {author} {\bibnamefont {Margenau},
  \bibfnamefont {H.}}, \ and\ \bibinfo {author} {\bibfnamefont {N.~R.}\
  \bibnamefont {Kestner}}} (\bibinfo {year} {1969}),\ \href@noop {} {\emph
  {\bibinfo {title} {Theory of Intermolecular Forces}}}\ (\bibinfo  {publisher}
  {Pergamon Press})\BibitemShut {NoStop}%
\bibitem [{\citenamefont {Maslov}(1992)}]{mas}%
  \BibitemOpen
  \bibfield  {author} {\bibinfo {author} {\bibnamefont {Maslov}, \bibfnamefont
  {D.~L.}}} (\bibinfo {year} {1992}),\ \href@noop {} {\bibfield  {journal}
  {\bibinfo  {journal} {Phys. Rev. B}\ }\textbf {\bibinfo {volume} {45}},\
  \bibinfo {pages} {1911}}\BibitemShut {NoStop}%
\bibitem [{\citenamefont {Matveev}(2013)}]{matveev}%
  \BibitemOpen
  \bibfield  {author} {\bibinfo {author} {\bibnamefont {Matveev}, \bibfnamefont
  {K.~A.}}} (\bibinfo {year} {2013}),\ \href@noop {} {\bibfield  {journal}
  {\bibinfo  {journal} {Zh. Eksp. Teor. Fiz.}\ }\textbf {\bibinfo {volume}
  {144}},\ \bibinfo {pages} {585}},\ \bibinfo {note} {[JETP {\bf 117}, 508
  (2013)]}\BibitemShut {NoStop}%
\bibitem [{\citenamefont {Mauser}\ \emph {et~al.}(2010)\citenamefont {Mauser},
  \citenamefont {Como}, \citenamefont {Baldauf}, \citenamefont {Rogach},
  \citenamefont {Huang}, \citenamefont {Talapin},\ and\ \citenamefont
  {Feldmann}}]{mau}%
  \BibitemOpen
  \bibfield  {author} {\bibinfo {author} {\bibnamefont {Mauser}, \bibfnamefont
  {C.}}, \bibinfo {author} {\bibfnamefont {E.~D.}\ \bibnamefont {Como}},
  \bibinfo {author} {\bibfnamefont {J.}~\bibnamefont {Baldauf}}, \bibinfo
  {author} {\bibfnamefont {A.~L.}\ \bibnamefont {Rogach}}, \bibinfo {author}
  {\bibfnamefont {J.}~\bibnamefont {Huang}}, \bibinfo {author} {\bibfnamefont
  {D.}~\bibnamefont {Talapin}}, \ and\ \bibinfo {author} {\bibfnamefont
  {J.}~\bibnamefont {Feldmann}}} (\bibinfo {year} {2010}),\ \href@noop {}
  {\bibfield  {journal} {\bibinfo  {journal} {Phys. Rev. B}\ }\textbf {\bibinfo
  {volume} {82}},\ \bibinfo {pages} {081306(R)}}\BibitemShut {NoStop}%
\bibitem [{\citenamefont {Micklitz}\ \emph {et~al.}(2010)\citenamefont
  {Micklitz}, \citenamefont {Rech},\ and\ \citenamefont {Matveev}}]{mick}%
  \BibitemOpen
  \bibfield  {author} {\bibinfo {author} {\bibnamefont {Micklitz},
  \bibfnamefont {T.}}, \bibinfo {author} {\bibfnamefont {J.}~\bibnamefont
  {Rech}}, \ and\ \bibinfo {author} {\bibfnamefont {K.~A.}\ \bibnamefont
  {Matveev}}} (\bibinfo {year} {2010}),\ \href {\doibase
  10.1103/PhysRevB.81.115313} {\bibfield  {journal} {\bibinfo  {journal} {Phys.
  Rev. B}\ }\textbf {\bibinfo {volume} {81}},\ \bibinfo {pages}
  {115313}}\BibitemShut {NoStop}%
\bibitem [{\citenamefont {Min}\ \emph {et~al.}(2008)\citenamefont {Min},
  \citenamefont {Bistritzer}, \citenamefont {Su},\ and\ \citenamefont
  {MacDonald}}]{bis}%
  \BibitemOpen
  \bibfield  {author} {\bibinfo {author} {\bibnamefont {Min}, \bibfnamefont
  {H.}}, \bibinfo {author} {\bibfnamefont {R.}~\bibnamefont {Bistritzer}},
  \bibinfo {author} {\bibfnamefont {J.-J.}\ \bibnamefont {Su}}, \ and\ \bibinfo
  {author} {\bibfnamefont {A.~H.}\ \bibnamefont {MacDonald}}} (\bibinfo {year}
  {2008}),\ \href {\doibase 10.1103/PhysRevB.78.121401} {\bibfield  {journal}
  {\bibinfo  {journal} {Phys. Rev. B}\ }\textbf {\bibinfo {volume} {78}},\
  \bibinfo {pages} {121401}}\BibitemShut {NoStop}%
\bibitem [{\citenamefont {Mink}\ \emph {et~al.}(2012)\citenamefont {Mink},
  \citenamefont {Stoof}, \citenamefont {Duine}, \citenamefont {Polini},\ and\
  \citenamefont {Vignale}}]{min}%
  \BibitemOpen
  \bibfield  {author} {\bibinfo {author} {\bibnamefont {Mink}, \bibfnamefont
  {M.~P.}}, \bibinfo {author} {\bibfnamefont {H.~T.~C.}\ \bibnamefont {Stoof}},
  \bibinfo {author} {\bibfnamefont {R.~A.}\ \bibnamefont {Duine}}, \bibinfo
  {author} {\bibfnamefont {M.}~\bibnamefont {Polini}}, \ and\ \bibinfo {author}
  {\bibfnamefont {G.}~\bibnamefont {Vignale}}} (\bibinfo {year} {2012}),\
  \href@noop {} {\bibfield  {journal} {\bibinfo  {journal} {Phys. Rev. Lett.}\
  }\textbf {\bibinfo {volume} {108}},\ \bibinfo {pages} {186402}}\BibitemShut
  {NoStop}%
\bibitem [{\citenamefont {Mink}\ \emph {et~al.}(2013)\citenamefont {Mink},
  \citenamefont {Stoof}, \citenamefont {Duine}, \citenamefont {Polini},\ and\
  \citenamefont {Vignale}}]{min2}%
  \BibitemOpen
  \bibfield  {author} {\bibinfo {author} {\bibnamefont {Mink}, \bibfnamefont
  {M.~P.}}, \bibinfo {author} {\bibfnamefont {H.~T.~C.}\ \bibnamefont {Stoof}},
  \bibinfo {author} {\bibfnamefont {R.~A.}\ \bibnamefont {Duine}}, \bibinfo
  {author} {\bibfnamefont {M.}~\bibnamefont {Polini}}, \ and\ \bibinfo {author}
  {\bibfnamefont {G.}~\bibnamefont {Vignale}}} (\bibinfo {year} {2013}),\ \href
  {\doibase 10.1103/PhysRevB.88.235311} {\bibfield  {journal} {\bibinfo
  {journal} {Phys. Rev. B}\ }\textbf {\bibinfo {volume} {88}},\ \bibinfo
  {pages} {235311}}\BibitemShut {NoStop}%
\bibitem [{\citenamefont {Mishchenko}\ \emph {et~al.}(2004)\citenamefont
  {Mishchenko}, \citenamefont {Reizer},\ and\ \citenamefont {Glazman}}]{reiz}%
  \BibitemOpen
  \bibfield  {author} {\bibinfo {author} {\bibnamefont {Mishchenko},
  \bibfnamefont {E.~G.}}, \bibinfo {author} {\bibfnamefont {M.~Y.}\
  \bibnamefont {Reizer}}, \ and\ \bibinfo {author} {\bibfnamefont {L.~I.}\
  \bibnamefont {Glazman}}} (\bibinfo {year} {2004}),\ \href {\doibase
  10.1103/PhysRevB.69.195302} {\bibfield  {journal} {\bibinfo  {journal} {Phys.
  Rev. B}\ }\textbf {\bibinfo {volume} {69}},\ \bibinfo {pages}
  {195302}}\BibitemShut {NoStop}%
\bibitem [{\citenamefont {Moldoveanu}\ and\ \citenamefont
  {Tanatar}(2009)}]{mol}%
  \BibitemOpen
  \bibfield  {author} {\bibinfo {author} {\bibnamefont {Moldoveanu},
  \bibfnamefont {V.}}, \ and\ \bibinfo {author} {\bibfnamefont
  {B.}~\bibnamefont {Tanatar}}} (\bibinfo {year} {2009}),\ \href@noop {}
  {\bibfield  {journal} {\bibinfo  {journal} {Europhys. Lett.}\ }\textbf
  {\bibinfo {volume} {86}},\ \bibinfo {pages} {67004}}\BibitemShut {NoStop}%
\bibitem [{\citenamefont {Morath}\ \emph {et~al.}(2009)\citenamefont {Morath},
  \citenamefont {Seamons}, \citenamefont {Reno},\ and\ \citenamefont
  {Lilly}}]{mor}%
  \BibitemOpen
  \bibfield  {author} {\bibinfo {author} {\bibnamefont {Morath}, \bibfnamefont
  {C.~P.}}, \bibinfo {author} {\bibfnamefont {J.~A.}\ \bibnamefont {Seamons}},
  \bibinfo {author} {\bibfnamefont {J.~L.}\ \bibnamefont {Reno}}, \ and\
  \bibinfo {author} {\bibfnamefont {M.~P.}\ \bibnamefont {Lilly}}} (\bibinfo
  {year} {2009}),\ \href@noop {} {\bibfield  {journal} {\bibinfo  {journal}
  {Phys. Rev. B}\ }\textbf {\bibinfo {volume} {79}},\ \bibinfo {pages}
  {041305(R)}}\BibitemShut {NoStop}%
\bibitem [{\citenamefont {Morimoto}\ \emph {et~al.}(2003)\citenamefont
  {Morimoto}, \citenamefont {Iwase}, \citenamefont {Aoki}, \citenamefont
  {Sasaki}, \citenamefont {Ochiai}, \citenamefont {Shailos}, \citenamefont
  {Bird}, \citenamefont {Lilly}, \citenamefont {Reno},\ and\ \citenamefont
  {Simmons}}]{mori}%
  \BibitemOpen
  \bibfield  {author} {\bibinfo {author} {\bibnamefont {Morimoto},
  \bibfnamefont {T.}}, \bibinfo {author} {\bibfnamefont {Y.}~\bibnamefont
  {Iwase}}, \bibinfo {author} {\bibfnamefont {N.}~\bibnamefont {Aoki}},
  \bibinfo {author} {\bibfnamefont {T.}~\bibnamefont {Sasaki}}, \bibinfo
  {author} {\bibfnamefont {Y.}~\bibnamefont {Ochiai}}, \bibinfo {author}
  {\bibfnamefont {A.}~\bibnamefont {Shailos}}, \bibinfo {author} {\bibfnamefont
  {J.~P.}\ \bibnamefont {Bird}}, \bibinfo {author} {\bibfnamefont {M.~P.}\
  \bibnamefont {Lilly}}, \bibinfo {author} {\bibfnamefont {J.~L.}\ \bibnamefont
  {Reno}}, \ and\ \bibinfo {author} {\bibfnamefont {J.~A.}\ \bibnamefont
  {Simmons}}} (\bibinfo {year} {2003}),\ \href {\doibase
  http://dx.doi.org/10.1063/1.1579851} {\bibfield  {journal} {\bibinfo
  {journal} {Applied Physics Letters}\ }\textbf {\bibinfo {volume}
  {82}}~(\bibinfo {number} {22}),\ \bibinfo {pages} {3952}}\BibitemShut
  {NoStop}%
\bibitem [{\citenamefont {Mortensen}\ \emph {et~al.}(2001)\citenamefont
  {Mortensen}, \citenamefont {Flensberg},\ and\ \citenamefont {Jauho}}]{mor1}%
  \BibitemOpen
  \bibfield  {author} {\bibinfo {author} {\bibnamefont {Mortensen},
  \bibfnamefont {N.~A.}}, \bibinfo {author} {\bibfnamefont {K.}~\bibnamefont
  {Flensberg}}, \ and\ \bibinfo {author} {\bibfnamefont {A.-P.}\ \bibnamefont
  {Jauho}}} (\bibinfo {year} {2001}),\ \href {\doibase
  10.1103/PhysRevLett.86.1841} {\bibfield  {journal} {\bibinfo  {journal}
  {Phys. Rev. Lett.}\ }\textbf {\bibinfo {volume} {86}},\ \bibinfo {pages}
  {1841}}\BibitemShut {NoStop}%
\bibitem [{\citenamefont {Mortensen}\ \emph
  {et~al.}(2002{\natexlab{a}})\citenamefont {Mortensen}, \citenamefont
  {Flensberg},\ and\ \citenamefont {Jauho}}]{mfj}%
  \BibitemOpen
  \bibfield  {author} {\bibinfo {author} {\bibnamefont {Mortensen},
  \bibfnamefont {N.~A.}}, \bibinfo {author} {\bibfnamefont {K.}~\bibnamefont
  {Flensberg}}, \ and\ \bibinfo {author} {\bibfnamefont {A.-P.}\ \bibnamefont
  {Jauho}}} (\bibinfo {year} {2002}{\natexlab{a}}),\ \href@noop {} {\bibfield
  {journal} {\bibinfo  {journal} {Phys. Rev. B}\ }\textbf {\bibinfo {volume}
  {65}},\ \bibinfo {pages} {085317}}\BibitemShut {NoStop}%
\bibitem [{\citenamefont {Mortensen}\ \emph
  {et~al.}(2002{\natexlab{b}})\citenamefont {Mortensen}, \citenamefont
  {Flensberg},\ and\ \citenamefont {Jauho}}]{mor2}%
  \BibitemOpen
  \bibfield  {author} {\bibinfo {author} {\bibnamefont {Mortensen},
  \bibfnamefont {N.~A.}}, \bibinfo {author} {\bibfnamefont {K.}~\bibnamefont
  {Flensberg}}, \ and\ \bibinfo {author} {\bibfnamefont {A.-P.}\ \bibnamefont
  {Jauho}}} (\bibinfo {year} {2002}{\natexlab{b}}),\ \href
  {http://stacks.iop.org/1402-4896/2002/i=T101/a=044} {\bibfield  {journal}
  {\bibinfo  {journal} {Physica Scripta}\ }\textbf {\bibinfo {volume}
  {2002}}~(\bibinfo {number} {T101}),\ \bibinfo {pages} {177}}\BibitemShut
  {NoStop}%
\bibitem [{\citenamefont {Moskalenko}(1962)}]{mosk}%
  \BibitemOpen
  \bibfield  {author} {\bibinfo {author} {\bibnamefont {Moskalenko},
  \bibfnamefont {S.~A.}}} (\bibinfo {year} {1962}),\ \href@noop {} {\bibfield
  {journal} {\bibinfo  {journal} {Fiz. Tverd. Tela.}\ }\textbf {\bibinfo
  {volume} {4}},\ \bibinfo {pages} {276}},\ \bibinfo {note} {[Sov. Phys. Solid
  State. {\bf 4}, 199 (1962)]}\BibitemShut {NoStop}%
\bibitem [{\citenamefont {Mo{\v{s}}ko}\ \emph {et~al.}(1992)\citenamefont
  {Mo{\v{s}}ko}, \citenamefont {Cambel},\ and\ \citenamefont
  {Mo{\v{s}}kov{\'a}}}]{mos}%
  \BibitemOpen
  \bibfield  {author} {\bibinfo {author} {\bibnamefont {Mo{\v{s}}ko},
  \bibfnamefont {M.}}, \bibinfo {author} {\bibfnamefont {V.}~\bibnamefont
  {Cambel}}, \ and\ \bibinfo {author} {\bibfnamefont {A.}~\bibnamefont
  {Mo{\v{s}}kov{\'a}}}} (\bibinfo {year} {1992}),\ \href@noop {} {\bibfield
  {journal} {\bibinfo  {journal} {Phys. Rev. B}\ }\textbf {\bibinfo {volume}
  {46}},\ \bibinfo {pages} {5012}}\BibitemShut {NoStop}%
\bibitem [{\citenamefont {Mott}\ and\ \citenamefont {Jones}(1936)}]{Mott}%
  \BibitemOpen
  \bibfield  {author} {\bibinfo {author} {\bibnamefont {Mott}, \bibfnamefont
  {N.~F.}}, \ and\ \bibinfo {author} {\bibfnamefont {H.}~\bibnamefont {Jones}}}
  (\bibinfo {year} {1936}),\ \href@noop {} {\emph {\bibinfo {title} {The Theory
  of the Properties of Metals and Alloys}}}\ (\bibinfo  {publisher} {Clarendon,
  Oxford})\BibitemShut {NoStop}%
\bibitem [{\citenamefont {M{\"u}ller}\ and\ \citenamefont
  {Sachdev}(2008)}]{mu1}%
  \BibitemOpen
  \bibfield  {author} {\bibinfo {author} {\bibnamefont {M{\"u}ller},
  \bibfnamefont {M.}}, \ and\ \bibinfo {author} {\bibfnamefont
  {S.}~\bibnamefont {Sachdev}}} (\bibinfo {year} {2008}),\ \href@noop {}
  {\bibfield  {journal} {\bibinfo  {journal} {Phys. Rev. B}\ }\textbf {\bibinfo
  {volume} {78}},\ \bibinfo {pages} {115419}}\BibitemShut {NoStop}%
\bibitem [{\citenamefont {M{\"u}ller}\ \emph {et~al.}(2009)\citenamefont
  {M{\"u}ller}, \citenamefont {Schmalian},\ and\ \citenamefont {Fritz}}]{kin1}%
  \BibitemOpen
  \bibfield  {author} {\bibinfo {author} {\bibnamefont {M{\"u}ller},
  \bibfnamefont {M.}}, \bibinfo {author} {\bibfnamefont {J.}~\bibnamefont
  {Schmalian}}, \ and\ \bibinfo {author} {\bibfnamefont {L.}~\bibnamefont
  {Fritz}}} (\bibinfo {year} {2009}),\ \href@noop {} {\bibfield  {journal}
  {\bibinfo  {journal} {Phys. Rev. Lett.}\ }\textbf {\bibinfo {volume} {103}},\
  \bibinfo {pages} {025301}}\BibitemShut {NoStop}%
\bibitem [{\citenamefont {Muraki}\ \emph {et~al.}(2004)\citenamefont {Muraki},
  \citenamefont {Lok}, \citenamefont {Kraus}, \citenamefont {Dietsche},
  \citenamefont {von Klitzing}, \citenamefont {Schuh}, \citenamefont
  {Bichler},\ and\ \citenamefont {W.Wegscheider}}]{mur}%
  \BibitemOpen
  \bibfield  {author} {\bibinfo {author} {\bibnamefont {Muraki}, \bibfnamefont
  {K.}}, \bibinfo {author} {\bibfnamefont {J.~G.~S.}\ \bibnamefont {Lok}},
  \bibinfo {author} {\bibfnamefont {S.}~\bibnamefont {Kraus}}, \bibinfo
  {author} {\bibfnamefont {W.}~\bibnamefont {Dietsche}}, \bibinfo {author}
  {\bibfnamefont {K.}~\bibnamefont {von Klitzing}}, \bibinfo {author}
  {\bibfnamefont {D.}~\bibnamefont {Schuh}}, \bibinfo {author} {\bibfnamefont
  {M.}~\bibnamefont {Bichler}}, \ and\ \bibinfo {author} {\bibnamefont
  {W.Wegscheider}}} (\bibinfo {year} {2004}),\ \href@noop {} {\bibfield
  {journal} {\bibinfo  {journal} {Phys. Rev. Lett.}\ }\textbf {\bibinfo
  {volume} {92}},\ \bibinfo {pages} {246801}}\BibitemShut {NoStop}%
\bibitem [{\citenamefont {Murphy}\ \emph {et~al.}(1994)\citenamefont {Murphy},
  \citenamefont {Eisenstein}, \citenamefont {Boebinger}, \citenamefont
  {Pfeiffer},\ and\ \citenamefont {West}}]{murph}%
  \BibitemOpen
  \bibfield  {author} {\bibinfo {author} {\bibnamefont {Murphy}, \bibfnamefont
  {S.~Q.}}, \bibinfo {author} {\bibfnamefont {J.~P.}\ \bibnamefont
  {Eisenstein}}, \bibinfo {author} {\bibfnamefont {G.~S.}\ \bibnamefont
  {Boebinger}}, \bibinfo {author} {\bibfnamefont {L.~N.}\ \bibnamefont
  {Pfeiffer}}, \ and\ \bibinfo {author} {\bibfnamefont {K.~W.}\ \bibnamefont
  {West}}} (\bibinfo {year} {1994}),\ \href {\doibase
  10.1103/PhysRevLett.72.728} {\bibfield  {journal} {\bibinfo  {journal} {Phys.
  Rev. Lett.}\ }\textbf {\bibinfo {volume} {72}},\ \bibinfo {pages}
  {728}}\BibitemShut {NoStop}%
\bibitem [{\citenamefont {Nandi}\ \emph {et~al.}(2012)\citenamefont {Nandi},
  \citenamefont {Finck}, \citenamefont {Eisenstein}, \citenamefont {Pfeiffer},\
  and\ \citenamefont {West}}]{nan}%
  \BibitemOpen
  \bibfield  {author} {\bibinfo {author} {\bibnamefont {Nandi}, \bibfnamefont
  {D.}}, \bibinfo {author} {\bibfnamefont {A.~D.~K.}\ \bibnamefont {Finck}},
  \bibinfo {author} {\bibfnamefont {J.~P.}\ \bibnamefont {Eisenstein}},
  \bibinfo {author} {\bibfnamefont {L.~N.}\ \bibnamefont {Pfeiffer}}, \ and\
  \bibinfo {author} {\bibfnamefont {K.~W.}\ \bibnamefont {West}}} (\bibinfo
  {year} {2012}),\ \href@noop {} {\bibfield  {journal} {\bibinfo  {journal}
  {Nature}\ }\textbf {\bibinfo {volume} {488}},\ \bibinfo {pages}
  {481}}\BibitemShut {NoStop}%
\bibitem [{\citenamefont {Narozhny}(2007)}]{me1}%
  \BibitemOpen
  \bibfield  {author} {\bibinfo {author} {\bibnamefont {Narozhny},
  \bibfnamefont {B.~N.}}} (\bibinfo {year} {2007}),\ \href@noop {} {\bibfield
  {journal} {\bibinfo  {journal} {Phys. Rev. B}\ }\textbf {\bibinfo {volume}
  {76}},\ \bibinfo {pages} {153409}}\BibitemShut {NoStop}%
\bibitem [{\citenamefont {Narozhny}\ and\ \citenamefont {Aleiner}(2000)}]{me0}%
  \BibitemOpen
  \bibfield  {author} {\bibinfo {author} {\bibnamefont {Narozhny},
  \bibfnamefont {B.~N.}}, \ and\ \bibinfo {author} {\bibfnamefont {I.~L.}\
  \bibnamefont {Aleiner}}} (\bibinfo {year} {2000}),\ \href@noop {} {\bibfield
  {journal} {\bibinfo  {journal} {Phys. Rev. Lett.}\ }\textbf {\bibinfo
  {volume} {84}},\ \bibinfo {pages} {5383}}\BibitemShut {NoStop}%
\bibitem [{\citenamefont {Narozhny}\ \emph {et~al.}(2001)\citenamefont
  {Narozhny}, \citenamefont {Aleiner},\ and\ \citenamefont {Stern}}]{me2}%
  \BibitemOpen
  \bibfield  {author} {\bibinfo {author} {\bibnamefont {Narozhny},
  \bibfnamefont {B.~N.}}, \bibinfo {author} {\bibfnamefont {I.~L.}\
  \bibnamefont {Aleiner}}, \ and\ \bibinfo {author} {\bibfnamefont
  {A.}~\bibnamefont {Stern}}} (\bibinfo {year} {2001}),\ \href@noop {}
  {\bibfield  {journal} {\bibinfo  {journal} {Phys. Rev. Lett.}\ }\textbf
  {\bibinfo {volume} {86}},\ \bibinfo {pages} {3610}}\BibitemShut {NoStop}%
\bibitem [{\citenamefont {Narozhny}\ \emph {et~al.}(2015)\citenamefont
  {Narozhny}, \citenamefont {Gornyi}, \citenamefont {Titov}, \citenamefont
  {Sch{\"u}tt},\ and\ \citenamefont {Mirlin}}]{mef}%
  \BibitemOpen
  \bibfield  {author} {\bibinfo {author} {\bibnamefont {Narozhny},
  \bibfnamefont {B.~N.}}, \bibinfo {author} {\bibfnamefont {I.~V.}\
  \bibnamefont {Gornyi}}, \bibinfo {author} {\bibfnamefont {M.}~\bibnamefont
  {Titov}}, \bibinfo {author} {\bibfnamefont {M.}~\bibnamefont {Sch{\"u}tt}}, \
  and\ \bibinfo {author} {\bibfnamefont {A.~D.}\ \bibnamefont {Mirlin}}}
  (\bibinfo {year} {2015}),\ \href {\doibase 10.1103/PhysRevB.91.035414}
  {\bibfield  {journal} {\bibinfo  {journal} {Phys. Rev. B}\ }\textbf {\bibinfo
  {volume} {91}},\ \bibinfo {pages} {035414}}\BibitemShut {NoStop}%
\bibitem [{\citenamefont {Narozhny}\ \emph {et~al.}(2012)\citenamefont
  {Narozhny}, \citenamefont {Titov}, \citenamefont {Gornyi},\ and\
  \citenamefont {Ostrovsky}}]{met}%
  \BibitemOpen
  \bibfield  {author} {\bibinfo {author} {\bibnamefont {Narozhny},
  \bibfnamefont {B.~N.}}, \bibinfo {author} {\bibfnamefont {M.}~\bibnamefont
  {Titov}}, \bibinfo {author} {\bibfnamefont {I.~V.}\ \bibnamefont {Gornyi}}, \
  and\ \bibinfo {author} {\bibfnamefont {P.~M.}\ \bibnamefont {Ostrovsky}}}
  (\bibinfo {year} {2012}),\ \href@noop {} {\bibfield  {journal} {\bibinfo
  {journal} {Phys. Rev. B}\ }\textbf {\bibinfo {volume} {85}},\ \bibinfo
  {pages} {195421}}\BibitemShut {NoStop}%
\bibitem [{\citenamefont {Narozhny}\ \emph {et~al.}(2002)\citenamefont
  {Narozhny}, \citenamefont {Zala},\ and\ \citenamefont {Aleiner}}]{zn2}%
  \BibitemOpen
  \bibfield  {author} {\bibinfo {author} {\bibnamefont {Narozhny},
  \bibfnamefont {B.~N.}}, \bibinfo {author} {\bibfnamefont {G.}~\bibnamefont
  {Zala}}, \ and\ \bibinfo {author} {\bibfnamefont {I.~L.}\ \bibnamefont
  {Aleiner}}} (\bibinfo {year} {2002}),\ \href@noop {} {\bibfield  {journal}
  {\bibinfo  {journal} {Phys. Rev. B}\ }\textbf {\bibinfo {volume} {65}},\
  \bibinfo {pages} {180202}}\BibitemShut {NoStop}%
\bibitem [{\citenamefont {Nazarov}\ and\ \citenamefont {Averin}(1998)}]{naz}%
  \BibitemOpen
  \bibfield  {author} {\bibinfo {author} {\bibnamefont {Nazarov}, \bibfnamefont
  {Y.~V.}}, \ and\ \bibinfo {author} {\bibfnamefont {D.~V.}\ \bibnamefont
  {Averin}}} (\bibinfo {year} {1998}),\ \href@noop {} {\bibfield  {journal}
  {\bibinfo  {journal} {Phys. Rev. Lett.}\ }\textbf {\bibinfo {volume} {81}},\
  \bibinfo {pages} {653}}\BibitemShut {NoStop}%
\bibitem [{\citenamefont {Noh}\ \emph {et~al.}(1998)\citenamefont {Noh},
  \citenamefont {Zelakiewicz}, \citenamefont {Feng}, \citenamefont {Gramila},
  \citenamefont {Pfeiffer},\ and\ \citenamefont {West}}]{no1}%
  \BibitemOpen
  \bibfield  {author} {\bibinfo {author} {\bibnamefont {Noh}, \bibfnamefont
  {H.}}, \bibinfo {author} {\bibfnamefont {S.}~\bibnamefont {Zelakiewicz}},
  \bibinfo {author} {\bibfnamefont {X.~G.}\ \bibnamefont {Feng}}, \bibinfo
  {author} {\bibfnamefont {T.~J.}\ \bibnamefont {Gramila}}, \bibinfo {author}
  {\bibfnamefont {L.~N.}\ \bibnamefont {Pfeiffer}}, \ and\ \bibinfo {author}
  {\bibfnamefont {K.~W.}\ \bibnamefont {West}}} (\bibinfo {year} {1998}),\
  \href@noop {} {\bibfield  {journal} {\bibinfo  {journal} {Phys. Rev. B}\
  }\textbf {\bibinfo {volume} {58}},\ \bibinfo {pages} {12621}}\BibitemShut
  {NoStop}%
\bibitem [{\citenamefont {Noh}\ \emph {et~al.}(1999)\citenamefont {Noh},
  \citenamefont {Zelakiewicz}, \citenamefont {Gramila}, \citenamefont
  {Pfeiffer},\ and\ \citenamefont {West}}]{noh}%
  \BibitemOpen
  \bibfield  {author} {\bibinfo {author} {\bibnamefont {Noh}, \bibfnamefont
  {H.}}, \bibinfo {author} {\bibfnamefont {S.}~\bibnamefont {Zelakiewicz}},
  \bibinfo {author} {\bibfnamefont {T.~J.}\ \bibnamefont {Gramila}}, \bibinfo
  {author} {\bibfnamefont {L.~N.}\ \bibnamefont {Pfeiffer}}, \ and\ \bibinfo
  {author} {\bibfnamefont {K.~W.}\ \bibnamefont {West}}} (\bibinfo {year}
  {1999}),\ \href@noop {} {\bibfield  {journal} {\bibinfo  {journal} {Phys.
  Rev. B}\ }\textbf {\bibinfo {volume} {59}},\ \bibinfo {pages}
  {13114}}\BibitemShut {NoStop}%
\bibitem [{\citenamefont {Nomura}\ and\ \citenamefont
  {MacDonald}(2006)}]{nom1}%
  \BibitemOpen
  \bibfield  {author} {\bibinfo {author} {\bibnamefont {Nomura}, \bibfnamefont
  {K.}}, \ and\ \bibinfo {author} {\bibfnamefont {A.~H.}\ \bibnamefont
  {MacDonald}}} (\bibinfo {year} {2006}),\ \href@noop {} {\bibfield  {journal}
  {\bibinfo  {journal} {Phys. Rev. Lett.}\ }\textbf {\bibinfo {volume} {96}},\
  \bibinfo {pages} {256602}}\BibitemShut {NoStop}%
\bibitem [{\citenamefont {Nomura}\ and\ \citenamefont
  {MacDonald}(2007)}]{nom2}%
  \BibitemOpen
  \bibfield  {author} {\bibinfo {author} {\bibnamefont {Nomura}, \bibfnamefont
  {K.}}, \ and\ \bibinfo {author} {\bibfnamefont {A.~H.}\ \bibnamefont
  {MacDonald}}} (\bibinfo {year} {2007}),\ \href@noop {} {\bibfield  {journal}
  {\bibinfo  {journal} {Phys. Rev. Lett.}\ }\textbf {\bibinfo {volume} {98}},\
  \bibinfo {pages} {076602}}\BibitemShut {NoStop}%
\bibitem [{\citenamefont {Onac}\ \emph {et~al.}(2006)\citenamefont {Onac},
  \citenamefont {Balestro}, \citenamefont {van Beveren}, \citenamefont
  {Hartmann}, \citenamefont {Nazarov},\ and\ \citenamefont
  {Kouwenhoven}}]{kou2}%
  \BibitemOpen
  \bibfield  {author} {\bibinfo {author} {\bibnamefont {Onac}, \bibfnamefont
  {E.}}, \bibinfo {author} {\bibfnamefont {F.}~\bibnamefont {Balestro}},
  \bibinfo {author} {\bibfnamefont {L.~H.~W.}\ \bibnamefont {van Beveren}},
  \bibinfo {author} {\bibfnamefont {U.}~\bibnamefont {Hartmann}}, \bibinfo
  {author} {\bibfnamefont {Y.~V.}\ \bibnamefont {Nazarov}}, \ and\ \bibinfo
  {author} {\bibfnamefont {L.~P.}\ \bibnamefont {Kouwenhoven}}} (\bibinfo
  {year} {2006}),\ \href@noop {} {\bibfield  {journal} {\bibinfo  {journal}
  {Phys. Rev. Lett.}\ }\textbf {\bibinfo {volume} {96}},\ \bibinfo {pages}
  {176601}}\BibitemShut {NoStop}%
\bibitem [{\citenamefont {von Oppen}\ \emph {et~al.}(2001)\citenamefont {von
  Oppen}, \citenamefont {Simon},\ and\ \citenamefont {Stern}}]{opp}%
  \BibitemOpen
  \bibfield  {author} {\bibinfo {author} {\bibnamefont {von Oppen},
  \bibfnamefont {F.}}, \bibinfo {author} {\bibfnamefont {S.~H.}\ \bibnamefont
  {Simon}}, \ and\ \bibinfo {author} {\bibfnamefont {A.}~\bibnamefont {Stern}}}
  (\bibinfo {year} {2001}),\ \href@noop {} {\bibfield  {journal} {\bibinfo
  {journal} {Phys. Rev. Lett.}\ }\textbf {\bibinfo {volume} {87}},\ \bibinfo
  {pages} {106803}}\BibitemShut {NoStop}%
\bibitem [{\citenamefont {Oreg}\ and\ \citenamefont {Halperin}(1999)}]{ore}%
  \BibitemOpen
  \bibfield  {author} {\bibinfo {author} {\bibnamefont {Oreg}, \bibfnamefont
  {Y.}}, \ and\ \bibinfo {author} {\bibfnamefont {B.~I.}\ \bibnamefont
  {Halperin}}} (\bibinfo {year} {1999}),\ \href@noop {} {\bibfield  {journal}
  {\bibinfo  {journal} {Phys. Rev. B}\ }\textbf {\bibinfo {volume} {60}},\
  \bibinfo {pages} {5679}}\BibitemShut {NoStop}%
\bibitem [{\citenamefont {Oreg}\ and\ \citenamefont {Kamenev}(1998)}]{or2}%
  \BibitemOpen
  \bibfield  {author} {\bibinfo {author} {\bibnamefont {Oreg}, \bibfnamefont
  {Y.}}, \ and\ \bibinfo {author} {\bibfnamefont {A.}~\bibnamefont {Kamenev}}}
  (\bibinfo {year} {1998}),\ \href@noop {} {\bibfield  {journal} {\bibinfo
  {journal} {Phys. Rev. Lett.}\ }\textbf {\bibinfo {volume} {80}},\ \bibinfo
  {pages} {2421}}\BibitemShut {NoStop}%
\bibitem [{\citenamefont {Orgad}\ and\ \citenamefont {Levit}(1996)}]{org}%
  \BibitemOpen
  \bibfield  {author} {\bibinfo {author} {\bibnamefont {Orgad}, \bibfnamefont
  {D.}}, \ and\ \bibinfo {author} {\bibfnamefont {S.}~\bibnamefont {Levit}}}
  (\bibinfo {year} {1996}),\ \href@noop {} {\bibfield  {journal} {\bibinfo
  {journal} {Phys. Rev. B}\ }\textbf {\bibinfo {volume} {53}},\ \bibinfo
  {pages} {7964}}\BibitemShut {NoStop}%
\bibitem [{\citenamefont {Ostrovsky}\ \emph {et~al.}(2006)\citenamefont
  {Ostrovsky}, \citenamefont {Gornyi},\ and\ \citenamefont {Mirlin}}]{ost}%
  \BibitemOpen
  \bibfield  {author} {\bibinfo {author} {\bibnamefont {Ostrovsky},
  \bibfnamefont {P.~M.}}, \bibinfo {author} {\bibfnamefont {I.~V.}\
  \bibnamefont {Gornyi}}, \ and\ \bibinfo {author} {\bibfnamefont {A.~D.}\
  \bibnamefont {Mirlin}}} (\bibinfo {year} {2006}),\ \href@noop {} {\bibfield
  {journal} {\bibinfo  {journal} {Phys. Rev. B}\ }\textbf {\bibinfo {volume}
  {74}},\ \bibinfo {pages} {235443}}\BibitemShut {NoStop}%
\bibitem [{\citenamefont {Ostrovsky}\ \emph {et~al.}(2007)\citenamefont
  {Ostrovsky}, \citenamefont {Gornyi},\ and\ \citenamefont {Mirlin}}]{mir}%
  \BibitemOpen
  \bibfield  {author} {\bibinfo {author} {\bibnamefont {Ostrovsky},
  \bibfnamefont {P.~M.}}, \bibinfo {author} {\bibfnamefont {I.~V.}\
  \bibnamefont {Gornyi}}, \ and\ \bibinfo {author} {\bibfnamefont {A.~D.}\
  \bibnamefont {Mirlin}}} (\bibinfo {year} {2007}),\ \href@noop {} {\bibfield
  {journal} {\bibinfo  {journal} {Eur. Phys. J. Special Topics}\ }\textbf
  {\bibinfo {volume} {148}},\ \bibinfo {pages} {63}}\BibitemShut {NoStop}%
\bibitem [{\citenamefont {Park}\ and\ \citenamefont {Das~Sarma}(2006)}]{park}%
  \BibitemOpen
  \bibfield  {author} {\bibinfo {author} {\bibnamefont {Park}, \bibfnamefont
  {K.}}, \ and\ \bibinfo {author} {\bibfnamefont {S.}~\bibnamefont
  {Das~Sarma}}} (\bibinfo {year} {2006}),\ \href {\doibase
  10.1103/PhysRevB.74.035338} {\bibfield  {journal} {\bibinfo  {journal} {Phys.
  Rev. B}\ }\textbf {\bibinfo {volume} {74}},\ \bibinfo {pages}
  {035338}}\BibitemShut {NoStop}%
\bibitem [{\citenamefont {Patel}\ \emph {et~al.}(1997)\citenamefont {Patel},
  \citenamefont {Linfield}, \citenamefont {Brown}, \citenamefont {Pepper},
  \citenamefont {Ritchie},\ and\ \citenamefont {Jones}}]{pat}%
  \BibitemOpen
  \bibfield  {author} {\bibinfo {author} {\bibnamefont {Patel}, \bibfnamefont
  {N.~K.}}, \bibinfo {author} {\bibfnamefont {E.~H.}\ \bibnamefont {Linfield}},
  \bibinfo {author} {\bibfnamefont {K.~M.}\ \bibnamefont {Brown}}, \bibinfo
  {author} {\bibfnamefont {M.}~\bibnamefont {Pepper}}, \bibinfo {author}
  {\bibfnamefont {D.~A.}\ \bibnamefont {Ritchie}}, \ and\ \bibinfo {author}
  {\bibfnamefont {G.~A.~C.}\ \bibnamefont {Jones}}} (\bibinfo {year} {1997}),\
  \href@noop {} {\bibfield  {journal} {\bibinfo  {journal} {Semicond. Sci.
  Technol.}\ }\textbf {\bibinfo {volume} {12}},\ \bibinfo {pages}
  {309}}\BibitemShut {NoStop}%
\bibitem [{\citenamefont {Pereira}\ and\ \citenamefont {Sela}(2010)}]{sela}%
  \BibitemOpen
  \bibfield  {author} {\bibinfo {author} {\bibnamefont {Pereira}, \bibfnamefont
  {R.~G.}}, \ and\ \bibinfo {author} {\bibfnamefont {E.}~\bibnamefont {Sela}}}
  (\bibinfo {year} {2010}),\ \href {\doibase 10.1103/PhysRevB.82.115324}
  {\bibfield  {journal} {\bibinfo  {journal} {Phys. Rev. B}\ }\textbf {\bibinfo
  {volume} {82}},\ \bibinfo {pages} {115324}}\BibitemShut {NoStop}%
\bibitem [{\citenamefont {Peres}\ \emph {et~al.}(2011)\citenamefont {Peres},
  \citenamefont {dos Santos},\ and\ \citenamefont {Neto}}]{net}%
  \BibitemOpen
  \bibfield  {author} {\bibinfo {author} {\bibnamefont {Peres}, \bibfnamefont
  {N.~M.~R.}}, \bibinfo {author} {\bibfnamefont {J.~M. B.~L.}\ \bibnamefont
  {dos Santos}}, \ and\ \bibinfo {author} {\bibfnamefont {A.~H.~C.}\
  \bibnamefont {Neto}}} (\bibinfo {year} {2011}),\ \href@noop {} {\bibfield
  {journal} {\bibinfo  {journal} {Europhys. Lett.}\ }\textbf {\bibinfo {volume}
  {95}},\ \bibinfo {pages} {18001}}\BibitemShut {NoStop}%
\bibitem [{\citenamefont {Pesin}\ and\ \citenamefont
  {MacDonald}(2011)}]{pesin}%
  \BibitemOpen
  \bibfield  {author} {\bibinfo {author} {\bibnamefont {Pesin}, \bibfnamefont
  {D.~A.}}, \ and\ \bibinfo {author} {\bibfnamefont {A.~H.}\ \bibnamefont
  {MacDonald}}} (\bibinfo {year} {2011}),\ \href {\doibase
  10.1103/PhysRevB.84.075308} {\bibfield  {journal} {\bibinfo  {journal} {Phys.
  Rev. B}\ }\textbf {\bibinfo {volume} {84}},\ \bibinfo {pages}
  {075308}}\BibitemShut {NoStop}%
\bibitem [{\citenamefont {Pikalov}\ and\ \citenamefont {Fil}(2012)}]{fil3}%
  \BibitemOpen
  \bibfield  {author} {\bibinfo {author} {\bibnamefont {Pikalov}, \bibfnamefont
  {A.~A.}}, \ and\ \bibinfo {author} {\bibfnamefont {D.~V.}\ \bibnamefont
  {Fil}}} (\bibinfo {year} {2012}),\ \href {\doibase 10.1186/1556-276X-7-145}
  {\bibfield  {journal} {\bibinfo  {journal} {Nanoscale Research Lett.}\
  }\textbf {\bibinfo {volume} {7}},\ \bibinfo {pages} {145}}\BibitemShut
  {NoStop}%
\bibitem [{\citenamefont {Pillarisetty}\ \emph {et~al.}(2002)\citenamefont
  {Pillarisetty}, \citenamefont {Noh}, \citenamefont {Tsui}, \citenamefont
  {Poortere}, \citenamefont {Tutuc},\ and\ \citenamefont {Shayegan}}]{pi2}%
  \BibitemOpen
  \bibfield  {author} {\bibinfo {author} {\bibnamefont {Pillarisetty},
  \bibfnamefont {R.}}, \bibinfo {author} {\bibfnamefont {H.}~\bibnamefont
  {Noh}}, \bibinfo {author} {\bibfnamefont {D.~C.}\ \bibnamefont {Tsui}},
  \bibinfo {author} {\bibfnamefont {E.~P.~D.}\ \bibnamefont {Poortere}},
  \bibinfo {author} {\bibfnamefont {E.}~\bibnamefont {Tutuc}}, \ and\ \bibinfo
  {author} {\bibfnamefont {M.}~\bibnamefont {Shayegan}}} (\bibinfo {year}
  {2002}),\ \href@noop {} {\bibfield  {journal} {\bibinfo  {journal} {Phys.
  Rev. Lett.}\ }\textbf {\bibinfo {volume} {89}},\ \bibinfo {pages}
  {016805}}\BibitemShut {NoStop}%
\bibitem [{\citenamefont {Pillarisetty}\ \emph {et~al.}(2003)\citenamefont
  {Pillarisetty}, \citenamefont {Noh}, \citenamefont {Tsui}, \citenamefont
  {Poortere}, \citenamefont {Tutuc},\ and\ \citenamefont {Shayegan}}]{pi22}%
  \BibitemOpen
  \bibfield  {author} {\bibinfo {author} {\bibnamefont {Pillarisetty},
  \bibfnamefont {R.}}, \bibinfo {author} {\bibfnamefont {H.}~\bibnamefont
  {Noh}}, \bibinfo {author} {\bibfnamefont {D.~C.}\ \bibnamefont {Tsui}},
  \bibinfo {author} {\bibfnamefont {E.~P.~D.}\ \bibnamefont {Poortere}},
  \bibinfo {author} {\bibfnamefont {E.}~\bibnamefont {Tutuc}}, \ and\ \bibinfo
  {author} {\bibfnamefont {M.}~\bibnamefont {Shayegan}}} (\bibinfo {year}
  {2003}),\ \href@noop {} {\bibfield  {journal} {\bibinfo  {journal} {Phys.
  Rev. Lett.}\ }\textbf {\bibinfo {volume} {90}},\ \bibinfo {pages}
  {226801}}\BibitemShut {NoStop}%
\bibitem [{\citenamefont {Pillarisetty}\ \emph
  {et~al.}(2005{\natexlab{a}})\citenamefont {Pillarisetty}, \citenamefont
  {Noh}, \citenamefont {Tsui}, \citenamefont {Poortere}, \citenamefont
  {Tutuc},\ and\ \citenamefont {Shayegan}}]{pi23}%
  \BibitemOpen
  \bibfield  {author} {\bibinfo {author} {\bibnamefont {Pillarisetty},
  \bibfnamefont {R.}}, \bibinfo {author} {\bibfnamefont {H.}~\bibnamefont
  {Noh}}, \bibinfo {author} {\bibfnamefont {D.~C.}\ \bibnamefont {Tsui}},
  \bibinfo {author} {\bibfnamefont {E.~P.~D.}\ \bibnamefont {Poortere}},
  \bibinfo {author} {\bibfnamefont {E.}~\bibnamefont {Tutuc}}, \ and\ \bibinfo
  {author} {\bibfnamefont {M.}~\bibnamefont {Shayegan}}} (\bibinfo {year}
  {2005}{\natexlab{a}}),\ \href@noop {} {\bibfield  {journal} {\bibinfo
  {journal} {Phys. Rev. Lett.}\ }\textbf {\bibinfo {volume} {94}},\ \bibinfo
  {pages} {016807}}\BibitemShut {NoStop}%
\bibitem [{\citenamefont {Pillarisetty}\ \emph {et~al.}(2004)\citenamefont
  {Pillarisetty}, \citenamefont {Noh}, \citenamefont {Tutuc}, \citenamefont
  {Poortere}, \citenamefont {Tsui},\ and\ \citenamefont {Shayegan}}]{pi21}%
  \BibitemOpen
  \bibfield  {author} {\bibinfo {author} {\bibnamefont {Pillarisetty},
  \bibfnamefont {R.}}, \bibinfo {author} {\bibfnamefont {H.}~\bibnamefont
  {Noh}}, \bibinfo {author} {\bibfnamefont {E.}~\bibnamefont {Tutuc}}, \bibinfo
  {author} {\bibfnamefont {E.~D.}\ \bibnamefont {Poortere}}, \bibinfo {author}
  {\bibfnamefont {D.}~\bibnamefont {Tsui}}, \ and\ \bibinfo {author}
  {\bibfnamefont {M.}~\bibnamefont {Shayegan}}} (\bibinfo {year} {2004}),\
  \href@noop {} {\bibfield  {journal} {\bibinfo  {journal} {Physica E}\
  }\textbf {\bibinfo {volume} {22}},\ \bibinfo {pages} {300}}\BibitemShut
  {NoStop}%
\bibitem [{\citenamefont {Pillarisetty}\ \emph
  {et~al.}(2005{\natexlab{b}})\citenamefont {Pillarisetty}, \citenamefont
  {Noh}, \citenamefont {Tutuc}, \citenamefont {Poortere}, \citenamefont {Lai},
  \citenamefont {Tsui},\ and\ \citenamefont {Shayegan}}]{pil}%
  \BibitemOpen
  \bibfield  {author} {\bibinfo {author} {\bibnamefont {Pillarisetty},
  \bibfnamefont {R.}}, \bibinfo {author} {\bibfnamefont {H.}~\bibnamefont
  {Noh}}, \bibinfo {author} {\bibfnamefont {E.}~\bibnamefont {Tutuc}}, \bibinfo
  {author} {\bibfnamefont {E.~P.~D.}\ \bibnamefont {Poortere}}, \bibinfo
  {author} {\bibfnamefont {K.}~\bibnamefont {Lai}}, \bibinfo {author}
  {\bibfnamefont {D.}~\bibnamefont {Tsui}}, \ and\ \bibinfo {author}
  {\bibfnamefont {M.}~\bibnamefont {Shayegan}}} (\bibinfo {year}
  {2005}{\natexlab{b}}),\ \href@noop {} {\bibfield  {journal} {\bibinfo
  {journal} {Phys. Rev. B}\ }\textbf {\bibinfo {volume} {71}},\ \bibinfo
  {pages} {115307}}\BibitemShut {NoStop}%
\bibitem [{\citenamefont {Pogrebinskii}(1977)}]{pog}%
  \BibitemOpen
  \bibfield  {author} {\bibinfo {author} {\bibnamefont {Pogrebinskii},
  \bibfnamefont {M.~B.}}} (\bibinfo {year} {1977}),\ \href@noop {} {\bibfield
  {journal} {\bibinfo  {journal} {Fiz. Tekh. Poluprovodn.}\ }\textbf {\bibinfo
  {volume} {11}},\ \bibinfo {pages} {637}},\ \bibinfo {note} {[Sov. Phys.
  Semicond. {\bf 11}, 372 (1977)]}\BibitemShut {NoStop}%
\bibitem [{\citenamefont {Ponomarenko}\ \emph {et~al.}(2011)\citenamefont
  {Ponomarenko}, \citenamefont {Geim}, \citenamefont {Zhukov}, \citenamefont
  {Jalil}, \citenamefont {Morozov}, \citenamefont {Novoselov}, \citenamefont
  {Grigorieva}, \citenamefont {Hill}, \citenamefont {Cheianov}, \citenamefont
  {Fal'ko}, \citenamefont {Watanabe}, \citenamefont {Taniguchi},\ and\
  \citenamefont {Gorbachev}}]{ge1}%
  \BibitemOpen
  \bibfield  {author} {\bibinfo {author} {\bibnamefont {Ponomarenko},
  \bibfnamefont {L.}}, \bibinfo {author} {\bibfnamefont {A.~K.}\ \bibnamefont
  {Geim}}, \bibinfo {author} {\bibfnamefont {A.~A.}\ \bibnamefont {Zhukov}},
  \bibinfo {author} {\bibfnamefont {R.}~\bibnamefont {Jalil}}, \bibinfo
  {author} {\bibfnamefont {S.~V.}\ \bibnamefont {Morozov}}, \bibinfo {author}
  {\bibfnamefont {K.~S.}\ \bibnamefont {Novoselov}}, \bibinfo {author}
  {\bibfnamefont {I.~V.}\ \bibnamefont {Grigorieva}}, \bibinfo {author}
  {\bibfnamefont {E.~H.}\ \bibnamefont {Hill}}, \bibinfo {author}
  {\bibfnamefont {V.~V.}\ \bibnamefont {Cheianov}}, \bibinfo {author}
  {\bibfnamefont {V.~I.}\ \bibnamefont {Fal'ko}}, \bibinfo {author}
  {\bibfnamefont {K.}~\bibnamefont {Watanabe}}, \bibinfo {author}
  {\bibfnamefont {T.}~\bibnamefont {Taniguchi}}, \ and\ \bibinfo {author}
  {\bibfnamefont {R.~V.}\ \bibnamefont {Gorbachev}}} (\bibinfo {year} {2011}),\
  \href@noop {} {\bibfield  {journal} {\bibinfo  {journal} {Nature Physics}\
  }\textbf {\bibinfo {volume} {7}},\ \bibinfo {pages} {958}}\BibitemShut
  {NoStop}%
\bibitem [{\citenamefont {Ponomarenko}(2013)}]{priv1}%
  \BibitemOpen
  \bibfield  {author} {\bibinfo {author} {\bibnamefont {Ponomarenko},
  \bibfnamefont {L.~A.}}} (\bibinfo {year} {2013}),\ \href@noop {} {}\bibinfo
  {howpublished} {private communication}\BibitemShut {NoStop}%
\bibitem [{\citenamefont {Ponomarenko}\ and\ \citenamefont
  {Averin}(2000)}]{poa}%
  \BibitemOpen
  \bibfield  {author} {\bibinfo {author} {\bibnamefont {Ponomarenko},
  \bibfnamefont {V.~V.}}, \ and\ \bibinfo {author} {\bibfnamefont {D.~V.}\
  \bibnamefont {Averin}}} (\bibinfo {year} {2000}),\ \href@noop {} {\bibfield
  {journal} {\bibinfo  {journal} {Phys. Rev. Lett.}\ }\textbf {\bibinfo
  {volume} {85}},\ \bibinfo {pages} {4928}}\BibitemShut {NoStop}%
\bibitem [{\citenamefont {Price}\ \emph {et~al.}(2008)\citenamefont {Price},
  \citenamefont {Savchenko}, \citenamefont {Allison},\ and\ \citenamefont
  {Ritchie}}]{sav}%
  \BibitemOpen
  \bibfield  {author} {\bibinfo {author} {\bibnamefont {Price}, \bibfnamefont
  {A.~S.}}, \bibinfo {author} {\bibfnamefont {A.~K.}\ \bibnamefont
  {Savchenko}}, \bibinfo {author} {\bibfnamefont {G.}~\bibnamefont {Allison}},
  \ and\ \bibinfo {author} {\bibfnamefont {D.~A.}\ \bibnamefont {Ritchie}}}
  (\bibinfo {year} {2008}),\ \href@noop {} {\bibfield  {journal} {\bibinfo
  {journal} {Physica E}\ }\textbf {\bibinfo {volume} {40}},\ \bibinfo {pages}
  {961}}\BibitemShut {NoStop}%
\bibitem [{\citenamefont {Price}\ \emph {et~al.}(2007)\citenamefont {Price},
  \citenamefont {Savchenko}, \citenamefont {Narozhny}, \citenamefont
  {Allison},\ and\ \citenamefont {Ritchie}}]{mes}%
  \BibitemOpen
  \bibfield  {author} {\bibinfo {author} {\bibnamefont {Price}, \bibfnamefont
  {A.~S.}}, \bibinfo {author} {\bibfnamefont {A.~K.}\ \bibnamefont
  {Savchenko}}, \bibinfo {author} {\bibfnamefont {B.~N.}\ \bibnamefont
  {Narozhny}}, \bibinfo {author} {\bibfnamefont {G.}~\bibnamefont {Allison}}, \
  and\ \bibinfo {author} {\bibfnamefont {D.~A.}\ \bibnamefont {Ritchie}}}
  (\bibinfo {year} {2007}),\ \href@noop {} {\bibfield  {journal} {\bibinfo
  {journal} {Science}\ }\textbf {\bibinfo {volume} {316}},\ \bibinfo {pages}
  {99}}\BibitemShut {NoStop}%
\bibitem [{\citenamefont {Price}\ \emph {et~al.}(2010)\citenamefont {Price},
  \citenamefont {Savchenko},\ and\ \citenamefont {Ritchie}}]{prs}%
  \BibitemOpen
  \bibfield  {author} {\bibinfo {author} {\bibnamefont {Price}, \bibfnamefont
  {A.~S.}}, \bibinfo {author} {\bibfnamefont {A.~K.}\ \bibnamefont
  {Savchenko}}, \ and\ \bibinfo {author} {\bibfnamefont {D.~A.}\ \bibnamefont
  {Ritchie}}} (\bibinfo {year} {2010}),\ \href@noop {} {\bibfield  {journal}
  {\bibinfo  {journal} {Phys. Rev. B}\ }\textbf {\bibinfo {volume} {81}},\
  \bibinfo {pages} {193303}}\BibitemShut {NoStop}%
\bibitem [{\citenamefont {Price}(1983)}]{pri}%
  \BibitemOpen
  \bibfield  {author} {\bibinfo {author} {\bibnamefont {Price}, \bibfnamefont
  {P.~J.}}} (\bibinfo {year} {1983}),\ \href@noop {} {\bibfield  {journal}
  {\bibinfo  {journal} {Physica (Amsterdam)}\ }\textbf {\bibinfo {volume}
  {117B}},\ \bibinfo {pages} {750}}\BibitemShut {NoStop}%
\bibitem [{\citenamefont {Price}(1984)}]{price}%
  \BibitemOpen
  \bibfield  {author} {\bibinfo {author} {\bibnamefont {Price}, \bibfnamefont
  {P.~J.}}} (\bibinfo {year} {1984}),\ \href@noop {} {\bibfield  {journal}
  {\bibinfo  {journal} {Solid State Communications}\ }\textbf {\bibinfo
  {volume} {51}}~(\bibinfo {number} {8}),\ \bibinfo {pages} {607}}\BibitemShut
  {NoStop}%
\bibitem [{\citenamefont {Price}(1988)}]{pr1}%
  \BibitemOpen
  \bibfield  {author} {\bibinfo {author} {\bibnamefont {Price}, \bibfnamefont
  {P.~J.}}} (\bibinfo {year} {1988}),\ in\ \href@noop {} {\emph {\bibinfo
  {booktitle} {The Physics of Submicron Semiconductor Devices}}},\ \bibinfo
  {editor} {edited by\ \bibinfo {editor} {\bibfnamefont {H.}~\bibnamefont
  {Grubin}}, \bibinfo {editor} {\bibfnamefont {D.~K.}\ \bibnamefont {Ferry}}, \
  and\ \bibinfo {editor} {\bibfnamefont {C.}~\bibnamefont {Jacoboni}}}\
  (\bibinfo  {publisher} {Plenum, New York})\BibitemShut {NoStop}%
\bibitem [{\citenamefont {Principi}\ \emph {et~al.}(2012)\citenamefont
  {Principi}, \citenamefont {Carrega}, \citenamefont {Asgari}, \citenamefont
  {Pellegrini},\ and\ \citenamefont {Polini}}]{ppp}%
  \BibitemOpen
  \bibfield  {author} {\bibinfo {author} {\bibnamefont {Principi},
  \bibfnamefont {A.}}, \bibinfo {author} {\bibfnamefont {M.}~\bibnamefont
  {Carrega}}, \bibinfo {author} {\bibfnamefont {R.}~\bibnamefont {Asgari}},
  \bibinfo {author} {\bibfnamefont {V.}~\bibnamefont {Pellegrini}}, \ and\
  \bibinfo {author} {\bibfnamefont {M.}~\bibnamefont {Polini}}} (\bibinfo
  {year} {2012}),\ \href@noop {} {\bibfield  {journal} {\bibinfo  {journal}
  {Phys. Rev. B}\ }\textbf {\bibinfo {volume} {86}},\ \bibinfo {pages}
  {085421}}\BibitemShut {NoStop}%
\bibitem [{\citenamefont {Profumo}\ \emph {et~al.}(2012)\citenamefont
  {Profumo}, \citenamefont {Asgari}, \citenamefont {Polini},\ and\
  \citenamefont {MacDonald}}]{plas2}%
  \BibitemOpen
  \bibfield  {author} {\bibinfo {author} {\bibnamefont {Profumo}, \bibfnamefont
  {R.~E.~V.}}, \bibinfo {author} {\bibfnamefont {R.}~\bibnamefont {Asgari}},
  \bibinfo {author} {\bibfnamefont {M.}~\bibnamefont {Polini}}, \ and\ \bibinfo
  {author} {\bibfnamefont {A.~H.}\ \bibnamefont {MacDonald}}} (\bibinfo {year}
  {2012}),\ \href@noop {} {\bibfield  {journal} {\bibinfo  {journal} {Phys.
  Rev. B}\ }\textbf {\bibinfo {volume} {85}},\ \bibinfo {pages}
  {085443}}\BibitemShut {NoStop}%
\bibitem [{\citenamefont {Prunnila}\ \emph {et~al.}(2008)\citenamefont
  {Prunnila}, \citenamefont {Laakso}, \citenamefont {Kivioja},\ and\
  \citenamefont {Ahopelto}}]{pru}%
  \BibitemOpen
  \bibfield  {author} {\bibinfo {author} {\bibnamefont {Prunnila},
  \bibfnamefont {M.}}, \bibinfo {author} {\bibfnamefont {S.~J.}\ \bibnamefont
  {Laakso}}, \bibinfo {author} {\bibfnamefont {J.~M.}\ \bibnamefont {Kivioja}},
  \ and\ \bibinfo {author} {\bibfnamefont {J.}~\bibnamefont {Ahopelto}}}
  (\bibinfo {year} {2008}),\ \href@noop {} {\bibfield  {journal} {\bibinfo
  {journal} {Appl. Phys. Lett.}\ }\textbf {\bibinfo {volume} {93}},\ \bibinfo
  {pages} {112113}}\BibitemShut {NoStop}%
\bibitem [{\citenamefont {Pustilnik}\ \emph {et~al.}(2003)\citenamefont
  {Pustilnik}, \citenamefont {Mishchenko}, \citenamefont {Glazman},\ and\
  \citenamefont {Andreev}}]{pus}%
  \BibitemOpen
  \bibfield  {author} {\bibinfo {author} {\bibnamefont {Pustilnik},
  \bibfnamefont {M.}}, \bibinfo {author} {\bibfnamefont {E.~G.}\ \bibnamefont
  {Mishchenko}}, \bibinfo {author} {\bibfnamefont {L.~I.}\ \bibnamefont
  {Glazman}}, \ and\ \bibinfo {author} {\bibfnamefont {A.~V.}\ \bibnamefont
  {Andreev}}} (\bibinfo {year} {2003}),\ \href@noop {} {\bibfield  {journal}
  {\bibinfo  {journal} {Phys. Rev. Lett.}\ }\textbf {\bibinfo {volume} {91}},\
  \bibinfo {pages} {126805}}\BibitemShut {NoStop}%
\bibitem [{\citenamefont {Pustilnik}\ \emph {et~al.}(2006)\citenamefont
  {Pustilnik}, \citenamefont {Mishchenko},\ and\ \citenamefont
  {Starykh}}]{sta}%
  \BibitemOpen
  \bibfield  {author} {\bibinfo {author} {\bibnamefont {Pustilnik},
  \bibfnamefont {M.}}, \bibinfo {author} {\bibfnamefont {E.~G.}\ \bibnamefont
  {Mishchenko}}, \ and\ \bibinfo {author} {\bibfnamefont {O.~A.}\ \bibnamefont
  {Starykh}}} (\bibinfo {year} {2006}),\ \href@noop {} {\bibfield  {journal}
  {\bibinfo  {journal} {Phys. Rev. Lett.}\ }\textbf {\bibinfo {volume} {97}},\
  \bibinfo {pages} {246803}}\BibitemShut {NoStop}%
\bibitem [{\citenamefont {Raichev}\ and\ \citenamefont
  {Vasilopoulos}(2000)}]{ra2}%
  \BibitemOpen
  \bibfield  {author} {\bibinfo {author} {\bibnamefont {Raichev}, \bibfnamefont
  {O.}}, \ and\ \bibinfo {author} {\bibfnamefont {P.}~\bibnamefont
  {Vasilopoulos}}} (\bibinfo {year} {2000}),\ \href@noop {} {\bibfield
  {journal} {\bibinfo  {journal} {Phys. Rev. B}\ }\textbf {\bibinfo {volume}
  {61}},\ \bibinfo {pages} {7511}}\BibitemShut {NoStop}%
\bibitem [{\citenamefont {Raichev}(1997)}]{raichev}%
  \BibitemOpen
  \bibfield  {author} {\bibinfo {author} {\bibnamefont {Raichev}, \bibfnamefont
  {O.~E.}}} (\bibinfo {year} {1997}),\ \href {\doibase
  http://dx.doi.org/10.1063/1.363909} {\bibfield  {journal} {\bibinfo
  {journal} {Journal of Applied Physics}\ }\textbf {\bibinfo {volume}
  {81}}~(\bibinfo {number} {3}),\ \bibinfo {pages} {1302}}\BibitemShut
  {NoStop}%
\bibitem [{\citenamefont {Raichev}\ and\ \citenamefont
  {Vasilopoulos}(1999)}]{ra21}%
  \BibitemOpen
  \bibfield  {author} {\bibinfo {author} {\bibnamefont {Raichev}, \bibfnamefont
  {O.~E.}}, \ and\ \bibinfo {author} {\bibfnamefont {P.}~\bibnamefont
  {Vasilopoulos}}} (\bibinfo {year} {1999}),\ \href {\doibase
  10.1103/PhysRevLett.83.3697} {\bibfield  {journal} {\bibinfo  {journal}
  {Phys. Rev. Lett.}\ }\textbf {\bibinfo {volume} {83}},\ \bibinfo {pages}
  {3697}}\BibitemShut {NoStop}%
\bibitem [{\citenamefont {Raikh}\ and\ \citenamefont {von Oppen}(2002)}]{rai}%
  \BibitemOpen
  \bibfield  {author} {\bibinfo {author} {\bibnamefont {Raikh}, \bibfnamefont
  {M.~E.}}, \ and\ \bibinfo {author} {\bibfnamefont {F.}~\bibnamefont {von
  Oppen}}} (\bibinfo {year} {2002}),\ \href@noop {} {\bibfield  {journal}
  {\bibinfo  {journal} {Phys. Rev. Lett.}\ }\textbf {\bibinfo {volume} {89}},\
  \bibinfo {pages} {106601}}\BibitemShut {NoStop}%
\bibitem [{\citenamefont {Read}(1990)}]{read}%
  \BibitemOpen
  \bibfield  {author} {\bibinfo {author} {\bibnamefont {Read}, \bibfnamefont
  {N.}}} (\bibinfo {year} {1990}),\ \href {\doibase
  10.1103/PhysRevLett.65.1502} {\bibfield  {journal} {\bibinfo  {journal}
  {Phys. Rev. Lett.}\ }\textbf {\bibinfo {volume} {65}},\ \bibinfo {pages}
  {1502}}\BibitemShut {NoStop}%
\bibitem [{\citenamefont {Reed}\ \emph {et~al.}(2010)\citenamefont {Reed},
  \citenamefont {Uchoa}, \citenamefont {Joe}, \citenamefont {Gan},
  \citenamefont {Casa}, \citenamefont {Fradkin},\ and\ \citenamefont
  {Abbamonte}}]{ree}%
  \BibitemOpen
  \bibfield  {author} {\bibinfo {author} {\bibnamefont {Reed}, \bibfnamefont
  {J.~P.}}, \bibinfo {author} {\bibfnamefont {B.}~\bibnamefont {Uchoa}},
  \bibinfo {author} {\bibfnamefont {Y.~I.}\ \bibnamefont {Joe}}, \bibinfo
  {author} {\bibfnamefont {Y.}~\bibnamefont {Gan}}, \bibinfo {author}
  {\bibfnamefont {D.}~\bibnamefont {Casa}}, \bibinfo {author} {\bibfnamefont
  {E.}~\bibnamefont {Fradkin}}, \ and\ \bibinfo {author} {\bibfnamefont
  {P.}~\bibnamefont {Abbamonte}}} (\bibinfo {year} {2010}),\ \href {\doibase
  10.1126/science.1190920} {\bibfield  {journal} {\bibinfo  {journal}
  {Science}\ }\textbf {\bibinfo {volume} {330}}~(\bibinfo {number} {6005}),\
  \bibinfo {pages} {805}}\BibitemShut {NoStop}%
\bibitem [{\citenamefont {Renn}(1992)}]{renn}%
  \BibitemOpen
  \bibfield  {author} {\bibinfo {author} {\bibnamefont {Renn}, \bibfnamefont
  {S.~R.}}} (\bibinfo {year} {1992}),\ \href {\doibase
  10.1103/PhysRevLett.68.658} {\bibfield  {journal} {\bibinfo  {journal} {Phys.
  Rev. Lett.}\ }\textbf {\bibinfo {volume} {68}},\ \bibinfo {pages}
  {658}}\BibitemShut {NoStop}%
\bibitem [{\citenamefont {Reznikov}\ \emph {et~al.}(1995)\citenamefont
  {Reznikov}, \citenamefont {Heiblum}, \citenamefont {Shtrikman},\ and\
  \citenamefont {Mahalu}}]{res}%
  \BibitemOpen
  \bibfield  {author} {\bibinfo {author} {\bibnamefont {Reznikov},
  \bibfnamefont {M.}}, \bibinfo {author} {\bibfnamefont {M.}~\bibnamefont
  {Heiblum}}, \bibinfo {author} {\bibfnamefont {H.}~\bibnamefont {Shtrikman}},
  \ and\ \bibinfo {author} {\bibfnamefont {D.}~\bibnamefont {Mahalu}}}
  (\bibinfo {year} {1995}),\ \href {\doibase 10.1103/PhysRevLett.75.3340}
  {\bibfield  {journal} {\bibinfo  {journal} {Phys. Rev. Lett.}\ }\textbf
  {\bibinfo {volume} {75}},\ \bibinfo {pages} {3340}}\BibitemShut {NoStop}%
\bibitem [{\citenamefont {Rieder}\ \emph {et~al.}(2014)\citenamefont {Rieder},
  \citenamefont {Micklitz}, \citenamefont {Levchenko},\ and\ \citenamefont
  {Matveev}}]{lev4}%
  \BibitemOpen
  \bibfield  {author} {\bibinfo {author} {\bibnamefont {Rieder}, \bibfnamefont
  {M.-T.}}, \bibinfo {author} {\bibfnamefont {T.}~\bibnamefont {Micklitz}},
  \bibinfo {author} {\bibfnamefont {A.}~\bibnamefont {Levchenko}}, \ and\
  \bibinfo {author} {\bibfnamefont {K.~A.}\ \bibnamefont {Matveev}}} (\bibinfo
  {year} {2014}),\ \href {\doibase 10.1103/PhysRevB.90.165405} {\bibfield
  {journal} {\bibinfo  {journal} {Phys. Rev. B}\ }\textbf {\bibinfo {volume}
  {90}},\ \bibinfo {pages} {165405}}\BibitemShut {NoStop}%
\bibitem [{\citenamefont {Rojo}(1999)}]{roj}%
  \BibitemOpen
  \bibfield  {author} {\bibinfo {author} {\bibnamefont {Rojo}, \bibfnamefont
  {A.~G.}}} (\bibinfo {year} {1999}),\ \href@noop {} {\bibfield  {journal}
  {\bibinfo  {journal} {J. Phys.: Condens. Matter}\ }\textbf {\bibinfo {volume}
  {11}},\ \bibinfo {pages} {R31}}\BibitemShut {NoStop}%
\bibitem [{\citenamefont {Rojo}\ and\ \citenamefont {Mahan}(1992)}]{rom}%
  \BibitemOpen
  \bibfield  {author} {\bibinfo {author} {\bibnamefont {Rojo}, \bibfnamefont
  {A.~G.}}, \ and\ \bibinfo {author} {\bibfnamefont {G.~D.}\ \bibnamefont
  {Mahan}}} (\bibinfo {year} {1992}),\ \href@noop {} {\bibfield  {journal}
  {\bibinfo  {journal} {Phys. Rev. Lett.}\ }\textbf {\bibinfo {volume} {68}},\
  \bibinfo {pages} {2074}}\BibitemShut {NoStop}%
\bibitem [{\citenamefont {Roth}\ \emph {et~al.}(2009)\citenamefont {Roth},
  \citenamefont {Brüne}, \citenamefont {Buhmann}, \citenamefont {Molenkamp},
  \citenamefont {Maciejko}, \citenamefont {Qi},\ and\ \citenamefont
  {Zhang}}]{molen}%
  \BibitemOpen
  \bibfield  {author} {\bibinfo {author} {\bibnamefont {Roth}, \bibfnamefont
  {A.}}, \bibinfo {author} {\bibfnamefont {C.}~\bibnamefont {Brüne}}, \bibinfo
  {author} {\bibfnamefont {H.}~\bibnamefont {Buhmann}}, \bibinfo {author}
  {\bibfnamefont {L.~W.}\ \bibnamefont {Molenkamp}}, \bibinfo {author}
  {\bibfnamefont {J.}~\bibnamefont {Maciejko}}, \bibinfo {author}
  {\bibfnamefont {X.-L.}\ \bibnamefont {Qi}}, \ and\ \bibinfo {author}
  {\bibfnamefont {S.-C.}\ \bibnamefont {Zhang}}} (\bibinfo {year} {2009}),\
  \href {\doibase 10.1126/science.1174736} {\bibfield  {journal} {\bibinfo
  {journal} {Science}\ }\textbf {\bibinfo {volume} {325}}~(\bibinfo {number}
  {5938}),\ \bibinfo {pages} {294}}\BibitemShut {NoStop}%
\bibitem [{\citenamefont {Rozhkov}(2008)}]{roz}%
  \BibitemOpen
  \bibfield  {author} {\bibinfo {author} {\bibnamefont {Rozhkov}, \bibfnamefont
  {A.~V.}}} (\bibinfo {year} {2008}),\ \href@noop {} {\bibfield  {journal}
  {\bibinfo  {journal} {Phys. Rev. B}\ }\textbf {\bibinfo {volume} {77}},\
  \bibinfo {pages} {125109}}\BibitemShut {NoStop}%
\bibitem [{\citenamefont {Rozhkov}(2009)}]{roz2}%
  \BibitemOpen
  \bibfield  {author} {\bibinfo {author} {\bibnamefont {Rozhkov}, \bibfnamefont
  {A.~V.}}} (\bibinfo {year} {2009}),\ \href@noop {} {\bibfield  {journal}
  {\bibinfo  {journal} {Phys. Rev. B}\ }\textbf {\bibinfo {volume} {79}},\
  \bibinfo {pages} {249903(E)}}\BibitemShut {NoStop}%
\bibitem [{\citenamefont {Rubel}\ \emph
  {et~al.}(1997{\natexlab{a}})\citenamefont {Rubel}, \citenamefont {Fischer},
  \citenamefont {Dietsche}, \citenamefont {J{\"o}rger}, \citenamefont
  {Klitzing},\ and\ \citenamefont {Eberl}}]{ru2}%
  \BibitemOpen
  \bibfield  {author} {\bibinfo {author} {\bibnamefont {Rubel}, \bibfnamefont
  {H.}}, \bibinfo {author} {\bibfnamefont {A.}~\bibnamefont {Fischer}},
  \bibinfo {author} {\bibfnamefont {W.}~\bibnamefont {Dietsche}}, \bibinfo
  {author} {\bibfnamefont {C.}~\bibnamefont {J{\"o}rger}}, \bibinfo {author}
  {\bibfnamefont {K.}~\bibnamefont {Klitzing}}, \ and\ \bibinfo {author}
  {\bibfnamefont {K.}~\bibnamefont {Eberl}}} (\bibinfo {year}
  {1997}{\natexlab{a}}),\ \href@noop {} {\bibfield  {journal} {\bibinfo
  {journal} {Physica E}\ }\textbf {\bibinfo {volume} {1}},\ \bibinfo {pages}
  {160}}\BibitemShut {NoStop}%
\bibitem [{\citenamefont {Rubel}\ \emph {et~al.}(1998)\citenamefont {Rubel},
  \citenamefont {Fischer}, \citenamefont {Dietsche}, \citenamefont
  {J{\"o}rger}, \citenamefont {Klitzing},\ and\ \citenamefont {Eberl}}]{ru22}%
  \BibitemOpen
  \bibfield  {author} {\bibinfo {author} {\bibnamefont {Rubel}, \bibfnamefont
  {H.}}, \bibinfo {author} {\bibfnamefont {A.}~\bibnamefont {Fischer}},
  \bibinfo {author} {\bibfnamefont {W.}~\bibnamefont {Dietsche}}, \bibinfo
  {author} {\bibfnamefont {C.}~\bibnamefont {J{\"o}rger}}, \bibinfo {author}
  {\bibfnamefont {K.}~\bibnamefont {Klitzing}}, \ and\ \bibinfo {author}
  {\bibfnamefont {K.}~\bibnamefont {Eberl}}} (\bibinfo {year} {1998}),\
  \href@noop {} {\bibfield  {journal} {\bibinfo  {journal} {Physica B}\
  }\textbf {\bibinfo {volume} {249-251}},\ \bibinfo {pages} {859}}\BibitemShut
  {NoStop}%
\bibitem [{\citenamefont {Rubel}\ \emph
  {et~al.}(1997{\natexlab{b}})\citenamefont {Rubel}, \citenamefont {Fischer},
  \citenamefont {Dietsche}, \citenamefont {von Klitzing},\ and\ \citenamefont
  {Eberl}}]{rub}%
  \BibitemOpen
  \bibfield  {author} {\bibinfo {author} {\bibnamefont {Rubel}, \bibfnamefont
  {H.}}, \bibinfo {author} {\bibfnamefont {A.}~\bibnamefont {Fischer}},
  \bibinfo {author} {\bibfnamefont {W.}~\bibnamefont {Dietsche}}, \bibinfo
  {author} {\bibfnamefont {K.}~\bibnamefont {von Klitzing}}, \ and\ \bibinfo
  {author} {\bibfnamefont {K.}~\bibnamefont {Eberl}}} (\bibinfo {year}
  {1997}{\natexlab{b}}),\ \href@noop {} {\bibfield  {journal} {\bibinfo
  {journal} {Phys. Rev. Lett.}\ }\textbf {\bibinfo {volume} {78}},\ \bibinfo
  {pages} {1763}}\BibitemShut {NoStop}%
\bibitem [{\citenamefont {Rubel}\ \emph {et~al.}(1995)\citenamefont {Rubel},
  \citenamefont {Linfield}, \citenamefont {Hill}, \citenamefont {Nicholls},
  \citenamefont {Ritchie}, \citenamefont {Brown}, \citenamefont {Pepper},\ and\
  \citenamefont {Jones}}]{ru3}%
  \BibitemOpen
  \bibfield  {author} {\bibinfo {author} {\bibnamefont {Rubel}, \bibfnamefont
  {H.}}, \bibinfo {author} {\bibfnamefont {E.~H.}\ \bibnamefont {Linfield}},
  \bibinfo {author} {\bibfnamefont {N.~P.~R.}\ \bibnamefont {Hill}}, \bibinfo
  {author} {\bibfnamefont {J.~T.}\ \bibnamefont {Nicholls}}, \bibinfo {author}
  {\bibfnamefont {D.~A.}\ \bibnamefont {Ritchie}}, \bibinfo {author}
  {\bibfnamefont {K.~M.}\ \bibnamefont {Brown}}, \bibinfo {author}
  {\bibfnamefont {M.}~\bibnamefont {Pepper}}, \ and\ \bibinfo {author}
  {\bibfnamefont {G.~A.~C.}\ \bibnamefont {Jones}}} (\bibinfo {year} {1995}),\
  \href@noop {} {\bibfield  {journal} {\bibinfo  {journal} {Semicond. Sci.
  Technol.}\ }\textbf {\bibinfo {volume} {10}},\ \bibinfo {pages}
  {1229}}\BibitemShut {NoStop}%
\bibitem [{\citenamefont {Rubel}\ \emph {et~al.}(1996)\citenamefont {Rubel},
  \citenamefont {Linfield}, \citenamefont {Hill}, \citenamefont {Nicholls},
  \citenamefont {Ritchie}, \citenamefont {Brown}, \citenamefont {Pepper},\ and\
  \citenamefont {Jones}}]{ru32}%
  \BibitemOpen
  \bibfield  {author} {\bibinfo {author} {\bibnamefont {Rubel}, \bibfnamefont
  {H.}}, \bibinfo {author} {\bibfnamefont {E.~H.}\ \bibnamefont {Linfield}},
  \bibinfo {author} {\bibfnamefont {N.~P.~R.}\ \bibnamefont {Hill}}, \bibinfo
  {author} {\bibfnamefont {J.~T.}\ \bibnamefont {Nicholls}}, \bibinfo {author}
  {\bibfnamefont {D.~A.}\ \bibnamefont {Ritchie}}, \bibinfo {author}
  {\bibfnamefont {K.~M.}\ \bibnamefont {Brown}}, \bibinfo {author}
  {\bibfnamefont {M.}~\bibnamefont {Pepper}}, \ and\ \bibinfo {author}
  {\bibfnamefont {G.~A.~C.}\ \bibnamefont {Jones}}} (\bibinfo {year} {1996}),\
  \href@noop {} {\bibfield  {journal} {\bibinfo  {journal} {Surf. Sci.}\
  }\textbf {\bibinfo {volume} {361-362}},\ \bibinfo {pages} {134}}\BibitemShut
  {NoStop}%
\bibitem [{\citenamefont {Sakhi}(1997)}]{sak}%
  \BibitemOpen
  \bibfield  {author} {\bibinfo {author} {\bibnamefont {Sakhi}, \bibfnamefont
  {S.}}} (\bibinfo {year} {1997}),\ \href@noop {} {\bibfield  {journal}
  {\bibinfo  {journal} {Phys. Rev. B}\ }\textbf {\bibinfo {volume} {56}},\
  \bibinfo {pages} {4098}}\BibitemShut {NoStop}%
\bibitem [{\citenamefont {S{\'a}nchez}\ \emph {et~al.}(2010)\citenamefont
  {S{\'a}nchez}, \citenamefont {L{\'o}pez}, \citenamefont {S{\'a}nchez},\ and\
  \citenamefont {B{\"u}ttiker}}]{sab}%
  \BibitemOpen
  \bibfield  {author} {\bibinfo {author} {\bibnamefont {S{\'a}nchez},
  \bibfnamefont {R.}}, \bibinfo {author} {\bibfnamefont {R.}~\bibnamefont
  {L{\'o}pez}}, \bibinfo {author} {\bibfnamefont {D.}~\bibnamefont
  {S{\'a}nchez}}, \ and\ \bibinfo {author} {\bibfnamefont {M.}~\bibnamefont
  {B{\"u}ttiker}}} (\bibinfo {year} {2010}),\ \href {\doibase
  10.1103/PhysRevLett.104.076801} {\bibfield  {journal} {\bibinfo  {journal}
  {Phys. Rev. Lett.}\ }\textbf {\bibinfo {volume} {104}},\ \bibinfo {pages}
  {076801}}\BibitemShut {NoStop}%
\bibitem [{\citenamefont {Scharf}\ and\ \citenamefont
  {Matos-Abiague}(2012)}]{sch}%
  \BibitemOpen
  \bibfield  {author} {\bibinfo {author} {\bibnamefont {Scharf}, \bibfnamefont
  {B.}}, \ and\ \bibinfo {author} {\bibfnamefont {A.}~\bibnamefont
  {Matos-Abiague}}} (\bibinfo {year} {2012}),\ \href@noop {} {\bibfield
  {journal} {\bibinfo  {journal} {Phys. Rev. B}\ }\textbf {\bibinfo {volume}
  {86}},\ \bibinfo {pages} {115425}}\BibitemShut {NoStop}%
\bibitem [{\citenamefont {Schliemann}\ \emph {et~al.}(2001)\citenamefont
  {Schliemann}, \citenamefont {Girvin},\ and\ \citenamefont {MacDonald}}]{sgm}%
  \BibitemOpen
  \bibfield  {author} {\bibinfo {author} {\bibnamefont {Schliemann},
  \bibfnamefont {J.}}, \bibinfo {author} {\bibfnamefont {S.~M.}\ \bibnamefont
  {Girvin}}, \ and\ \bibinfo {author} {\bibfnamefont {A.~H.}\ \bibnamefont
  {MacDonald}}} (\bibinfo {year} {2001}),\ \href {\doibase
  10.1103/PhysRevLett.86.1849} {\bibfield  {journal} {\bibinfo  {journal}
  {Phys. Rev. Lett.}\ }\textbf {\bibinfo {volume} {86}},\ \bibinfo {pages}
  {1849}}\BibitemShut {NoStop}%
\bibitem [{\citenamefont {Schlottmann}(2004{\natexlab{a}})}]{slm}%
  \BibitemOpen
  \bibfield  {author} {\bibinfo {author} {\bibnamefont {Schlottmann},
  \bibfnamefont {P.}}} (\bibinfo {year} {2004}{\natexlab{a}}),\ \href@noop {}
  {\bibfield  {journal} {\bibinfo  {journal} {Phys. Rev. B}\ }\textbf {\bibinfo
  {volume} {69}},\ \bibinfo {pages} {035110}}\BibitemShut {NoStop}%
\bibitem [{\citenamefont {Schlottmann}(2004{\natexlab{b}})}]{slm2}%
  \BibitemOpen
  \bibfield  {author} {\bibinfo {author} {\bibnamefont {Schlottmann},
  \bibfnamefont {P.}}} (\bibinfo {year} {2004}{\natexlab{b}}),\ \href@noop {}
  {\bibfield  {journal} {\bibinfo  {journal} {Phys. Rev. B}\ }\textbf {\bibinfo
  {volume} {70}},\ \bibinfo {pages} {115306}}\BibitemShut {NoStop}%
\bibitem [{\citenamefont {Schmult}\ \emph {et~al.}(2010)\citenamefont
  {Schmult}, \citenamefont {Tiemann}, \citenamefont {Dietsche},\ and\
  \citenamefont {von Klitzing}}]{std}%
  \BibitemOpen
  \bibfield  {author} {\bibinfo {author} {\bibnamefont {Schmult}, \bibfnamefont
  {S.}}, \bibinfo {author} {\bibfnamefont {L.}~\bibnamefont {Tiemann}},
  \bibinfo {author} {\bibfnamefont {W.}~\bibnamefont {Dietsche}}, \ and\
  \bibinfo {author} {\bibfnamefont {K.}~\bibnamefont {von Klitzing}}} (\bibinfo
  {year} {2010}),\ \href@noop {} {\bibfield  {journal} {\bibinfo  {journal} {J.
  Vac. Sci. Technol. B}\ }\textbf {\bibinfo {volume} {28}},\ \bibinfo {pages}
  {C3C1}}\BibitemShut {NoStop}%
\bibitem [{\citenamefont {Sch{\"u}tt}\ \emph {et~al.}(2011)\citenamefont
  {Sch{\"u}tt}, \citenamefont {Ostrovsky}, \citenamefont {Gornyi},\ and\
  \citenamefont {Mirlin}}]{pol}%
  \BibitemOpen
  \bibfield  {author} {\bibinfo {author} {\bibnamefont {Sch{\"u}tt},
  \bibfnamefont {M.}}, \bibinfo {author} {\bibfnamefont {P.~M.}\ \bibnamefont
  {Ostrovsky}}, \bibinfo {author} {\bibfnamefont {I.~V.}\ \bibnamefont
  {Gornyi}}, \ and\ \bibinfo {author} {\bibfnamefont {A.~D.}\ \bibnamefont
  {Mirlin}}} (\bibinfo {year} {2011}),\ \href@noop {} {\bibfield  {journal}
  {\bibinfo  {journal} {Phys. Rev. B}\ }\textbf {\bibinfo {volume} {83}},\
  \bibinfo {pages} {155441}}\BibitemShut {NoStop}%
\bibitem [{\citenamefont {Sch{\"u}tt}\ \emph {et~al.}(2013)\citenamefont
  {Sch{\"u}tt}, \citenamefont {Ostrovsky}, \citenamefont {Titov}, \citenamefont
  {Gornyi}, \citenamefont {Narozhny},\ and\ \citenamefont {Mirlin}}]{mem}%
  \BibitemOpen
  \bibfield  {author} {\bibinfo {author} {\bibnamefont {Sch{\"u}tt},
  \bibfnamefont {M.}}, \bibinfo {author} {\bibfnamefont {P.~M.}\ \bibnamefont
  {Ostrovsky}}, \bibinfo {author} {\bibfnamefont {M.}~\bibnamefont {Titov}},
  \bibinfo {author} {\bibfnamefont {I.~V.}\ \bibnamefont {Gornyi}}, \bibinfo
  {author} {\bibfnamefont {B.~N.}\ \bibnamefont {Narozhny}}, \ and\ \bibinfo
  {author} {\bibfnamefont {A.~D.}\ \bibnamefont {Mirlin}}} (\bibinfo {year}
  {2013}),\ \href@noop {} {\bibfield  {journal} {\bibinfo  {journal} {Phys.
  Rev. Lett.}\ }\textbf {\bibinfo {volume} {110}},\ \bibinfo {pages}
  {026601}}\BibitemShut {NoStop}%
\bibitem [{\citenamefont {Seamons}\ \emph {et~al.}(2009)\citenamefont
  {Seamons}, \citenamefont {Morath}, \citenamefont {Reno},\ and\ \citenamefont
  {Lilly}}]{sea}%
  \BibitemOpen
  \bibfield  {author} {\bibinfo {author} {\bibnamefont {Seamons}, \bibfnamefont
  {J.~A.}}, \bibinfo {author} {\bibfnamefont {C.~P.}\ \bibnamefont {Morath}},
  \bibinfo {author} {\bibfnamefont {J.~L.}\ \bibnamefont {Reno}}, \ and\
  \bibinfo {author} {\bibfnamefont {M.~P.}\ \bibnamefont {Lilly}}} (\bibinfo
  {year} {2009}),\ \href@noop {} {\bibfield  {journal} {\bibinfo  {journal}
  {Phys. Rev. Lett.}\ }\textbf {\bibinfo {volume} {102}},\ \bibinfo {pages}
  {026804}}\BibitemShut {NoStop}%
\bibitem [{\citenamefont {Seeger}(2002)}]{seeger2002}%
  \BibitemOpen
  \bibfield  {author} {\bibinfo {author} {\bibnamefont {Seeger}, \bibfnamefont
  {K.}}} (\bibinfo {year} {2002}),\ \href@noop {} {\emph {\bibinfo {title}
  {Semiconductor Physics}}}\ (\bibinfo  {publisher} {Springer})\BibitemShut
  {NoStop}%
\bibitem [{\citenamefont {Seradjeh}\ \emph {et~al.}(2009)\citenamefont
  {Seradjeh}, \citenamefont {Moore},\ and\ \citenamefont {Franz}}]{moo}%
  \BibitemOpen
  \bibfield  {author} {\bibinfo {author} {\bibnamefont {Seradjeh},
  \bibfnamefont {B.}}, \bibinfo {author} {\bibfnamefont {J.~E.}\ \bibnamefont
  {Moore}}, \ and\ \bibinfo {author} {\bibfnamefont {M.}~\bibnamefont {Franz}}}
  (\bibinfo {year} {2009}),\ \href {\doibase 10.1103/PhysRevLett.103.066402}
  {\bibfield  {journal} {\bibinfo  {journal} {Phys. Rev. Lett.}\ }\textbf
  {\bibinfo {volume} {103}},\ \bibinfo {pages} {066402}}\BibitemShut {NoStop}%
\bibitem [{\citenamefont {Shahbazyan}\ and\ \citenamefont
  {Ulloa}(1997{\natexlab{a}})}]{sha}%
  \BibitemOpen
  \bibfield  {author} {\bibinfo {author} {\bibnamefont {Shahbazyan},
  \bibfnamefont {T.~V.}}, \ and\ \bibinfo {author} {\bibfnamefont {S.~E.}\
  \bibnamefont {Ulloa}}} (\bibinfo {year} {1997}{\natexlab{a}}),\ \href@noop {}
  {\bibfield  {journal} {\bibinfo  {journal} {Phys. Rev. B}\ }\textbf {\bibinfo
  {volume} {55}},\ \bibinfo {pages} {13702}}\BibitemShut {NoStop}%
\bibitem [{\citenamefont {Shahbazyan}\ and\ \citenamefont
  {Ulloa}(1997{\natexlab{b}})}]{sha1}%
  \BibitemOpen
  \bibfield  {author} {\bibinfo {author} {\bibnamefont {Shahbazyan},
  \bibfnamefont {T.~V.}}, \ and\ \bibinfo {author} {\bibfnamefont {S.~E.}\
  \bibnamefont {Ulloa}}} (\bibinfo {year} {1997}{\natexlab{b}}),\ \href@noop {}
  {\bibfield  {journal} {\bibinfo  {journal} {Physica E}\ }\textbf {\bibinfo
  {volume} {1}},\ \bibinfo {pages} {259}}\BibitemShut {NoStop}%
\bibitem [{\citenamefont {Shen}(2013)}]{shen}%
  \BibitemOpen
  \bibfield  {author} {\bibinfo {author} {\bibnamefont {Shen}, \bibfnamefont
  {S.-Q.}}} (\bibinfo {year} {2013}),\ \href@noop {} {\emph {\bibinfo {title}
  {Topological Insulators: Dirac Equation in Condensed Matters}}}\ (\bibinfo
  {publisher} {Springer})\BibitemShut {NoStop}%
\bibitem [{\citenamefont {Shevchenko}(1976)}]{shev}%
  \BibitemOpen
  \bibfield  {author} {\bibinfo {author} {\bibnamefont {Shevchenko},
  \bibfnamefont {S.~I.}}} (\bibinfo {year} {1976}),\ \href@noop {} {\bibfield
  {journal} {\bibinfo  {journal} {Fiz. Nizk. Temp.}\ }\textbf {\bibinfo
  {volume} {2}},\ \bibinfo {pages} {505}},\ \bibinfo {note} {[Sov. J. Low-Temp.
  Phys. {\bf 2}, 251 (1976)]}\BibitemShut {NoStop}%
\bibitem [{\citenamefont {Shimshoni}(1995)}]{sh1}%
  \BibitemOpen
  \bibfield  {author} {\bibinfo {author} {\bibnamefont {Shimshoni},
  \bibfnamefont {E.}}} (\bibinfo {year} {1995}),\ \href@noop {} {\bibfield
  {journal} {\bibinfo  {journal} {Phys. Rev. B}\ }\textbf {\bibinfo {volume}
  {51}},\ \bibinfo {pages} {9415}}\BibitemShut {NoStop}%
\bibitem [{\citenamefont {Shimshoni}\ and\ \citenamefont {Sondhi}(1994)}]{shi}%
  \BibitemOpen
  \bibfield  {author} {\bibinfo {author} {\bibnamefont {Shimshoni},
  \bibfnamefont {E.}}, \ and\ \bibinfo {author} {\bibfnamefont {S.~L.}\
  \bibnamefont {Sondhi}}} (\bibinfo {year} {1994}),\ \href@noop {} {\bibfield
  {journal} {\bibinfo  {journal} {Phys. Rev. B}\ }\textbf {\bibinfo {volume}
  {49}},\ \bibinfo {pages} {11484}}\BibitemShut {NoStop}%
\bibitem [{\citenamefont {Shon}\ and\ \citenamefont {Ando}(1998)}]{and2}%
  \BibitemOpen
  \bibfield  {author} {\bibinfo {author} {\bibnamefont {Shon}, \bibfnamefont
  {N.~H.}}, \ and\ \bibinfo {author} {\bibfnamefont {T.}~\bibnamefont {Ando}}}
  (\bibinfo {year} {1998}),\ \href@noop {} {\bibfield  {journal} {\bibinfo
  {journal} {Journal of the Physical Society of Japan}\ }\textbf {\bibinfo
  {volume} {67}}~(\bibinfo {number} {7}),\ \bibinfo {pages} {2421}}\BibitemShut
  {NoStop}%
\bibitem [{\citenamefont {Shylau}\ and\ \citenamefont {Jauho}(2014)}]{shy}%
  \BibitemOpen
  \bibfield  {author} {\bibinfo {author} {\bibnamefont {Shylau}, \bibfnamefont
  {A.~A.}}, \ and\ \bibinfo {author} {\bibfnamefont {A.-P.}\ \bibnamefont
  {Jauho}}} (\bibinfo {year} {2014}),\ \href@noop {} {\bibfield  {journal}
  {\bibinfo  {journal} {Phys. Rev. B}\ }\textbf {\bibinfo {volume} {89}},\
  \bibinfo {pages} {165421}}\BibitemShut {NoStop}%
\bibitem [{\citenamefont {Simon}\ \emph {et~al.}(2003)\citenamefont {Simon},
  \citenamefont {Rezayi},\ and\ \citenamefont {Milovanovic}}]{sim}%
  \BibitemOpen
  \bibfield  {author} {\bibinfo {author} {\bibnamefont {Simon}, \bibfnamefont
  {S.~H.}}, \bibinfo {author} {\bibfnamefont {E.~H.}\ \bibnamefont {Rezayi}}, \
  and\ \bibinfo {author} {\bibfnamefont {M.~V.}\ \bibnamefont {Milovanovic}}}
  (\bibinfo {year} {2003}),\ \href {\doibase 10.1103/PhysRevLett.91.046803}
  {\bibfield  {journal} {\bibinfo  {journal} {Phys. Rev. Lett.}\ }\textbf
  {\bibinfo {volume} {91}},\ \bibinfo {pages} {046803}}\BibitemShut {NoStop}%
\bibitem [{\citenamefont {Singwi}\ \emph {et~al.}(1968)\citenamefont {Singwi},
  \citenamefont {Tosi}, \citenamefont {Land},\ and\ \citenamefont
  {Sj{\"o}lander}}]{stls}%
  \BibitemOpen
  \bibfield  {author} {\bibinfo {author} {\bibnamefont {Singwi}, \bibfnamefont
  {K.~S.}}, \bibinfo {author} {\bibfnamefont {M.~P.}\ \bibnamefont {Tosi}},
  \bibinfo {author} {\bibfnamefont {R.~H.}\ \bibnamefont {Land}}, \ and\
  \bibinfo {author} {\bibfnamefont {A.}~\bibnamefont {Sj{\"o}lander}}}
  (\bibinfo {year} {1968}),\ \href {\doibase 10.1103/PhysRev.176.589}
  {\bibfield  {journal} {\bibinfo  {journal} {Phys. Rev.}\ }\textbf {\bibinfo
  {volume} {176}},\ \bibinfo {pages} {589}}\BibitemShut {NoStop}%
\bibitem [{\citenamefont {Sirenko}\ and\ \citenamefont
  {Vasilopoulos}(1992)}]{si2}%
  \BibitemOpen
  \bibfield  {author} {\bibinfo {author} {\bibnamefont {Sirenko}, \bibfnamefont
  {Y.~M.}}, \ and\ \bibinfo {author} {\bibfnamefont {P.}~\bibnamefont
  {Vasilopoulos}}} (\bibinfo {year} {1992}),\ \href@noop {} {\bibfield
  {journal} {\bibinfo  {journal} {Phys. Rev. B}\ }\textbf {\bibinfo {volume}
  {46}},\ \bibinfo {pages} {1611}}\BibitemShut {NoStop}%
\bibitem [{\citenamefont {Sivan}\ \emph {et~al.}(1992)\citenamefont {Sivan},
  \citenamefont {Solomon},\ and\ \citenamefont {Shtrikman}}]{siv}%
  \BibitemOpen
  \bibfield  {author} {\bibinfo {author} {\bibnamefont {Sivan}, \bibfnamefont
  {U.}}, \bibinfo {author} {\bibfnamefont {P.~M.}\ \bibnamefont {Solomon}}, \
  and\ \bibinfo {author} {\bibfnamefont {H.}~\bibnamefont {Shtrikman}}}
  (\bibinfo {year} {1992}),\ \href@noop {} {\bibfield  {journal} {\bibinfo
  {journal} {Phys. Rev. Lett.}\ }\textbf {\bibinfo {volume} {68}},\ \bibinfo
  {pages} {1196}}\BibitemShut {NoStop}%
\bibitem [{\citenamefont {Smith}\ and\ \citenamefont
  {Jensen}(1989)}]{smith1989}%
  \BibitemOpen
  \bibfield  {author} {\bibinfo {author} {\bibnamefont {Smith}, \bibfnamefont
  {H.}}, \ and\ \bibinfo {author} {\bibfnamefont {H.~H.}\ \bibnamefont
  {Jensen}}} (\bibinfo {year} {1989}),\ \href@noop {} {\emph {\bibinfo {title}
  {Transport Phenomena}}}\ (\bibinfo  {publisher} {Oxford University
  Press})\BibitemShut {NoStop}%
\bibitem [{\citenamefont {Snoke}(2002)}]{sno}%
  \BibitemOpen
  \bibfield  {author} {\bibinfo {author} {\bibnamefont {Snoke}, \bibfnamefont
  {D.~W.}}} (\bibinfo {year} {2002}),\ \href@noop {} {\bibfield  {journal}
  {\bibinfo  {journal} {Science}\ }\textbf {\bibinfo {volume} {298}},\ \bibinfo
  {pages} {1368}}\BibitemShut {NoStop}%
\bibitem [{\citenamefont {Snoke}(2008)}]{snoke}%
  \BibitemOpen
  \bibfield  {author} {\bibinfo {author} {\bibnamefont {Snoke}, \bibfnamefont
  {D.~W.}}} (\bibinfo {year} {2008}),\ \href@noop {} {\emph {\bibinfo {title}
  {Solid State Physics. Essential Concepts}}}\ (\bibinfo  {publisher}
  {Addison-Wesley})\BibitemShut {NoStop}%
\bibitem [{\citenamefont {Sodemann}\ \emph {et~al.}(2012)\citenamefont
  {Sodemann}, \citenamefont {Pesin},\ and\ \citenamefont {MacDonald}}]{sode}%
  \BibitemOpen
  \bibfield  {author} {\bibinfo {author} {\bibnamefont {Sodemann},
  \bibfnamefont {I.}}, \bibinfo {author} {\bibfnamefont {D.~A.}\ \bibnamefont
  {Pesin}}, \ and\ \bibinfo {author} {\bibfnamefont {A.~H.}\ \bibnamefont
  {MacDonald}}} (\bibinfo {year} {2012}),\ \href {\doibase
  10.1103/PhysRevB.85.195136} {\bibfield  {journal} {\bibinfo  {journal} {Phys.
  Rev. B}\ }\textbf {\bibinfo {volume} {85}},\ \bibinfo {pages}
  {195136}}\BibitemShut {NoStop}%
\bibitem [{\citenamefont {S{\"o}derstr{\"o}m}\ \emph
  {et~al.}(1996)\citenamefont {S{\"o}derstr{\"o}m}, \citenamefont {Buyanov},\
  and\ \citenamefont {Sernelius}}]{sod}%
  \BibitemOpen
  \bibfield  {author} {\bibinfo {author} {\bibnamefont {S{\"o}derstr{\"o}m},
  \bibfnamefont {E.}}, \bibinfo {author} {\bibfnamefont {A.~V.}\ \bibnamefont
  {Buyanov}}, \ and\ \bibinfo {author} {\bibfnamefont {B.~E.}\ \bibnamefont
  {Sernelius}}} (\bibinfo {year} {1996}),\ \href@noop {} {\bibfield  {journal}
  {\bibinfo  {journal} {J. Phys.: Condens. Matter}\ }\textbf {\bibinfo {volume}
  {8}},\ \bibinfo {pages} {3705}}\BibitemShut {NoStop}%
\bibitem [{\citenamefont {Solomon}\ and\ \citenamefont
  {Laikhtman}(1991)}]{ex11}%
  \BibitemOpen
  \bibfield  {author} {\bibinfo {author} {\bibnamefont {Solomon}, \bibfnamefont
  {P.~M.}}, \ and\ \bibinfo {author} {\bibfnamefont {B.}~\bibnamefont
  {Laikhtman}}} (\bibinfo {year} {1991}),\ \href@noop {} {\bibfield  {journal}
  {\bibinfo  {journal} {Superlattices Microstruct.}\ }\textbf {\bibinfo
  {volume} {10}},\ \bibinfo {pages} {89}}\BibitemShut {NoStop}%
\bibitem [{\citenamefont {Solomon}\ \emph {et~al.}(1989)\citenamefont
  {Solomon}, \citenamefont {Price}, \citenamefont {Frank},\ and\ \citenamefont
  {Tulipe}}]{ex1}%
  \BibitemOpen
  \bibfield  {author} {\bibinfo {author} {\bibnamefont {Solomon}, \bibfnamefont
  {P.~M.}}, \bibinfo {author} {\bibfnamefont {P.~J.}\ \bibnamefont {Price}},
  \bibinfo {author} {\bibfnamefont {D.~J.}\ \bibnamefont {Frank}}, \ and\
  \bibinfo {author} {\bibfnamefont {D.~C.~L.}\ \bibnamefont {Tulipe}}}
  (\bibinfo {year} {1989}),\ \href@noop {} {\bibfield  {journal} {\bibinfo
  {journal} {Phys. Rev. Lett.}\ }\textbf {\bibinfo {volume} {63}},\ \bibinfo
  {pages} {2508}}\BibitemShut {NoStop}%
\bibitem [{\citenamefont {Son}(2007)}]{son}%
  \BibitemOpen
  \bibfield  {author} {\bibinfo {author} {\bibnamefont {Son}, \bibfnamefont
  {D.~T.}}} (\bibinfo {year} {2007}),\ \href@noop {} {\bibfield  {journal}
  {\bibinfo  {journal} {Phys. Rev. B}\ }\textbf {\bibinfo {volume} {75}},\
  \bibinfo {pages} {235423}}\BibitemShut {NoStop}%
\bibitem [{\citenamefont {Song}\ \emph {et~al.}(2013)\citenamefont {Song},
  \citenamefont {Abanin},\ and\ \citenamefont {Levitov}}]{sl3}%
  \BibitemOpen
  \bibfield  {author} {\bibinfo {author} {\bibnamefont {Song}, \bibfnamefont
  {J.~C.~W.}}, \bibinfo {author} {\bibfnamefont {D.~A.}\ \bibnamefont
  {Abanin}}, \ and\ \bibinfo {author} {\bibfnamefont {L.~S.}\ \bibnamefont
  {Levitov}}} (\bibinfo {year} {2013}),\ \href@noop {} {\bibfield  {journal}
  {\bibinfo  {journal} {Nanolett.}\ }\textbf {\bibinfo {volume} {13}},\
  \bibinfo {pages} {3631}}\BibitemShut {NoStop}%
\bibitem [{\citenamefont {Song}\ and\ \citenamefont {Levitov}(2012)}]{sl1}%
  \BibitemOpen
  \bibfield  {author} {\bibinfo {author} {\bibnamefont {Song}, \bibfnamefont
  {J.~C.~W.}}, \ and\ \bibinfo {author} {\bibfnamefont {L.~S.}\ \bibnamefont
  {Levitov}}} (\bibinfo {year} {2012}),\ \href@noop {} {\bibfield  {journal}
  {\bibinfo  {journal} {Phys. Rev. Lett.}\ }\textbf {\bibinfo {volume} {109}},\
  \bibinfo {pages} {236602}}\BibitemShut {NoStop}%
\bibitem [{\citenamefont {Song}\ and\ \citenamefont {Levitov}(2013)}]{sl2}%
  \BibitemOpen
  \bibfield  {author} {\bibinfo {author} {\bibnamefont {Song}, \bibfnamefont
  {J.~C.~W.}}, \ and\ \bibinfo {author} {\bibfnamefont {L.~S.}\ \bibnamefont
  {Levitov}}} (\bibinfo {year} {2013}),\ \href@noop {} {\bibfield  {journal}
  {\bibinfo  {journal} {Phys. Rev. Lett.}\ }\textbf {\bibinfo {volume} {111}},\
  \bibinfo {pages} {126601}}\BibitemShut {NoStop}%
\bibitem [{\citenamefont {Sothmann}\ \emph {et~al.}(2012)\citenamefont
  {Sothmann}, \citenamefont {S\'anchez}, \citenamefont {Jordan},\ and\
  \citenamefont {B\"uttiker}}]{soth}%
  \BibitemOpen
  \bibfield  {author} {\bibinfo {author} {\bibnamefont {Sothmann},
  \bibfnamefont {B.}}, \bibinfo {author} {\bibfnamefont {R.}~\bibnamefont
  {S\'anchez}}, \bibinfo {author} {\bibfnamefont {A.~N.}\ \bibnamefont
  {Jordan}}, \ and\ \bibinfo {author} {\bibfnamefont {M.}~\bibnamefont
  {B\"uttiker}}} (\bibinfo {year} {2012}),\ \href {\doibase
  10.1103/PhysRevB.85.205301} {\bibfield  {journal} {\bibinfo  {journal} {Phys.
  Rev. B}\ }\textbf {\bibinfo {volume} {85}},\ \bibinfo {pages}
  {205301}}\BibitemShut {NoStop}%
\bibitem [{\citenamefont {Spielman}\ \emph {et~al.}(2000)\citenamefont
  {Spielman}, \citenamefont {Eisenstein}, \citenamefont {Pfeiffer},\ and\
  \citenamefont {West}}]{spi2}%
  \BibitemOpen
  \bibfield  {author} {\bibinfo {author} {\bibnamefont {Spielman},
  \bibfnamefont {I.~B.}}, \bibinfo {author} {\bibfnamefont {J.~P.}\
  \bibnamefont {Eisenstein}}, \bibinfo {author} {\bibfnamefont {L.~N.}\
  \bibnamefont {Pfeiffer}}, \ and\ \bibinfo {author} {\bibfnamefont {K.~W.}\
  \bibnamefont {West}}} (\bibinfo {year} {2000}),\ \href {\doibase
  10.1103/PhysRevLett.84.5808} {\bibfield  {journal} {\bibinfo  {journal}
  {Phys. Rev. Lett.}\ }\textbf {\bibinfo {volume} {84}},\ \bibinfo {pages}
  {5808}}\BibitemShut {NoStop}%
\bibitem [{\citenamefont {Spielman}\ \emph {et~al.}(2004)\citenamefont
  {Spielman}, \citenamefont {Kellogg}, \citenamefont {Eisenstein},
  \citenamefont {Pfeiffer},\ and\ \citenamefont {West}}]{spi}%
  \BibitemOpen
  \bibfield  {author} {\bibinfo {author} {\bibnamefont {Spielman},
  \bibfnamefont {I.~B.}}, \bibinfo {author} {\bibfnamefont {M.}~\bibnamefont
  {Kellogg}}, \bibinfo {author} {\bibfnamefont {J.~P.}\ \bibnamefont
  {Eisenstein}}, \bibinfo {author} {\bibfnamefont {L.~N.}\ \bibnamefont
  {Pfeiffer}}, \ and\ \bibinfo {author} {\bibfnamefont {K.~W.}\ \bibnamefont
  {West}}} (\bibinfo {year} {2004}),\ \href@noop {} {\bibfield  {journal}
  {\bibinfo  {journal} {Phys. Rev. B}\ }\textbf {\bibinfo {volume} {70}},\
  \bibinfo {pages} {081303(R)}}\BibitemShut {NoStop}%
\bibitem [{\citenamefont {Spivak}\ and\ \citenamefont {Kivelson}(2005)}]{spk}%
  \BibitemOpen
  \bibfield  {author} {\bibinfo {author} {\bibnamefont {Spivak}, \bibfnamefont
  {B.}}, \ and\ \bibinfo {author} {\bibfnamefont {S.}~\bibnamefont {Kivelson}}}
  (\bibinfo {year} {2005}),\ \href {\doibase 10.1103/PhysRevB.72.045355}
  {\bibfield  {journal} {\bibinfo  {journal} {Phys. Rev. B}\ }\textbf {\bibinfo
  {volume} {72}},\ \bibinfo {pages} {045355}}\BibitemShut {NoStop}%
\bibitem [{\citenamefont {Spivak}\ \emph {et~al.}(2010)\citenamefont {Spivak},
  \citenamefont {Kravchenko}, \citenamefont {Kivelson},\ and\ \citenamefont
  {Gao}}]{spiv}%
  \BibitemOpen
  \bibfield  {author} {\bibinfo {author} {\bibnamefont {Spivak}, \bibfnamefont
  {B.}}, \bibinfo {author} {\bibfnamefont {S.~V.}\ \bibnamefont {Kravchenko}},
  \bibinfo {author} {\bibfnamefont {S.~A.}\ \bibnamefont {Kivelson}}, \ and\
  \bibinfo {author} {\bibfnamefont {X.~P.~A.}\ \bibnamefont {Gao}}} (\bibinfo
  {year} {2010}),\ \href {\doibase 10.1103/RevModPhys.82.1743} {\bibfield
  {journal} {\bibinfo  {journal} {Rev. Mod. Phys.}\ }\textbf {\bibinfo {volume}
  {82}},\ \bibinfo {pages} {1743}}\BibitemShut {NoStop}%
\bibitem [{\citenamefont {Stauber}\ and\ \citenamefont
  {G\'omez-Santos}(2012)}]{plas3}%
  \BibitemOpen
  \bibfield  {author} {\bibinfo {author} {\bibnamefont {Stauber}, \bibfnamefont
  {T.}}, \ and\ \bibinfo {author} {\bibfnamefont {G.}~\bibnamefont
  {G\'omez-Santos}}} (\bibinfo {year} {2012}),\ \href@noop {} {\bibfield
  {journal} {\bibinfo  {journal} {Phys. Rev. B}\ }\textbf {\bibinfo {volume}
  {85}},\ \bibinfo {pages} {075410}}\BibitemShut {NoStop}%
\bibitem [{\citenamefont {Stern}\ and\ \citenamefont {Halperin}(1995)}]{ady}%
  \BibitemOpen
  \bibfield  {author} {\bibinfo {author} {\bibnamefont {Stern}, \bibfnamefont
  {A.}}, \ and\ \bibinfo {author} {\bibfnamefont {B.~I.}\ \bibnamefont
  {Halperin}}} (\bibinfo {year} {1995}),\ \href {\doibase
  10.1103/PhysRevB.52.5890} {\bibfield  {journal} {\bibinfo  {journal} {Phys.
  Rev. B}\ }\textbf {\bibinfo {volume} {52}},\ \bibinfo {pages}
  {5890}}\BibitemShut {NoStop}%
\bibitem [{\citenamefont {Stern}\ and\ \citenamefont {Halperin}(2002)}]{ad2}%
  \BibitemOpen
  \bibfield  {author} {\bibinfo {author} {\bibnamefont {Stern}, \bibfnamefont
  {A.}}, \ and\ \bibinfo {author} {\bibfnamefont {B.~I.}\ \bibnamefont
  {Halperin}}} (\bibinfo {year} {2002}),\ \href@noop {} {\bibfield  {journal}
  {\bibinfo  {journal} {Phys. Rev. Lett.}\ }\textbf {\bibinfo {volume} {88}},\
  \bibinfo {pages} {106801}}\BibitemShut {NoStop}%
\bibitem [{\citenamefont {Stern}\ \emph {et~al.}(2000)\citenamefont {Stern},
  \citenamefont {Sarma}, \citenamefont {Fisher},\ and\ \citenamefont
  {Girvin}}]{ads}%
  \BibitemOpen
  \bibfield  {author} {\bibinfo {author} {\bibnamefont {Stern}, \bibfnamefont
  {A.}}, \bibinfo {author} {\bibfnamefont {S.~D.}\ \bibnamefont {Sarma}},
  \bibinfo {author} {\bibfnamefont {M.~P.~A.}\ \bibnamefont {Fisher}}, \ and\
  \bibinfo {author} {\bibfnamefont {S.~M.}\ \bibnamefont {Girvin}}} (\bibinfo
  {year} {2000}),\ \href@noop {} {\bibfield  {journal} {\bibinfo  {journal}
  {Phys. Rev. Lett.}\ }\textbf {\bibinfo {volume} {84}},\ \bibinfo {pages}
  {139}}\BibitemShut {NoStop}%
\bibitem [{\citenamefont {Stern}\ and\ \citenamefont {Ussishkin}(1997)}]{uss}%
  \BibitemOpen
  \bibfield  {author} {\bibinfo {author} {\bibnamefont {Stern}, \bibfnamefont
  {A.}}, \ and\ \bibinfo {author} {\bibfnamefont {I.}~\bibnamefont
  {Ussishkin}}} (\bibinfo {year} {1997}),\ \href@noop {} {\bibfield  {journal}
  {\bibinfo  {journal} {Physica E}\ }\textbf {\bibinfo {volume} {1}},\ \bibinfo
  {pages} {176}}\BibitemShut {NoStop}%
\bibitem [{\citenamefont {Stern}(1967)}]{ste}%
  \BibitemOpen
  \bibfield  {author} {\bibinfo {author} {\bibnamefont {Stern}, \bibfnamefont
  {F.}}} (\bibinfo {year} {1967}),\ \href@noop {} {\bibfield  {journal}
  {\bibinfo  {journal} {Phys. Rev. Lett.}\ }\textbf {\bibinfo {volume} {18}},\
  \bibinfo {pages} {546}}\BibitemShut {NoStop}%
\bibitem [{\citenamefont {Stormer}\ \emph {et~al.}(1990)\citenamefont
  {Stormer}, \citenamefont {Pfeiffer}, \citenamefont {Baldwin},\ and\
  \citenamefont {West}}]{storm}%
  \BibitemOpen
  \bibfield  {author} {\bibinfo {author} {\bibnamefont {Stormer}, \bibfnamefont
  {H.~L.}}, \bibinfo {author} {\bibfnamefont {L.~N.}\ \bibnamefont {Pfeiffer}},
  \bibinfo {author} {\bibfnamefont {K.~W.}\ \bibnamefont {Baldwin}}, \ and\
  \bibinfo {author} {\bibfnamefont {K.~W.}\ \bibnamefont {West}}} (\bibinfo
  {year} {1990}),\ \href@noop {} {\bibfield  {journal} {\bibinfo  {journal}
  {Phys. Rev. B}\ }\textbf {\bibinfo {volume} {41}},\ \bibinfo {pages}
  {1278}}\BibitemShut {NoStop}%
\bibitem [{\citenamefont {Su}\ and\ \citenamefont {MacDonald}(2008)}]{sumac}%
  \BibitemOpen
  \bibfield  {author} {\bibinfo {author} {\bibnamefont {Su}, \bibfnamefont
  {J.-J.}}, \ and\ \bibinfo {author} {\bibfnamefont {A.~H.}\ \bibnamefont
  {MacDonald}}} (\bibinfo {year} {2008}),\ \href@noop {} {\bibfield  {journal}
  {\bibinfo  {journal} {Nature Physics}\ }\textbf {\bibinfo {volume} {4}},\
  \bibinfo {pages} {799}}\BibitemShut {NoStop}%
\bibitem [{\citenamefont {Suen}\ \emph {et~al.}(1992)\citenamefont {Suen},
  \citenamefont {Engel}, \citenamefont {Santos}, \citenamefont {Shayegan},\
  and\ \citenamefont {Tsui}}]{suen}%
  \BibitemOpen
  \bibfield  {author} {\bibinfo {author} {\bibnamefont {Suen}, \bibfnamefont
  {Y.~W.}}, \bibinfo {author} {\bibfnamefont {L.~W.}\ \bibnamefont {Engel}},
  \bibinfo {author} {\bibfnamefont {M.~B.}\ \bibnamefont {Santos}}, \bibinfo
  {author} {\bibfnamefont {M.}~\bibnamefont {Shayegan}}, \ and\ \bibinfo
  {author} {\bibfnamefont {D.~C.}\ \bibnamefont {Tsui}}} (\bibinfo {year}
  {1992}),\ \href {\doibase 10.1103/PhysRevLett.68.1379} {\bibfield  {journal}
  {\bibinfo  {journal} {Phys. Rev. Lett.}\ }\textbf {\bibinfo {volume} {68}},\
  \bibinfo {pages} {1379}}\BibitemShut {NoStop}%
\bibitem [{\citenamefont {Suprunenko}\ \emph {et~al.}(2012)\citenamefont
  {Suprunenko}, \citenamefont {Cheianov},\ and\ \citenamefont {Fal'ko}}]{fal2}%
  \BibitemOpen
  \bibfield  {author} {\bibinfo {author} {\bibnamefont {Suprunenko},
  \bibfnamefont {Y.~F.}}, \bibinfo {author} {\bibfnamefont {V.}~\bibnamefont
  {Cheianov}}, \ and\ \bibinfo {author} {\bibfnamefont {V.~I.}\ \bibnamefont
  {Fal'ko}}} (\bibinfo {year} {2012}),\ \href {\doibase
  10.1103/PhysRevB.86.155405} {\bibfield  {journal} {\bibinfo  {journal} {Phys.
  Rev. B}\ }\textbf {\bibinfo {volume} {86}},\ \bibinfo {pages}
  {155405}}\BibitemShut {NoStop}%
\bibitem [{\citenamefont {Svintsov}\ \emph {et~al.}(2012)\citenamefont
  {Svintsov}, \citenamefont {Vyurkov}, \citenamefont {Yurchenko}, \citenamefont
  {Otsuji},\ and\ \citenamefont {Ryzhii}}]{ryz}%
  \BibitemOpen
  \bibfield  {author} {\bibinfo {author} {\bibnamefont {Svintsov},
  \bibfnamefont {D.}}, \bibinfo {author} {\bibfnamefont {V.}~\bibnamefont
  {Vyurkov}}, \bibinfo {author} {\bibfnamefont {S.}~\bibnamefont {Yurchenko}},
  \bibinfo {author} {\bibfnamefont {T.}~\bibnamefont {Otsuji}}, \ and\ \bibinfo
  {author} {\bibfnamefont {V.}~\bibnamefont {Ryzhii}}} (\bibinfo {year}
  {2012}),\ \href@noop {} {\bibfield  {journal} {\bibinfo  {journal} {Journal
  of Applied Physics}\ }\textbf {\bibinfo {volume} {111}}~(\bibinfo {number}
  {8})}\BibitemShut {NoStop}%
\bibitem [{\citenamefont {Swierkowski}\ \emph {et~al.}(1995)\citenamefont
  {Swierkowski}, \citenamefont {Szymanski},\ and\ \citenamefont
  {Gortel}}]{ssg}%
  \BibitemOpen
  \bibfield  {author} {\bibinfo {author} {\bibnamefont {Swierkowski},
  \bibfnamefont {L.}}, \bibinfo {author} {\bibfnamefont {J.}~\bibnamefont
  {Szymanski}}, \ and\ \bibinfo {author} {\bibfnamefont {Z.~W.}\ \bibnamefont
  {Gortel}}} (\bibinfo {year} {1995}),\ \href@noop {} {\bibfield  {journal}
  {\bibinfo  {journal} {Phys. Rev. Lett.}\ }\textbf {\bibinfo {volume} {74}},\
  \bibinfo {pages} {3245}}\BibitemShut {NoStop}%
\bibitem [{\citenamefont {Swierkowski}\ \emph {et~al.}(1996)\citenamefont
  {Swierkowski}, \citenamefont {Szymanski},\ and\ \citenamefont
  {Gortel}}]{ssg2}%
  \BibitemOpen
  \bibfield  {author} {\bibinfo {author} {\bibnamefont {Swierkowski},
  \bibfnamefont {L.}}, \bibinfo {author} {\bibfnamefont {J.}~\bibnamefont
  {Szymanski}}, \ and\ \bibinfo {author} {\bibfnamefont {Z.~W.}\ \bibnamefont
  {Gortel}}} (\bibinfo {year} {1996}),\ \href@noop {} {\bibfield  {journal}
  {\bibinfo  {journal} {Surf. Sci.}\ }\textbf {\bibinfo {volume} {361-362}},\
  \bibinfo {pages} {130}}\BibitemShut {NoStop}%
\bibitem [{\citenamefont {Swierkowski}\ \emph {et~al.}(1997)\citenamefont
  {Swierkowski}, \citenamefont {Szymanski},\ and\ \citenamefont
  {Gortel}}]{ssg3}%
  \BibitemOpen
  \bibfield  {author} {\bibinfo {author} {\bibnamefont {Swierkowski},
  \bibfnamefont {L.}}, \bibinfo {author} {\bibfnamefont {J.}~\bibnamefont
  {Szymanski}}, \ and\ \bibinfo {author} {\bibfnamefont {Z.~W.}\ \bibnamefont
  {Gortel}}} (\bibinfo {year} {1997}),\ \href@noop {} {\bibfield  {journal}
  {\bibinfo  {journal} {Phys. Rev. B}\ }\textbf {\bibinfo {volume} {55}},\
  \bibinfo {pages} {2280}}\BibitemShut {NoStop}%
\bibitem [{\citenamefont {Takashina}\ \emph {et~al.}(2009)\citenamefont
  {Takashina}, \citenamefont {Nishiguchi}, \citenamefont {Ono}, \citenamefont
  {Fujiwara}, \citenamefont {Fujisawa}, \citenamefont {Hirayama},\ and\
  \citenamefont {Muraki}}]{tak}%
  \BibitemOpen
  \bibfield  {author} {\bibinfo {author} {\bibnamefont {Takashina},
  \bibfnamefont {K.}}, \bibinfo {author} {\bibfnamefont {K.}~\bibnamefont
  {Nishiguchi}}, \bibinfo {author} {\bibfnamefont {Y.}~\bibnamefont {Ono}},
  \bibinfo {author} {\bibfnamefont {A.}~\bibnamefont {Fujiwara}}, \bibinfo
  {author} {\bibfnamefont {T.}~\bibnamefont {Fujisawa}}, \bibinfo {author}
  {\bibfnamefont {Y.}~\bibnamefont {Hirayama}}, \ and\ \bibinfo {author}
  {\bibfnamefont {K.}~\bibnamefont {Muraki}}} (\bibinfo {year} {2009}),\
  \href@noop {} {\bibfield  {journal} {\bibinfo  {journal} {Appl. Phys. Lett.}\
  }\textbf {\bibinfo {volume} {94}},\ \bibinfo {pages} {142104}}\BibitemShut
  {NoStop}%
\bibitem [{\citenamefont {Tan}\ \emph {et~al.}(2007)\citenamefont {Tan},
  \citenamefont {Zhang}, \citenamefont {Stormer},\ and\ \citenamefont
  {Kim}}]{tzs}%
  \BibitemOpen
  \bibfield  {author} {\bibinfo {author} {\bibnamefont {Tan}, \bibfnamefont
  {Y.-W.}}, \bibinfo {author} {\bibfnamefont {Y.}~\bibnamefont {Zhang}},
  \bibinfo {author} {\bibfnamefont {H.~L.}\ \bibnamefont {Stormer}}, \ and\
  \bibinfo {author} {\bibfnamefont {P.}~\bibnamefont {Kim}}} (\bibinfo {year}
  {2007}),\ \href@noop {} {\bibfield  {journal} {\bibinfo  {journal} {Eur.
  Phys. J. Special Topics}\ }\textbf {\bibinfo {volume} {148}},\ \bibinfo
  {pages} {15}}\BibitemShut {NoStop}%
\bibitem [{\citenamefont {Tanatar}(1996)}]{ta2}%
  \BibitemOpen
  \bibfield  {author} {\bibinfo {author} {\bibnamefont {Tanatar}, \bibfnamefont
  {B.}}} (\bibinfo {year} {1996}),\ \href@noop {} {\bibfield  {journal}
  {\bibinfo  {journal} {Solid State Comm.}\ }\textbf {\bibinfo {volume} {99}},\
  \bibinfo {pages} {1}}\BibitemShut {NoStop}%
\bibitem [{\citenamefont {Tanatar}(1998)}]{ta22}%
  \BibitemOpen
  \bibfield  {author} {\bibinfo {author} {\bibnamefont {Tanatar}, \bibfnamefont
  {B.}}} (\bibinfo {year} {1998}),\ \href@noop {} {\bibfield  {journal}
  {\bibinfo  {journal} {Phys. Rev. B}\ }\textbf {\bibinfo {volume} {58}},\
  \bibinfo {pages} {1154}}\BibitemShut {NoStop}%
\bibitem [{\citenamefont {Tanatar}\ and\ \citenamefont {Das}(1996)}]{ta3}%
  \BibitemOpen
  \bibfield  {author} {\bibinfo {author} {\bibnamefont {Tanatar}, \bibfnamefont
  {B.}}, \ and\ \bibinfo {author} {\bibfnamefont {A.~K.}\ \bibnamefont {Das}}}
  (\bibinfo {year} {1996}),\ \href@noop {} {\bibfield  {journal} {\bibinfo
  {journal} {Phys. Rev. B}\ }\textbf {\bibinfo {volume} {54}},\ \bibinfo
  {pages} {13827}}\BibitemShut {NoStop}%
\bibitem [{\citenamefont {Tiemann}\ \emph
  {et~al.}(2008{\natexlab{a}})\citenamefont {Tiemann}, \citenamefont
  {Dietsche}, \citenamefont {Hauser},\ and\ \citenamefont {von
  Klitzing}}]{tie}%
  \BibitemOpen
  \bibfield  {author} {\bibinfo {author} {\bibnamefont {Tiemann}, \bibfnamefont
  {L.}}, \bibinfo {author} {\bibfnamefont {W.}~\bibnamefont {Dietsche}},
  \bibinfo {author} {\bibfnamefont {M.}~\bibnamefont {Hauser}}, \ and\ \bibinfo
  {author} {\bibfnamefont {K.}~\bibnamefont {von Klitzing}}} (\bibinfo {year}
  {2008}{\natexlab{a}}),\ \href
  {http://stacks.iop.org/1367-2630/10/i=4/a=045018} {\bibfield  {journal}
  {\bibinfo  {journal} {New Journal of Physics}\ }\textbf {\bibinfo {volume}
  {10}}~(\bibinfo {number} {4}),\ \bibinfo {pages} {045018}}\BibitemShut
  {NoStop}%
\bibitem [{\citenamefont {Tiemann}\ \emph
  {et~al.}(2008{\natexlab{b}})\citenamefont {Tiemann}, \citenamefont {Lok},
  \citenamefont {Dietsche}, \citenamefont {von Klitzing}, \citenamefont
  {Muraki}, \citenamefont {Schuh},\ and\ \citenamefont {Wegscheider}}]{tie2}%
  \BibitemOpen
  \bibfield  {author} {\bibinfo {author} {\bibnamefont {Tiemann}, \bibfnamefont
  {L.}}, \bibinfo {author} {\bibfnamefont {J.~G.~S.}\ \bibnamefont {Lok}},
  \bibinfo {author} {\bibfnamefont {W.}~\bibnamefont {Dietsche}}, \bibinfo
  {author} {\bibfnamefont {K.}~\bibnamefont {von Klitzing}}, \bibinfo {author}
  {\bibfnamefont {K.}~\bibnamefont {Muraki}}, \bibinfo {author} {\bibfnamefont
  {D.}~\bibnamefont {Schuh}}, \ and\ \bibinfo {author} {\bibfnamefont
  {W.}~\bibnamefont {Wegscheider}}} (\bibinfo {year} {2008}{\natexlab{b}}),\
  \href {\doibase 10.1103/PhysRevB.77.033306} {\bibfield  {journal} {\bibinfo
  {journal} {Phys. Rev. B}\ }\textbf {\bibinfo {volume} {77}},\ \bibinfo
  {pages} {033306}}\BibitemShut {NoStop}%
\bibitem [{\citenamefont {Titov}\ \emph
  {et~al.}(2013{\natexlab{a}})\citenamefont {Titov}, \citenamefont {Gorbachev},
  \citenamefont {Narozhny}, \citenamefont {Tudorovskiy}, \citenamefont
  {Sch{\"u}tt}, \citenamefont {Ostrovsky}, \citenamefont {Gornyi},
  \citenamefont {Mirlin}, \citenamefont {Katsnelson}, \citenamefont
  {Novoselov}, \citenamefont {Geim},\ and\ \citenamefont {Ponomarenko}}]{meg}%
  \BibitemOpen
  \bibfield  {author} {\bibinfo {author} {\bibnamefont {Titov}, \bibfnamefont
  {M.}}, \bibinfo {author} {\bibfnamefont {R.~V.}\ \bibnamefont {Gorbachev}},
  \bibinfo {author} {\bibfnamefont {B.~N.}\ \bibnamefont {Narozhny}}, \bibinfo
  {author} {\bibfnamefont {T.}~\bibnamefont {Tudorovskiy}}, \bibinfo {author}
  {\bibfnamefont {M.}~\bibnamefont {Sch{\"u}tt}}, \bibinfo {author}
  {\bibfnamefont {P.~M.}\ \bibnamefont {Ostrovsky}}, \bibinfo {author}
  {\bibfnamefont {I.~V.}\ \bibnamefont {Gornyi}}, \bibinfo {author}
  {\bibfnamefont {A.~D.}\ \bibnamefont {Mirlin}}, \bibinfo {author}
  {\bibfnamefont {M.~I.}\ \bibnamefont {Katsnelson}}, \bibinfo {author}
  {\bibfnamefont {K.~S.}\ \bibnamefont {Novoselov}}, \bibinfo {author}
  {\bibfnamefont {A.~K.}\ \bibnamefont {Geim}}, \ and\ \bibinfo {author}
  {\bibfnamefont {L.}~\bibnamefont {Ponomarenko}}} (\bibinfo {year}
  {2013}{\natexlab{a}}),\ \href@noop {} {\bibfield  {journal} {\bibinfo
  {journal} {Phys. Rev. Lett.}\ }\textbf {\bibinfo {volume} {111}},\ \bibinfo
  {pages} {166601}}\BibitemShut {NoStop}%
\bibitem [{\citenamefont {Titov}\ \emph
  {et~al.}(2013{\natexlab{b}})\citenamefont {Titov}, \citenamefont {Narozhny},\
  and\ \citenamefont {Gornyi}}]{priv2}%
  \BibitemOpen
  \bibfield  {author} {\bibinfo {author} {\bibnamefont {Titov}, \bibfnamefont
  {M.}}, \bibinfo {author} {\bibfnamefont {B.~N.}\ \bibnamefont {Narozhny}}, \
  and\ \bibinfo {author} {\bibfnamefont {I.~V.}\ \bibnamefont {Gornyi}}}
  (\bibinfo {year} {2013}{\natexlab{b}}),\ \href@noop {} {}\bibinfo
  {howpublished} {unpublished}\BibitemShut {NoStop}%
\bibitem [{\citenamefont {Tomonaga}(1950)}]{tom}%
  \BibitemOpen
  \bibfield  {author} {\bibinfo {author} {\bibnamefont {Tomonaga},
  \bibfnamefont {S.}}} (\bibinfo {year} {1950}),\ \href@noop {} {\bibfield
  {journal} {\bibinfo  {journal} {Prog. Theor. Phys.}\ }\textbf {\bibinfo
  {volume} {5}},\ \bibinfo {pages} {544}}\BibitemShut {NoStop}%
\bibitem [{\citenamefont {Tse}\ and\ \citenamefont {{Das Sarma}}(2007)}]{tse}%
  \BibitemOpen
  \bibfield  {author} {\bibinfo {author} {\bibnamefont {Tse}, \bibfnamefont
  {W.-K.}}, \ and\ \bibinfo {author} {\bibfnamefont {S.}~\bibnamefont {{Das
  Sarma}}}} (\bibinfo {year} {2007}),\ \href@noop {} {\bibfield  {journal}
  {\bibinfo  {journal} {Phys. Rev. B}\ }\textbf {\bibinfo {volume} {75}},\
  \bibinfo {pages} {045333}}\BibitemShut {NoStop}%
\bibitem [{\citenamefont {Tse}\ \emph {et~al.}(2007)\citenamefont {Tse},
  \citenamefont {Hu},\ and\ \citenamefont {Sarma}}]{da1}%
  \BibitemOpen
  \bibfield  {author} {\bibinfo {author} {\bibnamefont {Tse}, \bibfnamefont
  {W.-K.}}, \bibinfo {author} {\bibfnamefont {B.~Y.-K.}\ \bibnamefont {Hu}}, \
  and\ \bibinfo {author} {\bibfnamefont {S.~D.}\ \bibnamefont {Sarma}}}
  (\bibinfo {year} {2007}),\ \href@noop {} {\bibfield  {journal} {\bibinfo
  {journal} {Phys. Rev. B}\ }\textbf {\bibinfo {volume} {76}},\ \bibinfo
  {pages} {081401}}\BibitemShut {NoStop}%
\bibitem [{\citenamefont {Tso}\ \emph {et~al.}(1998)\citenamefont {Tso},
  \citenamefont {Geldart},\ and\ \citenamefont {Vasilopoulos}}]{ts2}%
  \BibitemOpen
  \bibfield  {author} {\bibinfo {author} {\bibnamefont {Tso}, \bibfnamefont
  {H.~C.}}, \bibinfo {author} {\bibfnamefont {D.~J.}\ \bibnamefont {Geldart}},
  \ and\ \bibinfo {author} {\bibfnamefont {P.}~\bibnamefont {Vasilopoulos}}}
  (\bibinfo {year} {1998}),\ \href@noop {} {\bibfield  {journal} {\bibinfo
  {journal} {Phys. Rev. B}\ }\textbf {\bibinfo {volume} {57}},\ \bibinfo
  {pages} {6561}}\BibitemShut {NoStop}%
\bibitem [{\citenamefont {Tso}\ and\ \citenamefont {Vasilopoulos}(1992)}]{ts4}%
  \BibitemOpen
  \bibfield  {author} {\bibinfo {author} {\bibnamefont {Tso}, \bibfnamefont
  {H.~C.}}, \ and\ \bibinfo {author} {\bibfnamefont {P.}~\bibnamefont
  {Vasilopoulos}}} (\bibinfo {year} {1992}),\ \href@noop {} {\bibfield
  {journal} {\bibinfo  {journal} {Phys. Rev. B}\ }\textbf {\bibinfo {volume}
  {45}},\ \bibinfo {pages} {1333}}\BibitemShut {NoStop}%
\bibitem [{\citenamefont {Tso}\ \emph {et~al.}(1992)\citenamefont {Tso},
  \citenamefont {Vasilopoulos},\ and\ \citenamefont {Peeters}}]{tso}%
  \BibitemOpen
  \bibfield  {author} {\bibinfo {author} {\bibnamefont {Tso}, \bibfnamefont
  {H.~C.}}, \bibinfo {author} {\bibfnamefont {P.}~\bibnamefont {Vasilopoulos}},
  \ and\ \bibinfo {author} {\bibfnamefont {F.~M.}\ \bibnamefont {Peeters}}}
  (\bibinfo {year} {1992}),\ \href@noop {} {\bibfield  {journal} {\bibinfo
  {journal} {Phys. Rev. Lett.}\ }\textbf {\bibinfo {volume} {68}},\ \bibinfo
  {pages} {2516}}\BibitemShut {NoStop}%
\bibitem [{\citenamefont {Tso}\ \emph {et~al.}(1993)\citenamefont {Tso},
  \citenamefont {Vasilopoulos},\ and\ \citenamefont {Peeters}}]{tvp}%
  \BibitemOpen
  \bibfield  {author} {\bibinfo {author} {\bibnamefont {Tso}, \bibfnamefont
  {H.~C.}}, \bibinfo {author} {\bibfnamefont {P.}~\bibnamefont {Vasilopoulos}},
  \ and\ \bibinfo {author} {\bibfnamefont {F.~M.}\ \bibnamefont {Peeters}}}
  (\bibinfo {year} {1993}),\ \href@noop {} {\bibfield  {journal} {\bibinfo
  {journal} {Phys. Rev. Lett.}\ }\textbf {\bibinfo {volume} {70}},\ \bibinfo
  {pages} {2146}}\BibitemShut {NoStop}%
\bibitem [{\citenamefont {Tso}\ \emph {et~al.}(1994)\citenamefont {Tso},
  \citenamefont {Vasilopoulos},\ and\ \citenamefont {Peeters}}]{tso2}%
  \BibitemOpen
  \bibfield  {author} {\bibinfo {author} {\bibnamefont {Tso}, \bibfnamefont
  {H.~C.}}, \bibinfo {author} {\bibfnamefont {P.}~\bibnamefont {Vasilopoulos}},
  \ and\ \bibinfo {author} {\bibfnamefont {F.~M.}\ \bibnamefont {Peeters}}}
  (\bibinfo {year} {1994}),\ \href@noop {} {\bibfield  {journal} {\bibinfo
  {journal} {Surf. Sci.}\ }\textbf {\bibinfo {volume} {305}},\ \bibinfo {pages}
  {400}}\BibitemShut {NoStop}%
\bibitem [{\citenamefont {Tutuc}\ \emph {et~al.}(2009)\citenamefont {Tutuc},
  \citenamefont {Pillarisetty},\ and\ \citenamefont {Shayegan}}]{tut}%
  \BibitemOpen
  \bibfield  {author} {\bibinfo {author} {\bibnamefont {Tutuc}, \bibfnamefont
  {E.}}, \bibinfo {author} {\bibfnamefont {R.}~\bibnamefont {Pillarisetty}}, \
  and\ \bibinfo {author} {\bibfnamefont {M.}~\bibnamefont {Shayegan}}}
  (\bibinfo {year} {2009}),\ \href@noop {} {\bibfield  {journal} {\bibinfo
  {journal} {Phys. Rev. B}\ }\textbf {\bibinfo {volume} {79}},\ \bibinfo
  {pages} {041303(R)}}\BibitemShut {NoStop}%
\bibitem [{\citenamefont {Tutuc}\ \emph {et~al.}(2004)\citenamefont {Tutuc},
  \citenamefont {Shayegan},\ and\ \citenamefont {Huse}}]{tuthuse}%
  \BibitemOpen
  \bibfield  {author} {\bibinfo {author} {\bibnamefont {Tutuc}, \bibfnamefont
  {E.}}, \bibinfo {author} {\bibfnamefont {M.}~\bibnamefont {Shayegan}}, \ and\
  \bibinfo {author} {\bibfnamefont {D.~A.}\ \bibnamefont {Huse}}} (\bibinfo
  {year} {2004}),\ \href {\doibase 10.1103/PhysRevLett.93.036802} {\bibfield
  {journal} {\bibinfo  {journal} {Phys. Rev. Lett.}\ }\textbf {\bibinfo
  {volume} {93}},\ \bibinfo {pages} {036802}}\BibitemShut {NoStop}%
\bibitem [{\citenamefont {Ussishkin}\ and\ \citenamefont {Stern}(1997)}]{us1}%
  \BibitemOpen
  \bibfield  {author} {\bibinfo {author} {\bibnamefont {Ussishkin},
  \bibfnamefont {I.}}, \ and\ \bibinfo {author} {\bibfnamefont
  {A.}~\bibnamefont {Stern}}} (\bibinfo {year} {1997}),\ \href@noop {}
  {\bibfield  {journal} {\bibinfo  {journal} {Phys. Rev. B}\ }\textbf {\bibinfo
  {volume} {56}},\ \bibinfo {pages} {4013}}\BibitemShut {NoStop}%
\bibitem [{\citenamefont {Ussishkin}\ and\ \citenamefont {Stern}(1998)}]{ust}%
  \BibitemOpen
  \bibfield  {author} {\bibinfo {author} {\bibnamefont {Ussishkin},
  \bibfnamefont {I.}}, \ and\ \bibinfo {author} {\bibfnamefont
  {A.}~\bibnamefont {Stern}}} (\bibinfo {year} {1998}),\ \href@noop {}
  {\bibfield  {journal} {\bibinfo  {journal} {Phys. Rev. Lett.}\ }\textbf
  {\bibinfo {volume} {81}},\ \bibinfo {pages} {3932}}\BibitemShut {NoStop}%
\bibitem [{\citenamefont {Varma}\ \emph {et~al.}(1994)\citenamefont {Varma},
  \citenamefont {Larkin},\ and\ \citenamefont {Abrahams}}]{var}%
  \BibitemOpen
  \bibfield  {author} {\bibinfo {author} {\bibnamefont {Varma}, \bibfnamefont
  {C.~M.}}, \bibinfo {author} {\bibfnamefont {A.~I.}\ \bibnamefont {Larkin}}, \
  and\ \bibinfo {author} {\bibfnamefont {E.}~\bibnamefont {Abrahams}}}
  (\bibinfo {year} {1994}),\ \href@noop {} {\bibfield  {journal} {\bibinfo
  {journal} {Phys. Rev. B}\ }\textbf {\bibinfo {volume} {49}},\ \bibinfo
  {pages} {13999}}\BibitemShut {NoStop}%
\bibitem [{\citenamefont {Vignale}(2005)}]{vig}%
  \BibitemOpen
  \bibfield  {author} {\bibinfo {author} {\bibnamefont {Vignale}, \bibfnamefont
  {G.}}} (\bibinfo {year} {2005}),\ \href@noop {} {\bibfield  {journal}
  {\bibinfo  {journal} {Phys. Rev. B}\ }\textbf {\bibinfo {volume} {71}},\
  \bibinfo {pages} {125103}}\BibitemShut {NoStop}%
\bibitem [{\citenamefont {Vignale}\ and\ \citenamefont
  {MacDonald}(1996)}]{vi2}%
  \BibitemOpen
  \bibfield  {author} {\bibinfo {author} {\bibnamefont {Vignale}, \bibfnamefont
  {G.}}, \ and\ \bibinfo {author} {\bibfnamefont {A.~H.}\ \bibnamefont
  {MacDonald}}} (\bibinfo {year} {1996}),\ \href@noop {} {\bibfield  {journal}
  {\bibinfo  {journal} {Phys. Rev. Lett.}\ }\textbf {\bibinfo {volume} {76}},\
  \bibinfo {pages} {2786}}\BibitemShut {NoStop}%
\bibitem [{\citenamefont {Vitkalov}(1998)}]{vit}%
  \BibitemOpen
  \bibfield  {author} {\bibinfo {author} {\bibnamefont {Vitkalov},
  \bibfnamefont {S.}}} (\bibinfo {year} {1998}),\ \href@noop {} {\bibfield
  {journal} {\bibinfo  {journal} {Pis'ma Zh. Eksp. Teor. Fiz.}\ }\textbf
  {\bibinfo {volume} {67}},\ \bibinfo {pages} {276}},\ \bibinfo {note} {[JETP
  Lett. {\bf 67}, 295 (1998)]}\BibitemShut {NoStop}%
\bibitem [{\citenamefont {Walter}\ \emph {et~al.}(2011)\citenamefont {Walter},
  \citenamefont {Bostwick}, \citenamefont {Jeon}, \citenamefont {Speck},
  \citenamefont {Ostler}, \citenamefont {Seyller}, \citenamefont {Moreschini},
  \citenamefont {Chang}, \citenamefont {Polini}, \citenamefont {Asgari},
  \citenamefont {MacDonald}, \citenamefont {Horn},\ and\ \citenamefont
  {Rotenberg}}]{plasmaron2}%
  \BibitemOpen
  \bibfield  {author} {\bibinfo {author} {\bibnamefont {Walter}, \bibfnamefont
  {A.~L.}}, \bibinfo {author} {\bibfnamefont {A.}~\bibnamefont {Bostwick}},
  \bibinfo {author} {\bibfnamefont {K.-J.}\ \bibnamefont {Jeon}}, \bibinfo
  {author} {\bibfnamefont {F.}~\bibnamefont {Speck}}, \bibinfo {author}
  {\bibfnamefont {M.}~\bibnamefont {Ostler}}, \bibinfo {author} {\bibfnamefont
  {T.}~\bibnamefont {Seyller}}, \bibinfo {author} {\bibfnamefont
  {L.}~\bibnamefont {Moreschini}}, \bibinfo {author} {\bibfnamefont {Y.~J.}\
  \bibnamefont {Chang}}, \bibinfo {author} {\bibfnamefont {M.}~\bibnamefont
  {Polini}}, \bibinfo {author} {\bibfnamefont {R.}~\bibnamefont {Asgari}},
  \bibinfo {author} {\bibfnamefont {A.~H.}\ \bibnamefont {MacDonald}}, \bibinfo
  {author} {\bibfnamefont {K.}~\bibnamefont {Horn}}, \ and\ \bibinfo {author}
  {\bibfnamefont {E.}~\bibnamefont {Rotenberg}}} (\bibinfo {year} {2011}),\
  \href@noop {} {\bibfield  {journal} {\bibinfo  {journal} {Phys. Rev. B}\
  }\textbf {\bibinfo {volume} {84}},\ \bibinfo {pages} {085410}}\BibitemShut
  {NoStop}%
\bibitem [{\citenamefont {Wang}\ \emph {et~al.}(2005)\citenamefont {Wang},
  \citenamefont {Mishchenko},\ and\ \citenamefont {Demler}}]{wmd}%
  \BibitemOpen
  \bibfield  {author} {\bibinfo {author} {\bibnamefont {Wang}, \bibfnamefont
  {D.-W.}}, \bibinfo {author} {\bibfnamefont {E.~G.}\ \bibnamefont
  {Mishchenko}}, \ and\ \bibinfo {author} {\bibfnamefont {E.}~\bibnamefont
  {Demler}}} (\bibinfo {year} {2005}),\ \href@noop {} {\bibfield  {journal}
  {\bibinfo  {journal} {Phys. Rev. Lett.}\ }\textbf {\bibinfo {volume} {95}},\
  \bibinfo {pages} {086802}}\BibitemShut {NoStop}%
\bibitem [{\citenamefont {Wang}\ and\ \citenamefont
  {da~Cunha~Lima}(2001)}]{wan}%
  \BibitemOpen
  \bibfield  {author} {\bibinfo {author} {\bibnamefont {Wang}, \bibfnamefont
  {X.}}, \ and\ \bibinfo {author} {\bibfnamefont {I.~C.}\ \bibnamefont
  {da~Cunha~Lima}}} (\bibinfo {year} {2001}),\ \href@noop {} {\bibfield
  {journal} {\bibinfo  {journal} {Phys. Rev. B}\ }\textbf {\bibinfo {volume}
  {63}},\ \bibinfo {pages} {205312}}\BibitemShut {NoStop}%
\bibitem [{\citenamefont {Wei}\ \emph {et~al.}(2009)\citenamefont {Wei},
  \citenamefont {Bao}, \citenamefont {Pu}, \citenamefont {Lau},\ and\
  \citenamefont {Shi}}]{an2}%
  \BibitemOpen
  \bibfield  {author} {\bibinfo {author} {\bibnamefont {Wei}, \bibfnamefont
  {P.}}, \bibinfo {author} {\bibfnamefont {W.}~\bibnamefont {Bao}}, \bibinfo
  {author} {\bibfnamefont {Y.}~\bibnamefont {Pu}}, \bibinfo {author}
  {\bibfnamefont {C.~N.}\ \bibnamefont {Lau}}, \ and\ \bibinfo {author}
  {\bibfnamefont {J.}~\bibnamefont {Shi}}} (\bibinfo {year} {2009}),\ \href
  {\doibase 10.1103/PhysRevLett.102.166808} {\bibfield  {journal} {\bibinfo
  {journal} {Phys. Rev. Lett.}\ }\textbf {\bibinfo {volume} {102}},\ \bibinfo
  {pages} {166808}}\BibitemShut {NoStop}%
\bibitem [{\citenamefont {Wen}(1995)}]{wen1}%
  \BibitemOpen
  \bibfield  {author} {\bibinfo {author} {\bibnamefont {Wen}, \bibfnamefont
  {X.-G.}}} (\bibinfo {year} {1995}),\ \href {\doibase
  10.1080/00018739500101566} {\bibfield  {journal} {\bibinfo  {journal}
  {Advances in Physics}\ }\textbf {\bibinfo {volume} {44}}~(\bibinfo {number}
  {5}),\ \bibinfo {pages} {405}}\BibitemShut {NoStop}%
\bibitem [{\citenamefont {Wen}\ and\ \citenamefont {Zee}(1992)}]{zee}%
  \BibitemOpen
  \bibfield  {author} {\bibinfo {author} {\bibnamefont {Wen}, \bibfnamefont
  {X.-G.}}, \ and\ \bibinfo {author} {\bibfnamefont {A.}~\bibnamefont {Zee}}}
  (\bibinfo {year} {1992}),\ \href {\doibase 10.1103/PhysRevLett.69.1811}
  {\bibfield  {journal} {\bibinfo  {journal} {Phys. Rev. Lett.}\ }\textbf
  {\bibinfo {volume} {69}},\ \bibinfo {pages} {1811}}\BibitemShut {NoStop}%
\bibitem [{\citenamefont {Wiersma}\ \emph {et~al.}(2006)\citenamefont
  {Wiersma}, \citenamefont {Lok}, \citenamefont {Tiemann}, \citenamefont
  {Dietsche}, \citenamefont {von Klitzing}, \citenamefont {Schuh},\ and\
  \citenamefont {Wegscheider}}]{wie3}%
  \BibitemOpen
  \bibfield  {author} {\bibinfo {author} {\bibnamefont {Wiersma}, \bibfnamefont
  {R.}}, \bibinfo {author} {\bibfnamefont {J.}~\bibnamefont {Lok}}, \bibinfo
  {author} {\bibfnamefont {L.}~\bibnamefont {Tiemann}}, \bibinfo {author}
  {\bibfnamefont {W.}~\bibnamefont {Dietsche}}, \bibinfo {author}
  {\bibfnamefont {K.}~\bibnamefont {von Klitzing}}, \bibinfo {author}
  {\bibfnamefont {D.}~\bibnamefont {Schuh}}, \ and\ \bibinfo {author}
  {\bibfnamefont {W.}~\bibnamefont {Wegscheider}}} (\bibinfo {year} {2006}),\
  \href {\doibase http://dx.doi.org/10.1016/j.physe.2006.08.029} {\bibfield
  {journal} {\bibinfo  {journal} {Physica E: Low-dimensional Systems and
  Nanostructures}\ }\textbf {\bibinfo {volume} {35}}~(\bibinfo {number} {2}),\
  \bibinfo {pages} {320 }},\ \bibinfo {note} {proceedings of the 14th
  International Winterschool on New Developments in Solid State Physics
  -Charges and spins in nanostructures: basics and devices}\BibitemShut
  {NoStop}%
\bibitem [{\citenamefont {Wiersma}\ \emph {et~al.}(2004)\citenamefont
  {Wiersma}, \citenamefont {Lok}, \citenamefont {Kraus}, \citenamefont
  {Dietsche}, \citenamefont {von Klitzing}, \citenamefont {Schuh},
  \citenamefont {Bichler}, \citenamefont {Tranitz},\ and\ \citenamefont
  {Wegscheider}}]{wie2}%
  \BibitemOpen
  \bibfield  {author} {\bibinfo {author} {\bibnamefont {Wiersma}, \bibfnamefont
  {R.~D.}}, \bibinfo {author} {\bibfnamefont {J.~G.~S.}\ \bibnamefont {Lok}},
  \bibinfo {author} {\bibfnamefont {S.}~\bibnamefont {Kraus}}, \bibinfo
  {author} {\bibfnamefont {W.}~\bibnamefont {Dietsche}}, \bibinfo {author}
  {\bibfnamefont {K.}~\bibnamefont {von Klitzing}}, \bibinfo {author}
  {\bibfnamefont {D.}~\bibnamefont {Schuh}}, \bibinfo {author} {\bibfnamefont
  {M.}~\bibnamefont {Bichler}}, \bibinfo {author} {\bibfnamefont {H.-P.}\
  \bibnamefont {Tranitz}}, \ and\ \bibinfo {author} {\bibfnamefont
  {W.}~\bibnamefont {Wegscheider}}} (\bibinfo {year} {2004}),\ \href {\doibase
  10.1103/PhysRevLett.93.266805} {\bibfield  {journal} {\bibinfo  {journal}
  {Phys. Rev. Lett.}\ }\textbf {\bibinfo {volume} {93}},\ \bibinfo {pages}
  {266805}}\BibitemShut {NoStop}%
\bibitem [{\citenamefont {Wiersma}\ \emph {et~al.}(2007)\citenamefont
  {Wiersma}, \citenamefont {Lok}, \citenamefont {Tiemann}, \citenamefont
  {Dietsche}, \citenamefont {von Klitzing}, \citenamefont {Wegscheider},\ and\
  \citenamefont {Schuh}}]{wie}%
  \BibitemOpen
  \bibfield  {author} {\bibinfo {author} {\bibnamefont {Wiersma}, \bibfnamefont
  {R.~D.}}, \bibinfo {author} {\bibfnamefont {J.~G.~S.}\ \bibnamefont {Lok}},
  \bibinfo {author} {\bibfnamefont {L.}~\bibnamefont {Tiemann}}, \bibinfo
  {author} {\bibfnamefont {W.}~\bibnamefont {Dietsche}}, \bibinfo {author}
  {\bibfnamefont {K.}~\bibnamefont {von Klitzing}}, \bibinfo {author}
  {\bibfnamefont {W.}~\bibnamefont {Wegscheider}}, \ and\ \bibinfo {author}
  {\bibfnamefont {D.}~\bibnamefont {Schuh}}} (\bibinfo {year} {2007}),\ \href
  {\doibase 10.1142/S0217979207042719} {\bibfield  {journal} {\bibinfo
  {journal} {International Journal of Modern Physics B}\ }\textbf {\bibinfo
  {volume} {21}},\ \bibinfo {pages} {1256}}\BibitemShut {NoStop}%
\bibitem [{\citenamefont {Wunsch}\ \emph {et~al.}(2006)\citenamefont {Wunsch},
  \citenamefont {Stauber}, \citenamefont {Sols},\ and\ \citenamefont
  {Guinea}}]{plas0}%
  \BibitemOpen
  \bibfield  {author} {\bibinfo {author} {\bibnamefont {Wunsch}, \bibfnamefont
  {B.}}, \bibinfo {author} {\bibfnamefont {T.}~\bibnamefont {Stauber}},
  \bibinfo {author} {\bibfnamefont {F.}~\bibnamefont {Sols}}, \ and\ \bibinfo
  {author} {\bibfnamefont {F.}~\bibnamefont {Guinea}}} (\bibinfo {year}
  {2006}),\ \href@noop {} {\bibfield  {journal} {\bibinfo  {journal} {New
  Journal of Physics}\ }\textbf {\bibinfo {volume} {8}}~(\bibinfo {number}
  {12}),\ \bibinfo {pages} {318}}\BibitemShut {NoStop}%
\bibitem [{\citenamefont {Yamamoto}\ \emph {et~al.}(2002)\citenamefont
  {Yamamoto}, \citenamefont {Stopa}, \citenamefont {Tokura}, \citenamefont
  {Hirayama},\ and\ \citenamefont {Tarucha}}]{ya2}%
  \BibitemOpen
  \bibfield  {author} {\bibinfo {author} {\bibnamefont {Yamamoto},
  \bibfnamefont {M.}}, \bibinfo {author} {\bibfnamefont {M.}~\bibnamefont
  {Stopa}}, \bibinfo {author} {\bibfnamefont {Y.}~\bibnamefont {Tokura}},
  \bibinfo {author} {\bibfnamefont {Y.}~\bibnamefont {Hirayama}}, \ and\
  \bibinfo {author} {\bibfnamefont {S.}~\bibnamefont {Tarucha}}} (\bibinfo
  {year} {2002}),\ \href@noop {} {\bibfield  {journal} {\bibinfo  {journal}
  {Physica E}\ }\textbf {\bibinfo {volume} {12}},\ \bibinfo {pages}
  {726}}\BibitemShut {NoStop}%
\bibitem [{\citenamefont {Yamamoto}\ \emph {et~al.}(2006)\citenamefont
  {Yamamoto}, \citenamefont {Stopa}, \citenamefont {Tokura}, \citenamefont
  {Hirayama},\ and\ \citenamefont {Tarucha}}]{ya1}%
  \BibitemOpen
  \bibfield  {author} {\bibinfo {author} {\bibnamefont {Yamamoto},
  \bibfnamefont {M.}}, \bibinfo {author} {\bibfnamefont {M.}~\bibnamefont
  {Stopa}}, \bibinfo {author} {\bibfnamefont {Y.}~\bibnamefont {Tokura}},
  \bibinfo {author} {\bibfnamefont {Y.}~\bibnamefont {Hirayama}}, \ and\
  \bibinfo {author} {\bibfnamefont {S.}~\bibnamefont {Tarucha}}} (\bibinfo
  {year} {2006}),\ \href@noop {} {\bibfield  {journal} {\bibinfo  {journal}
  {Science}\ }\textbf {\bibinfo {volume} {313}},\ \bibinfo {pages}
  {204}}\BibitemShut {NoStop}%
\bibitem [{\citenamefont {Yamamoto}\ \emph {et~al.}(2012)\citenamefont
  {Yamamoto}, \citenamefont {Takagi}, \citenamefont {Stopa},\ and\
  \citenamefont {Tarucha}}]{yam}%
  \BibitemOpen
  \bibfield  {author} {\bibinfo {author} {\bibnamefont {Yamamoto},
  \bibfnamefont {M.}}, \bibinfo {author} {\bibfnamefont {H.}~\bibnamefont
  {Takagi}}, \bibinfo {author} {\bibfnamefont {M.}~\bibnamefont {Stopa}}, \
  and\ \bibinfo {author} {\bibfnamefont {S.}~\bibnamefont {Tarucha}}} (\bibinfo
  {year} {2012}),\ \href@noop {} {\bibfield  {journal} {\bibinfo  {journal}
  {Phys. Rev. B}\ }\textbf {\bibinfo {volume} {85}},\ \bibinfo {pages}
  {041308(R)}}\BibitemShut {NoStop}%
\bibitem [{\citenamefont {Yan}\ \emph {et~al.}(2012{\natexlab{a}})\citenamefont
  {Yan}, \citenamefont {Li}, \citenamefont {Chandra}, \citenamefont {Tulevski},
  \citenamefont {Wu}, \citenamefont {Freitag}, \citenamefont {Zhu},
  \citenamefont {Avouris},\ and\ \citenamefont {Xia}}]{plex2}%
  \BibitemOpen
  \bibfield  {author} {\bibinfo {author} {\bibnamefont {Yan}, \bibfnamefont
  {H.}}, \bibinfo {author} {\bibfnamefont {X.}~\bibnamefont {Li}}, \bibinfo
  {author} {\bibfnamefont {B.}~\bibnamefont {Chandra}}, \bibinfo {author}
  {\bibfnamefont {G.}~\bibnamefont {Tulevski}}, \bibinfo {author}
  {\bibfnamefont {Y.}~\bibnamefont {Wu}}, \bibinfo {author} {\bibfnamefont
  {M.}~\bibnamefont {Freitag}}, \bibinfo {author} {\bibfnamefont
  {W.}~\bibnamefont {Zhu}}, \bibinfo {author} {\bibfnamefont {P.}~\bibnamefont
  {Avouris}}, \ and\ \bibinfo {author} {\bibfnamefont {F.}~\bibnamefont {Xia}}}
  (\bibinfo {year} {2012}{\natexlab{a}}),\ \href@noop {} {\bibfield  {journal}
  {\bibinfo  {journal} {Nature Nanotech.}\ }\textbf {\bibinfo {volume} {7}},\
  \bibinfo {pages} {330}}\BibitemShut {NoStop}%
\bibitem [{\citenamefont {Yan}\ \emph {et~al.}(2012{\natexlab{b}})\citenamefont
  {Yan}, \citenamefont {Li}, \citenamefont {Li}, \citenamefont {Zhu},
  \citenamefont {Avouris},\ and\ \citenamefont {Xia}}]{magplas}%
  \BibitemOpen
  \bibfield  {author} {\bibinfo {author} {\bibnamefont {Yan}, \bibfnamefont
  {H.}}, \bibinfo {author} {\bibfnamefont {Z.}~\bibnamefont {Li}}, \bibinfo
  {author} {\bibfnamefont {X.}~\bibnamefont {Li}}, \bibinfo {author}
  {\bibfnamefont {W.}~\bibnamefont {Zhu}}, \bibinfo {author} {\bibfnamefont
  {P.}~\bibnamefont {Avouris}}, \ and\ \bibinfo {author} {\bibfnamefont
  {F.}~\bibnamefont {Xia}}} (\bibinfo {year} {2012}{\natexlab{b}}),\ \href@noop
  {} {\bibfield  {journal} {\bibinfo  {journal} {Nano Letters}\ }\textbf
  {\bibinfo {volume} {12}}~(\bibinfo {number} {7}),\ \bibinfo {pages}
  {3766}}\BibitemShut {NoStop}%
\bibitem [{\citenamefont {Yang}(1998)}]{ya3}%
  \BibitemOpen
  \bibfield  {author} {\bibinfo {author} {\bibnamefont {Yang}, \bibfnamefont
  {K.}}} (\bibinfo {year} {1998}),\ \href@noop {} {\bibfield  {journal}
  {\bibinfo  {journal} {Phys. Rev. B}\ }\textbf {\bibinfo {volume} {58}},\
  \bibinfo {pages} {R4246}}\BibitemShut {NoStop}%
\bibitem [{\citenamefont {Yang}\ and\ \citenamefont {MacDonald}(2001)}]{ma2}%
  \BibitemOpen
  \bibfield  {author} {\bibinfo {author} {\bibnamefont {Yang}, \bibfnamefont
  {K.}}, \ and\ \bibinfo {author} {\bibfnamefont {A.~H.}\ \bibnamefont
  {MacDonald}}} (\bibinfo {year} {2001}),\ \href@noop {} {\bibfield  {journal}
  {\bibinfo  {journal} {Phys. Rev. B}\ }\textbf {\bibinfo {volume} {63}},\
  \bibinfo {pages} {073301}}\BibitemShut {NoStop}%
\bibitem [{\citenamefont {Yang}\ \emph {et~al.}(1994)\citenamefont {Yang},
  \citenamefont {Moon}, \citenamefont {Zheng}, \citenamefont {MacDonald},
  \citenamefont {Girvin}, \citenamefont {Yoshioka},\ and\ \citenamefont
  {Zhang}}]{moon}%
  \BibitemOpen
  \bibfield  {author} {\bibinfo {author} {\bibnamefont {Yang}, \bibfnamefont
  {K.}}, \bibinfo {author} {\bibfnamefont {K.}~\bibnamefont {Moon}}, \bibinfo
  {author} {\bibfnamefont {L.}~\bibnamefont {Zheng}}, \bibinfo {author}
  {\bibfnamefont {A.~H.}\ \bibnamefont {MacDonald}}, \bibinfo {author}
  {\bibfnamefont {S.~M.}\ \bibnamefont {Girvin}}, \bibinfo {author}
  {\bibfnamefont {D.}~\bibnamefont {Yoshioka}}, \ and\ \bibinfo {author}
  {\bibfnamefont {S.-C.}\ \bibnamefont {Zhang}}} (\bibinfo {year} {1994}),\
  \href {\doibase 10.1103/PhysRevLett.72.732} {\bibfield  {journal} {\bibinfo
  {journal} {Phys. Rev. Lett.}\ }\textbf {\bibinfo {volume} {72}},\ \bibinfo
  {pages} {732}}\BibitemShut {NoStop}%
\bibitem [{\citenamefont {Yang}\ \emph {et~al.}(2011)\citenamefont {Yang},
  \citenamefont {Koralek}, \citenamefont {Orenstein}, \citenamefont {Tibbetts},
  \citenamefont {Reno},\ and\ \citenamefont {Lilly}}]{yan}%
  \BibitemOpen
  \bibfield  {author} {\bibinfo {author} {\bibnamefont {Yang}, \bibfnamefont
  {L.}}, \bibinfo {author} {\bibfnamefont {J.~D.}\ \bibnamefont {Koralek}},
  \bibinfo {author} {\bibfnamefont {J.}~\bibnamefont {Orenstein}}, \bibinfo
  {author} {\bibfnamefont {D.~R.}\ \bibnamefont {Tibbetts}}, \bibinfo {author}
  {\bibfnamefont {J.~L.}\ \bibnamefont {Reno}}, \ and\ \bibinfo {author}
  {\bibfnamefont {M.~P.}\ \bibnamefont {Lilly}}} (\bibinfo {year} {2011}),\
  \href@noop {} {\bibfield  {journal} {\bibinfo  {journal} {Phys. Rev. Lett.}\
  }\textbf {\bibinfo {volume} {106}},\ \bibinfo {pages} {246401}}\BibitemShut
  {NoStop}%
\bibitem [{\citenamefont {Yoon}\ \emph {et~al.}(2010)\citenamefont {Yoon},
  \citenamefont {Tiemann}, \citenamefont {Schmult}, \citenamefont {Dietsche},
  \citenamefont {von Klitzing},\ and\ \citenamefont {Wegscheider}}]{yoo}%
  \BibitemOpen
  \bibfield  {author} {\bibinfo {author} {\bibnamefont {Yoon}, \bibfnamefont
  {Y.}}, \bibinfo {author} {\bibfnamefont {L.}~\bibnamefont {Tiemann}},
  \bibinfo {author} {\bibfnamefont {S.}~\bibnamefont {Schmult}}, \bibinfo
  {author} {\bibfnamefont {W.}~\bibnamefont {Dietsche}}, \bibinfo {author}
  {\bibfnamefont {K.}~\bibnamefont {von Klitzing}}, \ and\ \bibinfo {author}
  {\bibfnamefont {W.}~\bibnamefont {Wegscheider}}} (\bibinfo {year} {2010}),\
  \href {\doibase 10.1103/PhysRevLett.104.116802} {\bibfield  {journal}
  {\bibinfo  {journal} {Phys. Rev. Lett.}\ }\textbf {\bibinfo {volume} {104}},\
  \bibinfo {pages} {116802}}\BibitemShut {NoStop}%
\bibitem [{\citenamefont {Yoshioka}\ \emph {et~al.}(1989)\citenamefont
  {Yoshioka}, \citenamefont {MacDonald},\ and\ \citenamefont {Girvin}}]{yosh}%
  \BibitemOpen
  \bibfield  {author} {\bibinfo {author} {\bibnamefont {Yoshioka},
  \bibfnamefont {D.}}, \bibinfo {author} {\bibfnamefont {A.~H.}\ \bibnamefont
  {MacDonald}}, \ and\ \bibinfo {author} {\bibfnamefont {S.~M.}\ \bibnamefont
  {Girvin}}} (\bibinfo {year} {1989}),\ \href {\doibase
  10.1103/PhysRevB.39.1932} {\bibfield  {journal} {\bibinfo  {journal} {Phys.
  Rev. B}\ }\textbf {\bibinfo {volume} {39}},\ \bibinfo {pages}
  {1932}}\BibitemShut {NoStop}%
\bibitem [{\citenamefont {Zala}\ \emph {et~al.}(2001)\citenamefont {Zala},
  \citenamefont {Narozhny},\ and\ \citenamefont {Aleiner}}]{zna}%
  \BibitemOpen
  \bibfield  {author} {\bibinfo {author} {\bibnamefont {Zala}, \bibfnamefont
  {G.}}, \bibinfo {author} {\bibfnamefont {B.}~\bibnamefont {Narozhny}}, \ and\
  \bibinfo {author} {\bibfnamefont {I.}~\bibnamefont {Aleiner}}} (\bibinfo
  {year} {2001}),\ \href@noop {} {\bibfield  {journal} {\bibinfo  {journal}
  {Phys. Rev. B}\ }\textbf {\bibinfo {volume} {64}},\ \bibinfo {pages}
  {214204}}\BibitemShut {NoStop}%
\bibitem [{\citenamefont {Zelakiewicz}\ \emph {et~al.}(2000)\citenamefont
  {Zelakiewicz}, \citenamefont {Noh}, \citenamefont {Gramila}, \citenamefont
  {Pfeiffer},\ and\ \citenamefont {West}}]{zel}%
  \BibitemOpen
  \bibfield  {author} {\bibinfo {author} {\bibnamefont {Zelakiewicz},
  \bibfnamefont {S.}}, \bibinfo {author} {\bibfnamefont {H.}~\bibnamefont
  {Noh}}, \bibinfo {author} {\bibfnamefont {T.~J.}\ \bibnamefont {Gramila}},
  \bibinfo {author} {\bibfnamefont {L.~N.}\ \bibnamefont {Pfeiffer}}, \ and\
  \bibinfo {author} {\bibfnamefont {K.~W.}\ \bibnamefont {West}}} (\bibinfo
  {year} {2000}),\ \href@noop {} {\bibfield  {journal} {\bibinfo  {journal}
  {Phys. Rev. Lett.}\ }\textbf {\bibinfo {volume} {85}},\ \bibinfo {pages}
  {1942}}\BibitemShut {NoStop}%
\bibitem [{\citenamefont {Zhang}\ and\ \citenamefont {Jin}(2013)}]{zhajin}%
  \BibitemOpen
  \bibfield  {author} {\bibinfo {author} {\bibnamefont {Zhang}, \bibfnamefont
  {C.}}, \ and\ \bibinfo {author} {\bibfnamefont {G.}~\bibnamefont {Jin}}}
  (\bibinfo {year} {2013}),\ \href
  {http://stacks.iop.org/0953-8984/25/i=42/a=425604} {\bibfield  {journal}
  {\bibinfo  {journal} {Journal of Physics: Condensed Matter}\ }\textbf
  {\bibinfo {volume} {25}}~(\bibinfo {number} {42}),\ \bibinfo {pages}
  {425604}}\BibitemShut {NoStop}%
\bibitem [{\citenamefont {Zhang}\ and\ \citenamefont {Takahashi}(1993)}]{zha}%
  \BibitemOpen
  \bibfield  {author} {\bibinfo {author} {\bibnamefont {Zhang}, \bibfnamefont
  {C.}}, \ and\ \bibinfo {author} {\bibfnamefont {Y.}~\bibnamefont
  {Takahashi}}} (\bibinfo {year} {1993}),\ \href@noop {} {\bibfield  {journal}
  {\bibinfo  {journal} {J. Phys.: Condens. Matter}\ }\textbf {\bibinfo {volume}
  {5}},\ \bibinfo {pages} {5009}}\BibitemShut {NoStop}%
\bibitem [{\citenamefont {Zhang}\ and\ \citenamefont {Joglekar}(2008)}]{jog}%
  \BibitemOpen
  \bibfield  {author} {\bibinfo {author} {\bibnamefont {Zhang}, \bibfnamefont
  {C.-H.}}, \ and\ \bibinfo {author} {\bibfnamefont {Y.~N.}\ \bibnamefont
  {Joglekar}}} (\bibinfo {year} {2008}),\ \href {\doibase
  10.1103/PhysRevB.77.233405} {\bibfield  {journal} {\bibinfo  {journal} {Phys.
  Rev. B}\ }\textbf {\bibinfo {volume} {77}},\ \bibinfo {pages}
  {233405}}\BibitemShut {NoStop}%
\bibitem [{\citenamefont {Zhang}\ and\ \citenamefont {Zhang}(2012)}]{zh2}%
  \BibitemOpen
  \bibfield  {author} {\bibinfo {author} {\bibnamefont {Zhang}, \bibfnamefont
  {S.~S.-L.}}, \ and\ \bibinfo {author} {\bibfnamefont {S.}~\bibnamefont
  {Zhang}}} (\bibinfo {year} {2012}),\ \href@noop {} {\bibfield  {journal}
  {\bibinfo  {journal} {Phys. Rev. Lett.}\ }\textbf {\bibinfo {volume} {109}},\
  \bibinfo {pages} {096603}}\BibitemShut {NoStop}%
\bibitem [{\citenamefont {Zheng}\ and\ \citenamefont {MacDonald}(1993)}]{mac}%
  \BibitemOpen
  \bibfield  {author} {\bibinfo {author} {\bibnamefont {Zheng}, \bibfnamefont
  {L.}}, \ and\ \bibinfo {author} {\bibfnamefont {A.~H.}\ \bibnamefont
  {MacDonald}}} (\bibinfo {year} {1993}),\ \href@noop {} {\bibfield  {journal}
  {\bibinfo  {journal} {Phys. Rev. B}\ }\textbf {\bibinfo {volume} {48}},\
  \bibinfo {pages} {8203}}\BibitemShut {NoStop}%
\bibitem [{\citenamefont {Zhou}\ and\ \citenamefont {Kim}(1999)}]{zho}%
  \BibitemOpen
  \bibfield  {author} {\bibinfo {author} {\bibnamefont {Zhou}, \bibfnamefont
  {F.}}, \ and\ \bibinfo {author} {\bibfnamefont {Y.~B.}\ \bibnamefont {Kim}}}
  (\bibinfo {year} {1999}),\ \href@noop {} {\bibfield  {journal} {\bibinfo
  {journal} {Phys. Rev. B}\ }\textbf {\bibinfo {volume} {59}},\ \bibinfo
  {pages} {R7825}}\BibitemShut {NoStop}%
\bibitem [{\citenamefont {Zhu}\ \emph {et~al.}(2013)\citenamefont {Zhu},
  \citenamefont {Badalyan},\ and\ \citenamefont {Peeters}}]{zhu}%
  \BibitemOpen
  \bibfield  {author} {\bibinfo {author} {\bibnamefont {Zhu}, \bibfnamefont
  {J.-J.}}, \bibinfo {author} {\bibfnamefont {S.~M.}\ \bibnamefont {Badalyan}},
  \ and\ \bibinfo {author} {\bibfnamefont {F.~M.}\ \bibnamefont {Peeters}}}
  (\bibinfo {year} {2013}),\ \href@noop {} {\bibfield  {journal} {\bibinfo
  {journal} {Phys. Rev. B}\ }\textbf {\bibinfo {volume} {87}},\ \bibinfo
  {pages} {085401}}\BibitemShut {NoStop}%
\bibitem [{\citenamefont {Ziman}(1965)}]{ziman1965}%
  \BibitemOpen
  \bibfield  {author} {\bibinfo {author} {\bibnamefont {Ziman}, \bibfnamefont
  {J.~M.}}} (\bibinfo {year} {1965}),\ \href@noop {} {\emph {\bibinfo {title}
  {Principles of the Theory of Solids}}}\ (\bibinfo  {publisher}
  {Cambridge})\BibitemShut {NoStop}%
\bibitem [{\citenamefont {Zou}\ \emph {et~al.}(2009)\citenamefont {Zou},
  \citenamefont {Refael},\ and\ \citenamefont {Yoon}}]{zou}%
  \BibitemOpen
  \bibfield  {author} {\bibinfo {author} {\bibnamefont {Zou}, \bibfnamefont
  {Y.}}, \bibinfo {author} {\bibfnamefont {G.}~\bibnamefont {Refael}}, \ and\
  \bibinfo {author} {\bibfnamefont {J.}~\bibnamefont {Yoon}}} (\bibinfo {year}
  {2009}),\ \href@noop {} {\bibfield  {journal} {\bibinfo  {journal} {Phys.
  Rev. B}\ }\textbf {\bibinfo {volume} {80}},\ \bibinfo {pages}
  {180503(R)}}\BibitemShut {NoStop}%
\bibitem [{\citenamefont {Zou}\ \emph {et~al.}(2010)\citenamefont {Zou},
  \citenamefont {Refael},\ and\ \citenamefont {Yoon}}]{zou2}%
  \BibitemOpen
  \bibfield  {author} {\bibinfo {author} {\bibnamefont {Zou}, \bibfnamefont
  {Y.}}, \bibinfo {author} {\bibfnamefont {G.}~\bibnamefont {Refael}}, \ and\
  \bibinfo {author} {\bibfnamefont {J.}~\bibnamefont {Yoon}}} (\bibinfo {year}
  {2010}),\ \href@noop {} {\bibfield  {journal} {\bibinfo  {journal} {Phys.
  Rev. B}\ }\textbf {\bibinfo {volume} {82}},\ \bibinfo {pages}
  {104515}}\BibitemShut {NoStop}%
\bibitem [{\citenamefont {Zuev}\ \emph {et~al.}(2009)\citenamefont {Zuev},
  \citenamefont {Chang},\ and\ \citenamefont {Kim}}]{zuev}%
  \BibitemOpen
  \bibfield  {author} {\bibinfo {author} {\bibnamefont {Zuev}, \bibfnamefont
  {Y.~M.}}, \bibinfo {author} {\bibfnamefont {W.}~\bibnamefont {Chang}}, \ and\
  \bibinfo {author} {\bibfnamefont {P.}~\bibnamefont {Kim}}} (\bibinfo {year}
  {2009}),\ \href@noop {} {\bibfield  {journal} {\bibinfo  {journal} {Phys.
  Rev. Lett.}\ }\textbf {\bibinfo {volume} {102}},\ \bibinfo {pages}
  {096807}}\BibitemShut {NoStop}%
\bibitem [{\citenamefont {Zyuzin}\ and\ \citenamefont {Fiete}(2010)}]{zuz}%
  \BibitemOpen
  \bibfield  {author} {\bibinfo {author} {\bibnamefont {Zyuzin}, \bibfnamefont
  {V.~A.}}, \ and\ \bibinfo {author} {\bibfnamefont {G.~A.}\ \bibnamefont
  {Fiete}}} (\bibinfo {year} {2010}),\ \href {\doibase
  10.1103/PhysRevB.82.113305} {\bibfield  {journal} {\bibinfo  {journal} {Phys.
  Rev. B}\ }\textbf {\bibinfo {volume} {82}},\ \bibinfo {pages}
  {113305}}\BibitemShut {NoStop}%
\end{thebibliography}%

\end{document}